\documentclass[titlepage,11pt,letterpaper,reqno]{amsbook}
\usepackage{amsfonts}
\usepackage{amsmath}
\usepackage{amstext}
\usepackage{amscd}
\usepackage{amsgen}
\usepackage{amsxtra}
\usepackage{amsbsy}
\usepackage{amsopn}
\usepackage{amsfonts}
\usepackage{amsthm}
\usepackage{amssymb}
\usepackage{graphicx}
\usepackage{float}
\usepackage{bm}
\usepackage{enumerate}
\usepackage{tabularx}

\allowdisplaybreaks[4]
\title{An invariant bayesian model selection principle for gaussian data in a sparse representation}
\author{Eirik Fossgaard}
\address{Department of Mathematics and Statistics, Faculty of Science, University of Troms{\o}, 9037 Troms{\o}, Norway.}
\email{eirikf@math.uit.no, efossgaard@gmail.com}
\newcommand{\mathdef}{\overset{\text{def}}{=}}
\newcommand{\littleo}{\ensuremath{o}}
\newcommand{\bigo}{\ensuremath{O}}
\newcommand{\sgn}{\text{sgn}\,}
\newcommand{\erf}{\text{erf}\,}

\newtheorem{theorem}{Theorem}[section]
\newtheorem{corollary}{Corollary}[section]
\newtheorem{definition}{Definition}[section]
\newtheorem{proposition}{Proposition}[section]

\begin{document}
%\maketitle

%Copyright \copyright{ 2004 by Eirik Fossgaard}

\begin{abstract}
We develop a code length principle which is invariant to the choice of parameterization on the model distributions.
An invariant approximation formula for easy computation of the marginal distribution is provided
for gaussian likelihood models. We provide invariant estimators of the model parameters and formulate conditions 
under which these estimators are essentially posteriori unbiased for gaussian models. 
An upper bound on the coarseness of discretization on the model parameters is deduced. 
We introduce a discrimination measure between probability distributions and use it to construct probability 
distributions on model classes. The total code length is shown to equal the NML code length of Rissanen 
to within an additive constant when choosing Jeffreys prior distribution on the model parameters together with a particular 
choice of prior distribution on the model classes. Our model selection principle is applied to a gaussian estimation problem 
for data in a wavelet representation and its performance is tested and compared 
to alternative wavelet-based estimation methods in numerical experiments.\\ \\
\end{abstract}

\maketitle
\vspace{6cm}\begin{center}Copyright \copyright{ 2004 by Eirik Fossgaard}\end{center}
%\newpage 

%\include{/home/eirikf/hovedfag97/thesis/side2mal}
\tableofcontents

%\thanks{}
 
%PAPER
%\input{/home/eirikf/LaTeX/Thesis/MDL_Basics}
%\include{Errata}
\chapter{Introduction}
This thesis describes the development of a codelength and model selection principle for gaussian likelihood models which is invariant
to the choice of parameterization of the model. We provide an invariant marginal approximation formula and 
invariant estimators which are shown to be essentially a posteriori unbiased under ''reasonable'' conditions on the signal 
to noise ratio and data generating model. An upper bound on the coarseness of discretization of model parameters is deduced. 
Also, we introduce the concept of a {\em model class prior distribution}, 
which enables us to discriminate quantitatively in terms of {\em code lengths} between different choices of 
prior distributions on the parameters that we want to estimate. The model class distribution 
may be interpreted as a quantitive measure of the amount of trust we have in our prior information of the data generating process. 
We show in numerical experiments that the choice of model class prior distribution may be of 
crucial importance to the performance of estimators when estimating parameters in additive white gaussian noise. 
The principle is compared to the NML-principle of Rissanen in both theory and numerical experiments.

\section{Wavelet-based recovering of data corrupted by noise}
We will rely on the properties of {\em discrete orthogonal wavelet bases} \cite{Daubechies:1992, Mallat:1998, Wickerhauser:1994}
to provide us with a {\em sparse} (most coefficients are ''almost'' zero) representation of the data sets. Empirical work
\cite{Moulin-Liu:1999} has shown that the family of Generalized Gaussian distributions (GGD) may be used to provide reasonable models
for natural image data when represented in the wavelet domain. Wavelets have through the last 15 years been used extensively in
problems of estimating data corrupted by additive noise (denoising). The wavelet based methods may all be divided into three main 
steps: Given a dataset $x\in\mathbb{R}^{n}$, do
\begin{enumerate}
\item Expand the data $x\in\mathbb{R}^{n}$ into an discrete orthogonal wavelet basis $W\in\mathbb{R}^{n\times n}$ by computing 
the linear orthogonal transform $w\mathdef W^{T}x$.
\item Process the transformed data $w$ in the wavelet domain to yield $\hat{w}$.
\item Inverse-transform the processed transformed data $\hat{w}$ back into the original space domain to yield the estimate
$\hat{x}\mathdef W\hat{w}$.
\end{enumerate}
Several denoising techniques have been developed for processing in the wavelet domain, 
\cite{DJ_Ideal:1994, DJ_Adapt:1995, Moulin-Liu:1999, BruceGao:1995b, Vidakovic:1998}, the 
differences between methods depending on the type data and modeling assumptions. 
Common to most wavelet-based denoising techniques are shrinkage-estimators operating in the wavelet domain, and among these,  
threshold estimators in particular. The most popular threshold operators take the form:
\begin{align}
&\text{Hard threshold estimator: }h_{t}^{(hard)}(x) = \left\{\begin{array}{l}0, \mbox{ if } |x|<t,\\
	x,\mbox{ if } |x| \geq t.\end{array}\right.\label{hard_threshold}\\
&\text{Soft threshold estimator: }h_{t}^{(soft)}(x) = \left\{\begin{array}{l}0, \mbox{ if } |x|<t,\\
	x-\sgn{(x)}t,\mbox{ if } |x| \geq t.\end{array}\right.\label{soft_threshold}\\
&\text{Firm threshold estimator: }h_{t_{1},t_{2}}^{(firm)}(x) = \left\{\begin{array}{l}0, \mbox{ if } |x|<t_{1},\\
	\frac{\sgn{(x)}t_{2}(|x|-t_{1})}{t_{2}-t_{1}},\mbox{ if } t_{1}\leq |x|\leq t_{2},\\
	x,\mbox{ if } |x| \geq t_{2}.\end{array}\right.\label{semisoft_threshold}
\end{align}
The generic case studied in the litterature referenced above is that of recovering an unknown function 
$g(t):[0,1]\longrightarrow\mathbb{R}$ at sample points $0\leq s_{i}\leq 1,\ 1\leq i\leq n$ by providing estimates of the discrete samples 
$\theta=\{g(s_{i})\}_{i=1}^{n}\in\mathbb{R}^{n}$ when corrupted by 
additive white gaussian noise $\eta\in\mathbb{R}^{n}$. The samples are all modelled as independently and identically 
distributed:
\begin{align}
& x_{i} = g(s_{i})+\eta_{i},\ 1\leq i\leq n\label{data_model}
\intertext{where}
&\eta_{i}\sim \mathcal{N}(0,\sigma), \ 1\leq i\leq n\nonumber
\intertext{and $g$ is the underlying unknown function:}
&g:[0,1]\longrightarrow\mathbb{R}.\nonumber
\intertext{At the sample points $s_{i}$ we define}
&\theta_{i}\mathdef g(s_{i}).\label{def_discrete_theta}
\end{align}
Hard and soft threshold estimators applied in the wavelet domain were studied in the work of Donoho and Johnstone 
\cite{DJ_Ideal:1994} and results on universal optimality of the estimators were reported: 
Let $\delta_{i}\in\{0,1\}$ be the ideal diagonal projection operator defined by
\begin{align}
&\delta_{i} = I_{\{|\theta_{i}|>\sigma\}}\label{ideal_digonal_projection}
\intertext{where $I$ is the indicator function. Supposing we have an {\em oracle} available providing us with the $\delta_{i}$, then the ideal risk 
$\mathcal{R}(\hat{\theta}^{(ideal)},\theta)$ $\mathdef E_{x}\|\hat{\theta}^{(ideal)}(x)-\theta\|_{2}^{2}$ of the ideal oracle estimator $\hat{\theta}^{(ideal)}$}
&\hat{\theta}_{i}^{(ideal)}(x_{i})\mathdef\delta_{i}x_{i}\label{ideal_projection_estimator} 
\intertext{becomes}
&\mathcal{R}(\hat{\theta}^{(ideal)},\theta)=\sum_{i=1}^{n}\min(|\theta_{i}|,\sigma)^{2}.\label{ideal_risk}
\intertext{The ideal risk in (\ref{ideal_risk}) is in general not attainable by any estimator without the aid of an oracle $\delta_{i}$, but the following
result on universal optimality of the estimator $\hat{\theta}^{(soft)}_{t_{n}}$ was shown in \cite{DJ_Ideal:1994}:}
&E\|\hat{\theta}^{(soft)}_{t_{n}}-\theta\|_{2}^{2}\leq (2\log{n}+1)\left(\sigma^{2}+\mathcal{R}(\hat{\theta}^{(ideal)},\theta)\right)\label{risk_bound}
\intertext{where $\hat{\theta}^{(soft)}_{t_{n}}$ is the soft threshold estimator (\ref{soft_threshold}) with threshold 
$t_{n}=\sigma\sqrt{2\log{n}}$. Furthermore, the result (\ref{risk_bound}) was shown to be asymptotically sharp in $n$, and
that no estimator can come closer to the ideal risk $\mathcal{R}(\hat{\theta}^{(ideal)},\theta)$ than this for all $\theta\in\mathbb{R}^{n}$ when forced to rely on the data $x$ alone.}\nonumber
\end{align}
These results were extended to the class of firm threshold estimators \cite{BruceGao:1995b, BruceGao:1995c}
and estimates on bias and variances of the estimators have also been provided \cite{BruceGao:1995a}. 

However, the universal threshold $t_{n}=\sigma\sqrt{2\log{n}}$ leads to an aggressive thresholding scheme on the data $x$
and the resulting estimates $\hat{\theta}(x)$ are often in experiments and applications found to suffer from oversmoothing
and loss of details, effects which are especially prominent in image denoising applications. 
Even though the result (\ref{risk_bound}) is universally optimal,
in most situations of practical interest the signal $\theta$ to be estimated is known to possess some degree of smoothness  
and this knowledge may be exploited to provide alternative (more sophisticated) wavelet shrinking estimators with 
better performance on this particular type of data. The percieved suboptimality of the universal thresholding scheme of Donoho and Johnstone
in particular cases could be expected, as their result on the universal optimality of the risk of the estimator was based purely on 
their new result in univariate normal decision theory, and did not presuppose anything concerning the wavelet representation of the data 
and/or the sparseness thereof. However, several minimax results on wavelet shrinkage estimators over wide ranges of 
Besov- and Triebel-type smoothness constraints were reported in \cite{DJ_minimax:1998}. 
Donoho and Johnstone in \cite{DJ_Adapt:1995} provided an {\em adaptive} {\em hybrid} thresholding scheme called SureShrink in the wavelet domain 
which was shown to be nearly minimax optimal when the underlying function $f$ belongs to a range of Besov spaces. 
The class of functions $f:[0,1]\rightarrow\mathbb{R}$ of total bounded variation $\|f\|_{tv}$ where 
\begin{align}
&\|f\|_{tv}\mathdef\sup\left\{\sum_{i=1}^{n}|f(s_{i+1})-f(s_{i})|:\ 0\leq s_{1}<\cdots<s_{n}\leq 1,\ n\in\mathbb{N}\right\}\label{def_tv}
\end{align}
are found to provide a reasonable class to embed ''most'' natural images in, \cite{MallatKalifa:2003_AnnStat}. 
Functions of total bounded variation belong on the scale of Besov spaces, \cite{DJ_Adapt:1995}. The SureShrink method
uses the Stein Unbiased Risk Estimate (SURE) \cite{Stein:1981} separately in each wavelet 
subband to compute the threshold minimizing the SURE-estimate. Renormalizing the data $x$ by the noise level $\sigma$ so that 
$x\sim\mathcal{N}(\theta,1)$ and letting $\hat{\theta}^{(t)}(x)$ denote the soft thresholding estimator
with threshold $t>0$, SURE states that
\begin{align}
&E_{x}\|\hat{\theta}^{(t)}(x)-\theta\|_{2}^{2}=E_{x}\text{ SURE}(x,t)\label{unbiased_sure}
\intertext{where}
&\text{SURE}(x,t)\mathdef n-2\sum_{i=1}^{n}I_{\{|x_{i}|<t\}}+\sum_{i=1}^{n}\min(|x_{i}|,t)^{2}\label{def_sure}
\intertext{and the SURE threshold $t_{S}$ is defined as}
&t_{S}\mathdef\text{arg min}_{0<t<t_{n}}\text{SURE}(x,t),\text{ where } t_{n}\mathdef\sqrt{2\log{n}}.\label{def_sure_threshold}
\end{align}
To circumvent issues of poor performance of SURE in cases of extreme sparsity
of the wavelet coefficients, a measure of sparseness of the wavelet representation of the data is computed in each subband, 
and if the representation within the subband is sparse ''enough'', the universal soft threshold estimator is used, 
otherwise the threshold $t\mathdef\min\left(t_{S},t_{n}\right)$ is used, thus making the method a hybrid between two different thresholding schemes.
This method has a fast $O(n\log{n})$ implementation. Moulin and Liu \cite{Moulin-Liu:1999}
found (empirically) the family of Generalized Gaussian Distributions (GGD) to be able to provide reasonable model distributions for the 
probability density distributions (pdf) of wavelet coefficients $\theta_{i}$ of natural image data, and estimators for different GGD distributions 
were investigated. Results from similar work were reported in \cite{ChangYuVetterli:2000}.

\section{Model selection, code lengths, prior information, invariance}

We briefly outline the connection between model selection, probability distributions 
and code length principles, for a thorough presentation on the theme we refer to \cite{CoverThomas:1991,Rissanen:1998b}. Let $X$ be a discrete 
random variable with range $A$ (finite or countably infinite) and pdf $p(x)$. Let $C(x)$ denote the codeword used to encode 
$x\in A$ in a binary representation
and let $L(x)$ denote the length (number of binary bits) of the codeword $C(x)$. 
The expected length $L(C)$ of the code $C(x)$ is then defined as $L(C)\mathdef E_{p}\{L(x)\}=\sum_{x\in A}p(x)L(x)$.
Furthermore, let $x^{n}\mathdef(x_{1},x_{2},...,x_{n})$ and define the codeword $C(x^{n})\mathdef C(x_{1})C(x_{2})\cdots C(x_{n})$
where $C(x_{1})C(x_{2})\cdots C(x_{n})$ denotes concatenation of codewords. We only want to consider decodeable codes $C$, i.e.
codes $C$ where $x_{i}\neq x_{j}\Rightarrow C(x_{i})\neq C(x_{j})$. An important class of such codes are the {\em prefix} codes which 
have the defining property that no codeword is a prefix of any other codeword. Any binary prefix code $C(x)$ with codeword lengths 
$L(x)$ satifies the {\em Kraft inequality}:
\begin{align}
&\sum_{x\in A}2^{-L(x)}\leq 1\label{KraftInequality}
\intertext{and conversely: For a given set of codeword lengths $L(x),\ x\in A$ satisfying (\ref{KraftInequality})
there exists a prefix code $C(x)$ with codeword lengths $L(x)$, \cite{Rissanen:1998b}. Then we note that for a given a pdf $q(x)$ 
on $x\in A$ we may define codeword lengths $L_{q}(x)\mathdef-\log_{2} q(x)$ and we then have}
&\sum_{x\in A}2^{-L_{q}(x)}=\sum_{x\in A}2^{\log{q(x)}} = \sum_{x\in A}q(x)=1\label{Kraft1}
\intertext{and conversely for given code $C^{\prime}(x)$ with codeword lengths $L^{\prime}(x)$ we may define a pdf $r(x)$ 
on $x\in A$ by}
&r(x)\mathdef \frac{2^{-L^{\prime}(x)}}{\sum_{x\in A}2^{-L^{\prime}(x)}}.\label{def_r_pdf}
\intertext{Let the entropy $H(X)$ of the random variable $X$ with range $A$ and pdf $p(x)$ be defined by}
&H(X)\mathdef -\sum_{x\in A}p(x)\log_{2}{p(x)}\label{def_entropy2}
\intertext{then the following inequality holds for any prefix code $C$}
&L(C)\geq H(X)\label{codelength_bound}
\intertext{with equality if and only if $L(x)=-\log_{2}{p(x)}, \forall x\in A$, \cite{Rissanen:1998b}. }\nonumber
\end{align}
That is, a prefix code $C$ with codeword lengths 
$L(x)=-\log_{2}{p(x)}$ is an optimal code in the sense that it minimizes the expected codeword length $L(C)$.
Assume $x^{n}$  a data set given to us and let the model class $\mathcal{M}=\{M_{1},M_{2},...\}$ be a set of models 
used to explain the data set $x^{n}$.
We may then construct a binary encoding scheme resulting in binary descriptions of both the model $M_{i}$ in question and the data 
$x^{n}$ in view of this model. In analogy with above notation, we let $L(s)$ denote the length of a binary description of an object $s$, we may write
\begin{align}
&L(x^{n},M_{i})\mathdef L(x^{n}|M_{i})+L(M_{i}).\label{description_length} 
\end{align}
We will use both the terms {\em code length} and {\em description length} of the data to mean the length of the encoded 
binary string representing the description of the data $x^{n}$. Because of (\ref{Kraft1}), (\ref{def_r_pdf}) we may restrict to considering 
{\em code lengths} and probability distributions rather than (the construction of) codes themselves. 
We use the term {\em code length principle} to denote the method of assigning a code length  
to a dataset $x^{n}$ and the model $M_{i}$ used to explain this dataset. A good model $M_{i}$ is one that leads to a short total code length 
$L(x^{n},M_{i})$. The term {\em minimum description length} refers to the principle of choosing the model 
$M^{*}\mathdef\text{arg min}_{M_{i}\in\mathcal{M}}L(x^{n},M_{i})$ as the model to be used to explain the data, that is the model providing
the shortest description of dataset {\em and} model together. 

Different code length principles have been proposed in the litterature, 
we will here point out two: The Minimum Description Length (MDL) principle (and in particular: the Normalized Maximum Likelihood (NML)-principle) 
of Rissanen \cite{Rissanen:1998a,Rissanen:1998b,Rissanen:2001} 
and the Minimum Message Length (MML) principle of Wallace \cite{WallaceFreeman:1987,OliverHand:1994,OliverBaxter:1994a}. 
The two principles are similar, but distinct, for a discussion of differences see \cite{OliverBaxter:1994b, Lanterman:2001}. An important difference
between these two principles stem from different views on the role of prior information on parameters. The following two citations 
provides some information on the MDL-view as Rissanen sees it:\\

\noindent ''(...) the suggestion that the (prior) distribution $\pi(\theta)$ (of parameters $\theta$) captures prior knowledge in an adequate manner is untenable and even totally unacceptable to many because of the interpretation difficulty whenever the parameter appears to be a contstant-albeit unknown. (...)'', 
\cite{Rissanen:1998b}, page 10.\\

\noindent And furthermore:\\
 
\noindent ''(...) In our view the parameter $\theta$ is generated by our selecting the model class, 
and it has no other 'inherent' meaning. (...)'', \cite{Rissanen:1998b} page 55. \\

\noindent On the other hand the MML-philosophy in the view of Wallace/Freeman states: \\

\noindent ''(...) there can be no substitute for careful specification of whatever prior knowledge is available (...)'', \cite{WallaceFreeman:1987}.\\

\noindent Our own opinion in this issue on the role of prior information and prior 
distributions on parameters is not quite as clear cut as in the statements cited above, 
but we may at least say this: On one hand we want to exploit and make the most of any prior information we have on 
the distribution of the noiseless data $\theta$ to help in 
providing a good estimate $\theta^{*}$, on the other hand we do not want to state claims on the prior distribution of 
the parameters $\theta$ which are too far from the ''truth'', whatever that may be. Introducing parametric prior distributions has a (heavy) price: 
It leads to the problem of providing ''sensible'' estimates of the parameters of the prior, a very difficult task in many cases, as indeed we 
experienced when applying our models and theory on the real world in the numerical work presented later in this thesis 
(the problem we experienced was basically of the kind of overfitting model to the data). 
This experience motivated us to introduce and construct a prior distribution on model classes: The model class prior distribution 
enables us to discriminate quantitatively between different choices of prior distributions on the model parameters $\theta$ which parameterize
the likelihood function. It may be used to provide a theoretically well-founded way of quantitatively penalizing over-optimistic 
judgements of robustness and/or ''truth'' and ''reasonability'' of prior knowledge of the data generating process as compared to some carefully chosen 
reference prior distribution. By careful we here mean that the chosen reference distribution should be not too informative, 
and (ideally) not too non-informative either. 

Also, we note that once (a prior distribution $\pi(\theta)$ on) 
parameters $\theta$ are introduced, the question of invariance \cite{Bala:1996,Bala:1997} arises: 
For given likelihood distribution $f(x|\theta)$, define the Fisher information matrix $F(\theta)$ by:
\begin{align}
%\intertext{For given likelihood distribution $f(x|\theta)$, define the Fisher information matrix $F(\theta)$ by}
&F(\theta)_{ij}\mathdef -E_{x}\frac{\partial}{\partial\theta_{i}\partial\theta_{j}}\log f(x|\theta),\label{def_fisher}
\intertext{observe that}
&m(x)\mathdef \int_{\theta\in\Theta} f(x|\theta)\pi(\theta)\ d\theta = \int_{\theta\in\Theta}\frac{f(x|\theta)\pi(\theta)}{|F(\theta)|^{1/2}} |F(\theta)|^{1/2}\ d\theta\label{Fisher_decompose_integral}
\intertext{and define}
& dV(\theta) \mathdef |F(\theta)|^{1/2}\ d\theta, \ \Phi(\theta)\mathdef -\log\left(\frac{f(x|\theta)\pi(\theta)}{|F(\theta)|^{1/2}}\right)\label{def_invariant_volume_and_Phi}
\intertext{then note that the MML estimator $\theta^{*}_{MML}$ is defined \cite{OliverBaxter:1994b} by}
&\theta_{MML}^{*}\mathdef\text{arg min}_{\theta\in\Theta}\Phi(\theta)\label{def_MML_estimator}
\intertext{and note that the integration measure $dV(\theta)$ is the Riemannian volume element 
which provides a reparameterization invariant integration measure on the parameter manifold $\Theta$ on which $\theta$ lives 
and furthermore: The choice of {\em Jeffreys distribution} $|F(\theta)|^{1/2}/\int |F(\beta)|^{1/2}\ d\beta$ as the prior $\pi(\theta)$ is 
equivalent to assuming equal prior likelihood of all distributions parameterized by $\theta\in\Theta$ as opposed to equal prior 
likelihood of parameters $\theta$, \cite{Bala:1996,Bala:1997}.
This choice of a non-informative prior distribution is what we will use when comparing our code length principle 
to the NML-principle of Rissanen.}\nonumber
\end{align}

\section{Connecting code length principles to wavelet-based denoising}
The observed ability of wavelet bases to provide sparse representations of several types of real world data sets of interest in diverse research 
fields (mammography, medical imaging, seismic data analysis) combined with results from the extensive empirical and theoretical research 
on properties of the wavelet expansions of data belonging to certain smoothness classes (Besov-scales, bounded total variation classes), 
provides information which may be exploited in building models, model selection and code length principles, 
for example in guiding the choice of prior distribution on the wavelet expansion coefficients of a dataset.
 
As pointed out in \cite{DJ_Ideal:1994}, the wavelet thresholding methods described previously may be viewed as model 
selection methods which pick a subset of the
wavelet basis vectors and fits a model to the data by optimizing some given criterion. In the case of the universal 
thresholding estimators $\theta_{t_{n}}^{(soft)}$, $\theta_{t_{n}}^{(hard)}$ the criterion is the least squares method.
In \cite{Saito:1994} a data adaptive model selection method for denoising data corrupted by additive white gaussian noise
was developed by using the Minimum Description Length (MDL) principle of Rissanen \cite{Rissanen:1996,Rissanen:1998b}
as the criterion to be optimized. The resulting denoising method consisted of thresholding the data in the wavelet 
domain with a hard thresholding estimator $h_{t}^{(hard)}$ with a data driven threshold $t$. However, the model selection principle 
presented in \cite{Saito:1994} was generally found in numerical experiments to result in large thresholds yielding very small models but
also a large degree of smoothing in the estimated data. The explanation for this lies in the crudeness of the coding assumptions made in 
this work: A constant budget of $\log_{2}{n}$ ($n$ is sample size) bits per wavelet coefficient included in the model was allocated for 
encoding the location of the coefficient inside the vector of wavelet expansion coefficients of the data, 
leading to an extra codelength term of $d\log_{2}{n}$ for model size $d$.  
This encoding of location of coefficients is in our view redundant in this case, as 
the rule for optimally selecting wavelet coefficients to include in the model is inherent to the model selection principle by
simply minimizing the codelength for given model size $d$.  In fact, it was shown in \cite{CohenMalahRaz:1998}
that for given deterministic noise variance $\sigma^{2}$ (that is $\sigma$ is given prior to the selection of the model), the coding assumptions 
in \cite{Saito:1994} leads to a hard thresholding scheme with threshold $t=\sigma\sqrt{3\log{n}}$ which is seen to  be larger than the 
universally optimal thresholds $t_{n}=\sigma\sqrt{2\log{n}}$ of \cite{DJ_Ideal:1994}. In \cite{Rissanen:2000} a MDL-based denoising method for 
data corrupted by additive white gaussian noise was deduced, resulting in a hard thresholding scheme with data driven threshold $t_{MDL}$. 
Furthermore it was argued that under reasonable and rather weak assumptions on the asymptotic (in sample size $n$) behaviour of the dataset $x$, 
the threshold $t_{MDL}\sim \hat{\sigma}_{ML}\sqrt{\log{n}}$ where $\hat{\sigma}_{ML}$ is the Maximum Likelihood estimate of the noise deviance 
$\sigma$. Another MDL-based (subband-dependent) method for simultaneous denoising and compression of image data in the wavelet domain 
was presented in \cite{Hansen-Yu:2000}. 

The model selection principle we will develop is based on minimizing the description length of 
the model and dataset when encoded in the binary code induced by our modelled marginal distribution $m(x)$ and a 
suitable model class prior distribution defined on the set of model classes in question. We will approximate 
the marginal distribution $m(x)$ in (\ref{Fisher_decompose_integral}) as follows:
\begin{enumerate}
\item Construct a reparameterization $\hat{\theta}=\psi^{-1}(\theta)$ with the property
that the reparameterized Fisher information $|\hat{F}(\hat{\theta})|$ is constant.
\item Use $\psi$ to reparameterize the marginal integral (\ref{Fisher_decompose_integral}). 
\item Letting $\hat{\Phi}(\hat{\theta})\mathdef\Phi(\psi(\hat{\theta}))$, expand the reparameterized marginal integral 
  around the MML estimate $\hat{\theta}^{*}_{MML}\mathdef \text{arg min}_{\hat{\theta}\in \hat{\Theta}}\hat{\Phi}(\hat{\theta})$
  by Taylor-expanding $\hat{\Phi}(\hat{\theta})$ around $\hat{\theta}^{*}_{MML}$.
\item Truncate the expansion of the integral to second order to yield the approximated marginal $\tilde{m}(x)$.
\end{enumerate}
Note that the approximation $\tilde{m}(x)$ of the marginal integral (\ref{Fisher_decompose_integral}) that results from the method outlined above
is invariant, in that it does not depend on our original more or less arbitrary choice of parameterization $\theta$. 
This independency of the approximation $\tilde{m}(x)$ of parameterization $\theta$ would in general not be the case 
(unlesss our original choice of parameterization $\theta$ was lucky enough to 
yield $|F(\theta)|=\text{constant}$) if we simply approximated the marginal integral directly by expanding 
$f(x|\theta)\pi(\theta)$ around the maximum posterior estimate $\theta^{*}_{MAP}$. 

Under some ''reasonable conditions'' on the data and prior distribution $\pi(\theta)$ which will be stated precisely later, 
we will show for gaussian likelihood models that the described second order approximation of the marginal integral has small error, 
and that the MML estimate $\theta^{*}_{MML}$ is ''very close'' to the posterior mean $\theta^{\sharp}$ 
\begin{align}
&\theta^{\sharp}\mathdef\frac{1}{m(x)}\int_{\theta\in\Theta}\theta f(x|\theta)\pi(\theta)\ d\theta\label{posterior_mean}
\end{align}
that is $\theta^{*}_{MML}$ is essentially unbiased in a posterior sense.
Furthermore, we will show that, for a gaussian likelihood and choosing Jeffreys distribution {\em both} as the prior distribution 
on the parameters {\em and} as the reference prior distribution for the model class distribution, the code length of the model and dataset 
is to within an additive constant equal to the NML code length developed in \cite{Rissanen:1996}.

\section{Organization of thesis}
In the second chapter (following the current chapter) we define the problem to be studied, describe the modeling assumptions, 
provide the necessary preliminaries on notation and theory and present the main theoretical results 
on the formula for the modelled data generating distribution. We present the development of the model class prior distribution and a
result on the coarsest possible choice of discretization of model parameters in a posterior perspective. 
The longest and computationally tedious proofs are put in the 
appendices to which we refer when appropriate. In the third chapter we apply the theory to the practical problem of denoising 
data in white gaussian noise and we present the results from our numerical experiments on the performance of our method. 
In the fourth chapter we extend our method to a case of non-white noise and present results from our numerical experiments.

\section{Contact information and documentation}
I may at the time of this writing be reached on the email addresses:\\ 
{\tt eirikf@math.uit.no} and: {\tt efossgaard@gmail.com}.
The code (Ansi C) developed to implement the theory in this thesis in the reported numerical experiments may be downloaded from:\\
{\tt http://www.math.uit.no/users/eirikf/}.

\section{Acknowledgements}
I want to express my sincerest thanks to my teacher and advisor, professor Tor Fl\aa, first for his suggestion of 
defining and applying for the support of a PhD-project for me at the University of Troms\o, then for his enthusiastic and 
enduring guidance during my contract period. I thank all my friends and colleagues in Troms\o (and Bod\o) for sharing some 
time with me both in and outside work: Hugues, Tormod, Truls, \O yvind, Olav, Robert, Tor-Arne, Vegard, Kurt and others: 
Please feel free to join the list ;-)
A very special thanks goes to my girlfriend Cathrine for her patience and support. The research reported in this thesis
was generously supported by the Norwegian Research Council under BeMatA project no. 135971 and the University of Troms\o.
\chapter{Development of an invariant code length principle}
\section{Definition of problem and data generating model}
Let $\mathbb{R}^{n}$ be euclidean $n$-dimensional space equipped with the euclidean inner product 
$\langle\cdot,\cdot\rangle:\mathbb{R}^{n}\times \mathbb{R}^{n}\longrightarrow \mathbb{R}$. 
Given data $\bm{x}=(x_{1},...,x_{n})\in\mathbb{R}^{n}$ modeled as 
\begin{equation}
\bm{x} = \bm{\theta}+\bm{\eta},\label{noise_model} 
\end{equation}
where  $\bm{\theta}=(\theta_{1},...,\theta_{n})\in\mathbb{R}^{n}$ is signal and $\bm{\eta}=(\eta_{1},...,\eta_{n})\in \mathbb{R}^{n}$ is noise, 
our goal is to estimate $\bm{\theta}$. We think of $\bm{\theta}$ as the sampled projection of some unknown real valued function 
$u:\mathbb{R}^{p}\longrightarrow\mathbb{R}$, $u\in X$, for some function space $X$, onto some $n$-dimensional orthogonal basis $\bm{W}$ spanning 
a $n$-dimensional subspace $V\subset X$. We will assume $X$ is some ''sufficiently nice'' subspace of $L^{2}(\mathbb{R}^{p})$ which members possess
some degree of smoothness. We model the noise coefficients $\{\eta_{i}\}_{i=1}^{n}$ as independently, identically distributed (IID) with mean zero, variance $\tau^{-1}$ and gaussian density function $f$. Thus, the data $\{x_{i}\}_{i=1}^{n}$ are independently distributed (ID) with $x_{i}\sim f(x_{i}|\theta_{i},\tau)$ where $\bm{\theta}=(\theta_{1},...,\theta_{n})^{T}$ are the mean values of the data $x_{i},\ i=1,...,n$, $\tau^{-1}$ is the 
variance of each $x_{i}$ and $f$ is a gaussian likelihood function. 
Define $f(\bm{x}|\bm{\theta},\tau)\mathdef f(x_{1}|\theta_{1},\tau)\cdots f(x_{n}|\theta_{n},\tau)$. 
Only $d<n$ of the parameters $\left\{\theta_{i}\right\}_{i=1}^{n}$ are considered to be free nonzero parameters which we are able 
to estimate ''reasonably'' accurate under the modeling assumption (\ref{noise_model}) and we will model
these $d$ parameters as independently identically distributed (IID). Thus the set of parameters $\bm{\theta}\in \mathbb{R}^{n}$ is a $d$-dimensional 
submanifold $\Theta_{d}$ of $\mathbb{R}^{n}$. In coordinates $\theta_{i}$ this may be expressed by a binary index vector
$\gamma_{d}=(\gamma_{d}(1),\gamma_{d}(2),...,\gamma_{d}(n))\in\{0,1\}^{n}$ where $\gamma_{d}$ has exactly $d$ nonzero elements. 
We define $\theta_{i}$ to be a model parameter if and only if $\gamma_{d}(i)=1$. Then we may write a prior density $\pi_{\lambda}(\theta_{i})$ on the form 
\begin{align}
&\pi_{\lambda}(\theta_{i})=\left\{\begin{array}{ll} h_{\lambda}(\theta_{i}), & \mbox{ if }\gamma_{d}(i)=1\\ 
g(\theta_{i}), & \mbox{ if } \gamma_{d}(i)=0,\end{array}\right.\label{prior}
\end{align}where $h_{\lambda}$ is some probability distribution parameterized by $\lambda$ centered in origo (zero first moment) 
and $\lambda^{-1/2}$ equals the second moment (deviance) and $g$ is some density.
We will restrict $\pi_{\lambda}$ to the class of priors which are everywhere smooth except possibly at the origin. 
We extend $h_{\lambda},g$ to densities on $\mathbb{R}^{d}$ and $\mathbb{R}^{n-d}$ respectively 
by assuming independence of the $\{\theta_{i}\}_{i=1}^{n}$. It is in most cases more difficult to have a clear a priori idea of what a 
suitable prior distribution $\varsigma(\tau)$ on the parameter $\tau$ should be. For reasons of simplicity in the computations to come, we will 
restrict the prior distribution $\varsigma(\tau)$ on $\tau$ to be the uniform distribution
\begin{align}
&\varsigma(\tau)=|I_{\tau}|^{-1},\ \forall\ \tau\in I_{\tau}\subset (0,\infty)\label{uniform_tau_prior}
\end{align} 
where $I_{\tau}$ is some bounded interval. We may reorder the index $i$ indexing $\bm{\theta}=(\theta_{1},...,\theta_{n})^{T}$ so that $\gamma_{d}(i)=1$ if and only if $1\leq i\leq d$ and zero otherwise. 
We reorder the data $\bm{x}=(x_{1},...,x_{n})^{T}$ by the same reordering performed on the parameters $\{\theta_{i}\}_{i=1}^{n}$.
We define $\bm{\theta}_{\parallel}\mathdef(\theta_{1},...,\theta_{d},\bm{0}_{1}^{T})\in\mathbb{R}^{n}$, 
$\bm{\theta}_{\perp}\mathdef(\bm{0}_{2}^{T},\theta_{d+1},...,\theta_{n})\in\mathbb{R}^{n}$, 
$\bm{x}_{\parallel}\mathdef(x_{1},...,x_{d},\bm{0}_{1}^{T})\in\mathbb{R}^{n}$, $\bm{x}_{\perp}\mathdef(\bm{0}_{2}^{T},x_{d+1},...,x_{n})$, where
$\bm{0}_{1}$ is the zero vector in $\mathbb{R}^{n-d}$ and $\bm{0}_{2}$ is the zero vector in $\mathbb{R}^{d}$.
We have then the orthogonal decompositions $\bm{x}=\bm{x}_{\parallel}+\bm{x}_{\perp}$, 
$\bm{\theta}=\bm{\theta}_{\parallel}+\bm{\theta}_{\perp}$ and through the set of model indices $\gamma_{d}$ and the basis $\bm{W}$ we get an 
induced orthogonal decomposition $V=V_{\parallel}\oplus V_{\perp}$. 
We model $g=\delta$, where $\delta$ is the  Dirac delta distribution, implying $\bm{\theta}_{\perp}=\bm{0}$ and thus $\bm{\theta}=\bm{\theta_{\parallel}}$ and
\begin{equation}
\pi_{\lambda}(\theta)=\delta\ast h_{\lambda}(\theta)=h_{\lambda}(\theta).\label{convolution_defi}
\end{equation}
To set up the proper definition of the marginal integral, a few words must be said on the status of the parameters
$\tau$, $\lambda$, i.e whether we consider them to be deterministic parameters defined prior to (and independent of) selection of 
model $\gamma_{d}$, or stochastic parameters depending on the model $\gamma_{d}$. In the litterature on model selection applied
to denoising there are examples on both approaches \cite{Hansen-Yu:2000}, \cite{Rissanen:2000}. We will here always consider
the parameter $\tau$ stochastic, uniformly distributed over some bounded interval 
$I_{\tau}\subset\mathbb{R}_{+}$, and its estimator $\tau^{*}$ to be determined in conjunction with the 
model $\gamma_{d}$. As for the parameter $\lambda$ we have deduced results on the marginal distribution for both 
cases. We will in the experiments section consider $\lambda$ to be stochastic and (for reasons of computational simplicity) 
uniformly distributed on some bounded interval.
The estimator $\lambda^{*}$  will therefore depend on the model $\gamma_{d}$. For now, however, we consider $\lambda$ deterministic.

Given $I_{\tau}$ and $\lambda$ we define the marginal density $m_{\gamma_{d}}(\bm{z}|I_{\tau},\lambda)$ by
\begin{align}
&m_{\gamma_{d}}(\bm{z}|I_{\tau},\lambda)\mathdef\frac{1}{|I_{\tau}|}\int_{\bm{\theta}\in\mathbb{R}^{d},\tau\in I_{\tau}}f(\bm{z}|\bm{\theta},\tau)\pi_{\lambda}(\bm{\theta})\ d\bm{\theta}\ d\tau.\label{def_marginal}
%\intertext{where $C_{\gamma_{d}}$ is a constant depending on both the model $\gamma_{d}$ and the given data set $\bm{x}$ and is to be 
%chosen so as to make $m_{\gamma_{d}}(\bm{z}|\tau^{*},\lambda)$ a proper density over a region $X$ containing the given data set $\bm{x}$, that is}
%&\int_{\bm{z}\in X}m_{\gamma_{d}}(\bm{z}|\tau^{*}(\bm{z}),\lambda)\ d\bm{z} = 1.\label{def_C_gamma_d}
\end{align}
%The region $X$ has to be chosen carefully and convergence issues will have to be adressed, 
%we will investigate this question in detail in the proof sections later.
We note that if $\lambda$ is considered stochastic, that is we consider it unknown to us prior to the model selection process, 
the integral in (\ref{def_marginal}) should also include an integration over a 
bounded $\lambda$-interval $I_{\lambda}\subset\mathbb{R}_{+}$. 
%containing the model dependent estimate $\lambda^{*}$. 
This is discussed in detail below, see Proposition \ref{proposition_prior_laplace_approximation_formula} and 
Corollary \ref{corollary_new_marginal}.
The subscript $\gamma_{d}$ in $m_{\gamma_{d}}$ is used to emphasize the dependence of the marginal density on the selected model indexed by $\gamma_{d}$. We consider $m_{{\gamma}_{d}}$ to be the data generating distribution in our model for the data, 
though it is not necessarily, and in most cases not, the true data-generating distribution $q$, say.

\section{Outline of motivation and strategy}
We will exploit the compression abilities of wavelets and wavelet packet bases on broad classes of natural signals and images to 
provide a sparse representation of the data in some (possibly data driven) wavelet domain. This will allow a smaller
data generating model (smaller model size $d$) and thus a more compact description of the data itself, parameterized by $\bm{\theta}$ and $\tau$. 
This will be essential to our use of the Minimum Desription Length Principle (MDL Principle) in constructing a posteriori unbiased estimators $\bm{\theta}^{\sharp}(\bm{x})$, $\tau^{\sharp}(\bm{x})$. 
%which are adapted to our estimation problem in the sense of adaptation to both model and data. 
The transforms we will consider are orthogonal transforms of wavelet-type.
Let $\bm{W}\in \mathbb{R}^{n\times n}$ be some discrete orthogonal basis of $\mathbb{R}^{n}$ consisting of discrete 
wavelet packet functions. We will consider the given data $\bm{x}$ to be the finest scale wavelet coefficients of the data available to us, 
so that $\bm{W}^{T}:\mathbb{R}^{n}\longrightarrow\mathbb{R}^{n}$ is a linear orthogonal operator on $\mathbb{R}^{n}$. Define $\bm{x}^{w}\mathdef\bm{W}^{T}\bm{x}$,
$\bm{\theta}^{w}\mathdef\bm{W}^{T}\bm{\theta}$. For notational 
simplicity we will drop the superscripts $w$, and assume that $\bm{x}$ and $\bm{\theta}$ are data and signal expanded in some
fixed suitable basis of wavelet type.

As we will se below, for many choices of ''realistic'' prior distributions for the parameters, 
our models will result in estimators $\bm{\theta}^{*}$ belonging to the class of thresholding estimators as 
have been described in \cite{DJ_Ideal:1994} and \cite{BruceGao:1995b}. 
Threshold estimators are MAP-estimators for the class of $\mbox{GGD}_{\nu}$ priors with shape parameter $0<\nu\leq 1$ as
demonstrated in \cite{Moulin-Liu:1999}.
%and Maximum Likelihood(ML)-estimators for Gaussian likelihood functions as demonstrated in \cite{Rissanen:2000}.
It is known from \cite{DJ_Ideal:1994}, that the MSE universally ideal
(meaning optimal over all $\bm{\theta}\in\mathbb{R}^{n}$)
threshold value $t_{n}$ grows like $\sigma\sqrt{2\log{n}}$ as $n\rightarrow\infty$ where $n$ is the sample size and $\sigma$ 
the noise deviance. 
%and that asymptotically in $n$ the MSE optimal threshold estimator has the form of a hard threshold estimator. 
Furthermore, note that the formula $t_{n}\sim \sqrt{2\log{n}}$ for the MSE ideal threshold value only applies for large $n$.
For smaller $n$ on the order of a few hundred the MSE optimal threshold values are significantly smaller than $\sigma\sqrt{2\log{n}}$, and
this remains true for an even larger range of sample sizes $n$ for the lower threshold $t_{1}$ in the firm threshold estimator 
(\ref{semisoft_threshold}).
The performance of the estimators in (\ref{hard_threshold})-(\ref{soft_threshold}) when using the universal MSE optimal 
threshold values $t_{n}$ is often found not to be satisfying on several
types of natural data encountered in problems of applied nature in that it leads to too much smoothing 
in the estimates. This lack of performance is mainly due to the fact that the universally optimal MSE value of the 
threshold $t$ is too large, in other words $t$ grows ''too fast'' with increasing dimension $n$ of the dataset.
Several refined/data adaptive threshold schemes 
as in \cite{BruceGao:1995c,DJ_Adapt:1995,ChangYuVetterli:2000,Moulin-Liu:1999} have been suggested. We will use model
selection in a wavelet basis to determine the relevant dimension $d<n$ of the dataset in this basis and the
compute the resulting data adaptive estimators $\bm{\theta}^{\sharp}$ and $\tau^{\sharp}$. We will seek to derive a model selection principle
and estimators $\bm{\theta}^{\sharp}$, $\tau^{\sharp}$ which are invariant to the choice of parameterization of our models.

The rationale behind the idea of decomposing the data $x$ into $\bm{x}=\bm{x}_{\parallel}+\bm{x}_{\perp}$ is 
the observation that the part of data $\bm{x}$ consisting of signal is efficiently compressed, meaning it can be accurately represented in the sense 
of small $\ell^{2}$ squared loss by a small subset of its expansion coefficients in a wavelet-type basis $\bm{W}$, whereas the noise is 
essentially not compressible in this type of basis. 
Thus, to some extent it is possible to choose the space $V_{\parallel}$ so that it contains most of the signal and 
therefore the space $V_{\perp}$ will contain mostly noise. We will make use of the Minimum Description Length Principle \cite{Rissanen:1998b}, 
\cite{Rissanen:1998a} to determine the ''best'' signal subspace $V_{\parallel}$ of the space $V$ where 
$V_{\parallel}=\text{Span}_{i:\gamma_{d}(i)=1}\{\bm{w}_{i}\}$ and $\{\bm{w}_{i}\}_{i:\gamma_{d}(i)=1}$ is 
some subset of the column vectors of the full basis matrix $\bm{W}=\{\bm{w}_{i}\}_{i=1}^{n}$. 

\section{Definition of the model class}
We need to know how to determine $\bm{x}_{\parallel}$. 
%For this we will use a Minimum Description Length Principle. 
As mentioned above, the marginal density $m_{\gamma_{d}}$ is likely not the {\em true} data generating distribution $q$. Depending on 
to which extent $m_{\gamma_{d}}$ is able to approximate $q$, we can expect $m_{\gamma_{d}}$ to approximate $q$ more or 
less closely in the space of probability distributions by optimizing the choice of the model index vector $\gamma_{d}$ under the modelled data 
generating distribution (\ref{def_marginal}). 
Beyond some subset of parameters $\{\theta_{i}\}_{i:\gamma_{d^{\prime}}(i)=1}$ of size $d^{\prime}\leq n$, it may be meaningless to try to estimate more parameters as 
these parameters do not capture more of the properties of the unknown underlying {\em true} data generating distribution $q$. 
That is, further adding of parameters to our model will result in overfitting $m_{\gamma_{d}}$ to the specific dataset $\bm{x}$ at hand, 
\cite{Bala:1996,Bala:1997}.

With this in mind, for given likelihood distribution $f(\bm{x}|\bm{\theta},\tau)$ and prior distribution $\pi_{\lambda}(\bm{\theta})$, 
let $M_{d}$ denote the class of all models with $d$ nonzero parameters $\theta_{i}$ as defined by index vectors $\gamma_{d}\in\left\{0,1\right\}^{n}$.
Because each index vector $\gamma_{d}$ index a different data generating distribution 
$m_{\gamma_{d}}$, we will say that $M_{d}$ is a model class for the modelled data generating distribution $m_{\gamma_{d}}$.
There are $\binom{n}{k}$ ways to pick $k$ elements out of $n$ elements. Therefore the number of distinct models inside
each model class $M_{k}$ is $\binom{n}{k}$. Letting $\mathcal{M}\mathdef\bigcup_{k=0}^{n}M_{k}$ denote the collection of all model classes 
under consideration, we have $|\mathcal{M}|=\sum_{k=0}^{n}|M_{k}|$$=\sum_{k=0}^{n}\binom{n}{k}=2^{n}$. 
This yields a total of $2^{n}$ different models.  

\section{Invariant Laplace-approximation of marginal density}
We will in this section develop a theory of a parameterization invariant approximation of the marginal distribution $m_{\gamma_{d}}(\bm{x})$ 
by expanding the defining integral (\ref{def_marginal}) about certain points $\bm{\theta}^{*}$ and $\tau^{*}$. 
We will start with the simplest case where we have complete knowledge of the prior distribution 
$\pi_{\lambda}(\bm{\theta})$, that is we know all its parameters. The result is shown in in Theorem \ref{theorem1}. Then we proceed to the 
case where an estimate of the parameters of the prior distribution $\pi_{\lambda}(\bm{\theta})$ has to be estimated from the 
given data set $\bm{x}$. The result is shown in Corollary \ref{corollary_new_marginal}.

\begin{definition}
The Fisher information matrix $\bm{F}(\bm{\beta})$ for a likelihood function $f(\bm{x}|\bm{\beta})$ parameterized by 
parameters $\bm{\beta}=(\beta_{1},...,\beta_{k})^{T}$ is defined as
\begin{align}
\bm{F}(\bm{\beta})\mathdef -\text{E}_{\bm{x}}\left\{\frac{\partial}{\partial\bm{\beta}}\log{f(\bm{x}|\bm{\beta})}\left(\frac{\partial}{\partial\bm{\beta}}\log{f(\bm{x}|\bm{\beta})}\right)^{T}\right\}.\label{FisherMatrix}
\end{align}
\end{definition}
As explained in \cite{Bala:1996} the Fisher information matrix induces a metric on the Riemannian parameter manifold in the space
of distributions parameterized by $\bm{\beta}$ and this metric is invariant to smooth transformations of the parameter vector $\bm{\beta}$.
We have therefore the following result:
\begin{proposition}
The integration measure $dV(\bm{\beta})=|\bm{F}(\bm{\beta})|^{1/2}d\bm{\beta}$ is a reparameterization invariant integration measure on the parameter manifold, where $|\bm{F}|$ denotes the absolute value of the determinant of the Fisher matrix $\bm{F}$.
\label{Invariant_Integration_Measure}
\end{proposition}
\begin{proof}
Let $\bm{\beta}=\bm{\psi}(\hat{\bm{\beta}})$ define a reparameterization of $\bm{\beta}$ with $g(\bm{z}|\hat{\bm{\beta}})\mathdef
f(\bm{z}|\bm{\psi}(\hat{\bm{\beta}}))$. The volume element $d\hat{V}(\hat{\bm{\beta}})$
in the reparameterized system is $d\hat{V}(\hat{\bm{\beta}})$ $=|\bm{J}_{\bm{\psi}}(\hat{\bm{\beta}})^{T}$ $\bm{F}(\bm{\psi}(\hat{\bm{\beta}}))\bm{J}_{\psi}(\hat{\bm{\beta}})|^{1/2}d\hat{\bm{\beta}}$, where $\bm{J}_{\bm{\psi}}(\hat{\bm{\beta}})_{ij}\mathdef\frac{\partial\beta_{i}}{\partial\hat{\beta}_{j}}$ is the jacobi matrix of the transformation $\bm{\psi}$.
Then the prior density $\pi_{\lambda}$ transforms as $\pi_{\lambda}(\bm{\beta})d\bm{\beta}\rightarrow\rho_{\hat{\lambda}}(\hat{\bm{\beta}})d\hat{\bm{\beta}}$ under $\bm{\beta}\rightarrow\bm{\psi}(\hat{\bm{\beta}})$ where $\rho_{\hat{\lambda}}(\hat{\bm{\beta}})\mathdef\pi_{\lambda}(\bm{\psi}(\hat{\bm{\beta}}))|\bm{J}_{\bm{\psi}}(\hat{\bm{\beta}})|$.
We have to show that $d\hat{V}(\hat{\bm{\beta}})=|\hat{\bm{F}}(\hat{\bm{\beta}})|^{1/2}d\hat{\bm{\beta}}$, where $\hat{\bm{F}}(\hat{\bm{\beta}})$ is the Fisher information matrix of the likelihood function $g$. We observe that
\begin{align}
&\hat{\bm{F}}_{ij}(\hat{\bm{\beta}})\mathdef-E_{\bm{z}\sim g}\left\{\frac{\partial^{2}\log g(\bm{z}|\hat{\bm{\beta}})}{\partial\hat{\beta}_{i}\partial\hat{\beta}_{j}}\right\}
=-E_{\bm{z}\sim f}\left\{\frac{\partial^{2}\log f(\bm{z}|\bm{\psi}({\hat{\bm{\beta}}}))}{\partial\hat{\beta}_{i}\partial\hat{\beta}_{j}}\right\},\nonumber\\
\intertext{by the chain rule we have }
&=-E_{\bm{z}\sim f}\left\{\sum_{k,l}\frac{\partial^{2} \log f(\bm{z}|\bm{\beta})}{\partial\beta_{l}\partial\beta_{k}}\frac{\partial\beta_{l}}{\partial\hat{\beta}_{i}}\frac{\partial\beta_{k}}{\partial\hat{\beta}_{j}}+\sum_{k}\frac{\partial \log f(\bm{z}|\bm{\beta})}{\partial\beta_{k}}\frac{\partial^{2}\beta_{k}}{\partial\hat{\beta}_{i}\partial{\hat{\beta}_{j}}}\right\},\label{Fisher_g}
\end{align}
now it is easily verified that $\text{E}_{\bm{z}\sim f}\left\{\frac{\partial \log f(\bm{z}|\bm{\beta})}{\partial\beta_{k}}\right\}=0,\ \forall\ k,$
and what remains is the $ij$ element of the matrix $\bm{J}_{\bm{\psi}}(\hat{\bm{\beta}})^{T}\bm{F}(\bm{\beta})\bm{J}_{\bm{\psi}}(\hat{\bm{\beta}})$,
thus we get $d\hat{V}(\hat{\bm{\beta}})=|\bm{J}_{\bm{\psi}}(\hat{\bm{\beta}})^{T}\bm{F}(\bm{\beta})\bm{J}_{\bm{\psi}}(\hat{\bm{\beta}})|^{1/2}\ d\hat{\bm{\beta}}=|\bm{F}(\bm{\beta})|^{1/2}|\bm{J}_{\bm{\psi}}(\hat{\bm{\beta}})|\ d\hat{\bm{\beta}}$ which proves the invariance of $dV(\bm{\beta})$
to smooth transformations of $\bm{\beta}$.
\end{proof}
\noindent Now rewrite the integral in (\ref{def_marginal}) as
\begin{align}
&m_{\gamma_{d}}(\bm{z}|I_{\tau},\lambda) = \frac{1}{|I_{\tau}|}\int_{\bm{\beta}\in\Theta\times T}f(\bm{z}|\bm{\beta})\pi_{\lambda}(\bm{\beta})\ d\bm{\beta}\nonumber\\
&=\frac{1}{|I_{\tau}|}\int_{\bm{\beta}\in\Theta\times T}f(\bm{z}|\bm{\beta})\frac{\pi_{\lambda}(\bm{\beta})}{|\bm{F}(\bm{\beta})|^{1/2}}\cdot\ |\bm{F}(\bm{\beta})|^{1/2}\ d\bm{\beta}\nonumber\\
&=\frac{1}{|I_{\tau}|}\int_{\bm{\beta}\in\Theta\times T}f(\bm{z}|\bm{\beta})\frac{\pi_{\lambda}(\bm{\beta})}{|\bm{F}(\bm{\beta})|^{1/2}}\ dV(\bm{\beta})\label{rewritten_marginal_p}
\end{align}
where $dV(\bm{\beta})$ is the reparametrization invariant integration measure discussed above. We have the following result
\begin{proposition}
The integrand $f(\bm{z}|\bm{\beta})\frac{\pi_{\lambda}(\bm{\beta})}{|\bm{F}(\bm{\beta})|^{1/2}}$ is invariant to reparameterizations $\bm{\beta}=\bm{\psi}(\hat{\bm{\beta}})$. 
\label{Invariant_Marginal_Integral}
\end{proposition}
\begin{proof}
To see this, simply observe that
\begin{align} 
&g(\bm{z}|\hat{\bm{\beta}})\frac{\rho(\hat{\bm{\beta}})}{|\hat{\bm{F}}(\hat{\bm{\beta}})|^{1/2}}=f(\bm{z}|\bm{\psi}(\hat{\bm{\beta}}))\frac{\pi_{\lambda}(\bm{\psi}(\hat{\bm{\beta}}))|\bm{J}_{\bm{\psi}}(\hat{\bm{\beta}})|}{|\bm{J}_{\bm{\psi}}(\hat{\bm{\beta}})^{T}\bm{F}(\bm{\beta})\bm{J}_{\bm{\psi}}(\hat{\bm{\beta}})|^{1/2}}\nonumber\\
&=f(\bm{z}|\bm{\beta})\frac{\pi_{\lambda}(\bm{\beta})}{|\bm{F}(\bm{\beta})|^{1/2}}.\label{invariance_of_integrand}
\end{align}
\end{proof}
\noindent We note that $-\log{\left(f(\bm{z}|\bm{\beta})\pi_{\lambda}(\bm{\beta})/|\bm{F}(\bm{\beta})|^{1/2}\right)}$ is, up to terms not depending on data $\bm{z}$ 
or parameters $\bm{\beta}$, the same expression one seeks to minimize in estimator and model selection by 
the Minimum Message Length (MML) principle in \cite{OliverBaxter:1994a}.

We may now proceed to calculate the integral in (\ref{def_marginal}) by a Laplace method which is invariant to reparameterizations. 
The Laplace method for evaluating marginal densities was investigated in \cite{TierneyKassKadane:1989}, \cite{TierneyKadane:1986}, \cite{KassKadaneTierney:1988} in the univariate case which may be straightforwardly extended to the the multivariate case of IID variables whereas in our case we face the problem of evaluating the marginal density in the multivariate case of ID variables which are not identically distributed, e.g different means ($E\{z_{i}\}=\theta_{i},\ 1\leq i\leq d$). 
Using the notation and definitions from above, we write
\begin{align}
&m_{\gamma_{d}}(\bm{z}|I_{\tau},\lambda)=\frac{1}{|I_{\tau}|}\int_{\bm{\theta}\in\Theta,\tau\in I_{\tau}}f(\bm{z}|\bm{\theta},\tau)\frac{\pi_{\lambda}(\bm{\theta})}{|\bm{F}(\bm{\theta},\tau)|^{1/2}}\ dV(\bm{\theta},\tau)
\label{invariant_marginal}\\
&=\frac{1}{|I_{\tau}|}\int_{\bm{\theta}\in\mathbb{R}^{d},\tau\in I_{\tau}}\exp\left[-\Phi(\bm{z},\tau,\bm{\theta})\right]\ dV(\bm{\theta},\tau)\label{exp_invariant_marginal}\\
&\text{where } -\Phi(\bm{z},\tau,\bm{\theta})\mathdef\log{\left[f(\bm{z}|\bm{\theta},\tau)\frac{\pi_{\lambda}(\bm{\theta})}{|\bm{F}(\bm{\theta},\tau)|^{1/2}}\right]_{v}}\label{def_Phi}
\intertext{and define the invariant MML-estimators by}
&\bm{\theta}^{*}\mathdef\text{arg min }_{\bm{\theta}\in\mathbb{R}^{d}}\ \Phi(\bm{z},\tau,\bm{\theta}),\ \tau^{*}\mathdef\text{arg min }_{\tau\in I_{\tau}}\ \Phi(\bm{z},\tau,\bm{\theta}),\label{def_invariant_estimators}
\end{align}
assuming the existence of extremal points $\bm{\theta}^{*}$ and $\tau^{*}$ 
where $\left.\frac{\partial\Phi}{\partial\bm{\theta}}(\bm{z},\tau,\bm{\theta})\right|_{\bm{\theta}=\bm{\theta}^{*}}$ $=$ $\bm{0}$
and $\left.\frac{\partial\Phi}{\partial\tau}(\bm{z},\tau,\bm{\theta})\right|_{\tau=\tau^{*}}=0$. It suffices that $\Phi(\bm{z},\tau,\bm{\theta})$
is a convex function in each of the parameter arguments $\tau$ and $\theta_{i},\ i=1,...,d$.
If we knew the exact form of the integration measure $dV(\bm{\theta},\tau)$, we could approximate the marginal density 
$m_{\gamma_{d}}(\bm{z}|I_{\tau},\lambda)$ 
by expanding the integral (\ref{exp_invariant_marginal}) around $\bm{\theta^{*}}$ and $\tau^{*}$ up to some order in $\bm{\theta}$ and $\tau$.
However, when doing such an expansion we want ''low order asymptotic 
convergence'' of the expansion series, to avoid both complex computations and complex
resulting formulas possibly difficult to analyse and implement. By ''low order asymptotic convergence'' we mean that second order Taylor approximations of 
$\Phi$ in (\ref{exp_invariant_marginal}) will be ''accurate enough'' for our purposes in the sense that asymptotically in the sample size $n$, our low order expansion of the integral will converge ''sufficiently fast'' to the exact value of the integral. 
%up to an error which depend on the likelihood $f$, prior distribution $\pi$, sample size $n$, the model size $d$ and the given data set $\bm{x}$. 
We will define ''accurate enough'' and ''sufficiently fast'' later. This ''low order asymptotic convergence'' 
may be difficult to achieve in arbitrary chosen parameterizations $\bm{\theta},\tau$. 
Also, the result would depend 
%both in form and accuracy 
on our more or less arbitrary choice of parameterization of the distributions $f$ and $\pi_{\lambda}$ in the first place.  
On this background we seek a reparameterization $\tau\mapsto\hat{\tau}$ and $\theta_{i}\mapsto\hat{\theta}_{i},\ 1\leq i\leq d$ 
yielding $dV(\bm{\theta},\tau)\rightarrow d\hat{V}(\hat{\tau},\hat{\bm{\theta}})=v_{0}\ d\hat{\bm{\theta}}\ d\hat{\tau}$ 
where $[v_{0}]_{u}$ is some positive real constant number.
To construct such a reparameterization we will limit our investigation to the case of a gaussian likelihood $f$. 
We then write
\begin{align}
&f(\bm{z}|\bm{\theta},\tau)=\left(\frac{\tau}{2\pi}\right)^{\frac{n}{2}}\exp\left(-\frac{\tau}{2}\|\bm{z}_{\perp}\|^{2}\right)\exp\left(-\frac{\tau}{2}\|\bm{z}_{\parallel}-\bm{\theta}\|^{2}\right).\label{split_likelihood}\\
\intertext{Let $\bar{\tau}$ be some real positive dimensionless constant number and let $\tau_{0}$ be some real positive constant with $[\tau_{0}]_{u}=[\tau]_{u}$. We choose}  
&\tau=\psi(\hat{\tau}),\ \psi(0)=\tau_{0},\ \theta_{i}=\phi(\hat{\theta}_{i},\hat{\tau})\mathdef\frac{\bar{\tau}^{1/2}}{\tau^{1/2}}\hat{\theta}_{i}
=\left(\frac{\bar{\tau}}{\psi(\hat{\tau})}\right)^{\frac{1}{2}}\hat{\theta}_{i},\ 1\leq i\leq d.\label{def_phi_psi}\\
\intertext{The Fisher matrix $\hat{\bm{F}}(\hat{\bm{\theta}},\hat{\tau})$ then evaluates to (see appendix)}
&\hat{\bm{F}}_{ij}(\hat{\bm{\theta}},\hat{\tau})=\left\{\begin{array}{ll}\left(\frac{1}{\psi(\hat{\tau})}\frac{d\psi(\hat{\tau})}{d\hat{\tau}}\right)^{2}
\left(\frac{n}{2}+\frac{\bar{\tau}}{4}\sum_{k=1}^{d}\hat{\theta}_{k}^{2}\right) &\text{ if } i=j=1,\\ 
\bar{\tau}, & \text{ if } i=j,\ 1< i,j\leq d,\\
-\frac{1}{2}\bar{\tau}^{\frac{1}{2}}\hat{\theta}_{j}\frac{1}{\psi(\hat{\tau})}\frac{d\psi(\hat{\tau})}{d\hat{\tau}} & \text{ if } i=1,\ 1<j\leq d,\\
-\frac{1}{2}\bar{\tau}^{\frac{1}{2}}\hat{\theta}_{i}\frac{1}{\psi(\hat{\tau})}\frac{d\psi(\hat{\tau})}{d\hat{\tau}} & \text{ if } j=1,\ 1<i\leq d,\\
0 & \text{ else.}\end{array}\right.\label{G_elements}\\
\intertext{As shown in the appendix, the determinant of $\hat{\bm{F}}(\hat{\bm{\theta}},\hat{\tau})$ as given in (\ref{G_elements}) above evaluates to}
&|\hat{\bm{F}}(\hat{\bm{\theta}},\hat{\tau})|=\frac{n}{2}\bar{\tau}^{d}\left(\frac{1}{\psi(\hat{\tau})}\frac{d\psi(\hat{\tau})}{d\hat{\tau}}\right)^{2}.\label{det_G}
\intertext{Now, our choice of reparameterization in (\ref{def_phi_psi}) implies $\hat{\theta}_{i},\ 1\leq i\leq d$ and $\hat{\tau}$ are dimensionless parameters. 
Therefore we may put $|\hat{\bm{F}}(\hat{\bm{\theta}},\hat{\tau})|^{1/2}=\bar{\bar{\tau}}^{d/2}$, where $\bar{\bar{\tau}}$ is some positive real dimensionless number.
This gives us together with (\ref{det_G}) and the initial condition in (\ref{def_phi_psi}) the equation}
&\frac{d\psi(\hat{\tau})}{d\hat{\tau}}=\pm\left(\frac{\bar{\bar{\tau}}}{\bar{\tau}}\right)^{\frac{d}{2}}\left(\frac{2}{n}\right)^{1/2}\psi(\hat{\tau}),\ \psi(0)=\tau_{0}.\label{diff_equation_psi}
\intertext{We choose the plus-sign in (\ref{diff_equation_psi}). This choice implies no loss of generality, as it is only a matter of sign convention on the parameter $\hat{\tau}$.  
Solving (\ref{diff_equation_psi}) then gives}
&\psi(\hat{\tau})=\tau_{0}\exp\left(\left(\frac{\bar{\bar{\tau}}}{\bar{\tau}}\right)^{\frac{d}{2}}\left(\frac{2}{n}\right)^{\frac{1}{2}}\hat{\tau}\right)\label{form_of_psi}
\intertext{where $\hat{\tau}$, $\bar{\tau}$, $\bar{\bar{\tau}}$ are dimensionless numbers and $[\tau_{0}]_{u}=[\tau]_{u}$. For notational convenience we define}
&\epsilon_{d}\mathdef\left(\frac{\bar{\bar{\tau}}}{\bar{\tau}}\right)^{\frac{d}{2}}\label{def_epsilon}
\intertext{and}
&\delta_{n}\mathdef\left(\frac{2}{n}\right)^{\frac{1}{2}}.\label{def_delta}
\intertext{Define}
&\hat{\Phi}(\bm{z},\hat{\tau},\hat{\bm{\theta}})\mathdef-\log{\left[\frac{g(\bm{z}|\hat{\bm{\theta}},\hat{\tau})\rho(\hat{\bm{\theta}})}{|\hat{\bm{F}}(\hat{\bm{\theta}},\hat{\tau})|^{1/2}}\right]_{v}}\nonumber\\
&=-\log{\left[\frac{f\left(\bm{z}|\bm{\phi}(\hat{\bm{\theta}},\hat{\tau}),\psi(\hat{\tau})\right)\pi_{\lambda}\left(\bm{\phi}(\hat{\bm{\theta}},\hat{\tau})\right)|\bm{J}_{\bm{\phi},\psi}|}{|\bm{J}_{\bm{\phi},\psi}^{T}\bm{F}(\bm{\phi}(\hat{\bm{\theta}},\hat{\tau}),\psi(\hat{\tau}))\bm{J}_{\bm{\phi},\psi}|^{1/2}}\right]_{v}}=\Phi(\bm{z},\psi(\hat{\tau}),\bm{\phi}(\hat{\bm{\theta}}))\label{def_Phi_hat}
\intertext{where we used (\ref{invariance_of_integrand}). Furthermore, define}
&\hat{\bm{\theta}}^{*}(\bm{z})\mathdef\text{arg min }_{\hat{\bm{\theta}}\in\hat{\Theta}}\ \hat{\Phi}(\bm{z},\hat{\tau},\hat{\bm{\theta}}),\ \hat{\tau}^{*}(\bm{z})\mathdef\text{arg min }_{\hat{\tau}\in \hat{I}_{\hat{\tau}}}\ \hat{\Phi}(\bm{z},\hat{\tau},\hat{\bm{\theta}})
\end{align}
where we have assumed $\hat{\Phi}(\bm{z},\hat{\tau},\hat{\bm{\theta}})$ is convex in each of its parameter arguments 
$\hat{\tau}$ and $\hat{\theta}_{i},\ i=1,...,d$, thus
the existence of $\hat{\tau}^{*}$ and $\hat{\bm{\theta}}^{*}$ is guaranteed. The integral in (\ref{exp_invariant_marginal}) 
defining $m_{\gamma_{d}}(\bm{z}|I_{\tau},\lambda)$ may then be rewritten as
\begin{align}
&m_{\gamma_{d}}(\bm{z}|I_{\tau},\lambda)=\bar{\bar{\tau}}^{\frac{d}{2}}\frac{1}{|I_{\tau}|}\int_{\hat{\bm{\theta}}\in\mathbb{R}^{d},\hat{\tau}\in \hat{I}_{\hat{\tau}}}\exp\left(-\hat{\Phi}(\bm{z},\hat{\tau},\hat{\bm{\theta}})\right)\ d\hat{\bm{\theta}}\ d\hat{\tau}.\label{reparameterized_integral}
\end{align}
Now, our constructed reparameterization above will provide us with the necessary means for approximating the marginal $m_{\gamma_{d}}(\bm{z}|I_{\tau},\lambda)$ to sufficient accuracy by a second order approximation which is invariant to reparameterizations. \\
\begin{theorem}
(Invariant second order approximation of marginal density) Let $\bm{x}\in\mathbb{R}^{n}$ be the given data set under the model (\ref{noise_model}) and 
an index vector $\gamma_{d}\in\{0,1\}^{n}$ of model indices with $d$ nonzero elements, $0<d<n$. Let $\bm{\theta}^{*}$ and $\tau^{*}$ be the invariant estimators defined in (\ref{def_invariant_estimators}). Let $I_{\tau}\subset(0,\infty)$ be a bounded closed interval containing the MML-estimate $\tau^{*}$. Let $f(\bm{x}|\bm{\theta},\tau)$ be a gaussian likelihood function of data $\bm{x}$ and $\pi_{\lambda}(\bm{\theta})$ a prior density on
the parameters $\bm{\theta}\in\mathbb{R}^{d}$ with a given variance $\lambda^{-1}$. Let $\bm{x}=\bm{x}_{\parallel}+\bm{x}_{\perp}$ be the orthogonal decomposition of the data induced by the selected model $\gamma_{d}$ of size $d$. Let $\bm{F}(\bm{\theta},\tau)$ denote the $(d+1)\times(d+1)$ Fisher matrix of the likelihood function $f(\bm{x}|\bm{\theta},\tau)$ with respect to parameters $\bm{\theta}$, $\tau$.
Let $\bm{H}(\bm{x},\bm{\theta},\tau)$ denote the $(d+1)\times(d+1)$ Hessian matrix of 
$\pi_{\lambda}(\bm{\theta})f(\bm{x}|\bm{\theta},\tau)\left|\bm{F}(\bm{\theta},\tau)\right|^{-\frac{1}{2}}$, let 
$P_{G}$ denote the Gaussian distribution function. Then the marginal $m_{\gamma_{d}}(\bm{x}|I_{\tau},\lambda)$ defined in (\ref{def_marginal}) may 
be expressed as follows:
\begin{align} 
&m_{\gamma_{d}}(\bm{x}|I_{\tau},\lambda)=\frac{(2\pi)^{\frac{d+1}{2}}}{|\bm{H}(\bm{x},\tau^{*},\bm{\theta}^{*})|^{\frac{1}{2}}}f(\bm{x}|\tau^{*},\bm{\theta}^{*})\pi_{\lambda}(\bm{\theta}^{*})|I_{\tau}|^{-1}\times\nonumber\\
&\left[\prod_{i=1}^{d}P_{G}\left((\tau^{*})^{\frac{1}{2}}|\theta_{i}^{*}|\left\{1+\littleo(\zeta)\right\}
\right)\right]\left\{1+\bigo\left(\kappa\right)\right\}\left\{1+\xi\right\}.\label{almost_invariant_marginal}
\intertext{
%where $C_{\gamma_{d}}^{-1}$ is a normalizing constant to ensure $\int_{\bm{z}\in X}m_{\gamma_{d}}(\bm{z}|I_{\tau},\lambda)d\bm{z}=1$ where 
%$X\ni \bm{x}$ is the region over which the marginal $m_{\gamma_{d}}(\bm{z}|I_{\tau},\lambda)$ is normalized. 
The formula (\ref{almost_invariant_marginal}) applies under the following sufficient conditions:}
\intertext{(1)\hspace{5mm}(Shape of prior) $\pi_{\lambda}(\theta)=C\cdot\lambda^{\frac{1}{2}}\exp(-h(\lambda^{\frac{1}{2}}\theta))$, some constant $C>0$, $\lambda>0$,  where $h$ is an integrable, symmetric function of $\theta$ such that}
&\lim_{|\theta|\rightarrow\infty}h(\lambda^{\frac{1}{2}}\theta)=\infty.\label{C1}\\
\intertext{(2)\hspace{5mm}(Heaviness of tails, integrability and smoothness on the prior) There exist constant real numbers $0<\nu<2$, $B_{\nu}^{\prime}\leq B_{\nu}$, $C_{\nu}>0$ so that the inequalities}
&B_{\nu}^{\prime}\leq h(\lambda^{\frac{1}{2}}\theta)\leq B_{\nu}+C_{\nu}\left|\lambda^{\frac{1}{2}}\theta\right|^{\nu},\ \forall\ [\theta]_{v}\in\mathbb{R},\ 1\leq i\leq d,\label{C2}\\
\intertext{and}
&0\leq\left|\left[\frac{\partial^{k}}{\partial\theta^{k}}h(\lambda^{\frac{1}{2}}\theta)\right]_{v}\right|\leq C_{\nu}\left|\left[\frac{\partial^{k}}{\partial\theta^{k}}\left|\lambda^{\frac{1}{2}}\theta\right|^{\nu}\right]_{v}\right|,\label{C3}\\
\intertext{hold for $\forall\ [\theta]_{v}\in\mathbb{R}$, and for all  $1\leq k<\infty,\ 1\leq i\leq d$.}\nonumber
\intertext{(3)\hspace{5mm} (SNR and model size) The number $0\leq\zeta<1$ is defined by}
&\zeta\mathdef \sup_{1\leq i\leq d}\left\{C_{\nu}\nu|\nu-1|\left(\frac{n}{d}\Omega(\lambda,\tau^{*})\right)^{-\frac{\nu}{2}}|(\tau^{*})^{\frac{1}{2}}\theta_{i}^{*}|^{\nu-2}\right\}<1\label{C4}
\intertext{where $\Omega(\lambda,\tau)$ is the signal to noise ratio (SNR)}
&\Omega(\lambda,\tau)\mathdef\frac{d\lambda^{-1}}{n\tau^{-1}}.\nonumber\\
%\intertext{(4)\hspace{5mm} (Bounds on model parameter estimates) We must have that}
%&\tau^{*}(\theta_{i}^{*})^{2}\geq 1,\ \forall\ i, \ 1\leq i\leq d.\label{C6}
%\intertext{(5)\hspace{5mm}(Scale of the second moment $\lambda^{-1}$) $\lambda^{-1}$ scales as}
%&\lambda^{-1}\propto d^{-1}\|\bm{\theta}^{*}\|_{2}^{2}.\label{C7}
\intertext{(4)\hspace{5mm} (The size and location of the interval $I_{\tau}$) The interval $I_{\tau}$ satisfies}
%&\left(\tau^{*}\exp\left[-\left(\frac{2\log{N(\lambda,\nu,\gamma_{d})}}{N(\lambda,\nu,\gamma_{d})}\right)^{\frac{1}{2}}\right],\tau^{*}\exp\left[\left(\frac{2\log{N(\lambda,\nu,\gamma_{d})}}{N(\lambda,\nu,\gamma_{d})}\right)^{\frac{1}{2}}\right]\right)\nonumber\\
%&\subset I_{\tau}\subset\left(\tau^{*}\exp\left[-\left(\frac{\log^{2}{N(\lambda,\nu,\gamma_{d})}}{N(\lambda,\nu,\gamma_{d})}\right)^{\frac{1}{2}}\right],\tau^{*}\exp\left[\left(\frac{2\log^{2}{N(\lambda,\nu,\gamma_{d})}}{N(\lambda,\nu,\gamma_{d})}\right)^{\frac{1}{2}}\right]\right)\label{C11}
&\tau^{*}\in I_{\tau}\subset\left(\tau^{*}\exp\left[-\left(\frac{\log^{2}{N(\lambda,\nu,\gamma_{d})}}{N(\lambda,\nu,\gamma_{d})}\right)^{\frac{1}{2}}\right],\tau^{*}\exp\left[\left(\frac{2\log^{2}{N(\lambda,\nu,\gamma_{d})}}{N(\lambda,\nu,\gamma_{d})}\right)^{\frac{1}{2}}\right]\right)\label{C11}
\intertext{where} 
&N(\lambda,\nu,\gamma_{d})\sim\left\{\begin{array}{ll}\frac{n-d+2}{2},&\text{ if } 0<\nu\leq 1\\
\frac{n-d+2}{2}-\frac{C_{\nu}\nu|\nu-1|}{4}d,&\text{ if } 1<\nu\leq 2.\end{array}\right.\nonumber
\intertext{We add that if the conditions (1)-(6) listed above are satisfied, one may then show the following bounds on $\kappa$ and $\xi$:\newline\newline
\noindent(5)\hspace{5mm}(The approximation error $\bigo\left(\kappa\right)$ from the Taylor terms above second order) $|\kappa|$ may be bounded from above by}
&|\kappa|< \frac{4}{3}(1+\zeta)\frac{C_{\nu}\nu|\nu-1|\cdot|\nu-2|}{\left(\frac{n}{d}\Omega(\tau^{*},\lambda)\right)^{\frac{\nu}{2}}}
\left|\sum_{j=1}^{d}\frac{\left|(\tau^{*})^{\frac{1}{2}}\theta_{j}^{*}\right|^{\nu-1}\sgn(\theta_{j}^{*})\left(1+\frac{2}{\tau^{*}(\theta_{j}^{*})^{2}}\right)}{\exp\left(\frac{1}{2}\tau^{*}(\theta^{*}_{j})^{2}\right)}\right|\nonumber\\
&+\frac{1}{N(\lambda,\nu,\gamma_{d})}\sum_{j=1}^{d}\frac{\tau^{*}(\theta_{j}^{*})^{2}}{\exp\left(\frac{1}{2}\tau^{*}(\theta_{j}^{*})^{2}\right)}\nonumber\\
&+\left|\frac{(2\pi)^{-\frac{1}{2}}}{N(\lambda,\nu,\gamma_{d})}\sum_{i,j=1}^{d}\frac{\tau^{*}(\bm{x}_{\parallel}(i)-\frac{1}{2}\theta_{i}^{*})(\bm{x}_{\parallel}(j)-\frac{1}{2}\theta_{j}^{*})}{\exp\left(\frac{1}{2}\tau^{*}\left[(\theta_{i}^{*})^{2}+(\theta_{j}^{*})^{2}\right]\right)}\right|.\label{C8}
\intertext{(6)\hspace{5mm}(The contribution $\xi$ from the integral of $\exp(-\Phi)$ over $\mathbb{R}^{d}\setminus\{\bigcup_{i=1}^{d}S_{i}\}$ where $S_{i}$ is the ''quadrant'' of $\mathbb{R}^{d}$ containing $\theta_{i}^{*}$.) The number $\xi$ may be bounded from above by}
&1<\xi+1<\nonumber\\
&\prod_{i=1}^{d}\left\{1+\left[2P_{G}\left(-\tau_{1}^{\frac{1}{2}}\left|\bm{x}_{\parallel}(i)\right|\right)\sup_{t\in\mathbb{R}}{\pi_{\lambda=1}(t)}/\pi_{\lambda=1}\left(u_{0}\left(\tau_{1}^{\frac{1}{2}}\bm{x}_{\parallel}(i)\right)\right)\right]\times\right.\nonumber\\
&\left.\left[1+\erf\left(\tau_{1}^{\frac{1}{2}}|\bm{x}_{\parallel}(i)|\right)\frac{\inf_{t\in\left(0,u_{0}\left(\tau_{1}^{\frac{1}{2}}\bm{x}_{\parallel}(i)\right)\right)}{\pi_{\lambda=1}(t)}}{\pi_{\lambda=1}\left(u_{0}\left(\tau_{1}^{\frac{1}{2}} \bm{x}_{\parallel}(i)\right)\right)}\right.\right.\nonumber\\
&\left.\left.-\frac{2C_{\nu}\nu}{(2\pi)^{\frac{1}{2}}}L_{\nu}(\tau_{1}^{\frac{1}{2}}\bm{x}_{\parallel}(i))\frac{\sup_{t\in\left(u_{0}\left(\tau_{1}^{\frac{1}{2}}\bm{x}_{\parallel}(i)\right),\infty\right)}{\pi_{\lambda=1}(t)}}{\pi_{\lambda=1}\left(u_{0}\left(\tau_{1}^{\frac{1}{2}}\bm{x}_{\parallel}(i)\right)\right)}\right]^{-1}\right\}\label{C9}
\intertext{if}
&\frac{\pi_{\lambda=1}\left(u_{0}\left(\tau_{1}^{\frac{1}{2}}\bm{x}_{\parallel}(i)\right)\right)}{\sup_{t\in\left(u_{0}\left(\tau_{1}^{1/2}\bm{x}_{\parallel}(i)\right),\infty\right)}{\pi_{\lambda=1}(t)}}> \frac{2C_{\nu}\nu}{(2\pi)^{\frac{1}{2}}}L_{\nu}\left(\tau_{1}^{\frac{1}{2}}\bm{x}_{\parallel}(i)\right),\ \forall\ i\in\gamma_{d}\nonumber
\intertext{where $\tau_{1}\in I_{\tau}$, $u_{0}(s)=\left(\frac{n}{d}\Omega(\lambda,\tau)\right)^{-\frac{1}{2}}|s|$ and}\nonumber\\
&L_{\nu}\left(\tau^{\frac{1}{2}}x\right)\mathdef\left\{\begin{array}{ll}\left|\tau^{\frac{1}{2}}x\right|^{\nu-1}\left(\frac{n}{d}\Omega(\lambda,\tau)\right)^{-\frac{\nu}{2}} & \text{ if }0<\nu\leq 1\\
\left(\frac{n}{d}\Omega(\lambda,\tau)\right)^{-\frac{1}{2}}\left(1+\left|\tau^{\frac{1}{2}}x\right|\left(\frac{n}{d}\Omega(\lambda,\tau)\right)^{-\frac{1}{2}}\right)
& \text{ if }1<\nu\leq 2.\end{array}\right.\label{C10}
\end{align}
\label{theorem1}
\end{theorem}
\begin{proof}
A proof is provided in the appendix.
\end{proof}
The invariant approximation of the marginal density $m_{\gamma_{d}}(\bm{x}|I_{\tau},\lambda)$ in (\ref{almost_invariant_marginal}) may now be fed into a code length principle to yield a best model size estimate $d^{\ast}$ and the best model $\gamma_{d^{\ast}}^{\ast}$
for a given dataset $\bm{x}\in\mathbb{R}^{n}$. This will yield a model selection principle invariant to reparameterizations in the sense
explained in above sections.

\section{Generalized Laplace-approximation of marginal density}
We now proceed to the case where the variance parameter $\lambda^{-1}$ of the prior distribution 
$\pi_{\lambda}(\bm{\theta})$ is unknown to us. We will then have to estimate the parameter $\lambda$ from the given data set. This implies that
the density $m_{\gamma}(\bm{x}|I_{\tau},\lambda)$ as written in (\ref{almost_invariant_marginal}) in Theorem \ref{theorem1} is not the 
marginal density for the data $\bm{x}$ as it contains the data dependent parameter $\lambda$. 
We must integrate out the parameter $\lambda\in I_{\lambda}$ from the formula in 
(\ref{def_marginal}), that is the marginal $m_{\gamma_{d}}(\bm{x}|I_{\tau},I_{\lambda})$ now becomes:
\begin{align}
&m_{\gamma_{d}}(\bm{x}|I_{\tau},I_{\lambda})\mathdef\frac{1}{|I_{\tau}|}\int_{\bm{\theta}\in\mathbb{R}^{d},\tau\in I_{\tau},\lambda\in I_{\lambda}}f(\bm{x}|\bm{\theta},\tau)\pi(\bm{\theta}|\lambda)l(\lambda)\ d\bm{\theta}\ d\tau\ d\lambda\label{def_marginal2}
\intertext{where $l(\lambda)$ is a prior distribution on the parameter $\lambda$. Let $I_{\lambda}\subset\mathbb{R}_{+}$ be a bounded interval, 
we model $\lambda$ as uniformly distributed on $I_{\lambda}$, and identically zero outside $I_{\lambda}$, that is}
&l(\lambda)=\left\{\begin{array}{ll}\frac{1}{|I_{\lambda}|},\text{ if } \lambda\in I_{\lambda}\\
0,\text{ otherwise }\end{array}\right.\label{prior_lambda}
\intertext{Now, by means of Theorem \ref{theorem1} we may write:}
&m_{\gamma_{d}}(\bm{x}|I_{\tau},I_{\lambda})=\frac{1}{|I_{\tau}|}\frac{1}{|I_{\lambda}|}\frac{(2\pi)^{\frac{d+1}{2}}f(\bm{x}|\bm{\theta}^{*},\tau^{*})}{|\bm{H}(\bm{x},\bm{\theta}^{*},\tau^{*})|^{\frac{1}{2}}}(1+\bigo(\kappa))(1+\xi)\times\nonumber\\
&\prod_{i=1}^{d}P_{G}\left((\tau^{*})^{\frac{1}{2}}|\theta_{i}^{*}|\left\{1+\littleo(\zeta)\right\}
\right)\int_{\lambda\in I_{\lambda}}\pi(\bm{\theta}^{*}|\lambda)\ d\lambda\label{marginal_integral2}
\intertext{where we have ignored any dependency of $\kappa$, $\xi$, $\bm{\theta}^{*}$, $\tau^{*}$ on $\lambda$ in the integration interval $I_{\lambda}$. This assumption will hold if we choose the location of the interval $I_{\lambda}$ properly and its width small enough as may be 
seen by examining the proof of Theorem \ref{theorem1}. 
We will preserve parameter invariance by following the same procedure of invariant Laplace-expansions as in sections above by expanding the desired integral in (\ref{marginal_integral2}) about a certain point $\lambda^{*}$. We need some definitions. Define}
&\bm{E}(\lambda)\mathdef-E_{\bm{\theta}}\left\{\frac{\partial^{2}}{\partial\lambda^{2}}\log{\pi(\bm{\theta}|\lambda)}\right\}\label{def_prior_fisher_matrix}
\intertext{and define}
&\Psi(\bm{\theta},\lambda)\mathdef-\log\left[\frac{\pi(\bm{\theta}|\lambda)}{|\bm{E}(\lambda)|^{\frac{1}{2}}}\right]\label{def_Psi}
\intertext{and define}
&\lambda^{*}\mathdef\text{arg inf}_{\lambda>0}\Psi(\bm{\theta},\lambda).\label{def_invariant_lambda_estimator}
\end{align}
We have the following result:
\begin{proposition}
Let $\pi(\bm{\beta}|\lambda)=\prod_{i=1}^{d}\pi(\beta_{i}|\lambda)$ be a density on $\bm{\beta}\in\mathbb{R}^{d}$ with variance $\lambda^{-1}$. 
Let $\bm{E}(\lambda)\mathdef-E_{\bm{\beta}}\left\{\frac{\partial^{2}}{\partial\lambda^{2}}\log{\pi(\bm{\beta}|\lambda)}\right\}$, let
$\Psi(\bm{\beta},\lambda)\mathdef-\log\left[\frac{\pi(\bm{\beta}|\lambda)}{|\bm{E}(\lambda)|^{\frac{1}{2}}}\right]$ and let
$\lambda^{*}\mathdef\text{arg inf}_{\lambda>0}\Psi(\bm{\beta},\lambda)$. Let $I_{\lambda}\subset\mathbb{R}_{+}$ be a bounded interval such that  
$\lambda^{*}\in I_{\lambda}$. Then we have:
\begin{align}
&\int_{\lambda\in I_{\lambda}}\pi(\bm{\beta}|\lambda)\ d\lambda = \frac{(2\pi)^{\frac{1}{2}}\pi(\bm{\beta}|\lambda^{*})}{|\Psi_{\lambda\lambda}(\bm{\beta},\lambda^{*})|^{\frac{1}{2}}}\left\{1+\bigo\left(\omega\right)\right\}.\label{lambda_integral}
\intertext{where}
&\omega\mathdef\frac{|\Psi_{\lambda\lambda\lambda}(\bm{\beta},\lambda^{*})|}{|\Psi_{\lambda\lambda}(\bm{\beta},\lambda^{*})|^{\frac{3}{2}}}=\bigo\left(\frac{1}{\sqrt{d}}\right).\label{def_omega}
\end{align}
Furthermore, the formula (\ref{lambda_integral}) is invariant to reparameterizations of the distribution $\pi$.
\label{proposition_prior_laplace_approximation_formula}
\end{proposition}
\begin{proof}
We define a map $\chi:\hat{\lambda}\rightarrow\lambda$ such that
\begin{align}
&\lambda=\chi(\hat{\lambda}),\ \chi(0)=\lambda_{0}\text{ and }\hat{\bm{E}}(\hat{\lambda})\mathdef-E_{\bm{\theta}}\left[\frac{\partial^{2}}{\partial\hat{\lambda}^{2}}\log\pi(\bm{\theta}|\chi(\hat{\lambda}))\right] 
= \bar{\lambda}^{-2}\in\mathbb{R}_{+}\label{def_chi}
\intertext{where $\bar{\lambda}>0$ is some constant number. The result follows by computing the Taylor-expansion 
  $\hat{T}(\hat{\lambda})$ of $\Psi(\bm{\beta},\chi(\hat{\lambda}))$ in $\hat{\lambda}$ 
  about the point $\hat{\lambda}^{*}\mathdef\chi^{-1}(\lambda^{*})$ and approximating the integral}
&\int_{\lambda\in I_{\lambda}}\exp\left(-\Psi(\bm{\beta},\lambda)\right)|\bm{E}(\lambda)|^{1/2}\ d\lambda=
\bar{\lambda}^{-1}\int_{\hat{\lambda}\in \hat{I}_{\hat{\lambda}}}\exp\left(-\Psi(\bm{\beta},\hat{\lambda})\right)\ d\hat{\lambda}\nonumber\\
&=\bar{\lambda}^{-1}\int_{\hat{\lambda}\in \hat{I}_{\hat{\lambda}}}\exp\left(-\hat{T}(\hat{\lambda})\right)\ d\hat{\lambda}\label{Psi_integral}
\intertext{to second order in $\hat{\lambda}$.}\nonumber
\end{align}
\end{proof} 
Using the result in (\ref{lambda_integral}) together with Theorem (\ref{theorem1}) we now have the following expression for the marginal density 
$m_{\gamma_{d}}(\bm{x}|I_{\tau},I_{\lambda})$:
\begin{corollary}
Given the conditions and notation in Theorem \ref{theorem1} and Proposition \ref{proposition_prior_laplace_approximation_formula}, we may state:
\begin{align}
&m_{\gamma_{d}}(\bm{x}|I_{\tau},I_{\lambda})=\frac{(2\pi)^{\frac{d+2}{2}}f(\bm{x}|\tau^{*},\bm{\theta}^{*})\pi(\bm{\theta}^{*}|\lambda^{*})|I_{\tau}|^{-1}|I_{\lambda}|^{-1}}{|\bm{H}(\bm{x},\tau^{*},\bm{\theta}^{*})|^{\frac{1}{2}}|\Psi_{\lambda\lambda}(\bm{\theta}^{*},\lambda^{*})|^{\frac{1}{2}}}\nonumber\\
&\times\left[\prod_{i=1}^{d}P_{G}\left((\tau^{*})^{\frac{1}{2}}|\theta_{i}^{*}|\left\{1+\littleo(\zeta)\right\}
\right)\right]\left\{1+\bigo(\kappa)\right\}\left\{1+\xi\right\}\left\{1+\bigo(\omega)\right\}.\label{almost_invariant_marginal2}
\end{align}
\label{corollary_new_marginal}
\end{corollary}
\begin{proof}
This is an immediate concequence of Theorem \ref{theorem1} and 
Proposition \ref{proposition_prior_laplace_approximation_formula}.
\end{proof}

\section{Marginal renormalization}
As pointed out in \cite{Rissanen:1998b}, using estimated values $\bm{\theta}^{*}(\bm{x})$, $\tau^{*}(\bm{x})$, $\lambda^{*}(\bm{\theta}^{*})$ for 
given data $\bm{x}$ instead of true parameter values $\bm{\theta}$, $\tau$, $\lambda$, does not yield an optimal code length for the data. 
That is, there is redundancy in the resulting code \cite{Rissanen:1998b}, and to remove this redundancy means to renormalize 
the marginal $m(\bm{x})$ in order to to get a proper density for use with the (IN)MDL Principle.
%Now, when replacing the parameters $\tau$, $\lambda$, $\bm{\theta}$ by their estimators $\tau^{*}(\bm{x})$, 
%$\lambda^{*}(\bm{x})$, $\bm{\theta}^{*}(\bm{x})$ as in (\ref{almost_invariant_marginal}), (\ref{almost_invariant_marginal2}) above, 
%then as pointed out in \cite{Rissanen:1998b}, $m_{\gamma_{d}}(\bm{x})$ is no longer a proper density, and has to be renormalized over some region $Y\subset\mathbb{R}^{n}$ containing the given dataset $\bm{x}$. 
 For given data set $\bm{x}$ and likelihood function 
$f(\bm{x}|\bm{\beta})$, Rissanen defined in \cite{Rissanen:2000} the normalized maximum likelihood (NML) marginal density $m_{NML}(\bm{x})$ by:
\begin{align}
&m_{NML}(\bm{x})\mathdef\frac{f\left(\bm{x}|\bm{\beta}^{*}\left(\bm{x}\right)\right)}{C_{NML}}\nonumber
\intertext{where $\bm{\beta}^{*}$ is the ML estimator and}
&C_{NML}\mathdef\int_{\bm{z}\in Y}f\left(\bm{z}|\bm{\beta}^{*}\left(\bm{z}\right)\right)\ d\bm{z}.
\label{NMDL_density}
\end{align}
The integration region $Y$ in the case of a gaussian likelihood $f$ was chosen through the ML parameter estimators to be the least possible 
{\em hyperspheres} containing the data $\bm{x}_{\perp}$ and $\bm{x}_{\parallel}$. While the ML esimator $\tau^{*}(\bm{x}_{\perp})$ for the noise
naturally imposes a spherical geometry on the part of the data space containing the noise, the same cannot be said of the ML estimator 
$\bm{\beta}^{*}$ for the parameters $\bm{\beta}$, which is simply: $\bm{\beta}^{*}(\bm{z}_{\parallel})=\bm{z}_{\parallel}$. 

It was shown in \cite{Rissanen:2001} that the density $m_{NML}(\bm{x})$ satisfies:
\begin{align}
m_{NML}(\bm{x})=\text{arg }\inf_{q\in Q}\sup_{g\in G}E_{\bm{x}\sim g}\left\{\log{\frac{f(\bm{x}|\bm{\beta}^{*}(\bm{x}))}{q(\bm{x})}}\right\}\label{def_NML_optimality}
\end{align}
where $G$ is the class of distributions $g(\bm{x})$ satisfying $E_{\bm{x}\sim g}\log\left(g(\bm{x})/\right.$$\left.f(\bm{x}|\bm{\beta}^{*})\right)$$<\infty$, $Q$ is the class of all densities and $\bm{\beta}^{*}$ is the ML-estimate of the parameters $\bm{\beta}$. This means that the code length $-\log{m_{NML}(\bm{x})}$ induced by the density $q(\bm{x})=m_{NML}(\bm{x})$ minimizes the expected difference between the the code lengths $-\log{f(\bm{x}|\bm{\beta}^{*}(\bm{x}))}$ and $-\log{q(\bm{x})}$, where expectation is taken with respect to the ''worst case'' data generating distribution $g$.
%The renormalization of the marginal  in our case is $C_{\gamma_{d}}$ defined in (\ref{def_C_gamma_d}). 
To compute the optimal code length, the domain $Y\ni\bm{x}$ on which the marginal density $m(\bm{x})$ is defined, 
has to be chosen properly, \cite{Rissanen:2000}. The expression (\ref{log_C_NML_bounds}) shows that the question for us is then for given data set $\bm{x}$ to choose the region 
$\Theta^{*}\ni\bm{\theta}^{*}$ properly. The choice of $\Theta^{*}$ may be of importance to our code length principle. This
region should not be chosen too big, neither too small. How to accomplish this? In \cite{Rissanen:2000}, the choice of $\Theta^{*}$ was taken to be 
the spherical region
\begin{align}
&\Theta^{*}_{NML}=\left\{\bm{z}\in\mathbb{R}^{d}:0<\|\bm{z}\|_{2}^{2}\leq\|\bm{x}_{\parallel}\|_{2}^{2}\right\}.\label{Rissanen_volume}
\end{align}
This is perhaps the most ''honest'' choice of integration region $\Theta^{*}$: In the absence of a prior distribution on the
parameters $\bm{\theta}\in\mathbb{R}^{d}$, the choice of a flat prior distribution on a domain with no preferred direction certainly
does not impose any prior constraints on the parameter $\bm{\theta}$, except that its expected norm is $\|\bm{x}_{\parallel}\|_{2}$. 
We will use the geometry imposed on the signal space $Y_{\parallel}$ by the prior distribution $\pi(\bm{\theta}|\lambda)$ through the invariant 
ML-estimator $\lambda^{*}$ defined in (\ref{def_invariant_lambda_estimator}). That is, for given data set $\bm{x}$ and model $\gamma_{d}$, we choose
\begin{align}
&\Theta^{*}=\left\{\bm{\theta}\in\mathbb{R}^{d}:\lambda^{*}(\bm{\theta})\in J_{\lambda}\ni\lambda^{*}(\bm{\theta}^{*}(\bm{x})) \right\}\label{Theta_star_region}
\end{align}
for some chosen interval $J_{\lambda}\subset\mathbb{R}_{+}$. This choice will ensure that $\bm{\theta}^{*}(\bm{x})\in\Theta^{*}$. 
\begin{figure}[h]
\begin{center}
\includegraphics[scale=0.6]{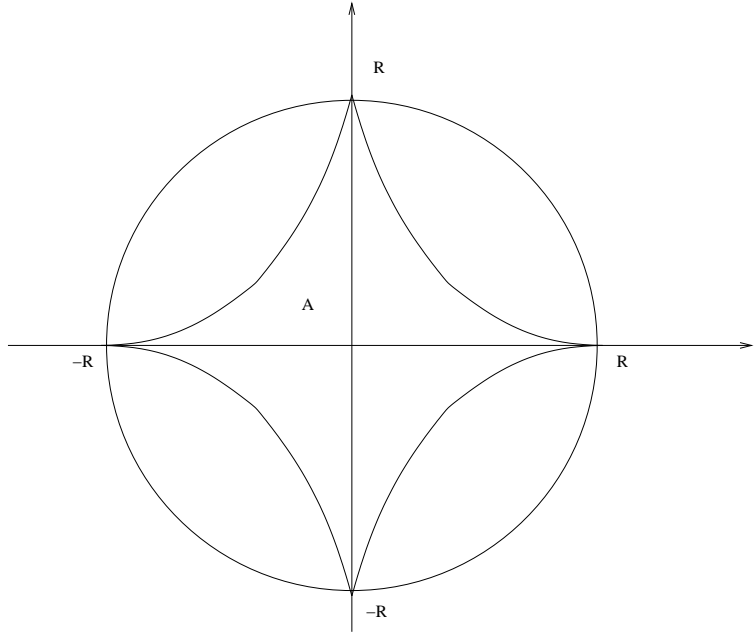}
\end{center}
\caption{Illustration of different geometries on the region $0<\sum_{i=1}^{d}|\theta_{i}|^{\nu}< R^{\nu}$ for the cases $\nu=2$ (circle) and $\nu=\nu_{0}<1$ (star) and model size $d=2$.}
\label{MAP_geometry}
\end{figure}

We define the renormalization $C_{\gamma_{d}}$ for the marginal $m_{\gamma_{d}}$ by:
\begin{align}
&C_{\gamma_{d}}\mathdef\int_{\bm{z}\in Y}\frac{(2\pi)^{\frac{d+2}{2}}f(\bm{z}|\tau^{*},\bm{\theta}^{*})\pi(\bm{\theta}^{*}|\lambda^{*})|I_{\tau}|^{-1}|I_{\lambda}|^{-1}}{|\bm{H}(\bm{z},\tau^{*},\bm{\theta}^{*})|^{\frac{1}{2}}|\Psi_{\lambda\lambda}(\bm{\theta}^{*},\lambda^{*})|^{\frac{1}{2}}}\nonumber\\
&\times\left[\prod_{i=1}^{d}P_{G}\left((\tau^{*})^{\frac{1}{2}}|\theta_{i}^{*}|\left\{1+\littleo(\zeta)\right\}
\right)\right]\ d\bm{z}.\label{def_C_gamma_d}
\end{align}
\noindent We have the following result on the marginal normalization $C_{\gamma_{d}}$:\\
\begin{proposition}
Under the conditions given in Theorem \ref{theorem1} and the following additional condition on the number $X(\nu,\lambda,\tau)$ defined below:
\begin{align}
&X(\nu,\lambda,\tau)\mathdef\frac{d}{n-d+2}\frac{2C_{\nu}^{2}\nu^{2}\left(\frac{n}{d}\Omega(\lambda,\tau)\right)^{-h(\nu)}}{1-\zeta}<1,\label{det_H_claim2}
\intertext{where}
&h(\nu)=\left\{\begin{array}{ll}\nu&\text{ if }0<\nu\leq 1\\
\nu/2 & \text{ if } 1<\nu\leq 2.\end{array}\right.\nonumber
\intertext{We then have}
&\log\left(\frac{n-d}{n-d+2}\right)-\frac{d}{2}\zeta-\frac{n-d}{2}\left\{\frac{dC_{\nu}^{2}\nu^{2}\left(\frac{n}{d}\Omega(\lambda^{*},\tau^{*})\right)^{-h(\nu)}}{n-d+2}\right.\nonumber\\
&\left.+\littleo\left(\left[\frac{dC_{\nu}^{2}\nu^{2}\left(\frac{n}{d}\Omega(\lambda^{*},\tau^{*})\right)^{-h(\nu)}}{n-d+2}\right]^{2}\right)\right\}+d\log{P_{G}\left((\tau^{*})^{1/2}\inf_{1\leq i\leq d}|\theta_{i}^{*}|\right)}\nonumber\\
&+\frac{1}{2}\log{(2\pi)}
+\log\int_{\bm{\theta}^{*}\in\Theta^{*},\tau^{*}\in J_{\tau^{*}}}\frac{d\bm{\theta}^{*}}{|I_{\lambda}|}\ \frac{d\tau^{*}}{|I_{\tau}|}\
\frac{\pi(\bm{\theta}^{*}|\lambda^{*})}{|\Psi_{\lambda\lambda}(\bm{\theta}^{*},\lambda^{*})|^{\frac{1}{2}}}\nonumber\\
&\leq\log{C_{\gamma_{d}}}\nonumber\\
&\leq\log\left(\frac{n-d}{n-d+2}\right)+\frac{d}{2}\zeta+\frac{1}{2}\log{(2\pi)}\nonumber\\
&+\log\int_{\bm{\theta}^{*}\in\Theta^{*},\tau^{*}\in J_{\tau^{*}}}\frac{d\bm{\theta}^{*}}{|I_{\lambda}|}\ \frac{d\tau^{*}}{|I_{\tau}|}\ \frac{\pi(\bm{\theta}^{*}|\lambda^{*})}{|\Psi_{\lambda\lambda}(\bm{\theta}^{*},\lambda^{*})|^{\frac{1}{2}}}.\label{log_C_NML_bounds}
\intertext{We note that}
&\zeta\mathdef\sup_{1\leq i\leq d}{C_{\nu}\nu|\nu-1|\left(\frac{n}{d}\Omega(\lambda^{*},\tau^{*})\right)^{-\frac{\nu}{2}}|(\tau^{*})^{\frac{1}{2}}\theta_{i}^{*}|^{\nu-2}}\nonumber
\intertext{and by (\ref{C_form5}) we have in the case of a prior distribution $\pi_{\lambda}(\bm{\theta})$ flat in $\bm{\theta}$ that}
&\log{C_{\gamma_{d}}}=\log\left(\frac{n-d}{n-d+2}\right)+\frac{1}{2}\log{(2\pi)}\nonumber\\
&+\log\int_{\bm{\theta}^{*}\in\Theta^{*},\tau^{*}\in J_{\tau^{*}}}
\frac{\pi(\bm{\theta}^{*}|\lambda^{*})}{|\Psi_{\lambda\lambda}(\bm{\theta}^{*},\lambda^{*})|^{\frac{1}{2}}}\ \frac{d\bm{\theta}^{*}}{|I_{\lambda}|}\ \frac{d\tau^{*}}{|I_{\tau}|}.\label{log_C_NML_bounds_flat_prior}
\end{align}
\label{Marginal_Renormalization_Constant}
\end{proposition}
\begin{proof}
A proof is given in the appendix.
\end{proof}
\subsection{Comments on Proposition \ref{Marginal_Renormalization_Constant}}
\begin{enumerate}
\item We note that using the expression (\ref{explicit_det_Hessian}) and retracing the steps leading up to (\ref{log_C_NML_bounds}), the inequality in 
(\ref{log_C_NML_bounds}) may be sharpened by replacing the term 
$\frac{d}{2}\zeta$ by: $\sum_{i=1}^{d}\mu_{\lambda^{*},\nu}(\tau^{*},\theta_{i}^{*})$ \\
$=C_{\nu}\nu|\nu-1|\left(\frac{n}{d}\Omega(\lambda^{*},\tau^{*})\right)^{-\nu/2}\sum_{i=1}^{d}|(\tau^{*})^{1/2}\theta_{i}^{*}|^{\nu-2}$.
\end{enumerate}

%\section{Finding the best model in a collection of model classes by use of the INMDL Principle}
\section{Discriminating between model classes}
We now proceed to find the best choice of model for our  estimation problem, where best
choice means choosing  the set of of model indices $\gamma^{\ast}_{d^{\ast}}$ yielding the shortest desciption in terms of 
code length in a binary alphabet of {\em both} model and data $\bm{x}$ when encoded under the modelled data generating distibution 
$m_{\gamma_{d}}(\bm{x})/C_{\gamma_{d}}$, that is
\begin{align}
\gamma^{\ast}_{d^{\ast}}\mathdef\text{arg inf}_{1\leq d\leq n,\gamma_{d}\in\{0,1\}^{n}}\left\{ -\log{\left(m_{\gamma_{d}}(\bm{x})/C_{\gamma_{d}}\right)}\right\}.\label{def_gamma_star_d_star}
\end{align} 
\noindent To find the optimal model, we must express the total code length $L(\bm{x},M_{d},\gamma_{d},d)$ needed to encode the data $\bm{x}$ for given model class $M_{d}$, model $\gamma_{d}$ and model size $d$. In previous work presented in \cite{Hansen-Yu:2000}, the process of encoding the data $\bm{x}$, the model index vector $\gamma_{d}$ and any model hyperparameters $\bm{\alpha}$ which are used in defining the model, was decomposed as follows:
\begin{equation}
L(\bm{x},\gamma_{d},\bm{\alpha})\mathdef L(\bm{x}|\bm{\alpha},\gamma_{d})+L(\gamma_{d}|\bm{\alpha})+L(\bm{\alpha}).\label{total_codelength_expression_YuHansen}
\end{equation}
\cite{Hansen-Yu:2000} then proceeded to address the question of how to select a suitable prior distribution for the model index vector $\gamma_{d}$. 
  In \cite{Hansen-Yu:2000} the $\gamma_{d}(i)$ were modelled as IID bernoulli distributed with parameter $p$, and a procedure for 
estimating the hyper-parameter $p$ was provided. However, the authors in \cite{Hansen-Yu:2000} noted that the estimation of the hyper-parameter 
$p$ is non-trivial, and some care had to be taken to avoid too large models. This is an experience we share from our own numerical experimental work as
well: Simply using the marginal formulas (\ref{almost_invariant_marginal}), (\ref{almost_invariant_marginal2}) and optimizing the resulting code length for the marginal distribution over the model size $d$, did in our numerical experiments more often than not lead to a code length expression with no minimum for $d<n/2$ or an optimal model size $d$ so large (comparable to $n/2$) that the stated sufficient conditions under which the asymptotic marginal expressions (\ref{almost_invariant_marginal}), (\ref{almost_invariant_marginal2}) are valid, are not satisfied. This suggests to us that we have been asking for too much in our use of the MDL principle: The extra degree of freedom introduced by the prior distribution $\pi_{\lambda}(\bm{\theta})$ through the parameter $\lambda$  has to be treated with care. However, we do not wish to introduce additional (hyper)parameters into our model classes $M_{d}$, as this will raise the problem of providing reasonable models and estimates for these parameters, which proved difficult to us: The resulting model selection principles and estimators performed poorly in experiments. 
We will therefore adopt a different strategy from that in \cite{Hansen-Yu:2000}. We observe that the model classes $M_{d}$ depend on the choice of prior distribution $\pi(\bm{\theta}|\lambda)$ and that there is (under the conditions and model given here for the data $\bm{x}$) no a priori reason to believe that all prior distributions are (or should be considered) equally likely for the given dataset $\bm{x}$. Therefore the code length measure induced by the marginal density $m_{\gamma_{d}}$  defined in (\ref{def_marginal}) on the collection $\mathcal{M}$ of model classes under consideration should be extended to a code length measure that in some way also quantifies our belief in a particular choice of prior distribution for given dataset $\bm{x}$ and model 
(\ref{noise_model}). Obviously, we cannot compare all possible choices of prior distributions. Also, we suspect the key to solving our problem described above of model overfitting the data, lies in the parameter $\lambda$ which is the only parameter discriminating between different models for given data $\bm{x}$, prior distribution $p$, model index vector $\gamma_{d}$, noise level estimate $\tau^{*}$ and parameter estimates $\bm{\theta}^{*}$. Therefore we will confine ourselves to constructing a measure for comparing our chosen prior distribution $p_{\lambda_{p}}$ with variance $\lambda_{p}^{-1}$ to some chosen reference distribution $q_{\lambda_{q}}$ with variance $\lambda_{q}^{-1}$. 
The distribution $p_{\lambda_{p}}$ is taken to be the best choice of model distribution for the unknown true distribution of $\bm{\theta}$ that we are able to come up with based on our prior knowledge (or our more or less qualified guesses) of the data and the data generating process. The distribution $q_{\lambda_{q}}$ is taken to be some kind of canonical reference prior distribution against which we will compare our choice $p_{\lambda_{p}}$. The problem is then to find a reasonable way to compare $p_{\lambda_{p}}$ and $q_{\lambda_{q}}$. For this we will make use of the entropy $S(p)$ of a distribution $p$, that is the expected (mean) code length for encoding data using $p$. 
Let $\mathcal{P}_{n}$ denote the collection of probability distributions defined on $\mathbb{R}^{n}$, then the entropy 
$S:\mathcal{P}_{n}\longrightarrow\mathbb{R}$ is defined as:
\begin{equation}
S(p)\mathdef E_{\bm{\theta}}\left\{-\log{p(\bm{\theta})}\right\}=-\int_{\bm{\theta}\in\mathbb{R}^{n}}p(\bm{\theta})\log p(\bm{\theta})\ d\bm{\theta}.
\label{def_entropy}
\end{equation} 
Now, we define 
\begin{align}
&\alpha\mathdef\frac{\lambda_{p}}{\lambda_{q}}\label{def_alpha}
\intertext{and}
&D(p_{\lambda_{p}},q_{\lambda_{q}})\mathdef \frac{\exp\left(S(p_{\lambda_{p}})-S(q_{\lambda_{q}})\right)}{\int_{\alpha\in I_{\alpha}}\exp\left(S(p_{\lambda_{p}})-S(q_{\lambda_{q}})\right)\ d\alpha}.\label{def_D}
\end{align}
We will call $D(p_{\lambda_{p}},q_{\lambda_{q}})$ a {\em model class prior distribution}.
We note that if the distributions $p$, $q$ live on the same parameter manifold, it is obvious that $S(p_{\lambda_{p}})-S(q_{\lambda_{q}})$ 
is parameterized by $\alpha$. If $p$ and $q$ live on different parameter manifolds, we may still parameterize  
$S(p_{\lambda_{p}})-S(q_{\lambda_{q}})$ by $\alpha=\lambda_{p}/\lambda_{q}$ if we ensure that $p$ and $q$ are normalized w.r.t 
an integration measure which is invariant to reparameterizations e.g the Fisher information measure. Some care will have to be taken in the choice of 
normalization interval $I_{\alpha}$ in (\ref{def_D}). This question will be further adressed below. The density $D(p,q)$  defined in (\ref{def_D}) may then be used to measure our prior belief in the distribution $p$ relative to the reference distribution $q$.

We proceed to compute $S(p)$ for the distributions of interest to us here, that is the GGD distribution and Jeffreys prior. In the case of a GGD distribution, we have
\begin{align}
&p_{\lambda}(\bm{\theta}) = \left(\frac{\nu\eta(\nu)}{2\Gamma(1/\nu)}\lambda^{1/2}\right)^{d}\exp\left(-\eta(\nu)^{\nu}\sum_{i=1}^{d}|\lambda^{1/2}\theta_{i}|^{\nu}\right)\label{def_p_GGD}\\
&S(p_{\lambda})\mathdef-\int_{\bm{\theta}\in\mathbb{R}^{d}}p_{\lambda}(\bm{\theta})\log{p_{\lambda}}(\bm{\theta})\ d\bm{\theta}\nonumber
\intertext{a trivial computation yields}
&S(p_{\lambda})=-d\log\left(\frac{\nu\eta(\nu)}{2\Gamma(1/\nu)}\lambda^{1/2}\right)+\frac{d}{\nu}.\label{S_GGD}
\intertext{Proceeding with Jeffreys prior distribution}
&q_{R}(\bm{\theta},\tau)=\frac{|\bm{F}(\bm{\theta},\tau)|^{1/2}}{\int_{\tau\in(0,\tau^{*}),\|\bm{\theta}\|_{2}\leq R}|\bm{F}(\bm{\theta},\tau)|^{1/2}d\bm{\theta}\ \frac{d\tau}{\tau}}\label{def_q_Jeffreys}
\intertext{where $\bm{F}$ is the Fisher matrix of the likelihood distribution. We note that the integration measure $\tau^{-1}d\tau$ in 
(\ref{def_q_Jeffreys}) is needed to make the normalization of $q_{R}$ invariant to reparameterizations. 
Using the gaussian distribution for the likelihood, a trivial computation yields}
&|\bm{F}(\bm{\theta},\tau)|^{1/2}=\sqrt{\frac{n}{2}}\tau^{d/2-1}\nonumber
\intertext{and some calculation then yields}
&S(q_{R})=-\int_{\|\bm{\theta}\|_{2}<R,\tau\in(0,\tau^{*})}\sqrt{\frac{n}{2}}\tau^{d/2-1}\log\left(\sqrt{\frac{n}{2}}\tau^{d/2-1}\right)\ d\bm{\theta}\ \frac{d\tau}{\tau}\nonumber\nonumber\\
&=\log\left(\sqrt{\frac{n}{2}}\frac{d-2}{2}\frac{\pi^{d/2}R^{d}}{\Gamma\left(\frac{d}{2}+1\right)}\right)+1-\frac{\frac{1}{2}\log\left(\frac{n}{2}\right)}{\sqrt{\frac{n}{2}}}\label{S_Jeffreys1}
\intertext{using Stirling approximation on the Gamma function yields}
&S(q_{R})=\frac{1}{2}\log\left(\frac{n}{2}\right)+\log\left(\frac{d-2}{2}\right)+1+\frac{d}{2}\log\left(\frac{2\pi eR}{d+2}\right)-\frac{\frac{1}{2}\log\left(\frac{n}{2}\right)}{\sqrt{\frac{n}{2}}}\nonumber\\
&+\log\left(\frac{e}{\sqrt{\pi}}(d+2)^{-1/2}\right)+\bigo(d^{-1}).\label{S_Jeffreys2}
\end{align}
We may now compute the measure $D(p,q)$ for the different $p$ and $q$ of interest to us in the current context. First, we consider the 
case where the reference distribution $q$ is taken to be Jeffreys prior distribution. This choice of reference distribution
may be interpreted as a very pessimistic one, in that this choice of a distribution flat in $\bm{\theta}$ states our complete lack of 
prior knowledge of the noiseless data $\bm{\theta}$, or rather our denial of imposing a more informative prior distribution on $\bm{\theta}$, 
that is a distribution with less entropy as reflected in (\ref{S_Jeffreys2}) where we see that the entropy of the distribution $q_{R}$ is 
up to an additive constant very close to the maximum entropy $d\log\sqrt{2\pi e\lambda^{-1}}$  
\cite{CoverThomas:1991} attained by a $d$-variate Gaussian distribution with variance $\lambda^{-1}=R^{2}/(d+2)$. Choosing a GGD distribution as the candidate for the true prior distribution $p_{\lambda_{p}}$ and setting
\begin{align}
&\lambda_{q}^{-1}\mathdef\frac{1}{d+2}R^{2},\ \alpha\mathdef\frac{\lambda_{p}}{\lambda_{q}},\ I_{\alpha}\mathdef(\alpha_{0},\alpha_{1}),\nonumber
\intertext{expression (\ref{def_D}) then becomes}
&D(p_{\lambda_{p}},q_{\lambda_{q}})=\frac{\exp\left(d\log\left(\frac{2\Gamma(1/\nu_{p})\exp(1/\nu_{p})}{\nu_{p}\eta(\nu_{p})\sqrt{2\pi e}}\right)-\frac{d}{2}\log(\lambda_{p}/\lambda_{q})\right)}{\int_{\alpha_{0}}^{\alpha_{1}}\exp\left(d\log\left(\frac{2\Gamma(1/\nu_{p})\exp(1/\nu_{p})}{\nu_{p}\eta(\nu_{p})\sqrt{2\pi e}}\right)-\frac{d}{2}\log(\lambda_{p}/\lambda_{q})\right)\ d\alpha}\nonumber\\
&=\frac{\alpha^{-d/2}}{\int_{\alpha_{0}}^{\alpha_{1}}\alpha^{-d/2}\ d\alpha}=\frac{d-2}{2}\left(\frac{\alpha_{0}}{\alpha}\right)^{\frac{d}{2}}
\frac{\alpha_{0}^{-1}}{1-\left(\frac{\alpha_{0}}{\alpha_{1}}\right)^{\frac{d}{2}-1}}.\label{D_form1}
\end{align}
We assume that $d\gg 1$, and/or $\alpha_{1}\gg\alpha_{0}$. We conclude that $D(p_{\lambda_{p}},q_{\lambda_{q}})$ for all practical purposes
only depends on the lower bound $\alpha_{0}$, and this dependence is very strong. Therefore, $\alpha_{0}$ has to be chosen carefully.
If we had a discretization $\Delta\alpha>0$ of the parameter $\alpha$, this would suggest a lower bound on our choice of $\alpha_{0}$, 
namely $\alpha_{0}\geq \Delta\alpha$. In lack of any prior information of how to choose $\alpha_{0}$, we settle for the most 
conservative choice $\alpha_{0}=\Delta\alpha$ as this choice of $\alpha_{0}$ will clearly make the code length contribution 
$-\log D(p,q)$ largest possible. By definition (\ref{def_alpha}) we may deduce the following connection between discretizations 
$\Delta\lambda_{p}$, $\Delta\lambda_{q}$ on parameters $\lambda_{p}$, $\lambda_{q}$ and the discretization $\Delta\alpha$, respectively:
\begin{align}
&\Delta\alpha\mathdef\sqrt{\left(\frac{\partial\alpha}{\partial\lambda_{p}}\Delta\lambda_{p}\right)^{2}+\left(\frac{\partial\alpha}{\partial\lambda_{q}}\Delta\lambda_{q}\right)^{2}}\nonumber\\
&=\sqrt{\left(\frac{1}{\lambda_{q}}\Delta\lambda_{p}\right)^{2}+\left(\frac{\lambda_{p}}{\lambda_{q}^{2}}\Delta\lambda_{q}\right)^{2}}
=\frac{\Delta\lambda_{p}}{\lambda_{q}}\sqrt{1+\frac{\lambda_{p}^{2}}{\lambda_{q}^{2}}\left(\frac{\Delta\lambda_{q}}{\Delta\lambda_{p}}\right)^{2}}.
\label{def_Delta_alpha}
\end{align}
Using the result shown in Proposition \ref{INMDL_optimal_quantization} on the posterior coarsest discretization of parameter $\lambda$, together with definitions 
(\ref{def_prior_fisher_matrix})-(\ref{def_Psi}) we get
\begin{equation}
\Delta\lambda_{p}=C_{\lambda_{p}}\frac{2}{\sqrt{\nu_{p}(d+2)}}\lambda_{p}^{*}, \text{ for $p$ GGD distribution in (\ref{def_p_GGD})}\label{Delta_lambda_p}
\end{equation}
\begin{equation}
\Delta\lambda_{q}=C_{\lambda_{q}}\frac{\sqrt{2}}{\sqrt{d+2}}\lambda_{q}^{*},\text{ for $q$ Jeffreys distribution in (\ref{def_q_Jeffreys})}\label{Delta_lambda_q}
\end{equation}
yielding
\begin{align}
&\Delta\alpha=C_{\lambda_{p}}2\alpha^{*}\sqrt{\frac{1+\frac{C_{\lambda_{q}}^{2}}{C_{\lambda_{p}}^{2}}\frac{\nu_{p}}{2}}{\nu_{p}(d+2)}}\label{Delta_alpha}
\intertext{where}
&\alpha^{*}\mathdef\frac{\lambda_{p}^{*}}{\lambda_{q}^{*}}\ \text{ and }\  0 < C_{\lambda_{p}},C_{\lambda_{q}}<1.\label{def_alpha_star}
\intertext{Setting $C_{\lambda_{p}}=C_{\lambda_{q}}$, this leads to}
&D(p_{\lambda_{p}},q_{\lambda_{q}})=\left(\frac{\alpha^{*}}{\alpha}\right)^{\frac{d}{2}}
\left(\frac{4C_{\lambda_{p}}\alpha^{*}}{d-2}\left(\frac{1+\frac{\nu_{p}}{2}}{\nu_{p}(d+2)}\right)^{\frac{1}{2}}\right)^{-1}\left(\frac{4C_{\lambda_{p}}^{2}(1+\frac{\nu_{p}}{2})}{\nu_{p}(d+2)}\right)^{\frac{d}{4}}\label{D_form_GGD_Jeffreys}
\intertext{in the case where the true distribution $p$ is taken to be the GGD distribution in (\ref{def_p_GGD}) and the reference distribution $q$ is taken to be Jeffreys distribution in (\ref{def_q_Jeffreys}). In the case where both $p$ and $q$ are taken to be GGD distributions with shape parameters
$\nu_{p}$ and $\nu_{q}$ respectively, using (\ref{Delta_lambda_p}) we get}
&D(p_{\lambda_{p}},q_{\lambda_{q}})=\left(\frac{\alpha^{*}}{\alpha}\right)^{\frac{d}{2}}
\left(\frac{4C_{\lambda_{p}}\alpha^{*}}{d-2}\left(\frac{1+\frac{\nu_{p}}{\nu_{q}}}{\nu_{p}(d+2)}\right)^{\frac{1}{2}}\right)^{-1}\left(\frac{4C_{\lambda_{p}}^{2}(1+\frac{\nu_{p}}{\nu_{q}})}{\nu_{p}(d+2)}\right)^{\frac{d}{4}}.\label{D_form_GGD_GGD}
\end{align} 
We note that in practice (\ref{D_form_GGD_Jeffreys}) and (\ref{D_form_GGD_GGD}) will be evaluated by plugging in the estimate $\alpha^{*}$ defined in (\ref{def_alpha_star}) for $\alpha$, and so we conclude that to leading order the contribution from the model class prior distribution $D(p,q)$ when the true prior distribution $p$ and reference distribution $q$ both are taken to be GGD, will be:
\begin{align} 
-\log{D(p,q)}\sim -\frac{d}{4}\log\left(\frac{4C_{\lambda_{p}}^{2}(1+\frac{\nu_{p}}{\nu_{q}})}{\nu_{p}(d+2)}\right)\label{leading_log_D_term} 
\end{align}
and when the reference distribution $q$ is taken to be Jeffreys distribution, $\nu_{q}$ is replaced by $\nu_{q}=2$ in 
(\ref{leading_log_D_term}) above. That is, the model class prior distribution $D(p,q)$ does not discriminate between $q$ a gaussian or $q$ a 
Jeffreys prior distribution when the likelihood for the data is gaussian. We note that the number $0<C_{\lambda_{p}}<1$ is connected to an estimate of 
an upper bound on the relative error of the posterior density through the relation (\ref{Delta_posterior}) in Proposition \ref{INMDL_optimal_quantization},
and we see from (\ref{leading_log_D_term}) that the code length contribution from $-\log{D(p,q)}$ will contain 
an additive term $-\frac{d}{2}\log{C_{\lambda_{p}}}$. We then end up with the following process for encoding the data $\bm{x}$
for given model class $M_{d}^{(p)}$, prior distribution $p$, 
reference prior distribution $q$, model size $d$ and model index vector $\gamma_{d}$:
\begin{align}
&L\left(\bm{x},M_{d}^{(p)},\gamma_{d},d\right)=L\left(\bm{x}|M_{d}^{(p)},\gamma_{d},d\right)
+L\left(M_{d}^{(p)}|,\gamma_{d},d\right)+L(\gamma_{d}|d)+L(d)\nonumber\\
&=-\log_{2}{\left(m_{\gamma_{d}}(\bm{x})/C_{\gamma_{d}}\right)}-\log_{2}{D(p,q)}+L(\gamma_{d}|d)+L(d).\label{total_codelength_expression_Fossgaard}
\end{align}
If we have no prior information on the optimal index vector $\gamma_{d}$ and model size $d$, then $L(\gamma_{d}|d)=\log_{2}{2^{n}}$ 
and $L(d)=\log_{2}(n)$ are constants, we will adopt this view here. The optimal model $\gamma_{d^{*}}^{*}$ is then found by computing
\begin{equation}
\gamma^{*}_{d^{*}}\mathdef\text{arg inf}_{0<d<n,\gamma_{d}\in\{0,1\}^{n}}L(\bm{x},M_{d}^{(p)},\gamma_{d},d).\label{best_gamma_d}
\end{equation}

\subsection{Comments on the term $D(p,q)$}
\begin{enumerate}
\item Using the prior distribution $D(p,q)$ on the model classes and choosing both the prior distribution $p$ and the reference distribution $q$
to be Jeffreys distributions with $\lambda_{p}$ independent of $\lambda_{q}$, we will end up with $-\log{D(p,q)}\approx -\frac{d}{4}\log\left(\frac{4C_{\lambda_{p}}^{2}}{d+2}\right)$ to leading order, as can be verified
by following the same steps as we did above for the case $p$ a GGD distribution and $q$ a Jeffreys distribution.
Since our numerical simulations reported in the experiments section below show that the NML code length principle of 
Rissanen \cite{Rissanen:2000} does not work well at all 
on datasets with signal to noise ratios below some level, whereas our INMDL principle in our simulations is seen to work as well as or better than NML-principle over broad regimes of SNR, it is tempting to suggest that when applied to denoising problems, the NML code length should be modified by adding an extra term of $-\frac{d}{4}\log\left(\frac{4C_{\lambda_{p}}^{2}}{d+2}\right)$. We think that more work in this area is needed as our reported numerical experiments seem to indicate that 
our suggested code length term of $-\log{D(p,q)}$ does not yield optimal model sizes, particularly not for the very small and the 
very high SNR values.
\item We emphazise that in the case where we choose $p\equiv q$ for prior distribution $p$ and reference distribution $q$, 
we have $\alpha\equiv 1$ and $D(p,q)$ becomes a constant. 
\end{enumerate}

%\section{Selecting the model for a given dataset and marginal density by the invariant NMDL Principle}
\section{Model selection by the INMDL Principle}
We must address in detail the question of how to actually find the optimal set $\gamma_{d}$ of model indices when given data $\bm{x}$.
Equation (\ref{best_gamma_d}) tells us
\begin{align}
&\gamma^{\ast}_{d^{\ast}}\mathdef\text{arg inf}_{0<d<n,\gamma_{d}\in\{0,1\}^{n}}L(\bm{x},M^{(\pi)}_{d},\gamma_{d},d).
\intertext{where $L(\bm{x},M_{d}^{(\pi)},\gamma_{d},d)$ is defined in (\ref{total_codelength_expression_Fossgaard}). Using formula (\ref{almost_invariant_marginal2}) together with (\ref{total_codelength_expression_Fossgaard}),(\ref{best_gamma_d}) we see we have to solve}
&\gamma_{d^{*}}^{*}=\text{arg inf}_{0<d<n,\gamma_{d}\in\{0,1\}^{n}}L(\bm{x},M_{d}^{(\pi)},\gamma_{d},d)\nonumber\\
&=\text{arg inf}_{0<d<n,\gamma_{d}\in\{0,1\}^{n}}\left\{L(M_{d}^{(\pi)}|\gamma_{d},d)\right.\nonumber\\
&\left.-\log \left[f(\bm{x}|\tau^{*},\bm{\theta}^{*})\right]_{v}-\log\left[\pi(\bm{\theta}^{*}|\lambda^{*})\right]_{v}-\log(2\pi)^{\frac{d+2}{2}}\right.\nonumber\\
&\left.+\log\left[|\bm{H}(\bm{x},\tau^{*},\bm{\theta}^{*})|^{\frac{1}{2}}\right]_{v}-\sum_{i=1}^{d}\log P_{G}\left((\tau^{*})^{\frac{1}{2}}|\theta_{i}^{*}|\left\{1+\littleo(\zeta)\right\}^{\frac{1}{2}}\right)\right.\nonumber\\
&\left.+\log\left[|\Psi_{\lambda\lambda}(\bm{\theta}^{*},\lambda^{*})|^{\frac{1}{2}}\right]_{v}+\log{C_{\gamma_{d}}}+\log{\left[|I_{\tau}|\cdot|I_{\lambda}|\right]_{v}}\right\}.\label{gamma_form1}
\intertext{Inserting the expressions (\ref{invariant_noise_estimator}), (\ref{explicit_det_Hessian2}) yields the expression}
&L(\bm{x},M_{d}^{(\pi)}\gamma_{d},d)=-\log D(\pi,q)+\log{C_{\gamma_{d}}}+\frac{n-d+2}{2}-\log\left[\pi(\bm{\theta}^{*}|\lambda^{*})\right]_{v}\nonumber\\
&-\frac{3}{2}\log\left(2\pi\right)-\frac{1}{2}\log{2}-\frac{n-d+1}{2}\log\left(\frac{n-d+2}{2\pi}\right)+\frac{d}{2}\littleo(\zeta)\nonumber\\
&+\log{\left[|I_{\tau}|\cdot|I_{\lambda}|\right]_{v}}+\frac{n-d+2}{2}\log\left[\|\bm{x}_{\perp}\|_{2}^{2}+\|\bm{x}_{\parallel}-\bm{\theta}^{*}\|_{2}^{2}\right]_{v}\nonumber\\
&+\log\left[|\Psi_{\lambda\lambda}(\bm{\theta}^{*},\lambda^{*})|^{\frac{1}{2}}\right]_{v}-\sum_{i=1}^{d}\log P_{G}\left((\tau^{*})^{\frac{1}{2}}|\theta_{i}^{*}|\left\{1+\littleo(\zeta)\right\}^{\frac{1}{2}}\right)\nonumber\\
&+\frac{1}{2}\log\left(1-\frac{2}{1+\littleo(\zeta)}\left\{1+\frac{\|\bm{x}_{\perp}\|_{2}^{2}}{\|\bm{x}_{\parallel}-\bm{\theta}^{*}\|_{2}^{2}}\right\}^{-1}\right).\label{gamma_form2}
%\intertext{We may find it useful to express (\ref{gamma_form1}) on the following alternative form. Using (\ref{def_Upsilon}) and (\ref{def_SNR}) we may write}
%&L(\bm{x},M_{d}^{(\pi)},\gamma_{d},d)=-\log D(\pi,q)+\log{C_{\gamma_{d}}}+\frac{n-2}{2}\log(2\pi)-\frac{1}{2}\log{2}\nonumber\\
%&+\frac{n-d+2}{2}+\sum_{i=1}^{d}f((\lambda^{*})^{\frac{1}{2}}\theta_{i}^{*})-d\log{K_{\pi_{\lambda}}}+\frac{d}{2}\log\left(\frac{n}{2\pi}\right)+\frac{d}{2}\log\left(\frac{\Omega(\lambda^{*},\tau^{*})}{d}\right)\nonumber\\
%&+\frac{d}{2}\littleo(\zeta)+\frac{n+2}{2}\log\left[\|\bm{x}_{\perp}\|_{2}^{2}+\|\bm{x}_{\parallel}-\bm{\theta}^{*}\|_{2}^{2}\right]_{v}-\frac{n+1}{2}\log(n-d+2)\nonumber\\
%&-\sum_{i=1}^{d}\log P_{G}\left((\tau^{*})^{\frac{1}{2}}|\theta_{i}^{*}|\left\{1+\littleo(\zeta)\right\}^{\frac{1}{2}}\right)+\log{\left[|I_{\tau}|\cdot|I_{\lambda}|\right]_{v}}\nonumber\\
%&+\log\left[|\Psi_{\lambda\lambda}(\bm{\theta}^{*},\lambda^{*})|^{\frac{1}{2}}\right]_{v}+\frac{1}{2}\log\left(1-\frac{2}{1+\littleo(\zeta)}\left\{1+\frac{\|\bm{x}_{\perp}\|_{2}^{2}}{\|\bm{x}_{\parallel}-\bm{\theta}^{*}\|_{2}^{2}}\right\}^{-1}\right),\label{gamma_form3}
%\intertext{where $K_{\pi_{\lambda}}$ is a normalization constant independent of both $\lambda$ and $d$, and $f$ is a symmetric function 
%with the properties stated in the proof of Theorem \ref{theorem1}. }
\intertext{Using the result in Proposition \ref{Marginal_Renormalization_Constant} we may write}
&L(\bm{x},M_{d}^{(\pi)},\gamma_{d},d)=Q\left(\bm{x},M_{d}^{(\pi)},\gamma_{d},d\right)+
Z\left(\bm{x},M_{d}^{(\pi)},\gamma_{d},d\right)\label{split_codelength_into_hairy_and_nice_parts}
\intertext{where}
&Z\left(\bm{x},M_{d}^{(\pi)},\gamma_{d},d\right)\mathdef\log{C_{\gamma_{d}}}-\frac{3}{2}\log{(2\pi)}-\frac{1}{2}\log{2}\nonumber\\
&-\log\int_{\bm{\theta}^{*}\in\Theta^{*},\tau^{*}\in J_{\tau^{*}}}\frac{\pi(\bm{\theta}^{*}|\lambda^{*})}{|\Psi_{\lambda\lambda}(\bm{\theta}^{*},\lambda^{*})|^{1/2}}\ \frac{d\bm{\theta}^{*}}{|I_{\lambda}|}\ \frac{d\tau^{*}}{|I_{\tau}|}\nonumber\\
&+\frac{n-d+2}{2}\log\left(1+\frac{\|\bm{x}_{\parallel}-\bm{\theta}^{*}\|_{2}^{2}}{\|\bm{x}_{\perp}\|_{2}^{2}}\right)
-\sum_{i=1}^{d}\log P_{G}\left((\tau^{*})^{\frac{1}{2}}|\theta_{i}^{*}|\left\{1+\littleo(\zeta)\right\}^{\frac{1}{2}}\right)\nonumber\\
&+\frac{1}{2}\log\left(1-\frac{2}{1+\littleo(\zeta)}\left\{1+\frac{\|\bm{x}_{\perp}\|_{2}^{2}}{\|\bm{x}_{\parallel}-\bm{\theta}^{*}\|_{2}^{2}}\right\}^{-1}\right)+\frac{d}{2}\littleo(\zeta)\label{hairy_codelength_part}
\intertext{and}
&Q\left(\bm{x},M_{d}^{(\pi)},\gamma_{d},d\right)\mathdef -\log D(\pi,q)+\frac{n-d+2}{2}+\log\left[|I_{\tau}||I_{\lambda}|\right]_{v}\nonumber\\
&-\frac{n-d+1}{2}\log\left(\frac{n-d+2}{2\pi}\right)+\frac{n-d+2}{2}\log\left[\|\bm{x}_{\perp}\|_{2}^{2}\right]_{v}\nonumber\\
&-\log\left[\left.\frac{\pi(\bm{\theta}^{*}|\lambda^{*})}{|\Psi_{\lambda\lambda}(\bm{\theta}^{*},\lambda^{*})|^{\frac{1}{2}}}\right/\int_{\bm{\theta}^{*}\in\Theta^{*},\tau^{*}\in J_{\tau^{*}}}
\frac{\pi(\bm{\theta}^{*}|\lambda^{*})}{|\Psi_{\lambda\lambda}(\bm{\theta}^{*},\lambda^{*})|^{\frac{1}{2}}}\ \frac{d\bm{\theta}^{*}}{|I_{\lambda}|}\ \frac{d\tau^{*}}{|I_{\tau}|}\right]_{v}.\label{def_nice_codelength_part}
\end{align}
We have the following result:
\begin{proposition}
Assume the conditions stated in Theorem \ref{theorem1} and let $q$ denote the chosen reference prior distribution.
Then, for given data $\bm{x}$ and prior distribution $\pi$, the optimal model class $M_{d^{*}}^{(\pi)}$ and model $\gamma_{d^{*}}^{*}$, is up to a code length precision of size $\Delta_{d}(\bm{x})$, selected as follows: Let
$S_{d}\subset A_{n}\mathdef\{1,2,3,...,n-1,n\}$ be of size $d\leq n$ and define $\gamma_{d}(i)=1$ if $i\in S_{d}$ and $\gamma_{d}(i)=0$ otherwise. The sets $S_{j}\mathdef \{l\}\cup S_{j-1}$, where $l\in A_{n}\setminus S_{j-1}$, 
are computed iteratively by minimizing the criterion $C(\bm{x}_{\parallel}(j)|S_{j-1})$ for each index $j:\ 1\leq j\leq d\leq n$ over the set of 
indices $l\in A_{n}\setminus S_{j-1}$ by putting $\bm{x}_{\parallel}(j)\mathdef\bm{x}(l)$ and defining
\begin{align}
&C(\bm{x}_{\parallel}(j)|S_{j-1})\mathdef -(n-d+2)\frac{|\bm{x}_{\parallel}(j)|}{\|\bm{x}\|_{2}^{2}-\|\bm{x}_{\parallel}\|_{2}^{2}}\nonumber\\
&-\frac{\partial}{\partial|\bm{x}_{\parallel}(j)|}\log\left[\left.\frac{\pi(\bm{\theta}^{*}|\lambda^{*})}{|\Psi_{\lambda\lambda}(\bm{\theta}^{*},\lambda^{*})|^{\frac{1}{2}}}\right/\int_{\bm{\theta}^{*}\in\Theta^{*},\tau^{*}\in J_{\tau}^{*}}\frac{\pi(\bm{\theta}^{*}|\lambda^{*})}{|\Psi_{\lambda\lambda}(\bm{\theta}^{*},\lambda^{*})|^{\frac{1}{2}}}\ \frac{d\bm{\theta}^{*}}{|I_{\lambda}|}\ \frac{d\tau^{*}}{|I_{\tau}|}\right]_{v},\label{def_model_selection_selection_criterion}\\
\intertext{where $\bm{x}_{\parallel}$, $\bm{\theta}^{*}$ and $\lambda^{*}$ are given by the model defined by the set $S_{j}\mathdef$$S_{j-1}\cup\{l\}$. }
&\Delta_{d}(\bm{x})\mathdef\log{C_{\gamma_{d}}}-\log\int_{\bm{\theta}^{*}\in\Theta^{*},\tau^{*}\in J_{\tau^{*}}}\frac{d\bm{\theta}^{*}}{|I_{\lambda}|}\ \frac{d\tau^{*}}{|I_{\tau}|}\ 
\frac{\pi(\bm{\theta}^{*}|\lambda^{*})}{|\Psi_{\lambda\lambda}(\bm{\theta}^{*},\lambda^{*})|^{\frac{1}{2}}}\nonumber\\
&+\frac{n-d+2}{2}\log\left(1+\frac{\|\bm{x}_{\parallel}-\bm{\theta}^{*}\|_{2}^{2}}{\|\bm{x}_{\perp}\|_{2}^{2}}\right)+\frac{d}{2}\littleo(\zeta)
-\frac{3}{2}\log{(2\pi)}-\frac{1}{2}\log{2}\nonumber\\
&+\frac{1}{2}\log\left(1-\frac{2}{1+\littleo(\zeta)}\left\{1+\frac{\|\bm{x}_{\perp}\|_{2}^{2}}{\|\bm{x}_{\parallel}-\bm{\theta}^{*}\|_{2}^{2}}\right\}^{-1}\right)\nonumber\\
&-\sum_{i=1}^{d}\log P_{G}\left((\tau^{*})^{\frac{1}{2}}|\theta_{i}^{*}|\left\{1+\littleo(\zeta)\right\}^{\frac{1}{2}}\right)\label{def_precision_simplified_codelength}
\intertext{and the value of $d^{\ast}$ is determined by minimizing the code length expression} 
&Q\left(\bm{x},M_{d}^{(\pi)},\gamma_{d},d\right)\mathdef -\log D(\pi,q)+\frac{n-d+2}{2}+\log\left[|I_{\tau}||I_{\lambda}|\right]_{v}\nonumber\\
&-\frac{n-d+1}{2}\log\left(\frac{n-d+2}{2\pi}\right)+\frac{n-d+2}{2}\log\left[\|\bm{x}_{\perp}\|_{2}^{2}\right]_{v}\nonumber\\
&-\log\left[\left.\frac{\pi(\bm{\theta}^{*}|\lambda^{*})}{|\Psi_{\lambda\lambda}(\bm{\theta}^{*},\lambda^{*})|^{\frac{1}{2}}}\right/\int_{\bm{\theta}^{*}\in\Theta^{*},\tau^{*}\in J_{\tau}^{*}}\frac{\pi(\bm{\theta}^{*}|\lambda^{*})}{|\Psi_{\lambda\lambda}(\bm{\theta}^{*},\lambda^{*})|^{\frac{1}{2}}}\ \frac{d\bm{\theta}^{*}}{|I_{\lambda}|}\ \frac{d\tau^{*}}{|I_{\tau}|}\right]_{v}\label{def_nice_codelength_part2}
\intertext{with respect to $d$, and $D(\pi,q)$ is given by (\ref{D_form_GGD_Jeffreys}) or (\ref{D_form_GGD_GGD}) and the total code length expression $L\left(\bm{x},M_{d}^{(\pi)}\gamma_{d},d\right)$ is given by}
&L\left(\bm{x},M_{d}^{(\pi)},\gamma_{d},d\right)=Q\left(\bm{x},M_{d}^{(\pi)},\gamma_{d},d\right)+\Delta_{d}(\bm{x}).\nonumber
\end{align}
\label{INMDL_Selection_Principle}
\end{proposition}
\begin{proof}
Under the given conditions the optimality of the selection algorithm defined by minimizing the criterion $C(\cdot)$ in 
(\ref{def_model_selection_selection_criterion}) follows by differentiating the expression (\ref{def_nice_codelength_part}). Then observe that the total code length expression is given by (\ref{split_codelength_into_hairy_and_nice_parts}). The result follows by recognizing that $\Delta_{d}=Z\left(\bm{x},M_{d}^{(\pi)},\gamma_{d},d\right)$.
\end{proof}
\subsection{Comments on Proposition \ref{INMDL_Selection_Principle}}
\begin{enumerate}
\item In the case we will be concentrating on: $\pi_{\lambda}$ is a GGD and $\lambda^{*}$ is the ML estimator the criterion 
$C(\cdot)$ in (\ref{def_model_selection_selection_criterion}) becomes
\begin{align}
&C(\bm{x}_{\parallel}(j)|S_{j-1})\mathdef -(n-d+2)\frac{|\bm{x}_{\parallel}(j)|}{\|\bm{x}\|_{2}^{2}-\|\bm{x}_{\parallel}\|_{2}^{2}}\nonumber\\
&+\frac{d-2}{\sum_{i=1}^{d}|\theta_{i}^{*}|^{\nu}}\frac{\partial|\theta_{j}^{*}|}{\partial|\bm{x}_{\parallel}(j)|}.\nonumber
\end{align}
Thus, the model selection process in this case may be implemented by a quicksort procedure.
\item Using the bounds in (\ref{log_C_NML_bounds}), (\ref{log_C_NML_bounds_flat_prior}) on $\log{C_{\gamma_{d}}}$ we may easily compute bounds on $\Delta_{d}$.
\item We note that if $\frac{\|\bm{x}_{\parallel}-\bm{\theta}^{*}\|_{2}^{2}}{\|\bm{x}_{\perp}\|_{2}^{2}}<1$, we have (by using the Taylor expansions centered in $y=0$ of $\log(1\pm y),\ 0\leq y<1$)
\begin{align}
&\log{C_{\gamma_{d}}}+\frac{n-d}{2}\frac{\|\bm{x}_{\parallel}-\bm{\theta}^{*}\|_{2}^{2}}{\|\bm{x}_{\perp}\|_{2}^{2}}-\frac{d}{2}\zeta
-\frac{3}{2}\log{(2\pi)}-\frac{1}{2}\log{2}\nonumber\\
&-\log\int_{\bm{\theta}^{*}\in\Theta^{*},\tau^{*}\in J_{\tau^{*}}}\frac{\pi(\bm{\theta}^{*}|\lambda^{*})}{|\Psi_{\lambda\lambda}(\bm{\theta}^{*},\lambda^{*})|^{\frac{1}{2}}}\ \frac{d\bm{\theta}^{*}}{|I_{\lambda}|}\ \frac{d\tau^{*}}{|I_{\tau}|}\nonumber\\
&\leq\Delta_{d}\nonumber\\
&\leq\log{C_{\gamma_{d}}}+\frac{n-d+14}{2}\frac{\|\bm{x}_{\parallel}-\bm{\theta}^{*}\|_{2}^{2}}{\|\bm{x}_{\perp}\|_{2}^{2}}+\frac{d}{2}\zeta\nonumber\\
&-d\log{P_{G}\left(\tau(\bm{x})^{\frac{1}{2}}\inf_{i\in\gamma_{d}}|\theta_{i}^{*}|\right)}-\frac{3}{2}\log{(2\pi)}-\frac{1}{2}\log{2}\nonumber\\
&-\log\int_{\bm{\theta}^{*}\in\Theta^{*},\tau^{*}\in J_{\tau^{*}}}\frac{\pi(\bm{\theta}^{*}|\lambda^{*})}{|\Psi_{\lambda\lambda}(\bm{\theta}^{*},\lambda^{*})|^{\frac{1}{2}}}\ \frac{d\bm{\theta}^{*}}{|I_{\lambda}|}\ \frac{d\tau^{*}}{|I_{\tau}|}.\label{Delta_bounds}
%\intertext{and we note that if the prior distribution is flat, $\|\bm{x}_{\parallel}-\bm{\theta}^{*}\|_{2}^{2}=0$ and $\zeta=0$ in (\ref{Delta_bounds}) thus implying $\Delta_{d}=0$.}\nonumber
\end{align}
\end{enumerate}

%\section{Comparison of the INMDL\--codelength to the NML\--code\-length for gaussian likelihood functions}
\section{The INMDL- versus NML-principle for gaussian likelihood}
It may be of interest to know how the code length-principle we have developed in the previous sections defers from the code length principle developed
by Rissanen in \cite{Rissanen:1998b} and \cite{Rissanen:1996} in the special case of a gaussian likelihood function. Starting out from the expression in (\ref{almost_invariant_marginal}), recalling the initial definitions of prior distributions on the parameters in (\ref{prior}) and (\ref{uniform_tau_prior}) we define the joint prior distribution $\pi_{\lambda}(\bm{\theta})\varsigma(\tau)$ as follows
\begin{align}
&\pi_{\lambda}(\bm{\theta})\varsigma(\tau)\mathdef\frac{\left|\bm{F}(\bm{\theta},\tau)\right|^{\frac{1}{2}}}{\int_{\bm{\theta}\in\Theta^{*},\tau\in I_{\tau}}\left|\bm{F}(\bm{\theta},\tau)\right|^{\frac{1}{2}}\ d\bm{\theta}\ d\tau}\label{Jeffreys_joint_prior}
\end{align}
Now, it is easy to verify that this choice of Jeffreys prior in (\ref{Jeffreys_joint_prior}) as a joint prior distribution satisfies the conditions on the prior $\pi_{\lambda}(\theta)$ stated in Theorem \ref{theorem1} when the likelihood function $f$ is gaussian. However, the Theorem \ref{theorem1} was deduced under a flat prior distribution on the parameter $\tau$. But by the proof of Theorem \ref{theorem1} we see that the special choice of joint prior distribution $\varsigma(\tau)\pi_{\lambda}(\bm{\theta})$ in (\ref{Jeffreys_joint_prior}) transforms under the chosen reparameterizations given in the proof of Theorem \ref{theorem1} to a constant and therefore does only contribute as a constant in any of the integrals discussed in the proof of Theorem \ref{theorem1}. Because of this fact, the formula (\ref{almost_invariant_marginal}) is still valid if we replace $\pi_{\lambda}(\bm{\theta}^{*})|I_{\tau}|^{-1}$  in (\ref{almost_invariant_marginal}) by the expression for $\pi_{\lambda}(\bm{\theta}^{*})\varsigma(\tau^{*})$ given in (\ref{Jeffreys_joint_prior}).
We observe that the prior distribution in (\ref{Jeffreys_joint_prior}) in the case of a gaussian likelihood is smooth in $\bm{\theta}$ and therefore the 
term $\prod_{i=1}^{d}P_{G}\left((\tau^{*})^{\frac{1}{2}}\theta_{i}^{*}\right)$ in (\ref{almost_invariant_marginal}) is to be replaced by $1$. By 
Corollary \ref{corollary_new_marginal} the code length defined in (\ref{total_codelength_expression_Fossgaard}) now becomes
\begin{align}
&L\left(\bm{x},M_{d}^{(\pi)},\gamma_{d},d\right)=-\log D(\pi,q)+\log{C_{\gamma_{d}}}-\log\left[f(\bm{x}|\bm{\theta}^{*},\tau^{*})\right]_{v}\nonumber\\
&-\log\left(2\pi\right)^{\frac{d+1}{2}}+\frac{1}{2}\log\left(\frac{\left|\bm{H}(\bm{x},\tau^{*},\bm{\theta}^{*})\right|}{\left|\bm{F}(\bm{\theta}^{*},\tau^{*})\right|}\right)+\log\left(|\Psi_{\lambda\lambda}(\bm{\theta}^{*},\lambda^{*})|^{1/2}|I_{\lambda}|\right)\nonumber\\
&+\log\left(\int_{\bm{\theta}\in\Theta^{*},\tau\in I_{\tau}}\left|\bm{F}(\bm{\theta},\tau)\right|^{\frac{1}{2}}\ d\bm{\theta}\ d\tau\right).\label{L_form}
\intertext{Now, by the fact that $|\Psi_{\lambda\lambda}(\bm{\theta}^{*},\lambda^{*})|^{1/2}$ $=\sqrt{(d+2)/2}/\lambda^{*}$ is a constant (independent of $\bm{\theta^{*}}$ with $1/\lambda^{*}=(d+2)^{-1}\|\bm{x}_{\parallel}\|_{2}^{2}$) and may therefore 
be taken out of the integral in (\ref{C_form5}) we get}
&\log{C_{\gamma_{d}}}=\log\left(\frac{n-d}{n-d+2}\right)+\frac{1}{2}\log{(2\pi)}-\log\left(|\Psi_{\lambda\lambda}(\bm{\theta}^{*},\lambda^{*})|^{1/2}|I_{\lambda}|\right).\label{log_C_gamma_d}
\intertext{Furthermore, in this case we have that $\pi\equiv q=$ Jeffreys distribution and so the parameter $\alpha$ defined in (\ref{def_alpha}) is identically $1$, i.e deterministic, and therefore the discretization $\Delta\alpha=0$. By (\ref{def_D}) we get 
$D(p,q)=\exp(0)/\int_{0}^{\alpha_{1}}\exp(0)\ d\alpha=\alpha_{1}^{-1}$ which is a constant, and may therefore be omitted from 
the code length expression. 
Letting $L^{\prime}\left(\bm{x},\gamma_{d}\right)$ denote the NML code length as developed in \cite{Rissanen:1996} we have}
&L^{\prime}\left(\bm{x},\gamma_{d}\right)=-\log\left[f(\bm{x}|\bm{\beta}^{*})\right]_{v}+\frac{d+1}{2}\log\left(\frac{n}{2\pi}\right)\nonumber\\
&+\log\left(\int_{\bm{\beta}\in\Theta^{*}\times I_{\tau}}\left|\bm{I}(\bm{\beta})\right|^{\frac{1}{2}}\ d\bm{\beta}\right)+\littleo(1)\label{L_prime_form}
\intertext{where}
&\bm{I}(\bm{\beta})\mathdef-n^{-1}E_{\bm{x}}\left\{\frac{\partial^{2}\log{f(\bm{x}|\bm{\beta})}}{\partial\beta_{i}\partial\beta_{j}}\right\},\ 1\leq i,j\leq n,\ \bm{\beta}\mathdef(\tau,\bm{\theta}^{T})^{T}\in\mathbb{R}^{d+1}\label{Rissanen_Fisher}
\intertext{and thus (\ref{L_prime_form}) may be rewritten as}
&L^{\prime}(\bm{x},\gamma_{d})=-\log\left[f(\bm{x}|\bm{\theta}^{*},\tau^{*})\right]_{v}-\log\left(2\pi\right)^{\frac{d+1}{2}}\nonumber\\
&+\log\left(\int_{\bm{\theta}\in\Theta^{*},\tau\in I_{\tau}}\left|\bm{F}(\bm{\theta},\tau)\right|^{\frac{1}{2}}\ d\bm{\theta}\ d\tau\right)+\littleo(1).\label{L_prime_form2}
\intertext{By (\ref{L_form}), (\ref{log_C_gamma_d}) and (\ref{L_prime_form2}) and observing that $\bm{H}(\bm{x},\tau^{*},\bm{\theta}^{*})$$=\bm{F}(\bm{\theta}^{*},\tau^{*})$ we conclude}
&L\left(\bm{x},M_{d}^{(\pi)},\gamma_{d},d\right)-L^{\prime}\left(\bm{x},\gamma_{d}\right)=\log\left(\frac{n-d}{n-d+2}\right)+\text{ constants }+\littleo(1).\label{diff_L_and_L_prime}
\end{align}
We note that the difference $L\left(\bm{x},M_{d}^{(\pi)},\gamma_{d},d\right)-L^{\prime}\left(\bm{x},\gamma_{d}\right)$$=\Delta_{d}(\bm{x})+\littleo(1)$, where $\Delta_{d}(\bm{x})$ is the code length precision in our INMDL model selection principle as given in Proposition \ref{INMDL_Selection_Principle}. Thus, in model selection, we may expect our INMDL principle to yield results very close to the NML-principle of Rissanen \cite{Rissanen:1996}, \cite{Rissanen:2000} in the case of a gaussian likelihood function $f(x|\theta,\tau)$ and Jeffreys prior (\ref{Jeffreys_joint_prior}) as a joint prior distribution $\pi_{\lambda}(\theta)\varsigma(\tau)$. We summarize our findings:\\ 

\begin{corollary}
Assume the conditions given in Theorem \ref{theorem1} and Proposition \ref{Marginal_Renormalization_Constant} and Proposition \ref{INMDL_Selection_Principle}. Let $L\left(\bm{x},M_{d}^{(\pi)},\gamma_{d},d\right)$ denote the code length as presented in Proposition \ref{INMDL_Selection_Principle} and let $L^{\prime}\left(\bm{x},\gamma_{d}\right)$ denote the NML code length as developed in \cite{Rissanen:1996}. Then 
\begin{align}
&L\left(\bm{x},M_{d}^{(\pi)},\gamma_{d},d\right)-L^{\prime}\left(\bm{x},\gamma_{d}\right)=\log\left(\frac{n-d}{n-d+2}\right)+\text{ constants }+\littleo(1).\label{diff_our_and_Rissanen_codelengths}
\end{align}
\end{corollary}
\begin{proof}
See discussion above.
\end{proof}

%\section{Invariant Laplace approximation of the posterior mean of parameters}
\section{The posterior mean of parameters}
Given the model $\gamma_{d}$ we want to compute the posterior means $\bm{\theta}^{\sharp}$ and $\tau^{\sharp}$ 
of the parameters $\bm{\theta}$ and $\tau$, that is 
\begin{align}
&\bm{\theta}^{\sharp}\mathdef E_{\bm{\theta},\tau}\left\{\bm{\theta}\right\}\label{def_posterior_theta}\\
&\tau^{\sharp}\mathdef E_{\bm{\theta},\tau}\left\{\tau\right\}\label{def_posterior_tau}
\intertext{The posterior $p_{\gamma_{d}}(\bm{\theta},\tau|\bm{x})$ is defined by Bayes rule:}
&p_{\gamma_{d}}(\tau,\bm{\theta}|\bm{x})\mathdef\frac{1}{m_{\gamma_{d}}(\bm{x})}f(\bm{x}|\tau,\bm{\theta})\pi_{\lambda}(\bm{\theta}).\label{def_posterior}
\intertext{Let $(\tau,\bm{\theta}^{T})^{T}= \bm{\Upsilon}((\hat{\tau},\hat{\bm{\theta}}^{T})^{T})$ 
denote the reparameterization induced by the mappings $\phi:(\hat{\tau},\hat{\theta})\rightarrow\theta$ and 
$\psi:\hat{\tau}\rightarrow\tau$ given in (\ref{def_phi_psi}) and (\ref{form_of_psi}). Let $\bm{J}_{\bm{\Upsilon}}$
be the jacobian of $\bm{\Upsilon}$ as computed in (\ref{form_of_J}), let $\bm{\Gamma}_{\bm{\Upsilon}}$ be the 3-tensor of second derivatives of $\bm{\Upsilon}$
and for notational simplicity define $\bm{\beta}\mathdef(\tau,\bm{\theta}^{T})^{T}$, 
$\hat{\bm{\beta}}\mathdef(\hat{\tau},\hat{\bm{\theta}}^{T})^{T}$. Then we may write to leading order}
&\bm{\beta} = \bm{\Upsilon}(\hat{\bm{\beta}}^{*})+\bm{J}_{\bm{\Upsilon}}(\hat{\bm{\beta}}^{*})(\hat{\bm{\beta}}-\hat{\bm{\beta}}^{*})+
\bm{\Gamma}(\hat{\bm{\beta}}^{*})(\hat{\bm{\beta}}-\hat{\bm{\beta}}^{*})(\hat{\bm{\beta}}-\hat{\bm{\beta}}^{*})^{T}\label{PSI_expansion}
\intertext{and thus}
&\bm{\beta}^{\sharp}=\bm{\beta}^{*}+\bm{J}_{\bm{\Upsilon}}(\hat{\bm{\beta}}^{*})\int_{\hat{\bm{\beta}}\in\hat{I}_{\hat{\tau}}\times\mathbb{R}^{d}}(\hat{\bm{\beta}}-\hat{\bm{\beta}}^{*})\frac{\exp\left(-\hat{\Phi}(\bm{x},\hat{\bm{\beta}})\right)}{m_{\gamma_{d}}(\bm{x})}|\hat{\bm{F}}(\hat{\bm{\beta}})|^{1/2}\ d\hat{\bm{\beta}}\nonumber\\
&+\bm{\Gamma}_{\bm{\Upsilon}}(\hat{\bm{\beta}}^{*})\int_{\hat{\bm{\beta}}\in\hat{I}_{\hat{\tau}}\times\mathbb{R}^{d}}(\hat{\bm{\beta}}-\hat{\bm{\beta}}^{*})(\hat{\bm{\beta}}-\hat{\bm{\beta}}^{*})^{T}\frac{\exp\left(-\hat{\Phi}(\bm{x},\hat{\bm{\beta}})\right)}{m_{\gamma_{d}}(\bm{x})}|\hat{\bm{F}}(\hat{\bm{\beta}})|^{1/2}\ d\hat{\bm{\beta}}.\label{expanded_posterior_means}
\end{align}
Proceeding as we did in the proof of Theorem \ref{theorem1} by claiming $\hat{\bm{\theta}}^{*}\in\mathbb{R}_{+}^{d}$ and 
splitting up the integrals 
in (\ref{expanded_posterior_means}) into integration over the two disjoint domains $\mathbb{R}_{+}^{d}$, $\mathbb{R}_{-}^{d}$ 
in the $\hat{\bm{\theta}}$-variable, the following result is a straightforward consequence of the proof of Theorem \ref{theorem1} together with
the observation (consider $\bm{J}_{\bm{\Upsilon}}$ in (\ref{form_of_J})) 
that the integrated contributions from the $\bm{\Gamma}_{\bm{\Upsilon}}$-part of the expansion 
for this particular $\bm{\Upsilon}$ does not contribute to leading order and may be neglected:

\begin{corollary}
(Corollary of the proof of Theorem \ref{theorem1}). The posterior bias of the estimators $\bm{\theta}^{*}$ and $\tau^{*}$ under the conditions in Theorem \ref{theorem1} may be written on the form:
\begin{align}
&E_{\bm{\theta},\tau}\left\{\theta_{i}\right\}=\theta_{i}^{*}+(\tau^{*})^{-\frac{1}{2}}\frac{(2\pi)^{-\frac{1}{2}}}{6}\frac{C_{\nu}\nu(\nu-1)(\nu-2)}{\left(\frac{n}{d}\Omega\left(\lambda,\tau^{*}\right)\right)^{\frac{\nu}{2}}}\exp\left(-\frac{1}{2}\tau^{*}(\theta_{i}^{*})^{2}\right)\nonumber\\
&\times\left\{\sgn(\theta_{i}^{*})+\sum_{j=1,j\neq i}^{d}\frac{\left|(\tau^{*})^{\frac{1}{2}}\theta_{j}^{*}\right|^{\nu-1}\sgn(\theta_{j}^{*})}{\exp\left(\frac{1}{2}\tau^{*}(\theta_{j}^{*})^{2}\right)}\right.\nonumber\\
&\left.+\bigo\left(\sum_{j=1}^{d}\frac{\tau^{*}(\theta_{j}^{*})^{2}}{n-d}\exp\left(-\frac{1}{2}\tau^{*}(\theta_{j}^{*})^{2}\right)\right)\right\},\ 1\leq i\leq d.\label{posterior_theta}\\
&E_{\bm{\theta},\tau}\left\{\tau\right\}=\tau^{*}\left\{1+\right.\nonumber\\
&\left.+\bigo\left(\nu(\nu-1)(\nu-2)\right.\right.\nonumber\\
&\left.\left.\times\frac{\sum_{j=1}^{d}\left|(\tau^{*})^{\frac{1}{2}}\theta_{j}^{*}\right|^{\nu-2}\left\{\sgn(\theta_{j}^{*})+(\tau^{*})^{\frac{1}{2}}\theta_{j}^{*}\exp\left(-\frac{1}{2}\tau^{*}(\theta_{j}^{*})^{2}\right)\right\}}{(n-d)\left(\frac{n}{d}\Omega\left(\lambda,\tau^{*}\right)\right)^{\frac{\nu}{2}}}\right)\right\}.\label{posterior_tau}
\end{align}
\label{Posterior_Unbiased_Estimators}
\end{corollary}
\begin{proof}
This result follows directly from the proof of Theorem \ref{theorem1}, by equations (\ref{def_L}) through (\ref{M_tau_int})
and by inspection of the jacobian $\bm{J}_{\bm{\Upsilon}}$ which is evaluated in (\ref{form_of_J}).
\end{proof}

\section{Discretization of model parameters}
\noindent We will in this section investigate how the INMDL-principle may be applied to deduce an upper bound on the discretization on the model parameters. We will use the result both to compute a sufficient mesh size on the grid on which we solve the nonlinear equation which determines the 
estimator $\theta^{*}$ in the experiments section below, and in our deduction of a model class prior distribution. In the previous sections, by means of Theorem \ref{theorem1} and its proof and Corollary \ref{corollary_new_marginal}, we have established the following formula for the posterior density $\hat{p}_{\gamma_{d}}$ as %(should perhaps include the differentials $d\hat{\bm{\theta}}$ ,$d\hat{\tau}$, $d\hat{\lambda}$ in formula below to make it clearer how the posterior density $p$ transforms under: $\bm{\theta}\rightarrow\hat{\bm{\theta}}$, $\tau\rightarrow\hat{\tau}$, $\lambda\rightarrow\hat{\lambda}$).
\begin{align}
&\hat{p}_{\gamma_{d}}\left(\hat{\bm{\theta}},\hat{\tau},\hat{\lambda}|\bm{x}\right)
=\frac{|I_{\tau}|^{-1}|I_{\lambda}|^{-1}}{m_{\gamma_{d}}(\bm{x})}\rho(\hat{\bm{\theta}}|\hat{\lambda})f\left(\bm{x}\left|\bm{\phi}(\hat{\bm{\theta}},\hat{\tau}),\psi(\hat{\tau})\right)\right.\nonumber\\
&=\frac{|I_{\tau}|^{-1}|I_{\lambda}|^{-1}}{m_{\gamma_{d}}(\bm{x})}\frac{\rho(\hat{\bm{\theta}}|\hat{\lambda})f\left(\bm{x}\left|\bm{\phi}(\hat{\bm{\theta}},\hat{\tau}),\psi(\hat{\tau})\right)\right.}{|\hat{\bm{F}}(\hat{\bm{\theta}}^{*},\hat{\tau}^{*})|^{1/2}}\cdot|\hat{\bm{F}}(\hat{\bm{\theta}}^{*},\hat{\tau}^{*})|^{1/2}\nonumber\\
&\approx\frac{|\hat{\bm{F}}(\hat{\bm{\theta}}^{*},\hat{\tau}^{*})|^{\frac{1}{2}}}{m_{\gamma_{d}}(\bm{x})}|I_{\tau}|^{-1}|I_{\lambda}|^{-1}\exp\left(-\hat{\Phi}(\bm{x},\hat{\bm{\beta}}^{*})\right)\nonumber\\
&\times\exp\left(-\frac{1}{2}(\hat{\bm{\beta}}-\hat{\bm{\beta}}^{*})^{T}\hat{\bm{H}}(\bm{x},\hat{\bm{\beta}}^{*})(\hat{\bm{\beta}}-\hat{\bm{\beta}}^{*})-\frac{1}{2}\hat{\Psi}_{\hat{\lambda}\hat{\lambda}}(\hat{\bm{\theta}}^{*},\hat{\lambda}^{*})(\hat{\lambda}-\hat{\lambda}^{*})^{2}\right)\nonumber\\
&=\frac{|\hat{\bm{F}}(\hat{\bm{\theta}}^{*},\hat{\tau}^{*})|^{1/2}}{|I_{\tau}|\cdot|I_{\lambda}|}\left[\frac{(2\pi)^{(d+2)/2}(\bm{x})|I_{\tau}|^{-1}|I_{\lambda}|^{-1}}{|\bm{H}(\bm{x},\bm{\theta}^{*},\tau^{*})|^{1/2}|\Psi_{\lambda\lambda}(\bm{\theta}^{*},\lambda^{*})|^{1/2}}\times\right.\nonumber\\
&\left.\pi(\bm{\theta}^{*}|\lambda^{*})f(\bm{x}|\bm{\theta}^{*},\tau^{*})\prod_{i=1}^{d}P_{G}\left((\tau^{*})^{\frac{1}{2}}|\theta_{i}^{*}|\right)\right]^{-1}\nonumber\\
&\times\exp\left(-\hat{\Phi}(\bm{x},\hat{\bm{\theta}}^{*},\hat{\tau}^{*})\right)\exp\left(-\frac{1}{2}(\hat{\bm{\beta}}-\hat{\bm{\beta}}^{*})^{T}\hat{\bm{H}}(\bm{x},\hat{\bm{\beta}}^{*})(\hat{\bm{\beta}}-\hat{\bm{\beta}}^{*})\right)\nonumber\\
&\times\exp\left(-\frac{1}{2}\hat{\Psi}_{\hat{\lambda}\hat{\lambda}}(\hat{\bm{\theta}}^{*},\hat{\lambda}^{*})(\hat{\lambda}-\hat{\lambda}^{*})^{2}\right)\nonumber\\
&=|\bm{H}(\bm{x},\bm{\theta}^{*},\tau^{*})|^{1/2}\frac{|\bm{J}_{\phi,\psi}^{T}\bm{F}(\bm{\theta}^{*},\tau^{*})\bm{J}_{\phi,\psi}|^{1/2}}{\pi(\bm{\theta}^{*}|\lambda^{*})f(\bm{x}|\bm{\theta}^{*},\tau^{*})}
|\hat{\Psi}_{\hat{\lambda}\hat{\lambda}}(\hat{\bm{\theta}}^{*},\hat{\lambda}^{*})|^{1/2}\nonumber\\
&\times(2\pi)^{-\frac{d+2}{2}}\exp\left(-\hat{\Phi}(\bm{x},\hat{\bm{\theta}}^{*},\hat{\tau}^{*})-\frac{1}{2}(\hat{\bm{\beta}}-\hat{\bm{\beta}}^{*})^{T}\hat{\bm{H}}(\bm{x},\hat{\bm{\beta}}^{*})(\hat{\bm{\beta}}-\hat{\bm{\beta}}^{*})\right)\nonumber\\
&\times\exp\left(-\frac{1}{2}\hat{\Psi}_{\hat{\lambda}\hat{\lambda}}(\hat{\bm{\theta}}^{*},\hat{\lambda}^{*})(\hat{\lambda}-\hat{\lambda}^{*})^{2}\right)\left[\prod_{i=1}^{d}P_{G}\left((\tau^{*})^{\frac{1}{2}}|\theta_{i}^{*}|\right)\right]^{-1}\nonumber\\
&=\left[\prod_{i=1}^{d}P_{G}\left((\tau^{*})^{\frac{1}{2}}|\theta_{i}^{*}|\right)\right]^{-1}|\hat{\bm{H}}(\bm{x},\hat{\bm{\theta}}^{*},\hat{\tau}^{*})|^{1/2}|\hat{\Psi}_{\hat{\lambda}\hat{\lambda}}(\hat{\bm{\theta}}^{*},\hat{\lambda}^{*})|^{1/2}\nonumber\\
&\times(2\pi)^{-\frac{d+2}{2}}\exp\left(-\frac{1}{2}(\hat{\bm{\beta}}-\hat{\bm{\beta}}^{*})^{T}\hat{\bm{H}}(\bm{x},\hat{\bm{\beta}}^{*})(\hat{\bm{\beta}}-\hat{\bm{\beta}}^{*})\right)\nonumber\\
&\times\exp\left(-\frac{1}{2}\hat{\Psi}_{\hat{\lambda}\hat{\lambda}}(\hat{\bm{\theta}}^{*},\hat{\lambda}^{*})(\hat{\lambda}-\hat{\lambda}^{*})^{2}\right)\label{hat_INMDL_approximated_posterior}
\end{align}
where $\hat{\bm{\beta}}\mathdef(\hat{\tau},\hat{\bm{\theta}}^{T})^{T}$ with $\hat{\bm{\theta}}$, $\hat{\tau}$, $\hat{\bm{\theta}}^{*}$, $\hat{\tau}^{*}$ defined in (\ref{def_phi_psi}), (\ref{def_invariant_estimators}), respectively, and $\hat{\lambda}$, $\hat{\lambda^{*}}$ defined in Proposition \ref{proposition_prior_laplace_approximation_formula}. The $\approx$-relation between the lefthand side and righthandside in (\ref{hat_INMDL_approximated_posterior}) is due to our omitting terms of order three and higher in the Taylor expansion of $\hat{\Phi}$ and $\hat{\Psi}$ used in formula (\ref{hat_INMDL_approximated_posterior}). These terms may be found by going through the proof of Theorem \ref{theorem1} and Proposition \ref{proposition_prior_laplace_approximation_formula}, but assuming the conditions in Theorem \ref{theorem1} under which the marginal approximation is valid, they may be omitted here. Now, 
since we do not know the exact form of the prior distribution $\pi(\bm{\theta}|\lambda)$, we do not know the exact forms of 
$\Psi(\bm{\theta}^{*},\lambda^{*})$, $\hat{\Psi}(\hat{\bm{\theta}}^{*},\hat{\lambda}^{*})$ and the map 
$\chi:\hat{\lambda}\rightarrow\lambda$. But we do know that the transformed Fisher ''matrix'' $\hat{\bm{E}}(\hat{\lambda})$ is constant 
and we may expect that the transformed Hessian $\hat{\Psi}_{\hat{\lambda}\hat{\lambda}}(\hat{\bm{\theta}},\hat{\lambda})$ of 
$-\log\left(\pi(\hat{\bm{\theta}}|\hat{\lambda})|\hat{\bm{E}}(\hat{\lambda})|^{-1/2}\right)$ satisfies
\begin{align}
&\hat{\Psi}_{\hat{\lambda}\hat{\lambda}}(\hat{\bm{\theta}}^{*},\hat{\lambda}^{*})\approx\hat{\bm{E}}(\hat{\lambda}^{*})=\bar{\lambda} = \text{ constant}.\label{det_hat_Hessian_hat_lambda}
\intertext{By means of (\ref{hat_Phi_theta_tau2})-(\ref{hat_Phi_theta_theta2}) we evaluate the Hessian $\hat{\bm{H}}$ to be}
&\hat{\bm{H}}(\bm{x},\hat{\bm{\theta}}^{*},\hat{\tau}^{*})=\left(\begin{array}{ccccc}a & b_{1} & b_{2} &\cdots & b_{d}\\
b_{1} & c_{1} & 0 & \cdots & 0\\
b_{2} & 0 & c_{2} & \cdots & 0\\
\vdots &\vdots &\vdots & \ddots &\vdots\\
b_{d} & 0 & 0 & \cdots & c_{d} \end{array}\right)\label{form_of_hat_H}
\intertext{where}
&a=\epsilon_{d}^{2}\left(\frac{n-d+2}{n}+\frac{1}{2n}\|(\tau^{*})^{\frac{1}{2}}\bm{\theta}^{*}\|_{2}^{2}+\frac{2}{n}\sum_{i=1}^{d}\tau^{*}(\theta_{i}^{*})^{2}\littleo(\mu_{\lambda,\nu}(\tau^{*},\theta_{i}^{*}))+\right.\nonumber\\
&\left.+\frac{2}{n}\sum_{i=1}^{d}\tau^{*}(\bm{x}_{\parallel}(i)-\theta_{i}^{*})\theta_{i}^{*}\right)\label{def_a2}\\
&b_{i}=-\epsilon_{d}\left(\frac{2}{n}\right)^{\frac{1}{2}}\bar{\tau}^{\frac{1}{2}}\left((\tau^{*})^{\frac{1}{2}}(\bm{x}_{\parallel}(i)-\frac{1}{2}\theta_{i}^{*})+\right.\nonumber\\
&\left.+\frac{1}{2}\littleo(\mu_{\lambda,\nu}(\tau^{*},\theta_{i}^{*}))(\tau^{*})^{\frac{1}{2}}\theta_{i}^{*}\right),\ 1\leq i\leq d\label{def_b_i}\\
&c_{i}=\bar{\tau}(1+\littleo(\mu_{\lambda,\nu}(\tau^{*},\theta_{i}^{*}))),\ 1\leq i\leq d.\label{def_c_i}
\intertext{Using (\ref{form_of_hat_H})-(\ref{def_c_i}) and the determinant formula (\ref{det_form_of_general_G}) we may write}
&|\hat{\bm{H}}(\bm{x},\hat{\bm{\theta}}^{*},\hat{\tau}^{*})|=\left(a-\sum_{j=1}^{d}\frac{b_{j}^{2}}{c_{j}}\right)\prod_{l=1}^{d}c_{l}\nonumber\\
&\approx\bar{\tau}^{d}\exp\left[\littleo\left(\left(\frac{n}{d}\Omega(\lambda^{*},\tau^{*})\right)^{-\frac{\nu}{2}}\sum_{i=1}^{d}|(\tau^{*})^{\frac{1}{2}}\theta_{i}^{*}|^{\nu-2}\right)\right]\times\nonumber\\
&\left[\epsilon_{d}^{2}\left(\frac{n-d+2}{n}+\frac{1}{2n}\|(\tau^{*})^{\frac{1}{2}}\bm{\theta}^{*}\|_{2}^{2}\right)-\epsilon_{d}^{2}\frac{2}{n}\sum_{i=1}^{d}
\frac{\left(\frac{1}{2}\bar{\tau}^{\frac{1}{2}}(\tau^{*})^{\frac{1}{2}}\theta_{i}^{*}\right)^{2}}{\bar{\tau}}\right]\nonumber\\
&=\epsilon_{d}^{2}\bar{\tau}^{d}\exp\left[\littleo\left(\left(\frac{n}{d}\Omega(\lambda^{*},\tau^{*})\right)^{-\frac{\nu}{2}}\sum_{i=1}^{d}|(\tau^{*})^{\frac{1}{2}}\theta_{i}^{*}|^{\nu-2}\right)\right]\nonumber\\
&\sim\bar{\bar{\tau}}^{d}=|\hat{\bm{F}}(\hat{\bm{\theta}}^{*},\hat{\tau}^{*})|,\label{det_hat_H}
\intertext{where $\sim$ means asymptotical equality as $n\rightarrow\infty$ and $\frac{d}{n}\rightarrow 0$. We want to determine the half-axes of the reduced quadratic form associated with $\hat{\bm{H}}$. We proceed with estimating the eigenvalues of $\hat{\bm{H}}$. Letting $\bm{Id}$ denote the identity matrix we have}
&\det\left(\hat{\kappa}\bm{Id}-\hat{\bm{H}}\right)=\det\left(\begin{array}{ccccc}\hat{\kappa}-a & -b_{1} & -b_{2} &\cdots & -b_{d}\\
-b_{1} & \hat{\kappa}-c_{1} & 0 & \cdots & 0\\
-b_{2} & 0 & \hat{\kappa}-c_{2} & \cdots & 0\\
\vdots &\vdots &\vdots & \ddots &\vdots\\
-b_{d} & 0 & 0 & \cdots & \hat{\kappa}-c_{d} \end{array}\right)\nonumber\\
&=\left(\hat{\kappa}-a-\sum_{j=1}^{d}\frac{b_{j}^{2}}{\hat{\kappa}-c_{j}}\right)\prod_{l=1}^{d}(\hat{\kappa}-c_{l})\label{char_polynomial}
\intertext{where we used the determinant formula (\ref{det_form_of_general_G}). The first eigenvalue $\hat{\kappa}=\hat{\kappa}_{1}$ may be found by solving}
&\hat{\kappa}_{1}-a-\sum_{i=1}^{d}\frac{b_{i}^{2}}{\hat{\kappa}_{1}-c_{i}}=\hat{\kappa}_{1}-a-\sum_{i=1}^{d}\frac{b_{i}^{2}}{\hat{\kappa}_{1}-\bar{\tau}(1+\littleo(\zeta))}=0.\nonumber
\intertext{We introduce the approximation $c_{i}\approx\bar{\tau},\ \forall\ i$ founded on the asssumption $\zeta\ll 1$ and get the equation}
&(\hat{\kappa}_{1}-a)(\hat{\kappa}_{1}-\bar{\tau})=\sum_{i=1}^{d}b_{i}^{2}\nonumber\\
\intertext{which yields}
&\hat{\kappa}_{1}=\frac{a+\bar{\tau}}{2}\left[1\pm\left(1+4\frac{\sum_{i=1}^{d}b_{i}^{2}-a\bar{\tau}}{(a+\bar{\tau})^{2}}\right)^{\frac{1}{2}}\right].\nonumber\\
\intertext{Using (\ref{def_a2}), (\ref{def_b_i}) yields}
&\hat{\kappa}_{1}=\frac{a+\bar{\tau}}{2}\left[1\pm\left(1+4\frac{\epsilon_{d}^{2}\bar{\tau}\left(-1+\frac{2}{n}\sum_{i=1}^{d}\littleo(\mu_{\lambda,\nu}(\tau^{*},\theta_{i}^{*}))\tau^{*}(\theta_{i}^{*})^{2}\right)}{\left(\epsilon_{d}^{2}\left(1+(2n)^{-1}\|(\tau^{*})^{\frac{1}{2}}\bm{\theta}^{*}\|_{2}^{2}\right)+\bar{\tau}\right)^{2}}\right)^{1/2}\right].\label{eigenvalue1}
%\intertext{Now, because of (\ref{C7}) in Theorem \ref{theorem1} we know how $n^{-1}\|(\tau^{*})^{\frac{1}{2}}\bm{\theta}^{*}\|_{2}^{2}$ scales:}
\intertext{Now, using:}
&n^{-1}\|(\tau^{*})^{\frac{1}{2}}\bm{\theta}^{*}\|_{2}^{2}\propto\frac{d\frac{1}{\lambda}}{n\frac{1}{\tau^{*}}}=\Omega(\lambda,\tau^{*})\label{scaling_law1}
\intertext{we may write (\ref{eigenvalue1}) as}
&\hat{\kappa}_{1}=\frac{a+\bar{\tau}}{2}\left(1\pm\left(1+\epsilon_{d}^{-2}\bar{\tau}\frac{-1+\Omega(\lambda^{*},\tau^{*})\littleo(\zeta)}{\left(1+\frac{1}{2}\Omega(\lambda^{*},\tau^{*})+\bar{\tau}\right)^{2}}\right)^{\frac{1}{2}}\right).\nonumber
\intertext{Assuming $\Omega(\lambda,\tau^{*})\gg 1$ or $\zeta\ll 1$ or alternatively $\epsilon_{d}\gg 1$ we may write}
&\hat{\kappa}_{1}\approx\frac{a+\bar{\tau}}{2}(1\pm 1).\nonumber
\intertext{Observing that $\kappa=0$ does not solve (\ref{char_polynomial}) we finally get}
&\kappa_{1}\approx a+\bar{\tau}=a(1+a^{-1}\bar{\tau})\approx a\left(1+\frac{\bar{\tau}}{\frac{n-d+2}{n}+(2n)^{-1}\|(\tau^{*})^{\frac{1}{2}}\bm{\theta}^{*}\|_{2}^{2}}\right)\nonumber\\
&\approx a\left(1+\frac{\bar{\tau}}{1+\frac{1}{2}\Omega(\lambda^{*},\tau^{*})}\right)\approx a\label{first_eigenvalue}
\intertext{where the $\approx$ here is taken to mean ''almost equality'' if $n$ is sufficiently large, $d/n$ sufficiently small, $\bar{\tau}\ll \Omega(\lambda^{*},\tau^{*})$ and we use 
$(\lambda^{*})^{-1}\sim d^{-1}\|\bm{\theta}^{*}\|_{2}^{2}$. Now, we estimate the rest of the eigenvalues $\hat{\kappa}_{i},\ 2\leq i\leq d+1$.
By (\ref{det_hat_H}) and (\ref{first_eigenvalue}) we have that}
&\epsilon_{d}^{2}\bar{\tau}^{d}\approx|\hat{\bm{H}}|=\prod_{i=1}^{d+1}\hat{\kappa}_{i}\approx a\prod_{i=2}^{d+1}\hat{\kappa}_{i}
\approx\epsilon_{d}^{2}\left(\frac{n-d+2}{n}+\frac{1}{2}\Omega(\lambda^{*},\tau^{*})\right)\prod_{i=2}^{d+1}\hat{\kappa}_{i}.\label{det_hat_H_approx}
\intertext{Now, because $c_{i}\approx\bar{\tau},\ \forall\ i$, we conclude that $c_{i}\approx c_{j},\ 1\leq i,j\leq d$ and from (\ref{char_polynomial}) we may then conclude that $\hat{\kappa}_{i}\approx\hat{\kappa}_{j},\ 2\leq i,j\leq d+1$. Then by (\ref{det_hat_H_approx}) we may write}
&\hat{\kappa}_{i}\approx\bar\tau\left(\frac{n-d+2}{n}+\frac{1}{2}\Omega(\lambda^{*},\tau^{*})\right)^{-\frac{1}{d}}\approx\bar{\tau},\ 2\leq i\leq d+1,\label{last_eigenvalues}
\intertext{where in the last $\approx$ above we made the assumption that $0<\log{\Omega}\ll d$. Letting $\hat{\bm{M}}$ denote the orthogonal matrix such that $\hat{\bm{M}}^{T}\hat{\bm{H}}\hat{\bm{M}}=\text{\bf{diag}}(\hat{\kappa}_{i})$ and define the orthogonal transformation of variables $\hat{\bm{\alpha}}\mathdef\hat{\bm{M}}^{T}\hat{\bm{\beta}}$. Plugging this change of variables into (\ref{hat_INMDL_approximated_posterior}) 
we see that $\hat{\bm{\alpha}}\sim$ $\mathcal{N}(\hat{\bm{M}}^{T}\hat{\bm{\beta}}^{*},\text{\bf{diag}}(\hat{\kappa_{i}}))$.  Now comparing the eigenvalues 
$\left\{\hat{\kappa}_{i}\right\}_{i=1}^{d+1}$ we have estimated above with the elements of $\hat{\bm{H}}$, we conclude that 
$\hat{\bm{M}}^{T}\hat{\bm{H}}\hat{\bm{M}}=$ $\text{\bf{diag}}\left\{\hat{\kappa}_{i}\right\}_{i=1}^{d+1}\approx$ $\text{\bf{diag}}(\hat{\bm{H}})$ up to our accuarcy of estimation of the $\hat{\kappa}_{i}$ above. It follows that the half axes of the reduced quadratic form (hyper-ellipsoid) associated with $\hat{\bm{H}}$ are approximately given by 
$\left\{\hat{\kappa}_{i}\right\}_{i=1}^{d+1}$. Expanding the model parameters $\lambda$, $\tau$ and $\theta_{i},\ 1\leq i\leq d$ into their differentials $\Delta\lambda$, $\Delta\tau$, $\Delta\theta_{i}$ we get}
&\Delta\lambda =\chi^{\prime}(\hat{\lambda})\Delta\hat{\lambda}\label{def_Delta_lambda}\\
&\Delta\tau=\frac{d\psi(\hat{\tau})}{d\hat{\tau}}\Delta\hat{\tau}=\frac{\partial}{\partial\hat{\tau}}\tau_{0}
\exp\left(\epsilon_{d}\left(\frac{2}{n}\right)^{\frac{1}{2}}\hat{\tau}\right)\Delta\hat{\tau}=\epsilon_{d}\left(\frac{2}{n}\right)^{\frac{1}{2}}\psi(\hat{\tau})\Delta\hat{\tau}\label{def_Delta_tau}\\
&\Delta\theta_{i}=\frac{\partial\phi(\hat{\theta}_{i},\hat{\tau})}{\partial\hat{\theta}_{i}}\Delta\hat{\theta}_{i}+\frac{\partial\phi(\hat{\theta}_{i},\hat{\tau})}{\partial\hat{\tau}}\Delta\hat{\tau}\nonumber\\
&=\bar{\tau}^{\frac{1}{2}}\psi^{-\frac{1}{2}}(\hat{\tau})\Delta\hat{\theta}_{i}-\frac{1}{2}\bar{\tau}^{\frac{1}{2}}\psi^{-\frac{1}{2}}(\hat{\tau})\hat{\theta}_{i}\epsilon_{d}\left(\frac{2}{n}\right)^{\frac{1}{2}}\Delta\hat{\tau}.\label{def_Delta_theta_i}
\intertext{Now, denoting the discretization size of $\hat{\theta}_{i}$ 
by $\Delta\hat{\theta}_{i}$, and the discretization sizes of $\hat{\tau}$, $\hat{\lambda}$ by $\Delta\hat{\tau}$ and $\Delta\hat{\lambda}$, respectively, we may write the relative uncertainty $\Delta\hat{p}/\hat{p}$ of the posterior density $\hat{p}(\hat{\bm{\theta}},\hat{\tau},\hat{\lambda}|\bm{x})$ due to the discretizations $\Delta\hat{\lambda},\Delta\hat{\tau},\Delta\hat{\theta}_{i}$ of parameters $\hat{\lambda},\hat{\tau},\hat{\theta}_{i},\ 1\leq i\leq d$ respectively, as follows:}
&\frac{\Delta\hat{p}}{\hat{p}}\mathdef\left[\left(\frac{1}{\hat{p}}\frac{\partial\hat{p}\left(\hat{\bm{\theta}},\hat{\lambda},\hat{\tau}|\bm{x}\right)}{\partial\hat{\lambda}}\Delta\hat{\lambda}\right)^{2}+\left(\frac{1}{\hat{p}}\frac{\partial\hat{p}\left(\hat{\bm{\theta}},\hat{\lambda},\hat{\tau}|\bm{x}\right)}{\partial\hat{\tau}}\Delta\hat{\tau}\right)^{2}+\right.\nonumber\\
&\left.+\sum_{i=1}^{d}\left(\frac{1}{\hat{p}}\frac{\partial\hat{p}\left(\hat{\bm{\theta}},\hat{\lambda},\hat{\tau}|\bm{x}\right)}{\partial\hat{\theta}_{i}}\Delta\hat{\theta}_{i}\right)^{2}\right]^{1/2}.\label{total_differential_expansion_posterior}
\intertext{We want to bound the relative error defined in (\ref{total_differential_expansion_posterior}) over the cell $\hat{C}_{\hat{\bm{\theta}}^{*},\hat{\tau}^{*},\hat{\lambda}^{*}}$ defined by}
&\hat{C}_{\hat{\bm{\theta}}^{*},\hat{\tau}^{*},\hat{\lambda}^{*}}\mathdef [\hat{\theta}_{1}^{*}-\Delta\hat{\theta}_{1},\hat{\theta}_{1}^{*}+\Delta\hat{\theta}_{1}]
\times[\hat{\theta}_{2}^{*}-\Delta\hat{\theta}_{2},\hat{\theta}_{2}^{*}+\Delta\hat{\theta}_{2}]\times\cdots\nonumber\\
&\times[\hat{\theta}_{d}^{*}-\Delta\hat{\theta}_{d},\hat{\theta}_{d}^{*}+\Delta\hat{\theta}_{d}]\times[\hat{\tau}^{*}-\Delta\hat{\tau},\hat{\tau}^{*}+\Delta\hat{\tau}]\times[\hat{\lambda}^{*}-\Delta\hat{\lambda},\hat{\lambda}^{*}+\Delta\hat{\lambda}].\label{quantized_parameter_unit_cell}
\intertext{When we consider (\ref{hat_INMDL_approximated_posterior}) together with (\ref{total_differential_expansion_posterior}) we see that $\Delta\hat{\lambda}$ should not scale coarser than}
&\Delta\hat{\lambda}\sim\left[\hat{\Psi}_{\hat{\lambda}\hat{\lambda}}(\hat{\bm{\theta}}^{*},\hat{\lambda}^{*})\right]^{-1/2}\cdot c_{\hat{\lambda}}\approx\bar{\lambda}^{-1/2}\cdot c_{\hat{\lambda}}\label{hat_lambda_scaling}
\intertext{where $0<c_{\hat{\lambda}}<1$ is some constant. Considering (\ref{hat_INMDL_approximated_posterior}) together with (\ref{form_of_hat_H}),
 (\ref{total_differential_expansion_posterior}) and the argument of approximating posterior covariances above which justifies treating $\hat{\bm{H}}(\bm{x},\hat{\bm{\theta}}^{*},\hat{\tau}^{*})$ as diagonal matrix 
%(SHOULD WE APPEAL TO OUR FREEDOM IN CHOOSING EPSILON OR BAR TAU TO MAKE HESSIAN MATRIX H APPROXIMATELY DIAGONAL???)
, we see that $\Delta\hat{\tau}$ should scale no coarser than}
&\Delta\hat{\tau}\sim a^{-\frac{1}{2}}\cdot c_{\hat{\tau}}=\epsilon_{d}^{-1}\left(\frac{n-d+2}{n}+\frac{1}{2n}\|(\tau^{*})^{\frac{1}{2}}\bm{\theta}^{*}\|_{2}^{2}\right)^{-\frac{1}{2}}\cdot c_{\hat{\tau}}\nonumber\\
&=_{E_{\theta}}\epsilon_{d}^{-1}\left(\frac{n-d+2}{n}+\frac{1}{2}\Omega(\lambda^{*},\tau^{*})\right)^{-\frac{1}{2}}\cdot c_{\hat{\tau}}\nonumber\\
&=\epsilon_{d}^{-1}\sqrt{2}\Omega^{-\frac{1}{2}}(\lambda^{*},\tau^{*})\left(1+2\frac{n-d+2}{n\Omega(\lambda^{*},\tau^{*})}\right)^{-\frac{1}{2}}\cdot c_{\hat{\tau}}\nonumber\\
&\approx\sqrt{2}\epsilon_{d}^{-1}\Omega^{-\frac{1}{2}}(\lambda^{*},\tau^{*})\cdot c_{\hat{\tau}}.\label{hat_tau_scaling}
\intertext{where $0<c_{\hat{\tau}}<1$ is some constant number. We have omitted terms of non-leading order in (\ref{def_a2}) and we assumed 
$\Omega(\lambda^{*},\tau^{*})\gg 1$ when writing the last $\approx$ above. Now, choosing the scaling on $\Delta\hat{\theta}$ as}
&\Delta\hat{\theta}_{i}\sim -\sgn(\hat{\theta}_{i})\cdot \frac{\bar{\tau}^{-\frac{1}{2}}}{\sqrt{d}}\cdot c_{\hat{\theta}},\label{hat_theta_scaling}
\intertext{where $0<c_{\hat{\theta}}<1$ is some constant number (see (\ref{hat_INMDL_approximated_posterior})), 
%should also bring in the error from the Laplace approximation step here) 
and the sign convention $\sgn(\Delta\hat{\theta}_{i})=-\sgn(\hat{\theta}_{i})$ is just a trick to make $|\Delta\theta_{i}|$ 
symmetric w.r.t sign of $\theta_{i}$, see (\ref{def_Delta_theta_i}). The  relations (\ref{total_differential_expansion_posterior})-(\ref{hat_theta_scaling}) now yields} 
&\sup_{\hat{\bm{\theta}},\hat{\tau},\hat{\lambda}\in \hat{C}_{\hat{\bm{\theta}}^{*},\hat{\tau}^{*},\hat{\lambda}^{*}}}\frac{\Delta\hat{p}(\hat{\bm{\theta}},\hat{\tau},\hat{\lambda}|\bm{x})}{\hat{p}(\hat{\bm{\theta}},\hat{\tau},\hat{\lambda}|\bm{x})} \leq\left(c_{\hat{\lambda}}^{2}+c_{\hat{\tau}}^{2}+\sum_{i=1}^{d}\left(\frac{c_{\hat{\theta}}}{\sqrt{d}}\right)^{2}\right)^{1/2}\nonumber\\
&=\sqrt{c_{\hat{\lambda}}^{2}+c_{\hat{\tau}}^{2}+c_{\hat{\theta}}^{2}}\label{sup_relative_error_on_posterior_density}
\intertext{Plugging (\ref{hat_lambda_scaling}), (\ref{hat_tau_scaling}), (\ref{hat_theta_scaling})  into (\ref{def_Delta_lambda}), (\ref{def_Delta_tau}) and (\ref{def_Delta_theta_i}) we get}
&\Delta\tau\sim\tau\frac{2}{\sqrt{n}}\Omega^{-\frac{1}{2}}(\lambda^{*},\tau^{*})\cdot c_{\hat{\tau}}\label{Delta_tau_form1}\\
&\Delta\theta_{i}\sim-\sgn(\theta_{i})\left(\frac{\tau^{-\frac{1}{2}}}{\sqrt{d}}\cdot c_{\hat{\theta}}+|\theta_{i}|\frac{1}{\sqrt{n}}\Omega^{-\frac{1}{2}}(\lambda^{*},\tau^{*})\cdot c_{\hat{\tau}}\right)\label{Delta_theta_form1}\\
&\Delta\lambda\sim\chi^{\prime}(\hat{\lambda})\left[\hat{\Psi}_{\hat{\lambda}\hat{\lambda}}(\hat{\bm{\theta}}^{*},\hat{\lambda}^{*})\right]^{-1/2}\cdot c_{\hat{\lambda}}.\label{Delta_lambda_form1}
\intertext{We note that (\ref{Delta_lambda_form1}) may be simplified by observing that}
&\hat{\Psi}_{\hat{\lambda}\hat{\lambda}}(\hat{\bm{\theta}},\hat{\lambda})\mathdef\frac{\partial^{2}}{\partial\hat{\lambda}^{2}}\Psi(\hat{\bm{\theta}},\chi(\hat{\lambda}))=\chi^{\prime\prime}(\hat{\lambda})\Psi_{\lambda}(\bm{\theta},\lambda)+\chi^{\prime}(\hat{\lambda})^{2}\Psi_{\lambda\lambda}(\bm{\theta},\lambda)\label{hat_Hessian_prior}
\intertext{and noting that $\Psi_{\lambda}(\bm{\theta}^{*},\lambda^{*})=0$ by definition of $\lambda^{*}$ we may by means of (\ref{hat_Hessian_prior}) 
write (\ref{Delta_lambda_form1}) as}
&\Delta\lambda\sim\chi^{\prime}(\hat{\lambda}^{*})\left[\chi^{\prime}(\hat{\lambda})^{2}\Psi_{\lambda\lambda}(\bm{\theta}^{*},\lambda^{*})\right]^{-1/2}\cdot c_{\hat{\lambda}}=|\Psi_{\lambda\lambda}(\bm{\theta}^{*},\lambda^{*})|^{-1/2}\cdot c_{\hat{\lambda}}.\label{Delta_lambda_form2}
\end{align}
The expressions (\ref{Delta_tau_form1}), (\ref{Delta_theta_form1}), (\ref{Delta_lambda_form1}) may be used to deduce an upper bound on the discretization to use in encoding the estimated parameters $\tau^{*}$, $\bm{\theta}^{*},\lambda^{*}$ while yielding the posterior distribution to within a prescribed precision. We note that in \cite{Rissanen:1998b} it is shown that the MDL-optimal choice of discretization of parameters scales like $n^{-1/2}$ (asymptotically in $n$). This should not be confused with the discretization given 
in (\ref{hat_theta_scaling}): We want a discretization which is fine enough to enable us to evaluate posterior probabilities to within 
some specified precision whereas Rissanen want a discretization yielding the shortest code length \cite{Rissanen:1998b}, \cite{Rissanen:1996}.
\begin{proposition}
The discretization $\Delta\tau$, $\Delta\lambda$, $\Delta\theta_{i}$ on the parameters $\tau$, $\lambda$, $\theta_{i},$ $\ 1\leq i\leq d$, respectively, given by 
\begin{align}
&\Delta\tau=\tau\frac{2}{\sqrt{n}}\Omega^{-\frac{1}{2}}(\lambda^{*},\tau^{*})\cdot c_{\hat{\tau}},\ 0< c_{\hat{\tau}}<1.\label{Delta_tau_form3}\\
&\Delta\theta_{i}=\frac{\tau^{-\frac{1}{2}}}{\sqrt{d}}\cdot c_{\hat{\theta}}\cdot\left(1+\left(\frac{d}{n}\frac{\tau|\theta_{i}|^{2}}{\Omega(\lambda^{*},\tau^{*})}\right)^{1/2}\frac{c_{\hat{\tau}}}{c_{\hat{\theta}}}\right),\ 0<c_{\hat{\theta}}<1.\label{Delta_theta_form3}\\
&\Delta\lambda=\chi^{\prime}(\hat{\lambda})\left[\hat{\Psi}_{\hat{\lambda}\hat{\lambda}}(\hat{\bm{\theta}}^{*},\hat{\lambda}^{*})\right]^{-1/2}\cdot c_{\hat{\lambda}}=|\Psi_{\lambda\lambda}(\bm{\theta}^{*},\lambda^{*})|^{-1/2}\cdot c_{\hat{\lambda}},\nonumber\\
&0<c_{\hat{\lambda}}<1.\label{Delta_lambda_form3}
\intertext{yields the following precision $\Delta\hat{p}(\hat{\bm{\theta}},\hat{\tau},\hat{\lambda}|\bm{x})$ on the posterior density 
$\hat{p}_{\gamma_{d}}(\hat{\bm{\theta}},\hat{\tau},\hat{\lambda}|\bm{x})$:}
&\frac{\Delta\hat{p}(\hat{\bm{\theta}},\hat{\tau},\hat{\lambda}|\bm{x})}{\hat{p}(\hat{\bm{\theta}},\hat{\tau},\hat{\lambda}|\bm{x})}\leq \sqrt{c_{\hat{\theta}}^{2}+c_{\hat{\lambda}}^{2}+c_{\hat{\tau}}^{2}}.\label{Delta_posterior}
\end{align}
\label{INMDL_optimal_quantization}
\end{proposition}
\begin{proof}
See discussion above.
\end{proof}
\subsection{Comments on Proposition \ref{INMDL_optimal_quantization}}
\begin{enumerate}
\item The discretization scheme given above should not be confused with the optimal $1/\sqrt{n}$ discretization given in \cite{Rissanen:1998b}, which is optimal in the sense of minimizing the expected difference w.r.t the worst data generating distribution $g$ between code lengths using 
the code length induced by any distribution $q(\bm{x})$ on data and the code length induced by $f(\bm{x};\bm{\beta}^{*}(\bm{x}))$, see 
\cite{Rissanen:2001} and (\ref{def_NML_optimality}). 
The discretization shown in Proposition \ref{INMDL_optimal_quantization} was developed to be the coarsest possible yielding
the posterior distribution to within a prescribed precision. 
\\

\item We see that the discretization of $\theta_{i}$ given above is data driven and implying a discretization that may well be finer or coarser 
than the MDL-optimal discretization of $\tau^{-1/2}/\sqrt{n}$, \cite{Rissanen:1998b}. It will generally lead to a finer discretization if $c_{\hat{\theta}}<\sqrt{d/n}$ and a coarser discretization if $c_{\hat{\theta}}>\sqrt{d/n}$. Also, we get coarser discretization for those indices $i$ where $\left(\frac{d}{n}\right)^{1/2}\frac{2\tau^{1/2}|\theta_{i}|}{\Omega^{1/2}(\lambda^{*},\tau^{*})}\cdot\frac{c_{\hat{\tau}}}{c_{\hat{\theta}}}>\sqrt{d/n}$.
\end{enumerate}

%\section{Unresolved questions}
%\begin{itemize}
%\item Whether to consider $\log\left(\frac{[|I_{\tau}|]_{v}}{2\pi}\right)$ part of the ''hairy part'' $Z_{\gamma_{d}}(\bm{x})$ or the ''nice'' part $Q_{\gamma_{d}}(\bm{x})$ of the codelength $L_{\gamma_{d}}(\bm{x})$. Check claims on $I_{\tau}$ originating from the proof of Theorem \ref{theorem1}.
%\end{itemize}
%Things to clean up:\\
%\begin{itemize}
%\item Possibly confusing use of notation: Use of $\Phi,\phi,\bm{\phi},\psi,\bm{\psi}$.
%\item Bad use of notation: $\littleo(\cdot)$. Have to improve this....
%\item Should introduce notation to compare order expressions.
%\item Check the details of Corollary \ref{Posterior_Unbiased_Estimators}, especially the $E\{\tau\}$-part.
%\item Check under which conditions on the prior distribution $\pi_{\lambda}(\theta)$ the estimator $\theta^{*}_{i}$ may be said to be monotone increasing function of 
%$\bm{x}_{\parallel}(i)$.
%\end{itemize}
\section{A formal approximative generalization to non-gaussian models}
The results we have obtained so far were deduced for models with IID gaussian likelihood distributions. However, it is possible 
to generalize the results to the case of non-gaussian IID likelihood models under some (smoothness) conditions on the distribution. 
The argument goes as follows: Given a IID non-gaussian likelihood
$f(\bm{x}|\bm{\theta},\bm{\alpha})$ $=\prod_{i=1}^{n}f(x_{i}|\theta_{i},\bm{\alpha})$, where $E_{x_{i}}[x_{i}]=\theta_{i}$, and $\bm{\alpha}=(\alpha_{1},...,\alpha_{s})$ are parameters of the distribution $f$, compute the Taylor expansion of 
$Q_{\bm{\alpha}}(\theta|x)\mathdef-\log{f(x|\theta,\bm{\alpha})}$ about $\theta=\theta_{0}=x$:
\begin{align}
&T_{Q_{\bm{\alpha}}}(\bm{\theta}|\bm{x})=\sum_{i=1}^{n}Q_{\bm{\alpha}}(x_{i}|x_{i})+\sum_{i=1}^{n}a(x_{i}|\bm{\alpha})(\theta_{i}-x_{i})\nonumber\\
&+\frac{1}{2}\sum_{i=1}^{n}b(x_{i}|\bm{\alpha})(\theta_{i}-x_{i})^{2}+R_{\bm{\alpha}}(\bm{\theta}|\bm{x})\nonumber\\
\intertext{where}
&R_{\bm{\alpha}}(\bm{\theta}|\bm{x})\mathdef\frac{1}{6}\sum_{i=1}^{n}\int_{x_{i}}^{\theta_{i}}c(z_{i}|\bm{\alpha})(\theta_{i}-z_{i})^{3}\ d{z_{i}},\ 
a(x_{i}|\bm{\alpha})\mathdef\left.\frac{\partial Q_{\bm{\alpha}}(\theta|x_{i})}{\partial \theta}\right|_{\theta=x_{i}},\nonumber\\ 
&b(x_{i}|\bm{\alpha})\mathdef\left.\frac{\partial^{2}Q_{\bm{\alpha}}(\theta|x_{i})}{\partial \theta^{2}}\right|_{\theta_{i}=x_{i}},\ 
c(x_{i}|\bm{\alpha})\mathdef\left.\frac{\partial^{3}Q_{\bm{\alpha}}(\theta|x_{i})}{\partial \theta^{3}}\right|_{\theta=x_{i}}.\label{Q_expansion2}
\intertext{Truncating the expansion $T_{Q_{\bm{\alpha}}}(\bm{\theta}|\bm{x})$ to second order in $\bm{\theta}$ will yield an approximation 
$g(\bm{x}|\bm{\theta},\bm{\alpha})$, which is a gaussian function of $\bm{\theta}$, to the likelihood model $f(\bm{x}|\bm{\theta},\bm{\alpha})$ and we may write}
&f(\bm{x}|\bm{\theta},\bm{\alpha})=g(\bm{x}|\bm{\theta},\bm{\alpha})Z_{\bm{\alpha}}(\bm{\theta}|\bm{x})\nonumber\\
\intertext{where}
&g(\bm{x}|\bm{\theta},\bm{\alpha})\mathdef\exp\left(-\sum_{i=1}^{n}Q_{\bm{\alpha}}(x_{i}|x_{i})+\sum_{i=1}^{n}\frac{a(x_{i}|\bm{\alpha})^{2}}{2b(x_{i}|\bm{\alpha})}\right)\nonumber\\
&\times\exp\left(-\frac{1}{2}\sum_{i=1}^{n}b(x_{i}|\bm{\alpha})\left(\theta_{i}-x_{i}+\frac{a(x_{i}|\bm{\alpha})}{b(x_{i}|\bm{\alpha})}\right)^{2}\right)\label{gaussian_part_likelihood}
\intertext{and}
&Z_{\bm{\alpha}}(\bm{\theta}|\bm{x})\mathdef\exp\left(-R_{\bm{\alpha}}(\bm{\theta}|\bm{x})\right).\label{error_part_likelihood}
\intertext{Although the second order approximation $g(\bm{x}|\bm{\theta},\bm{\alpha})$ in general will be a poor pointwise approximation to the 
density $f(\bm{x}|\bm{\theta},\bm{\alpha})$, it may {\em locally} in a vicinity of $\bm{\theta}=\bm{x}$ be sufficiently accurate to be used to compute the marginal integral 
$\int f(\bm{x}|\bm{\theta},\bm{\alpha})\pi_{\lambda}(\bm{\theta})\rho(\lambda)\zeta(\bm{\alpha})\ d\bm{\theta}\ d\bm{\alpha}\ d\lambda$ to within the desired accuracy. An analysis of the remainder term $Z_{\bm{\alpha}}(\bm{\theta}|\bm{x})$ will have to be carried out for the given likelihood $f$ to decide if this is the case. If so, we may define an approximative Fisher matrix $\bm{F}$ to the likelihood $f(\bm{x}|\bm{\theta},\bm{\alpha})$ by} 
&\bm{F}_{ij}(\bm{\beta})\mathdef -E_{\bm{x}}\left[\frac{\partial^{2}}{\partial\beta_{i}\partial\beta_{j}}\log{g(\bm{x}|\bm{\beta})}\right],\ \text{ where } \bm{\beta}\mathdef (\bm{\alpha}^{T},\bm{\theta}^{T})^{T}.\label{approximative_Fisher}
\intertext{Then we may proceed similar to the steps taken in (\ref{def_phi_psi})-(\ref{form_of_psi}) to find the reparameterizations 
$\phi: \hat{\bm{\theta}}\rightarrow\bm{\theta}$, $\psi:\hat{\bm{\alpha}}\rightarrow\bm{\alpha}$
which makes the reparameterized Fisher information $|\hat{\bm{F}}(\hat{\bm{\theta}},\hat{\bm{\alpha}})|$ a constant. In at least some cases of 
interest the reparameterizations defined in (\ref{def_phi_psi})-(\ref{form_of_psi}) should still apply with minor modifications and so would 
the (proof of) result in Theorem \ref{theorem1}.}\nonumber
\end{align}
\chapter{Applying the INMDL-principle to GGD-modelled data}
\section{Preliminaries}
We will investigate the performance of the INMDL-principle as developed in previous sections
when applied to GGD-modelled data. The GGD-model is frequently used when representing natural images in wavelet bases \cite{Moulin-Liu:1999}. Having found the invariant noise estimator $\tau^{*}$ in (\ref{invariant_noise_estimator}), 
we need to compute the invariant estimator $\theta^{*}$ defined in (\ref{def_invariant_estimators}) under the GGD-model.
The GGD family of distributions is a two-parameter family governed by the variance-parameter $\frac{1}{\lambda}>0$ and a shape parameter $\nu>0$ 
and has the form \cite{Moulin-Liu:1999}
\begin{align}
&\pi_{\lambda,\nu}(\theta)=\frac{\nu\eta(\nu)}{2\Gamma(1/\nu)}\lambda^{\frac{1}{2}}\exp\left(-\left[\eta(\nu)\lambda^{\frac{1}{2}}|\theta|\right]^{\nu}\right),\nonumber 
\intertext{where} 
&\eta(\nu)\mathdef\left(\frac{\Gamma(3/\nu)}{\Gamma(1/\nu)}\right)^{\frac{1}{2}}.\label{form_GGD_density}
\intertext{Under the assumption of IID additive white gaussian noise (WGN) the problem to solve is}
&\theta^{*} = \text{arg min}_{[\theta]_{v}\in\mathbb{R}}\left\{\frac{\tau}{2}(x-\theta)^{2}-\log\pi_{\lambda,\nu}(\theta)\right\}\nonumber\\
&=\text{arg min}_{[\theta]_{v}\in\mathbb{R}}\left\{\frac{\tau}{2}(x-\theta)^{2}+\left[\eta(\nu)\lambda^{\frac{1}{2}}|\theta|\right]^{\nu}\right\}\nonumber\\
&=\text{arg min}_{[\theta]_{v}\in\mathbb{R}}\left\{[x-\theta]_{v}^{2}+2\eta(\nu)^{\nu}\frac{[\lambda]_{v}^{\frac{\nu}{2}}}{[\tau]_{v}}|[\theta]_{v}|^{\nu}\right\}.\label{MAP_estimator_theta1}
\intertext{We will in the following consider the case $0<\nu<2$. We define}
&\Lambda(\nu,\lambda,\tau)\mathdef 2\eta(\nu)^{\nu}\frac{[\lambda]_{v}^{\frac{\nu}{2}}}{[\tau]_{v}}\label{def_Lambda}\\
&\theta\mathdef\Lambda^{\frac{1}{2-\nu}}(\nu,\lambda,\tau)\bar{\theta}\label{def_bar_theta}\\
&x\mathdef\Lambda^{\frac{1}{2-\nu}}(\nu,\lambda,\tau)\bar{x}.\label{def_bar_x}
\intertext{We see that the problem to solve may be written}
&{\bar{\theta}}^{*}=\text{arg min}_{[\bar{\theta}]_{v}\in\mathbb{R}}\left\{[\bar{x}-\bar{\theta}]_{v}^{2}+|[\bar{\theta}]_{v}|^{\nu}\right\}
\label{MAP_estimator_theta}
\intertext{The equation (\ref{MAP_estimator_theta}) may be solved numerically by means of standard numerical software or simply by linear interpolation as follows. Define}
&R(\bar{\theta})\mathdef [\bar{x}-\bar{\theta}]_{v}^{2}+|[\bar{\theta}]_{v}|^{\nu}\label{def_R}
\intertext{assuming $\bar{\theta}\neq 0$ we may then write}
&\frac{dR(\bar{\theta})}{d\bar{\theta}}=-2[\bar{x}-\bar{\theta}]_{v}+\nu\cdot\sgn([\bar{\theta}]_{v})|[\bar{\theta}]_{v}|^{\nu-1},\ [\bar{\theta}]_{v}\neq 0.\label{MAP_equation}
\intertext{We observe by (\ref{def_R}) that $R(\bar{\theta})$ is a convex function of $\bar{\theta}$ for $1\leq\nu<2$ and therefore $\bar{\theta}^{*}$ is given by $\frac{dR(\bar{\theta})}{d\bar{\theta}}=0$. It was shown in \cite{Moulin-Liu:1999} that in the case $0<\nu\leq 1$ there exists a threshold $t_{\nu}>0$ such that $|x|<t_{\nu}\Leftrightarrow\theta^{*}=0$ with}
&t_{\nu}\mathdef D_{\nu}\cdot\frac{[\lambda]_{v}^{\frac{\nu/2}{2-\nu}}}{[\tau]_{v}^{\frac{1}{2-\nu}}},\ 0<\nu\leq 1\label{threshold_size}\\
\intertext{where}
&D_{\nu}\mathdef(2-\nu)(2-2\nu)^{-\frac{1-\nu}{2-\nu}}\eta(\nu)^{\frac{\nu}{2-\nu}}.\label{def_D_nu}
\intertext{This yields}
&[\bar{\theta}^{*}]_{v} = 0 \Leftrightarrow |[\bar{x}]_{v}|<\bar{t}_{\nu}\mathdef\Lambda^{-\frac{1}{2-\nu}}t_{\nu}=
2^{-\frac{1}{2-\nu}}(2-\nu)(2-2\nu)^{-\frac{1-\nu}{2-\nu}}.\label{bar_threshold}
\intertext{We note that one may show that $0<\nu\leq 1\Rightarrow 0<\bar{t}_{\nu}<1$. We observe that}
&0=\left.\frac{dR(\bar{\theta})}{d\bar{\theta}}\right|_{\bar{\theta}=\bar{\theta}^{*}}\Rightarrow[\bar{x}]_{v}=[\bar{\theta}^{*}]_{v}+\frac{\nu}{2}\sgn([\bar{\theta}^{*}]_{v})|[\bar{\theta}^{*}]_{v}|^{\nu-1},\ [\bar{\theta}^{*}]_{v}\neq 0.\label{MAP_bar_x_of_bar_theta}
\intertext{The expression (\ref{MAP_bar_x_of_bar_theta}) applies to $|[\bar{x}]_{v}|\geq\bar{t}_{\nu}$ if $0<\nu\leq 1$ and (\ref{MAP_bar_x_of_bar_theta}) applies to 
all $[\bar{x}]_{v}$ if $1<\nu<2$. We further observe that the GGD-MAP estimator $\theta^{*}(x)$ and $\bar{\theta}^{*}(\bar{x})$ exhibit step discontinuities at 
$x=\pm t_{\nu}$, $\bar{x}=\pm\bar{t}_{\nu}$, respectively, when $0<\nu<1$: By (\ref{MAP_bar_x_of_bar_theta}) we see that}
&\bar{t}_{\nu}=\lim_{\bar{x}\rightarrow \bar{t}_{\nu}^{+}}[\bar{x}]_{v}=\lim_{\bar{x}\rightarrow\bar{t}_{\nu}^{+}}\left\{[\bar{\theta}^{*}(\bar{x})]_{v}+\frac{\nu}{2}\sgn([\bar{\theta}^{*}(\bar{x})]_{v})|[\bar{\theta}^{*}(\bar{x})]_{v}|^{\nu-1}\right\}\label{GGD_MAP_discontinuity}
\intertext{and while the lefthand side of (\ref{GGD_MAP_discontinuity}) is finite, the righthand side increases to $+\infty$ as $\bar{\theta}^{*}\rightarrow 0^{+}$, if $0<\nu<1$. Therefore, if $0<\nu<1$, there must exist a number $\bar{s}_{\nu}>0$ depending on $\nu$ such that $|\bar{x}|>\bar{t}_{\nu}\Rightarrow |\bar{\theta}^{*}(x)|\geq \bar{s}_{\nu}>0$. The size $\bar{s}_{\nu}$ of the step discontinuity may be computed (numerically) for given $0<\nu<1$ by solving}
&\bar{t}_{\nu}=\bar{s}_{\nu}+\frac{\nu}{2}\bar{s}_{\nu}^{\nu-1}.\label{GGD_MAP_estimator_step_size}
\intertext{We note that by (\ref{GGD_MAP_estimator_step_size}), (\ref{bar_threshold}) we have $\bar{s}_{\nu}\rightarrow\bar{t}_{\nu}\rightarrow 1$ as $\nu\rightarrow 0^{+}$ and $\bar{s}_{\nu}\rightarrow(\bar{t}_{\nu}-1/2)\rightarrow 0$ as $\nu\rightarrow 1^{-}$. 
%SAY SOMETHING OF THE IMPLICATIONS OF THIS FACT TO THE SIZE OF ZETA HERE. ALSO, COULD THE SIZE OF THE STEP BE ANALYTICALLY ESTIMATED FROM THE ESTIMATOR EQUATION?
}\nonumber
\end{align}
One can compile lookup tables of pairs of corresponding values $(\bar{x}$,$\bar{\theta}^{*})$ to the equation (\ref{MAP_estimator_theta}) by 
discretizing $\bar{\theta}^{*}$ to some specific precision $\Delta\bar{\theta}^{*}$ and then use equations (\ref{MAP_bar_x_of_bar_theta}), (\ref{bar_threshold}) to compute corresponding 
pairs of values $(\bar{x},\bar{\theta}^{*})$. Since we ultimately want 
the estimated value $\theta^{*}$ to some precision $\Delta\theta^{*}$, we have to ensure that the lookup table of pairs of values $(\bar{x}$,$\bar{\theta}^{*})$ 
is computed on a sufficiently fine grid with stepsize $\Delta\bar{\theta}^{*}$ yielding a sufficient precision $\Delta\theta^{*}=\Lambda^{\frac{1}{2-\nu}}(\nu,\lambda,\tau)\Delta\bar{\theta}^{*}$ when transforming by the formula (\ref{def_bar_theta}). Letting $\delta>0$ denote the desired precision on the parameters $\theta^{*}$, then it suffices to demand
\begin{align}
&\delta\geq\Lambda^{\frac{1}{2-\nu}}(\nu,\lambda,\tau)\Delta\bar{\theta}^{*}.\label{upper_bound_delta_bar_theta1}
\intertext{Rissanen in \cite{Rissanen:1998b} computed the asymptotically MDL-optimal discretization $\delta^{*}$ on the parameters which parameterize a $n$-variate distribution. In Proposition \ref{INMDL_optimal_quantization} in a previous section we presented a result on the {\em posterior} optimal discretization of parameters which deviates from the MDL-optimal $\delta^{*}$ in that it suggests a data-driven, possibly coarser discretization of the parameters. However, Proposition \ref{INMDL_optimal_quantization} shows that MDL-optimal discretization $\delta^{*}$ is a lower bound on the posterior optimal discretization $\delta$ (since $\text{E}_{\theta}[\lambda\theta^{2}]=1$ and $d/n<1$), and so  
in the $n$-variate IID case of a gaussian likelihood with deviation $\sigma$ we will use}
&\delta^{*}=\frac{\sigma}{\sqrt{n}}\label{optimal_parameter_precicion}
\intertext{and by (\ref{upper_bound_delta_bar_theta1}) we then find an upper bound for $\Delta\bar{\theta}^{*}$ to be}
&\Delta\bar{\theta}^{*}\leq\frac{\tau^{-\frac{1}{2}}}{\sqrt{n}}\cdot\Lambda^{-\frac{1}{2-\nu}}(\nu,\lambda,\tau)=\frac{\tau^{-\frac{1}{2}}}{\sqrt{n}}\cdot\left(
2\eta(\nu)^{\nu}\frac{[\lambda]_{v}^{\frac{\nu}{2}}}{[\tau]_{v}}\right)^{-\frac{1}{2-\nu}}\nonumber\\
&=\frac{1}{\sqrt{n}}\cdot\left(2\eta(\nu)^{\nu}\frac{[\lambda]_{v}^{\frac{\nu}{2}}}{[\tau]^{\frac{\nu}{2}}_{v}}\right)^{-\frac{1}{2-\nu}}
=\frac{1}{\sqrt{n}}\cdot\left(\frac{\frac{n}{d}\Omega(\lambda,\tau)}{2^{\frac{2}{\nu}}\eta(\nu)^{2}}\right)^{\frac{\nu/2}{2-\nu}}.\label{upper_bound_delta_bar_theta2}
\intertext{For most datasets of interest we may bound $\Omega(\lambda,\tau)$ from below by $1$ (which means that we exclude data models where the noise in the data has greater power than the signal part of the data). We define}
&\Delta \bar{x}\mathdef \frac{d \bar{x}}{d\bar{\theta}^{*}}\Delta\bar{\theta}^{*}\label{def_Delta_bar_x}
\intertext{then by differentiating (\ref{MAP_bar_x_of_bar_theta}) we get}
&\Delta\bar{x}=\left(1+\frac{\nu(\nu-1)}{2}\left|[\bar{\theta}^{*}]_{v}\right|^{\nu-2}\right)\Delta\bar{\theta}^{*},\ [\bar{\theta}^{*}]_{v}\neq 0\label{delta_bar_x_by_delta_bar_theta}
\intertext{which corresponds to}
&\Delta x=\left(1+\frac{\nu(\nu-1)}{2}\Lambda(\nu,\lambda,\tau)\left|[\theta^{*}]_{v}\right|^{\nu-2}\right)\Delta\theta^{*},\ [\theta^{*}]_{v}\neq 0\label{delta_x_by_delta_theta}
\intertext{Given a data value $\hat{\bar{x}}$, gridpoint pairs $(\bar{x}_{i},\bar{\theta}_{i}^{*})$ and 
$(\bar{x}_{i+1},\bar{\theta}_{i+1}^{*})$ with $\bar{x}_{i}\leq \hat{\bar{x}}\leq\bar{x}_{i+1}$, we define the estimated parameter value $\hat{\bar{\theta}}^{*}(\hat{\bar{x}})$ by the linear interpolation}
&\hat{\bar{\theta}}^{*}\mathdef\bar{\theta}_{i}^{*}+(\hat{\bar{x}}-\bar{x}_{i})\cdot\frac{\bar{\theta}_{i+1}^{*}-\bar{\theta}_{i}^{*}}{\bar{x}_{i+1}-\bar{x}_{i}}\label{interpolation_scheme_bar_theta}
\intertext{Expression (\ref{delta_bar_x_by_delta_bar_theta})  can be used to compute a bound on the interpolation error for $\hat{\bar{\theta}}^{*}$ for given gridsize $\Delta\bar{\theta}^{*}$. The interpolation error $\Delta\hat{\bar{\theta}}^{*}$ in the linear interpolation estimate $\hat{\bar{\theta}}^{*}(\hat{\bar{x}})$ may by equation (\ref{delta_bar_x_by_delta_bar_theta}) be bounded as follows}
&|\Delta\hat{\bar{\theta}}^{*}|\leq\min\left(\left|\hat{\bar{x}}-\bar{x}_{i}\right|,\left|\hat{\bar{x}}-\bar{x}_{i+1}\right|\right)\sup_{\bar{\theta}^{*}\in(\bar{\theta}_{i}^{*},\bar{\theta}_{i+1}^{*})}\left|1+\frac{\nu(\nu-1)}{2}\left|[\bar{\theta}^{*}]_{v}\right|^{\nu-2}\right|^{-1}.\label{error_bound_interpolated_bar_theta}
\end{align}

\section{The marginal normalization $C_{\gamma_{d}}$ for GGD priors}
We need to calulate the Fisher matrix $\bm{E}(\lambda)$ defined in (\ref{def_prior_fisher_matrix}) and $\Psi(\bm{\theta},\lambda)$ defined in 
(\ref{def_Psi}) and the invariant estimator 
$\lambda^{*}$ defined in (\ref{def_invariant_lambda_estimator}) to be able to compute $\Psi_{\lambda\lambda}(\bm{\theta}^{*},\lambda^{*})$
which is part of the formula for the marginal distribution given in Corollary \ref{corollary_new_marginal}. Plugging the definition 
(\ref{form_GGD_density}) into the defining formulas we get
\begin{align}
&\bm{E}(\lambda)\mathdef-E_{\bm{\theta}}\frac{\partial^{2}\log\pi(\bm{\theta}|\lambda)}{\partial\lambda^{2}}\nonumber\\
&=-E_{\bm{\theta}}\left\{-\frac{d/2}{\lambda^{2}}-\frac{\nu}{2}\left(\frac{\nu}{2}-1\right)\eta(\nu)^{\nu}\lambda^{\nu/2-2}\sum_{i=1}^{d}|\theta_{i}|^{\nu}\right\}\nonumber\\
\intertext{Now, a straightforward calculation yields:}
&E_{\theta}\{|\theta|^{\nu}\}=\frac{\Gamma\left(\frac{1}{\nu}+1\right)}{\Gamma\left(\frac{1}{\nu}\right)\eta(\nu)^{\nu}}\lambda^{-\nu/2}\label{GGD_side_calculation1}
\intertext{and we may then write}
&\bm{E}(\lambda)=\frac{d/2}{\lambda^{2}}+\frac{\nu}{2}\left(\frac{\nu}{2}-1\right)\eta(\nu)^{\nu}\lambda^{\nu/2-2}\sum_{i=1}^{d}\frac{\Gamma\left(\frac{1}{\nu}+1\right)}{\Gamma\left(\frac{1}{\nu}\right)\eta(\nu)^{\nu}}\lambda^{-\nu/2}=\frac{\nu d/4}{\lambda^{2}}.\label{GGD_fisher_matrix}
\intertext{We may then calculate $\Psi(\bm{\theta},\lambda)$ as}
&\Psi(\bm{\theta},\lambda)\mathdef-\log\left[\frac{\pi(\bm{\theta}|\lambda)}{|\bm{E}(\lambda)|^{1/2}}\right]_{v}=\frac{1}{2}\log{\left(\frac{\nu d}{4}\right)}-\log[\lambda]_{v}-\log{\left[\pi(\bm{\theta}|\lambda)\right]_{v}}\label{GGD_Psi_form1}
\intertext{and $\lambda^{*}$ then becomes}
&\lambda^{*}\mathdef\text{arg inf}_{\lambda>0}\Psi(\bm{\theta},\lambda)=\frac{(d+2)^{2/\nu}}{\nu^{2/\nu}\eta(\nu)^{2}\left(\sum_{i=1}^{d}|\theta_{i}|^{\nu}\right)^{2/\nu}}.\label{GGD_lambda_estimator_form1}
\intertext{For notational convenience, we define}
&R_{\nu}^{\nu}(\bm{\theta})\mathdef\sum_{i=1}^{d}|\theta_{i}|^{\nu}.\label{def_R_nu}
\intertext{We may now proceed to calculate}
&\Psi_{\lambda\lambda}(\bm{\theta},\lambda)=\frac{\partial^{2}}{\partial\lambda^{2}}\left(-\frac{d+2}{2}\log[\lambda]_{v}+\eta(\nu)^{\nu}[\lambda]_{v}^{\nu/2}R_{\nu}^{\nu}(\bm{\theta})\right)\nonumber\\
&=\frac{d+2}{2}\lambda^{-2}\left(1+\frac{\nu}{d+2}\left(\frac{\nu}{2}-1\right)\left[\eta(\nu)\lambda^{1/2}R_{\nu}(\bm{\theta})\right]^{\nu}\right)
\label{diff_Psi}
\intertext{and we may now by means of (\ref{GGD_lambda_estimator_form1}) evaluate}
&|\Psi_{\lambda\lambda}(\bm{\theta},\lambda^{*})|^{1/2}=\frac{1}{\lambda^{*}}\left(\frac{\nu(d+2)}{4}\right)^{1/2}\label{Hessian_prior}
\intertext{We may now calculate the quantization induced by the mapping $\chi(\hat{\lambda})$ on the parameter $\lambda$ as described in Proposition 
\ref{INMDL_optimal_quantization}. We have}
&\Delta\lambda=\chi^{\prime}(\hat{\lambda})|\hat{\Psi}(\hat{\bm{\theta}}^{*},\hat{\lambda}^{*})|^{-1/2}\cdot c_{\hat{\lambda}}=|\Psi_{\lambda\lambda}(\bm{\theta},\lambda^{*})|^{-1/2}\cdot c_{\hat{\lambda}}\nonumber\\
&=\frac{2}{\sqrt{\nu(d+2)}}\cdot\lambda^{*}\cdot c_{\hat{\lambda}}.\label{GGD_lambda_quant}
\intertext{We may now calculate the map $\chi:\hat{\lambda}\rightarrow\lambda$ which defines the invariant parameterization $\hat{\lambda}$.
Define the log likelihood $\hat{L}$ by}
&\hat{L}(\hat{\bm{\theta}})\mathdef-\log\left[\pi\left(\bm{\theta}\left|\right.\chi(\hat{\lambda})\right)\right]_{v}\nonumber\\
&=\frac{d}{2}\log\left[\chi(\hat{\lambda})\right]_{v}
+d\log\left(\frac{\nu\eta(\nu)}{2\Gamma\left(\frac{1}{\nu}\right)}\right)-\chi^{\nu/2}(\hat{\lambda})\eta(\nu)^{\nu}\sum_{i=1}^{d}|\theta_{i}|^{\nu}\label{def_hat_L}
\intertext{Now, the Fisher matrix $\hat{\bm{E}}(\hat{\bm{\theta}})$ in the invariant parameterization $\hat{\lambda}$ is defined by}
&\hat{\bm{E}}(\hat{\lambda})\mathdef -E_{\hat{\theta}}\left\{\frac{\partial^{2}}{\partial\hat{\lambda}^{2}}\hat{L}(\hat{\bm{\theta}})\right\}\label{def_invariant_fisher}
\intertext{To make the parameterization $\hat{\lambda}$ invariant, we have to demand}
&|\hat{\bm{E}}(\hat{\lambda})|=\bar{\lambda}^{-2}
\intertext{where $\bar{\lambda}>0$ is some constant number. This yields the equation}
&-E_{\hat{\theta}}\left\{\hat{L}(\hat{\bm{\theta}})\right\}=r(\hat{\lambda})\mathdef\hat{c}_{0}+\hat{c}_{1}\hat{\lambda}+\frac{1}{2}\bar{\lambda}^{-2}\hat{\lambda}^{2}\label{L_hat_eq1}
\intertext{for some real constants $\hat{c}_{0}$, $\hat{c}_{1}$. Plugging in the expression $\hat{L}(\hat{\bm{\theta}})$ from (\ref{def_hat_L}) into (\ref{L_hat_eq1}) yields}
&r(\hat{\lambda})=-\frac{d}{2}\log\left[\chi(\hat{\lambda})\right]_{v}-d\log\left(\frac{\nu\eta(\nu)}{2\Gamma\left(\frac{1}{\nu}\right)}\right)
+\chi^{\nu/2}(\hat{\lambda})\eta(\nu)^{\nu}\sum_{i=1}^{d}E_{\hat{\theta}}\left\{|\theta_{i}|^{\nu}\right\}\nonumber\\
&=-\frac{d}{2}\log\left[\chi(\hat{\lambda})\right]_{v}-d\log\left(\frac{\nu\eta(\nu)}{2\Gamma\left(\frac{1}{\nu}\right)}\right)
+d\cdot\frac{\chi^{\nu/2}(\hat{\lambda})\eta(\nu)^{\nu}\Gamma\left(\frac{1}{\nu}+1\right)}{\Gamma\left(\frac{1}{\nu}\right)\eta(\nu)^{\nu}}\chi^{-\nu/2}(\hat{\lambda}).\nonumber
\intertext{Now, solving for $\chi(\hat{\lambda})$ yields}
&\chi(\hat{\lambda})=\left(\frac{\nu\eta(\nu)}{2\Gamma\left(\frac{1}{\nu}\right)}\right)^{-2}\exp\left(2/\nu\right)\exp\left(-\frac{\left(\hat{\lambda}+\hat{c}_{1}\bar{\lambda}^{2}\right)^{2}}{d\bar{\lambda}^{2}}-\frac{2\hat{c}_{0}-\hat{c}_{1}^{2}\bar{\lambda}^{2}}{d}\right).\nonumber
\intertext{Specifying the initial condition $\chi(0)=\lambda_{0}$, we get}
&\chi(\hat{\lambda})=\lambda_{0}\exp\left(-\frac{\left(\hat{\lambda}+\hat{c}_{1}\bar{\lambda}^{2}\right)^{2}}{d\bar{\lambda}^{2}}+\frac{\hat{c}_{1}^{2}\bar{\lambda}^{2}}{d}\right).\label{chi_map}
\intertext{Using (\ref{GGD_lambda_estimator_form1}) we now evaluate the integral}
&\int_{\bm{\theta}\in\Theta}\frac{\pi(\bm{\theta}|\lambda^{*})}{|\Psi_{\lambda\lambda}(\bm{\theta},\lambda^{*})|^{1/2}}\ d\bm{\theta}
=\left(\frac{\nu(d+2)}{4}\right)^{-1/2}\left(\frac{\nu\eta(\nu)}{2\Gamma(1/\nu)}\right)^{d}\nonumber\\
&\times\int_{\bm{\theta}\in\Theta}(\lambda^{*})^{\frac{d+2}{2}}
\exp\left(-\eta(\nu)^{\nu}(\lambda^{*})^{\nu/2}\sum_{i=1}^{d}|\theta_{i}|^{\nu}\right)\ d\bm{\theta}\nonumber\\
&=\left(\frac{\nu(d+2)}{4}\right)^{-1/2}\left(\frac{\nu\eta(\nu)}{2\Gamma(1/\nu)}\right)^{d}\exp\left(-\frac{d+2}{\nu}\right)\left(\frac{(d+2)^{\frac{2}{\nu}}}{\nu^{\frac{2}{\nu}}\eta(\nu)^{2}}\right)^{\frac{d+2}{2}}\nonumber\\
&\times\int_{\bm{\theta}\in\Theta}\left(\sum_{i=1}^{d}|\theta_{i}|^{\nu}\right)^{-\frac{d+2}{\nu}}\ d\bm{\theta}.\label{C_integral1}
\intertext{Now, using}
&\Theta=\left\{\bm{\theta}\in\mathbb{R}^{d}\left.\right|0<(r_{\nu})^{\nu}<\sum_{i=1}^{d}|\theta_{i}|^{\nu}<\sum_{i=1}^{d}|\theta_{i}^{*}|^{\nu}\mathdef (R_{\nu})^{\nu}\right\}\label{def_Theta_region}
\intertext{and performing a suitable change of coordinates (see \cite{Gradshteyn:2000}, page 610) the integral (\ref{C_integral1}) evaluates to}
&\int_{\bm{\theta}\in\Theta}\frac{\pi(\bm{\theta}|\lambda^{*})}{|\Psi_{\lambda\lambda}(\bm{\theta},\lambda^{*})|^{1/2}}\ d\bm{\theta}\nonumber\\
&=\left(\frac{\nu(d+2)}{4}\right)^{-1/2}\left(\frac{\nu\eta(\nu)}{2\Gamma\left(1/\nu\right)}\right)^{d}\exp\left(-\frac{d+2}{\nu}\right)\left(\frac{(d+2)^{\frac{2}{\nu}}}{\nu^{\frac{2}{\nu}}\eta(\nu)^{2}}\right)^{\frac{d+2}{2}}\nonumber\\
&\times2^{d}\frac{\Gamma\left(\frac{1}{\nu}\right)^{d}}{\nu^{d}\Gamma\left(\frac{d}{\nu}\right)}\left(r_{\nu}^{-2}\int_{1}^{\infty}x^{-\frac{2}{\nu}-1}\ dx
-R_{\nu}^{-2}\int_{1}^{\infty}x^{-\frac{2}{\nu}-1}\ dx\right)\nonumber\\
&=(2\pi)^{-1/2}(d+2)^{\frac{2}{\nu}-\frac{1}{2}}d^{\frac{1}{2}}\eta(\nu)^{-1}\nu^{-\frac{2}{\nu}}
r_{\nu}^{-2}\left(1-\left(\frac{r_{\nu}}{R_{\nu}}\right)^{2}\right).\label{C_integral2}
\end{align}
This result may be plugged into Proposition \ref{Marginal_Renormalization_Constant} to yield the precise codelength contribution from 
the term $\log{C_{\gamma_{d}}}$. We observe that it will only contribute constant terms plus a $(2/\nu)\log{d}$ term.

%\section{The model selection algorithm for IID GGD distributed parameters}
\section{The model selection algorithm for GGD distributed parameters}
By Proposition \ref{INMDL_Selection_Principle} we see that we will have to investigate the behaviour of $-\log{\pi(\bm{\theta}^{*}|\lambda^{*})}$.
Choosing the ML estimator (\ref{GGD_lambda_estimator_form1}) for $\lambda^{*}$, we get
\begin{align}
&C(\bm{x}_{\parallel}(i)|S_{i-1})\mathdef-(n-d+2)\frac{|\bm{x}_{\parallel}(i)|}{\|\bm{x}\|_{2}^{2}-\|\bm{x}_{\parallel}\|_{2}^{2}}\nonumber\\
&-\frac{\partial}{\partial|\bm{x}_{\parallel}(i)|}\log\left(\frac{\pi(\bm{\theta}^{*}|\lambda^{*})}{|\Psi_{\lambda\lambda}(\bm{\theta}^{*},\lambda^{*})|^{\frac{1}{2}}}\right)\nonumber\\
&=-(n-d+2)\frac{|\bm{x}_{\parallel}(i)|}{\|\bm{x}\|_{2}^{2}-\|\bm{x}_{\parallel}\|_{2}^{2}}\nonumber\\
&+(d+2)\frac{|\theta_{i}^{*}|^{\nu-1}}{\sum_{j=1}^{d}|\theta_{j}^{*}|^{\nu}}\frac{\partial|\theta_{i}^{*}(\bm{x}_{\parallel}(i))|}{\partial|\bm{x}_{\parallel}(i)|},\ 1\leq i\leq d\label{GGD_evaluated_selection_criterion}
\intertext{It is easy to see that with a possible exception for the derivative term, all terms in (\ref{GGD_evaluated_selection_criterion}) are decreasing functions of $|\bm{x}_{\parallel}(i)|$. As for the derivative term, we see from (\ref{MAP_bar_x_of_bar_theta}) that this term is positive and bounded by 1 for $0<\nu\leq 2$ for $|\bm{x}_{\parallel}(i)|$ sufficiently large and so we may conclude that $C(\bm{x}_{\parallel}(i)|S_{i-1})$ is
a decreasing function of $|\bm{x}_{\parallel}(i)|$. Therefore, the $d$ nonzero elements of $\bm{x}_{\parallel}\in\mathbb{R}^{n}$
are the $d$ largest $|\bm{x}(i)|$ in the dataset $\bm{x}\in\mathbb{R}^{n}$.}\nonumber
\end{align}

%\section{The approximation errors in using the Laplace formula on the GGD model}
\section{The approximation errors for the GGD model}
We need to control the approximation error terms $\kappa$ and $\xi$ as defined in the proof of Theorem \ref{theorem1}. 
An easily computable upper bound for the error term $\kappa$ is given in Theorem \ref{theorem1}. The upper bound for 
the error term $\xi$ as shown in Theorem \ref{theorem1} may be considerably simplified in the case of a GGD prior 
distribution on the noiseless data. We have the following result:
\begin{proposition}
One may verify that the GGD distributions satisfy the conditions in Theorem \ref{theorem1}. Using the notation from Theorem \ref{theorem1}
we may state the following upper bound on the number $\xi$ for GGD prior distributions
\begin{align}
&\xi+1\leq\exp\exp\left\{d\left(\frac{1}{2}+\log{2}\right)-\frac{1}{2}\tau_{1}\|\bm{x}_{\parallel}\|_{2}^{2}-\sum_{i=1}^{d}\log\left|\tau_{1}^{\frac{1}{2}}\bm{x}_{\parallel}(i)\right|\right.\nonumber\\
&\left.+\sum_{i=1}^{d}\left[\eta(\nu)^{2}\frac{\tau_{1} \bm{x}_{\parallel}^{2}(i)}{\frac{n}{d}\Omega(\lambda,\tau_{1})}\right]^{\frac{\nu}{2}}
+\frac{2\eta(\nu)^{\nu}\nu}{(2\pi)^{\frac{1}{2}}}\sum_{i=1}^{d}L_{\nu}\left(\tau_{1}^{\frac{1}{2}}\bm{x}_{\parallel}(i)\right)\right\}\label{xi_GGD_upper_bound}
\intertext{where}
%&\tau_{1}=\tau^{*}\exp\left[-\left(\frac{\log^{2}{N(\lambda,\nu,\gamma_{d})}}{N(\lambda,\nu,\gamma_{d})}\right)^{1/2}\right],\nonumber\\
%&\tau_{2}=\tau^{*}\exp\left[\left(\frac{\log^{2}{N(\lambda,\nu,\gamma_{d})}}{N(\lambda,\nu,\gamma_{d})}\right)^{1/2}\right],\nonumber\\
&\tau^{*}\exp\left[-\left(\frac{\log^{2}{N(\lambda,\nu,\gamma_{d})}}{N(\lambda,\nu,\gamma_{d})}\right)^{1/2}\right]\nonumber\\
&\leq\tau_{1}\leq \exp\left[\left(\frac{\log^{2}{N(\lambda,\nu,\gamma_{d})}}{N(\lambda,\nu,\gamma_{d})}\right)^{1/2}\right],\nonumber\\
&L_{\nu}\left(\tau^{\frac{1}{2}}x\right)=\left\{\begin{array}{ll}\left|\tau^{\frac{1}{2}}x\right|^{\nu-1}\left(\frac{n}{d}\Omega(\lambda,\tau)\right)^{-\frac{\nu}{2}} & \text{ if }0<\nu\leq 1\\
\left(\frac{n}{d}\Omega(\lambda,\tau)\right)^{-\frac{1}{2}}\left(1+\left|\tau^{\frac{1}{2}}x\right|\left(\frac{n}{d}\Omega(\lambda,\tau)\right)^{-\frac{1}{2}}\right)& \text{ if }1<\nu<2.\end{array}\right.\nonumber
\end{align}
%We note that if the result of evaluating (\ref{xi_GGD_upper_bound}) with these values of $\tau_{1},\tau_{2}$ turns out too large to be of any use, we will have to evaluate (\ref{xi_GGD4}) shown in the proof below more accurately.
\end{proposition}
\begin{proof}
First, by (\ref{C9}) and the fact that $\pi_{\lambda}(x)$ is a monotone decreasing function of $|x|$, we observe that
\begin{align}
&\xi+1\leq\prod_{i=1}^{d}\left\{1+\frac{2P_{G}\left(-\tau_{1}^{\frac{1}{2}}\left|\bm{x}_{\parallel}(i)\right|\right)\pi_{\lambda=1}(0)/\pi_{\lambda=1}\left(\left[\frac{\tau_{1} \bm{x}_{\parallel}^{2}(i)}{\frac{n}{d}\Omega(\lambda,\tau_{1})}\right]^{\frac{1}{2}}\right)}{1+\erf(\tau_{1}^{\frac{1}{2}}|\bm{x}_{\parallel}(i)|)-\frac{2C_{\nu}\nu}{(2\pi)^{\frac{1}{2}}}L_{\nu}\left(\tau_{1}^{\frac{1}{2}}\bm{x}_{\parallel}(i)\right)}\right\}.\label{xi_GGD1}
\intertext{Plugging (\ref{form_GGD_density}) into (\ref{xi_GGD1}) we get}
&\xi+1\leq\prod_{i=1}^{d}\left\{1+\frac{2P_{G}\left(-\tau_{1}^{\frac{1}{2}}\left|\bm{x}_{\parallel}(i)\right|\right)\exp\left(\eta(\nu)^{\nu}\left[\frac{\tau_{1} \bm{x}_{\parallel}^{2}(i)}{\frac{n}{d}\Omega(\lambda,\tau_{1})}\right]^{\frac{\nu}{2}}\right)}{1+\erf(\tau_{1}^{\frac{1}{2}}|\bm{x}_{\parallel}(i)|)-\frac{2\eta(\nu)^{\nu}\nu}{(2\pi)^{\frac{1}{2}}}L_{\nu}\left(\tau_{1}^{\frac{1}{2}}\bm{x}_{\parallel}(i)\right)}\right\}.\label{xi_GGD2}
\intertext{Using the bound $P_{G}(-t)\leq t^{-1}\frac{1}{\sqrt{2\pi}}\exp\left(-\frac{1}{2}t^{2}\right),\ \forall\ t\neq 0$, we may write}
&\xi+1\leq\prod_{i=1}^{d}\left\{1+\frac{2\exp\left(-\frac{1}{2}\tau_{1}\bm{x}^{2}_{\parallel}(i)-\log\left|\tau_{1}^{\frac{1}{2}}\bm{x}_{\parallel}(i)\right|+\left[\eta(\nu)^{2}\frac{\tau_{1} \bm{x}_{\parallel}^{2}(i)}{\frac{n}{d}\Omega(\lambda,\tau_{1})}\right]^{\frac{\nu}{2}}\right)}{1+\erf(\tau_{1}^{\frac{1}{2}}|\bm{x}_{\parallel}(i)|)-\frac{2\eta(\nu)^{\nu}\nu}{(2\pi)^{\frac{1}{2}}}L_{\nu}\left(\tau_{1}^{\frac{1}{2}}\bm{x}_{\parallel}(i)\right)}\right\}\nonumber\\
&\leq\exp\left(\sum_{i=1}^{d}\left\{\frac{2\exp\left(-\frac{1}{2}\tau_{1}\bm{x}^{2}_{\parallel}(i)-\log\left|\tau_{1}^{\frac{1}{2}}\bm{x}_{\parallel}(i)\right|+\left[\eta(\nu)^{2}\frac{\tau_{1} \bm{x}_{\parallel}^{2}(i)}{\frac{n}{d}\Omega(\lambda,\tau_{1})}\right]^{\frac{\nu}{2}}\right)}{1+\erf(\tau_{1}^{\frac{1}{2}}|\bm{x}_{\parallel}(i)|)-\frac{2\eta(\nu)^{\nu}\nu}{(2\pi)^{\frac{1}{2}}}L_{\nu}\left(\tau_{1}^{\frac{1}{2}}\bm{x}_{\parallel}(i)\right)}\right\}\right)\label{xi_GGD3}
\intertext{where}
&L_{\nu}\left(\tau^{\frac{1}{2}}x\right)\mathdef\left\{\begin{array}{ll}\left(\tau^{\frac{1}{2}}|x|\right)^{\nu-1}\left(\frac{n}{d}\Omega(\lambda,\tau)\right)^{-\frac{\nu}{2}} & \text{ if }0<\nu\leq 1\\
\left(\frac{n}{d}\Omega(\lambda,\tau)\right)^{-\frac{1}{2}}& \text{ if }1<\nu<2\end{array}\right.\label{def_L_nu2}
\intertext{and}
&\tau_{1}\in I_{\tau}=\left(\tau^{*}\exp\left[-\left(\frac{2\log{N(\lambda,\nu,\gamma_{d})}}{N(\lambda,\nu,\gamma_{d})}\right)^{\frac{1}{2}}\right],\right.\nonumber\\
&\left.\tau^{*}\exp\left[\left(\frac{2\log{N(\lambda,\nu,\gamma_{d})}}{N(\lambda,\nu,\gamma_{d})}\right)^{\frac{1}{2}}\right]\right).\label{bounds_on_taus}
\intertext{We see that the righthandside of (\ref{xi_GGD3}) makes no sense when $\nu\rightarrow 0^{+}$ because $\lim_{\nu\rightarrow 0^{+}}\eta(\nu)^{\nu}=+\infty$ and also the validity of expression (\ref{xi_GGD3}) depends on}
&\frac{2\eta(\nu)^{\nu}\nu}{(2\pi)^{\frac{1}{2}}}L_{\nu}\left(\tau_{1}^{\frac{1}{2}}\bm{x}_{\parallel}(i)\right)\leq 1,\ \forall\ i\in\gamma_{d}.\label{GGD_model_claim1}
\intertext{This lack of generality is due to our choice of technique for estimating $\xi$ in the proof of Theorem \ref{theorem1} where we implicitely assumed}
&\pi_{\lambda=1}\left(\left(\frac{n}{d}\Omega(\lambda,\tau)\right)^{-\frac{1}{2}}\tau^{\frac{1}{2}}x\right)/(\sup_{t\in\mathbb{R}}{\pi_{\lambda=1}(t)})> 2\eta(\nu)^{\nu}\nu(2\pi)^{-\frac{1}{2}}L_{\nu}(\tau^{\frac{1}{2}}\bm{x}_{\parallel}(i)),\nonumber\\
&\forall\ i\in\gamma_{d},\ \forall \tau\in I_{\tau}, \label{implicit_assumption1}
\intertext{and is therefore not due to an intrinsic property of the model. Now, because of (\ref{GGD_model_claim1}) we have}
&-1<-2\eta(\nu)^{\nu}\nu(2\pi)^{-\frac{1}{2}}L_{\nu}(\tau_{1}^{\frac{1}{2}}\bm{x}_{\parallel})+\erf\left(\tau_{1}^{\frac{1}{2}}|\bm{x}_{\parallel}(i)|\right)<1\label{xi_denominator_inequality}
\intertext{and by the inequality $\log(1+x)>x-\frac{1}{2}x^{2},\ \forall\ |x|<1$ we then have}
&\log\left(1-2\eta(\nu)^{\nu}\nu(2\pi)^{-\frac{1}{2}}L_{\nu}(\tau_{1}^{\frac{1}{2}}\bm{x}_{\parallel})+\erf\left(\tau_{1}^{\frac{1}{2}}|\bm{x}_{\parallel}(i)|\right)\right)\nonumber\\
&>2\eta(\nu)^{\nu}\nu(2\pi)^{-\frac{1}{2}}L_{\nu}\left(\tau_{1}^{\frac{1}{2}}\bm{x}_{\parallel}(i)\right)+\erf\left(\tau_{1}^{\frac{1}{2}}|\bm{x}_{\parallel}(i)|\right)\nonumber\\
&-\frac{1}{2}\left(-2\eta(\nu)^{\nu}\nu(2\pi)^{-\frac{1}{2}}L_{\nu}\left(\tau_{1}^{\frac{1}{2}}\bm{x}_{\parallel}(i)\right)+\erf\left(\tau_{1}^{\frac{1}{2}}|\bm{x}_{\parallel}(i)|\right)\right)^{2}.\label{log_xi_denominator_inequality}
\intertext{Using (\ref{log_xi_denominator_inequality}) on the expression (\ref{xi_GGD3}) enables us to write}
&\xi+1\leq\exp\exp\left\{-\frac{1}{2}\tau_{1}\|\bm{x}_{\parallel}\|_{2}^{2}-\sum_{i=1}^{d}\log\left|\tau_{1}^{\frac{1}{2}}\bm{x}_{\parallel}(i)\right|+\sum_{i=1}^{d}\left[\eta(\nu)^{2}\frac{\tau_{1} \bm{x}_{\parallel}^{2}(i)}{\frac{n}{d}\Omega(\lambda,\tau_{1})}\right]^{\frac{\nu}{2}}\right.\nonumber\\
&\left.+d\log{2}+2\eta(\nu)^{\nu}\nu(2\pi)^{-\frac{1}{2}}\sum_{i=1}^{d}L_{\nu}\left(\tau_{1}^{\frac{1}{2}}\bm{x}_{\parallel}(i)\right)-\sum_{i=1}^{d}\erf\left(\tau_{1}^{\frac{1}{2}}|\bm{x}_{\parallel}(i)|\right)\right.\nonumber\\
&\left.+\frac{1}{2}\sum_{i=1}^{d}\left[-2\eta(\nu)^{\nu}\nu(2\pi)^{-\frac{1}{2}}\sum_{i=1}^{d}L_{\nu}\left(\tau_{1}^{\frac{1}{2}}\bm{x}_{\parallel}(i)\right)+\sum_{i=1}^{d}\erf\left(\tau_{1}^{\frac{1}{2}}|\bm{x}_{\parallel}(i)|\right)\right]^{2}\right\}\label{xi_GGD4}\\
&\leq\exp\exp\left\{d\left(\frac{1}{2}+\log{2}\right)-\frac{1}{2}\tau_{1}\|\bm{x}_{\parallel}\|_{2}^{2}-\sum_{i=1}^{d}\log\left|\tau_{1}^{\frac{1}{2}}\bm{x}_{\parallel}(i)\right|\right.\nonumber\\
&\left.+\sum_{i=1}^{d}\left[\eta(\nu)^{2}\frac{\tau_{1} \bm{x}_{\parallel}^{2}(i)}{\frac{n}{d}\Omega(\lambda,\tau_{1})}\right]^{\frac{\nu}{2}}
+2\eta(\nu)^{\nu}\nu(2\pi)^{-\frac{1}{2}}\sum_{i=1}^{d}L_{\nu}\left(\tau_{1}^{\frac{1}{2}}\bm{x}_{\parallel}(i)\right)\right\}\label{xi_GGD5}
\intertext{and we may easily evaluate an upper bound on the righthandside of (\ref{xi_GGD5}) by evaluation with}
&\tau_{1}=\tau^{*}\exp\left[-\left(\frac{\log^{2}{N(\lambda,\nu,\gamma_{d})}}{N(\lambda,\nu,\gamma_{d})}\right)^{1/2}\right],\nonumber\\
\intertext{and}
&\tau_{1}=\tau^{*}\exp\left[\left(\frac{\log^{2}{N(\lambda,\nu,\gamma_{d})}}{N(\lambda,\nu,\gamma_{d})}\right)^{1/2}\right]\nonumber
\intertext{with $N(\lambda,\nu,\gamma_{d})$ as given in Theorem \ref{theorem1}.}\nonumber
\end{align}
\end{proof} 

\section{Numerical methods and experiments}
\begin{figure}[h]
\begin{center}
\includegraphics[scale=0.75]{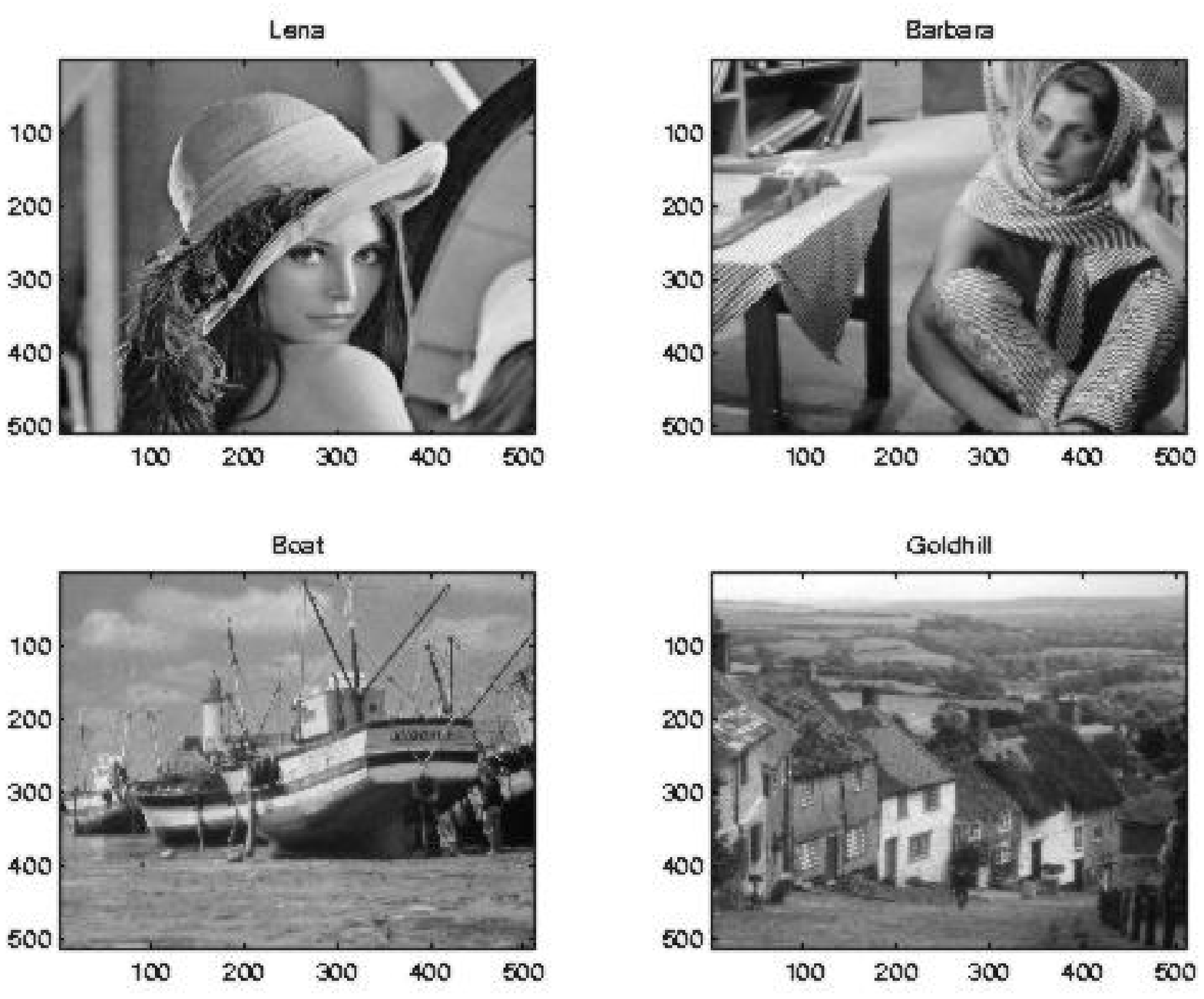}
\end{center}
\caption{Test images.}
\label{TestImages1}
\end{figure}

\begin{figure}[h]
\begin{center}
\includegraphics[scale=0.75]{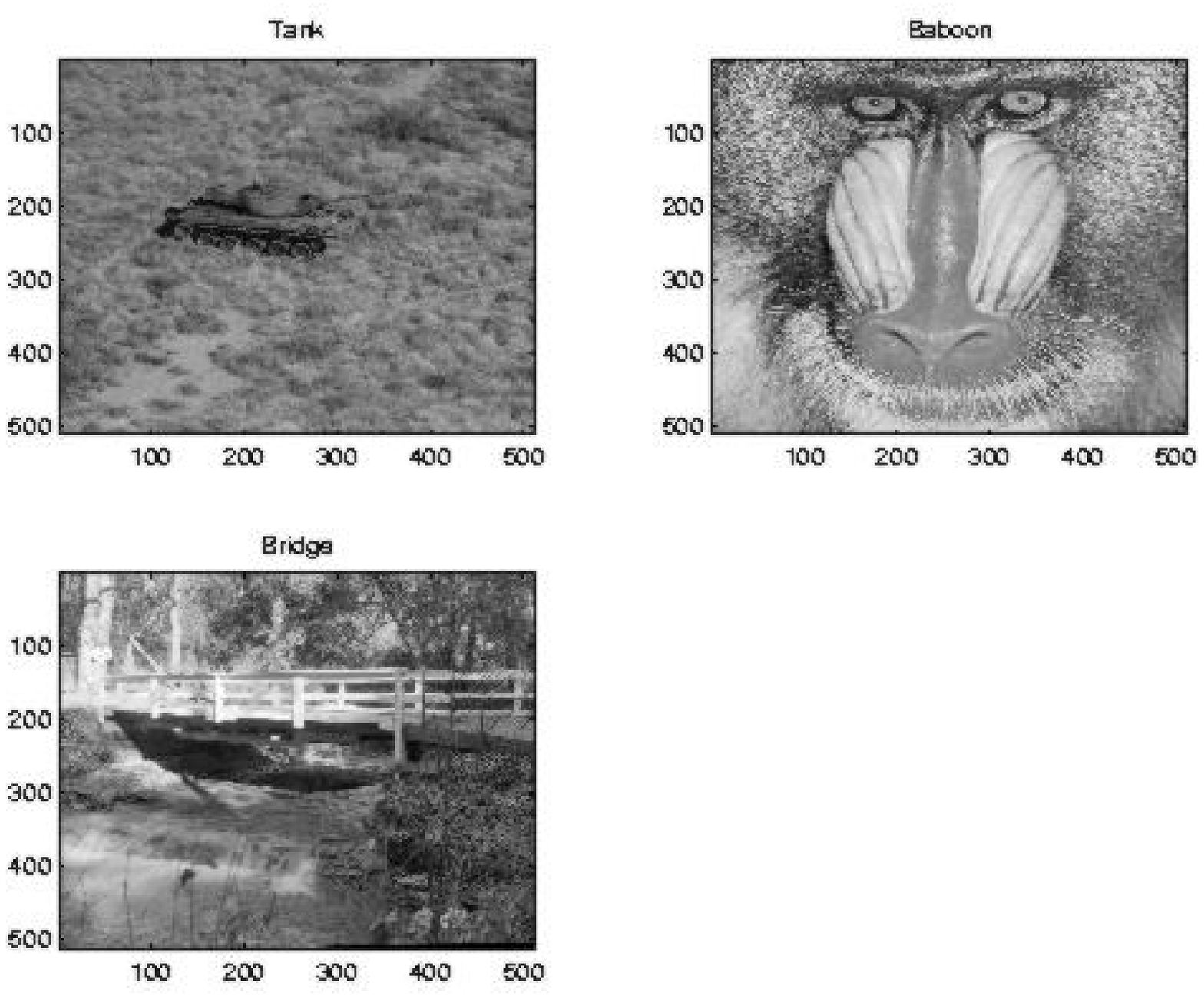}
\end{center}
\caption{Test images.}
\label{TestImages2}
\end{figure}

\begin{figure}[h]
\begin{center}
\includegraphics[scale=0.75]{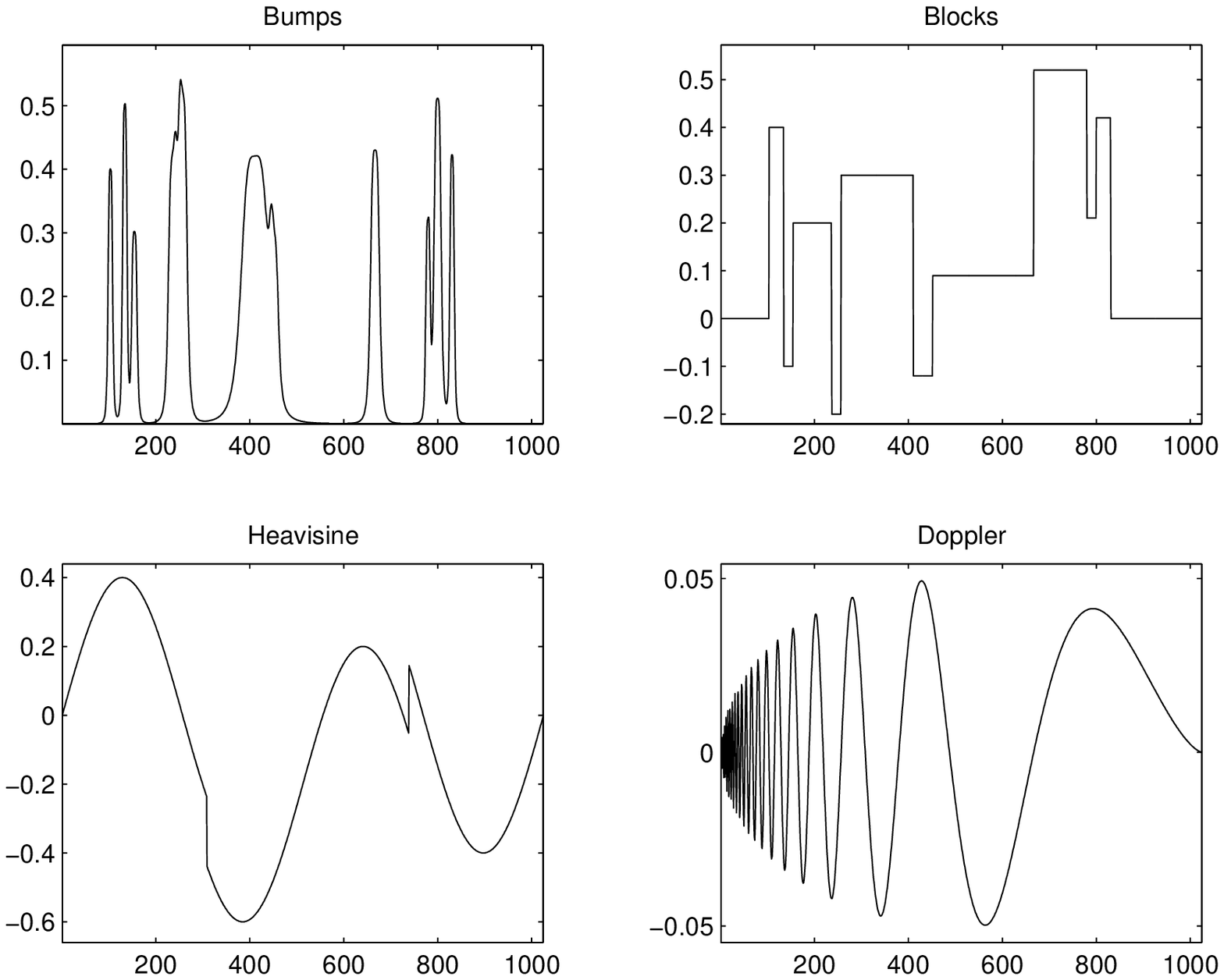}
\end{center}
\caption{Test signals.}
\label{TestSignals1}
\end{figure}

\begin{figure}[h]
\begin{center}
\includegraphics[scale=0.75]{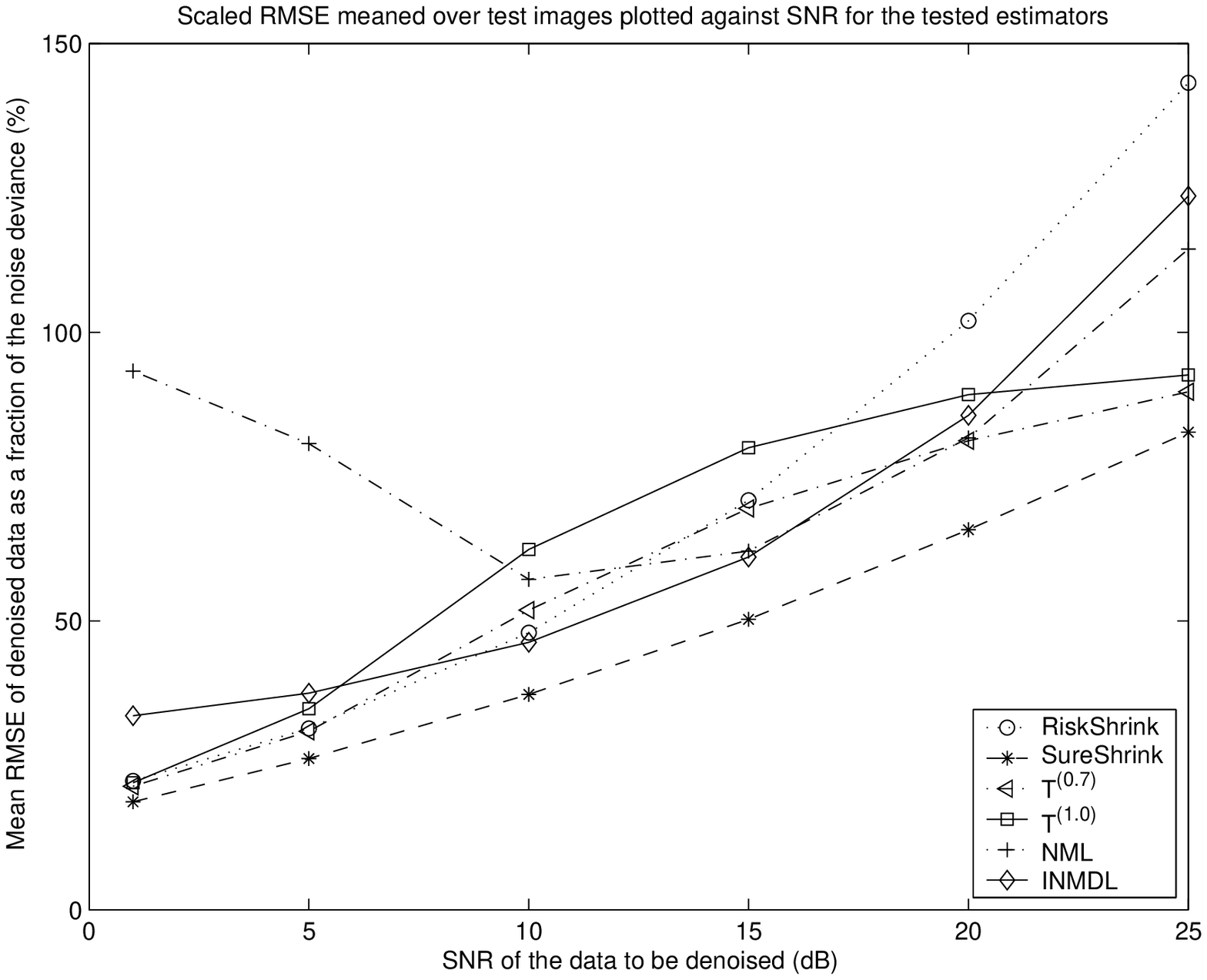}
\end{center}
\caption{Mean estimator performance on the test images.}
\label{RelativeEstimatorPerformance2}
\end{figure}

\begin{figure}[h]
\begin{center}
\includegraphics[scale=0.75]{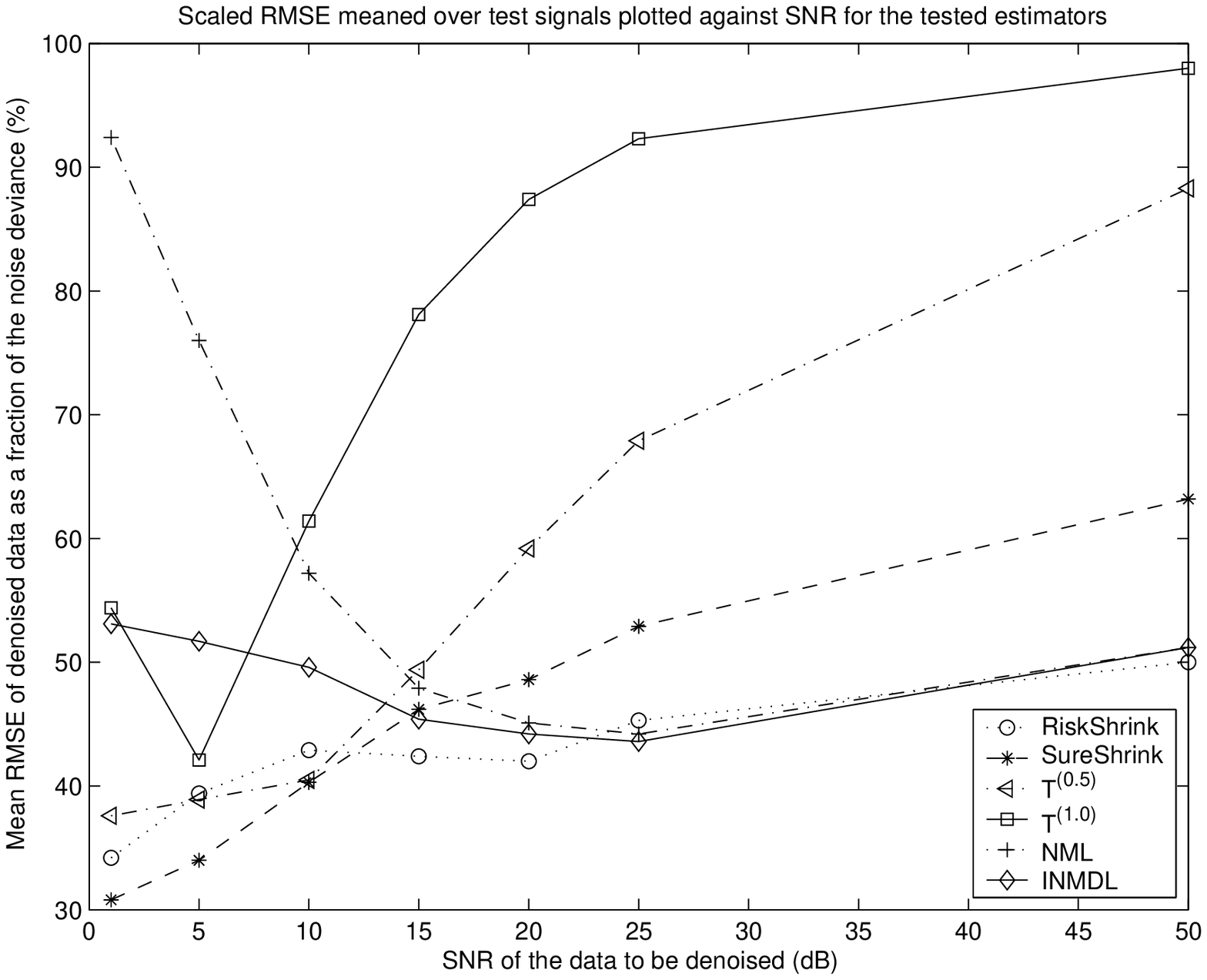}
\end{center}
\caption{Mean estimator performance on the 1-dimensional test signals.}
\label{RelativeEstimatorPerformance1}
\end{figure}

\begin{figure}[h]
\begin{center}
\includegraphics[scale=0.75]{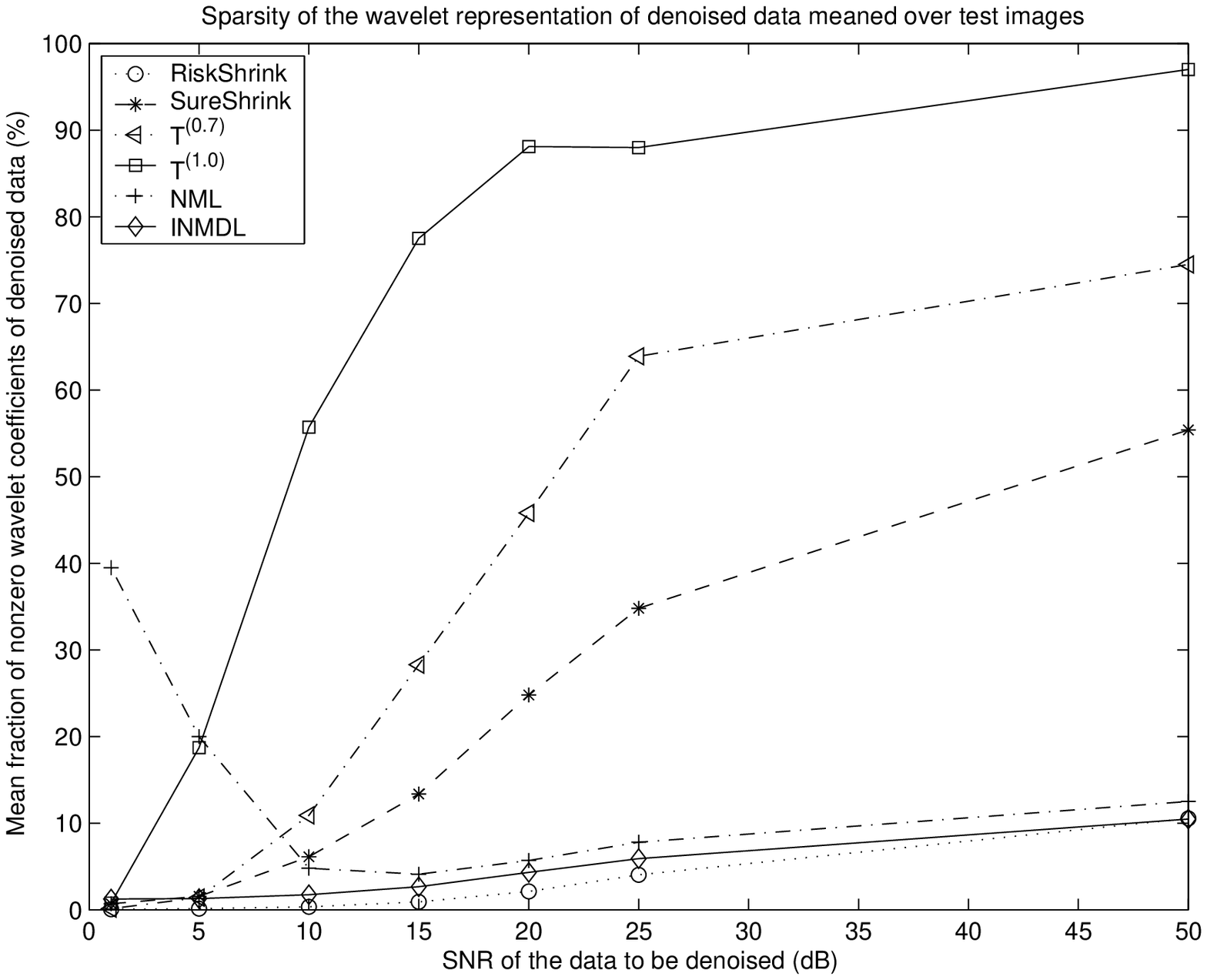}
\end{center}
\caption{The sparsity of the wavelet representation of the denoised test images measured as the fraction of nonzero wavelet 
  coefficient estimates.}
\label{WaveletSparsity2}
\end{figure}

\begin{figure}[h]
\begin{center}
\includegraphics[scale=0.75]{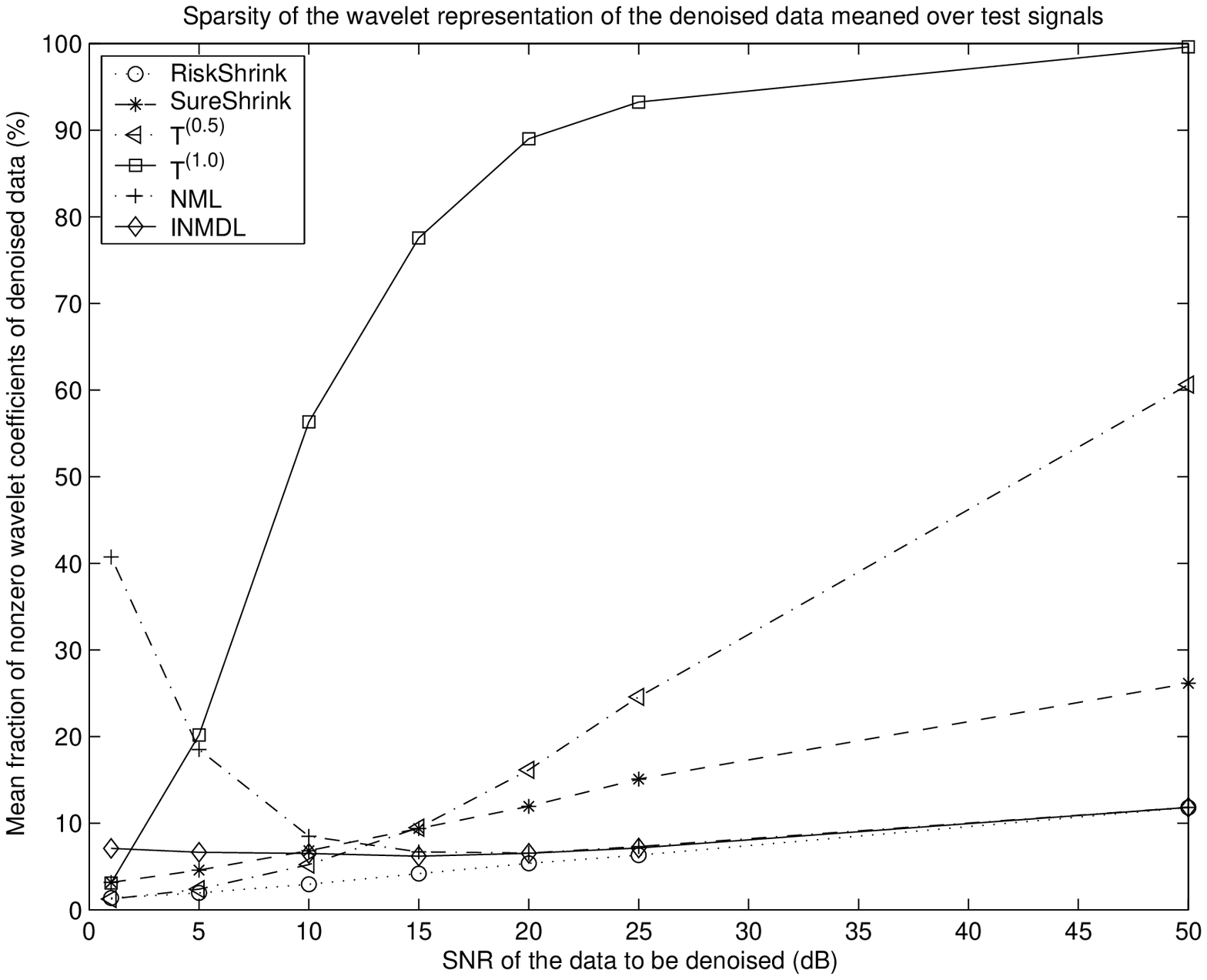}
\end{center}
\caption{The sparsity of the wavelet representation of the denoised 1-dimensional test signals measured as the fraction of nonzero wavelet coefficient estimates.}
\label{WaveletSparsity1}
\end{figure}

In this section we show the performance of our INMDL-algorithm when applied to the problem of estimating various kinds of 1-dimensional data (signals)
and 2-dimensional data (images) embedded in IID gaussian noise (the ''denoising''-problem) and we will compare the performance of INMDL-principle developed in the previous sections to various other kinds of denoising algorithms. Detailed numerical results are shown in the appendix while a graphical overview of estimator performance is shown in Figures \ref{RelativeEstimatorPerformance2}-\ref{WaveletSparsity1}. We will here focus on the NML-principle of Rissanen as presented in \cite{Rissanen:2000}, the RiskShrink-thresholding algorithms as presented in \cite{DJ_Ideal:1994}, \cite{BruceGao:1995a}, \cite{BruceGao:1995c}, (that is the universial hard thresholding scheme with threshold $\sigma\sqrt{2\log{N}}$ where $\sigma^{2}$ is the noise variance),the SureShrink-thresholding algorithm given in \cite{DJ_Adapt:1995} and the MAP-estimator deduced from an IID GGD model applied to the {\em full} data set 
\cite{Moulin-Liu:1999}, that is
\begin{align}
&\bm{x}=\bm{\theta}+\bm{\eta},\ \bm{\theta},\bm{\eta},\bm{x}\in\mathbb{R}^{N}
\intertext{and}
&\theta_{i}\sim \pi_{\lambda,\nu}(\theta_{i}),\ 1\leq i\leq N,\ \eta_{i}\sim\mathcal{N}(0,\tau^{-1/2}),\ 1\leq i\leq N. 
\end{align}
where $\pi_{\lambda,\nu}$ is a GGD distribution with mean zero, second moment $\lambda^{-1/2}$ and shape parameter $\nu$. Note the difference
from the model defined in (\ref{noise_model})-(\ref{prior}) from which we deduced our INMDL principle.
This MAP estimator equals the estimator called $T_{MAP}$ defined in \cite{Hansen-Yu:2000}, except that we use the exact MAP estimator  
(up to interpolation errors in the numerical approximation of this estimator, see expressions 
(\ref{upper_bound_delta_bar_theta2})-(\ref{error_bound_interpolated_bar_theta})) 
for general values on the shape parameter $\nu$, whereas in \cite{Hansen-Yu:2000} they use the MAP estimator for $\nu=1$ which is the soft 
thresholding operator (\ref{soft_threshold}). We adopt similar notation for this estimator: We write 
$T_{MAP}^{(\nu)}$ where $\nu$ signifies the shape parameter in the GGD distribution.
To make the conditions under which our reported numerical experiments were conducted, as clear as possible, we list some remarks:
\begin{enumerate}

\item The image data used in our experiments were mostly collected from the USC-SIPI Image database at\\ 
{\tt http://sipi.usc.edu/services/database/Database.html},\\ see Figure \ref{TestImages1} and Figure \ref{TestImages2}. 
The one dimensional signals used here are the standard examples used and defined in \cite{DJ_Adapt:1995}, see Figure \ref{TestSignals1}. 

\item All {\em images} used in experiments are bitdepth 8 gray level images of size $N = n\times n=512\times 512$ unless otherwise is specified. The one dimensional test signals are of length $N=1024$ unless otherwise is specified.

\item In the tables shown in the appendix, results obtained from datasets $\bm{x}\in\mathbb{R}^{N}$ with computer generated noise 
are shown. The definition of signal to noise ratio (SNR) of the dataset $\bm{x}=\bm{\theta}+\bm{\eta}$ used for signal $\bm{\theta}\in\mathbb{R}^{N}$ and 
noise $\bm{\eta}\in\mathbb{R}^{N}$ when generating datasets $\bm{x}$ with different SNR values is:
\begin{align}
&\text{SNR}=10\log_{10}\left(\frac{\|\bm{\theta}\|_{2}^{2}}{E_{\eta}\|\bm{\eta}\|_{2}^{2}}\right).\label{SNR_experiments}
\end{align}
where $f$ signifies the gaussian likelihood distribution.
\item The SNR measure used when reporting signal to noise ratios in the estimated signals $\bm{\theta}^{*}$ in the tables in appendix is:
\begin{align}
&\widehat{\text{SNR}}=10\log_{10}\left(\frac{\|\bm{\theta}^{*}\|_{2}^{2}}{\|\bm{\theta}-\bm{\theta}^{*}\|_{2}^{2}}\right).\label{SNR_experiments2}
\end{align}

\item The error measure used in tables below will be a scaled version of the root mean square error (RMSE) defined by
\begin{align}
&RMSE = \sqrt{\frac{1}{N}\|\bm{\theta}-\bm{\theta}^{*}\|_{2}^{2}\cdot\tau}\label{def_RMSE}
\end{align}
where $\bm{\theta}$ is the estimate of the signal $\bm{\theta}$ and $\tau^{-1}$ is the variance of the noise $\bm{\eta}$.
\item For the RiskShrink, SureShrink and $T_{MAP}$ algorithms the noise variance $\tau^{-1}$ was estimated from the highpass band using 
  the median estimator, see \cite{DJ_Adapt:1995}. Also for the $T_{MAP}$ algorithm we estimated the signal variance $\lambda^{-1}$
  by the moment estimator $\lambda^{*}$ defined by:
  \begin{align}
    \frac{1}{\lambda^{*}}\mathdef\max\left(0,\left(\frac{1}{N}\sum_{i=1}^{N}\bm{x}(i)^{2}\right)-\tau^{-1}\right)\label{def_lambda_estimator_exp}
\end{align}

\item The INMDL principle was implemented by an iterative scheme in our numerical experiments as follows: 
The NML principle of \cite{Rissanen:2000} is used to provide an initial estimate of the best model $\gamma_{d^{*}}^{*}$ 
from which we compute initial estimates $\tau^{*}$, $\lambda^{*}$ of variance parameters $\tau$ and $\lambda$ and then an initial estimate 
$\bm{\theta}^{*}$ of the wavelet coefficients $\bm{\theta}$ of the data is computed. 
These parameter estimates are then fed into the model selection principle as 
defined in Proposition \ref{INMDL_Selection_Principle} and a new estimate of the best model $\gamma_{d^{*}}^{*}$ may then be computed
and the iteration process continues with new updated estimates $\tau^{*}$, $\lambda^{*}$, $\bm{\theta}^{*}$ and so on.
The GGD shape parameter $\nu$ is also estimated in each iteration step using the estimate $\bm{\theta}^{*}$ and the estimator $\nu^{*}$
provided in \cite{DoVetterli:2002}. This whole model selection iteration procedure continues 
until changes in the estimates of the optimal model size $d^{*}$ between two iterations falls within 5\%. 
We also note that the number $C_{\lambda_{p}}$ in the model class prior distribution $D(p,q)$ in (\ref{D_form_GGD_Jeffreys}), (\ref{D_form_GGD_GGD}) 
was set to $C_{\lambda_{p}}=1.0$ in all our experiments reported below.

\item For image experiments, we show results from the GGD MAP estimator $T_{MAP}^{(\nu)}$  for values $\nu=1.0$ and $\nu=0.7$ 
on the GGD shape parameter. The reason for our choice of these values, are that extensive empirical investigation \cite{Moulin-Liu:1999}
show that a GGD model with $\nu\in(0.5,1.0)$ provides a reasonable prior model for many if not ''most'' natural images. Also, 
the choice of $\nu=1$ yields the  Laplace distribution which is very often used as a model distribution in the image denoising community because 
one then can obtain closed form analytical solutions to estimator and risk equations in the case of gaussian noise. 

\item For the experiments with 1-dimensional signals, we show results from the GGD MAP estimator $T_{MAP}^{(\nu)}$  
for values $\nu=0.5$ and $\nu=1.0$. Unlike the case of image data, we have in this case no prior knowledge which supports a choice of 
a GGD model for the data. However, the wavelet basis is known to yield sparse representations of piecewise smooth signals 
\cite{DJ_Ideal:1994}, so a GGD distribution with $\nu\leq 1$ could be worth a try. The choice $\nu=2$ yields a gaussian model 
distribution, which maximizes the entropy for a given variance, but this choice turned out to yield a very poorly performing 
estimator $T_{MAP}^{(2.0)}$, so we omit it.

\item The ordinary full depth periodic wavelet basis with a symmlet of filter length 16 (Symmlet 16) was used as the wavelet basis in all the 
  image experiments. 

\item All numerical experiments reported in this thesis were carried out on a 2.0 GHz Pentium4-Mobile PC with 768 MB RAM running FreeBSD-4.9 as operating system. 
The experiments were all implemented in the C programming language except a few cases were we have been using the NAG Fortran Library Mark 16 
for some standard mathematical functions and random number generators. The C compiler used was Intel C$++$ compiler version 7.1 (build 20030922Z).

\item We note that even though our implemented version of the INMDL procedure is quite computing intensive, it runs in $\bigo(N\log{N})$ time, 
and typically on our 2.0 GHz Pentium4 PC with image data with $N= n\times n=512\times 512$, the run time is about 50-90 seconds when the source 
code is compiled with full optimization. The computational bottleneck by far is the computation of the GGD-MAP estimate $\bm{\theta}^{*}$ by
linear interpolations. However, we have not gone to any effort in optimizing our implementation for speed. 
Considerable speed improvements may be possible.

\item The same noise realization was of course used when comparing the different algorithms shown in tables below. We only report results obtained
from a single realization of the noise because we found that the both SNR and RMSE results for all the denoising algorithms in the tables 
deviated by less than 1\% over 3 different noise realizations when used on test image ''Barbara''.

\item In the experiments on images below, we checked the validity of our asymptotic marginal 
formula in Theorem \ref{theorem1}, Corollary \ref{corollary_new_marginal} and the marginal renormalization constant in 
Proposition \ref{Marginal_Renormalization_Constant}, by checking (the upper bounds of) the numbers 
$\kappa<T_{\kappa}$, $\xi<T_{\xi}$, $\zeta$, $\omega<T_{\omega}$, $X$. These were found to vary as: 
$1.0E$$-3<T_{\kappa}<9.0E$$-2$, $1.0E2\leq T_{\xi}\leq 4.5E9$, 
$1.0E$$-3\leq \log_{2}{(1+T_{\xi})}/d\leq 4.0E-2$, $1.0E$$-2<T_{\omega}<9.0E$$-2$, $1.0E$$-3<X<1.0E$$-2$, $1.0E$$-3<\zeta<3.0E$$-2$.
Furthermore we observed that $2.4<\inf_{1\leq i\leq d}|\sqrt{\tau^{*}}\theta_{i}^{*}|<4.0$ always for the test images used. 
The posterior biases shown in Corollary \ref{Posterior_Unbiased_Estimators} were found to be of insignificant size: < 0.01\% 
of the estimator values $\theta_{i}^{*},\ 1\leq i\leq d$ and $\tau^{*}$, for all of the test images. We emphasize that
although the numerical values of $\xi$ were found to be large, the contribution from the term $(1+\xi)$ to the codelength
is given by: $-\log_{2}(1+\xi)/d$ per model sample in the mean, and this is found to be of the same order per model sample as the 
uncertainty $\pm 0.5\zeta$ in the codelength contribution from the marginal normalization $-\log_{2}C_{\gamma_{d}}$ 
(see Proposition \ref{Marginal_Renormalization_Constant}) which we have explicitely neglected.

\item For the experiments on 1-dimensional data below, we checked the validity of our asymptotic marginal 
formula in Theorem \ref{theorem1}, Corollary \ref{corollary_new_marginal} and the marginal renormalization constant in 
Proposition \ref{Marginal_Renormalization_Constant}, by checking (the upper bounds of) the numbers 
$\kappa<T_{\kappa}$, $\xi<T_{\xi}$, $\zeta$, $\omega<T_{\omega}$, $X$. These were found to vary as: 
$1.0E$$-3<T_{\kappa}<5.0E$$-2$, $1.0E$$0\leq T_{\xi}\leq 4.0E$$0$, $3.0E$$-3\leq \log_{2}{(1+T_{\xi})}/d\leq 2.0E$$-2$,
$1.0E$$-2<T_{\omega}<5.0E$$-1$, $1.0E$$-3<X<2.0E$$-2$, $1.0E$$-3\leq\zeta\leq 5.0E$$-2$.
Furthermore we observed that $1.96<\inf_{1\leq i\leq d}|\sqrt{\tau^{*}}\theta_{i}^{*}|<3.7$ always for the test signals used. 
The posterior biases: $E_{\theta,\tau}(\theta_{i}^{*}-\theta_{i})$ and:$E_{\theta,\tau}(\tau^{*}-\tau)$ shown in Corollary \ref{Posterior_Unbiased_Estimators} were found to be of insignificant size: < 0.001\% 
of the estimator values $\theta_{i}^{*},\ 1\leq i\leq d$ and $\tau^{*}$, for all of the test signals.

\end{enumerate}

%\clearpage

\newpage

\begin{figure}[h]
\begin{center}
\includegraphics[scale=0.75]{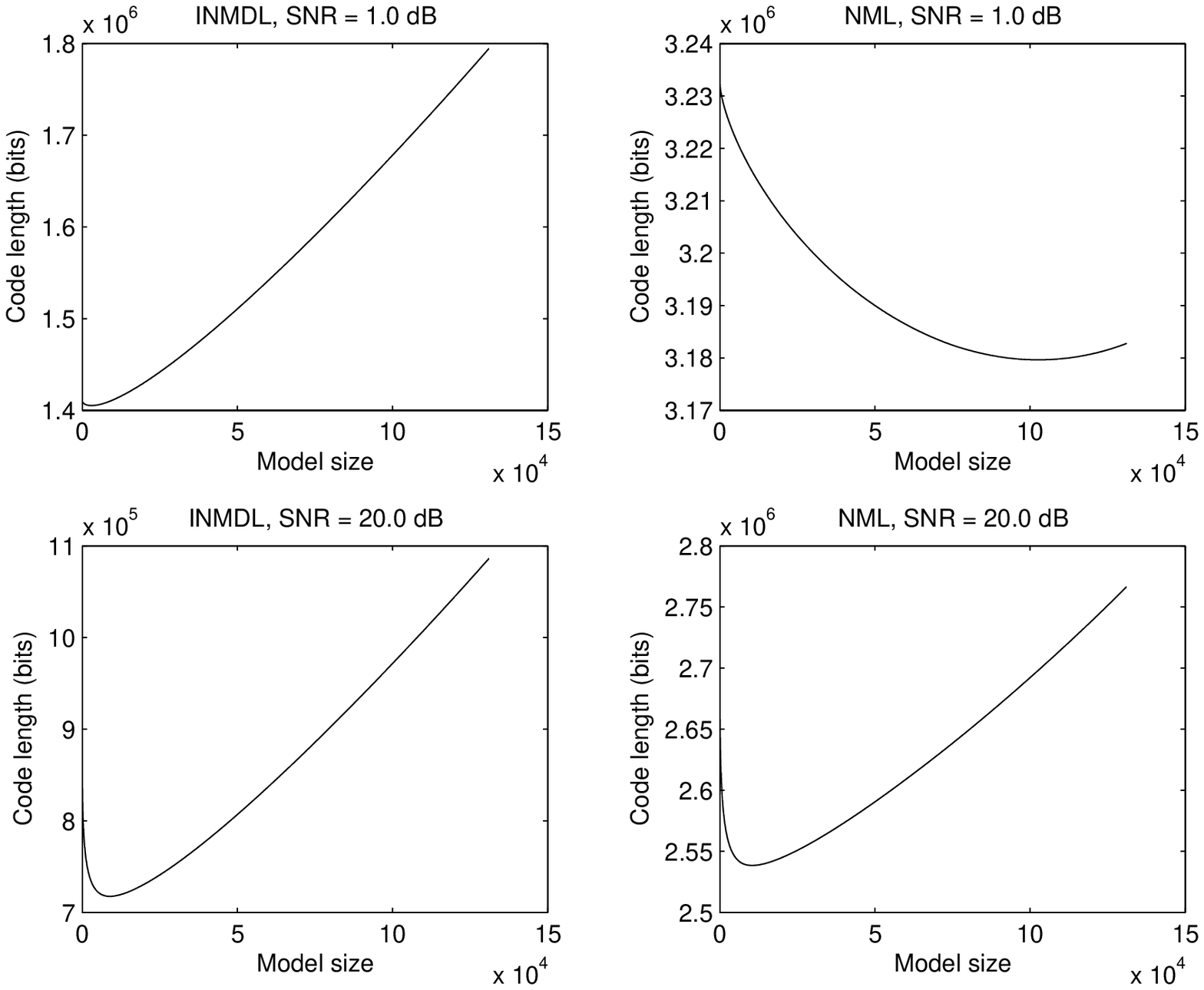}
\end{center}
\caption{The code length as a function of model size for the test image boat.}
\label{Code lengths}
\end{figure}

\subsection{Discussion of experimental results}

When applied to image data, the SureShrink method \cite{DJ_Adapt:1995} clearly outperforms all of the tested estimators over the whole range 
of tested SNR values as seen from Figure \ref{RelativeEstimatorPerformance2}. 
The SureShrink method is a hybrid method between a soft universial thresholding scheme as given in \cite{DJ_Ideal:1994}
and an adaptive thresholding scheme given by adapting the thresholds to minimize a risk estimate using Steins unbiased risk estimate (SURE) 
given in \cite{Stein:1981}. The hybrid scheme of SureShrink decides in each wavelet subband whether the signal is sparsely represented
in the subband. In sparse situations the universial thresholding scheme is used, otherwise the SURE method is used to provide risk 
estimates in each wavelet subband. Thus, different adaptive thresholds are used in 
each subband by the SureShrink, whereas the other methods use a global (identical in all subbands), although data adaptive, thresholding scheme.

Comparing the NML and INMDL-principle
we note that the NML-principle does not have a robust performance for the datasets tested here, 
it fails badly compared to all the other methods as the SNR falls below 10 on the dB scale as may be seen
from Figure \ref{RelativeEstimatorPerformance2} and Figure \ref{RelativeEstimatorPerformance1}. The performance of the INMDL-principle 
in the region of low SNR is the second worst method measured in RMSE for SNR $<5$ dB , but it does not fail as bad as the NML-principle.
Coupling these observations to the information in Figure \ref{WaveletSparsity1} and Figure \ref{WaveletSparsity2}, we conclude that the main 
explanation for the observed weak performance of NML and INMDL in the low SNR region, is that the sizes of the optimal models 
as predicted by these model selection principles are too large, this behaviour is especially clear for the NML-principle. 
The Figure \ref{RelativeEstimatorPerformance2}
shows that the INMDL-based estimator has second best performance of the tested estimators for image datasets in the SNR range $10 \text{ dB} <$ SNR $<15 \text{ dB}$. 
For image data in the high SNR region we see that performances of both NML and INMDL weakens as the SNR increases when compared to the 
GGD-MAP estimators and the SureShrink principle. Figure \ref{WaveletSparsity2} explains why: The predicted optimal model sizes are too small
in this SNR region for image data. However, for the 1-dimensional test data the situation is reversed: As the SNR increases the performance 
of NML and INMDL based estimators improves and outperform the SureShrink and the GGD-MAP estimators. The Figure \ref{WaveletSparsity2}
explains why: The SureShrink and the GGD-MAP estimators keeps too many wavelet coefficients for this type of data 
whereas the model sizes as predicted by the NML and the INMDL principles yields a smaller number of nonzero wavelet coefficient 
estimates which closely match the RiskShrink estimator both in performance and sparseness of the wavelet coefficient estimates.
We note that the RiskShrink estimator $\theta^{(RS)}_{i}$ is known to be universally near-optimal in the sense that to within a logarithmic factor 
it achives the ideal risk obtained with an oracle estimator \cite{DJ_Ideal:1994}, that is 
\begin{align}
&E_{\bm{x}}\|\bm{\theta}^{(RS)}(\bm{x})-\bm{\theta}\|_{2}^{2}\leq \left(2\log{N}+1\right)\left(\sigma^{2}+\sum_{i=1}^{N}\min(\theta_{i}^{2},\sigma^{2})\right),\forall\ \bm{\theta}\in\mathbb{R}^{N}\label{RiskShrink_riskbound} 
\end{align}
and no estimator can come closer to the ideal risk than $\theta^{(RS)}$ for all $\bm{\theta}\in\mathbb{R}^{N}$ without relying on an oracle.

Figure \ref{RelativeEstimatorPerformance2} and Figure \ref{RelativeEstimatorPerformance1} indicates that the tested estimators perform
quite differently relative to each other for a given SNR level, 
depending on whether the data belongs to the 2-dimensional test datasets or 1-dimensional test data in our experiments. The explanation may 
depend on several factors: The sample size $N$ which in the experiments here defer by two orders of magnitude between the 
two-dimensional and one-dimensional datasets. However, we verified (see remarks above) that the parameters controlling the error on 
our marginal approximation formula are well inside the required intervals for both sample sizes $N=2^{18}$ and $N=2^{10}$ in all our experiments.
Therefore we do not believe the observed differences 
in performance are primarily due to differences in sample size here. 
The ability of the wavelet basis to sparsely represent the data in the 
wavelet domain (''few'' large and many ''small'' wavelet expansion coefficients) is important. In this respect we note that 
wavelet bases are known to optimally (in a certain strictly defined sense) \cite{Mallat:1998b} represent data inside a ''ball'' 
of bounded total variation functions and a large class of ''natural'' images belong to this class of functions 
\cite{DJ_Adapt:1995}, \cite{MallatKalifa:2003_AnnStat}.
%\chapter{Applying the INMDL-principle to GGD-modeled data in an inverse problem}
\chapter{The INMDL-principle applied to an inverse problem}
\section{Definition of problem and data generating model}
We will investigate the performance of the INMDL-principle when applied to the problem of estimating signals or images $\bm{\theta}$ which have gone through a degradation process modelled as
\begin{align}
&y=u\ast\theta+\eta\label{deconv_problem}
\end{align}
where $u$ is a known lowpass filter and $\eta$ is IID gaussian noise and $\ast$ denotes the convolution operator.
We will rely on and use as reference work presented in \cite{MallatKalifa:2003_IEEE} and \cite{MallatKalifa:2003_AnnStat}, 
in particular we will use the {\em mirror wavelet} basis constructed in the cited papers, see appendix. 
The motivation behind our investigation into applying the INMDL-principle
to the deconvolution problem (\ref{deconv_problem}) is our experience from numerical simulations concerning the 
denoising problem in the previous section that the INMDL-principle as developed in previous sections seem to be very robust 
against high noise, and so one could expect that the INMDL-principle would eliminate the need for ''hard'' regularization
techniques like the cutoff-frequencies $k_{c}$ in the Fourier domain introduced in \cite{MallatKalifa:2003_AnnStat} or the modified threshold
estimators in \cite{MallatKalifa:2003_IEEE} demanding the a priori knowledge of the numbers $s_{B}[m]$ which are 
$s_{B}[m]\mathdef\sup_{f\in\Theta}|\langle f,b_{m}\rangle|$ where $b_{m}$ are elements in an orthogonal basis $B$
and $f$ belongs to a predefined set of signals (datasets) $\Theta$. 
Also the INMDL principle does not need to know the noise level $\tau^{-1}\mathdef E(\eta^{2})$ beforehand. 
In addition, the use of a prior distribution $\pi$ in the INMDL-principle allow for a more sophisticated modeling of the 
wavelet coefficients than in the papers \cite{MallatKalifa:2003_IEEE} and \cite{MallatKalifa:2003_AnnStat}, 
this may enable a better reconstruction of the degraded data. Formally deconvolving the data $\bm{x}$ in expression 
(\ref{deconv_problem}) yields
\begin{equation}
x\mathdef u^{-1}\ast y=\theta+u^{-1}\ast\eta\label{deconv_solution}
\end{equation}
where the inverse $u^{-1}$ is defined by
\begin{equation}
u^{-1}\mathdef\mathcal{F}^{-1}\left(\frac{1}{\hat{u}(\omega)}\right)\label{def_inverse_u}
\end{equation}
where 
\begin{equation}
\hat{u}(\omega)\mathdef\mathcal{F}(u)(\omega)\label{def_u_hat}
\end{equation}
and $\mathcal{F}$ denotes the Fourier transform. When the Fourier transform of the inverse filter 
$1/\hat{u}(\omega)$ is not bounded in the high frequencies, the noise $Z\mathdef u^{-1}\ast\eta$ resulting from the 
deconvolution of data $y$ in (\ref{deconv_solution}) is amplified by a factor that tends to infinity. Therefore, in general the
deconvolution problem (\ref{deconv_problem}) is an ill-posed inverse problem, and solutions to this type of problem must include some 
kind of regularization procedure for removing the worst part of the deconvolved noise $Z$. The INMDL-principle naturally provides a regularization
through a model selection process and we will now investigate how the INMDL principle may be adapted to deconvolution problems.
Assuming the convolution is circular, we may write the discretization of (\ref{deconv_problem}) on the form
\begin{align}
&\bm{y}=\bm{U}\bm{\theta}+\bm{\eta}\label{discrete_convolution}
\intertext{where $\bm{U}\in\mathbb{R}^{n\times n}$ is the matrix representation of the smoothing operation by
convolution by the lowpass filter $u$. We define}
&\bm{x}\mathdef\bm{U}^{-1}\bm{y}=\bm{\theta}+\bm{U}^{-1}\bm{\eta}\label{discrete_deconvolution}\\
&\bm{Z}\mathdef\bm{U}^{-1}\bm{\eta}\label{dicrete_deconvolved_noise}\\
&\bm{K}\mathdef E[\bm{Z}\bm{Z}^{T}]=\tau^{-1}\bm{U}^{-1}\bm{U}^{-T}\label{covariance_deconv_noise}
\intertext{The mirror wavelet basis $\widetilde{\bm{W}}\in\mathbb{R}^{n\times n}$ \cite{MallatKalifa:2003_AnnStat}, see appendix, 
approximately diagonalizes the covariance $\bm{K}$
of the deconvolved noise $\bm{Z}$, that is for $\widetilde{\bm{w}}_{k}\in\widetilde{\bm{W}}$, $1\leq k\leq n$ we have}
&\widetilde{\bm{K}}\mathdef\widetilde{\bm{W}}^{T}\bm{K}\widetilde{\bm{W}}\sim\text{\bf{diag}}\left(\widetilde{\bm{W}}^{T}\bm{K}\widetilde{\bm{W}}\right)\sim
\text{\bf{diag}}\left(\left\langle\bm{K}\widetilde{\bm{w}}_{k},\widetilde{\bm{w}}_{k}\right\rangle\right)_{1\leq k\leq n}\nonumber\\
&=\text{\bf{diag}}\left(\left\langle\widehat{\bm{K}}\widehat{\widetilde{\bm{w}}}_{k},\widehat{\widetilde{\bm{w}}}_{k}\right\rangle\right)_{1\leq k\leq n}\nonumber\\
&=\text{\bf{diag}}\left(\tau^{-1}\sum_{i=1}^{n}\frac{\left|\widehat{\widetilde{\bm{w}}}_{k}[i]\right|^{2}}{|\widehat{u}[i]|^{2}}\right)_{1\leq k\leq n}\mathdef\widetilde{\bm{K}}_{D}\label{deconv_noise_variances}
\end{align}
where $\widehat{\widetilde{\bm{w}}}_{k}\mathdef\bm{W}^{T}_{F}\widetilde{\bm{w}}_{k}$ 
is the discrete Fourier transform of $\widetilde{\bm{w}}_{k}$ and
$\widehat{\bm{K}}\mathdef\tau^{-1}\bm{W}^{T}_{F}\bm{U}^{-1}\bm{U}^{-T}\bm{W}_{F}$ and $\bm{W}_{F}\in\mathbb{C}^{n\times n}$ is 
the discrete Fourier basis on $\mathbb{C}^{n}$. 
%\section{A model selection algorithm for IID GGD distributed parameters embedded in non-white additive gaussian noise}
\section{The model selection algorithm}
We would like to be able to use our previous results to compute an approximation to the marginal
density of the deconvolved data $\bm{x}$. However, in the current case of a non-constant diagonal covariance matrix $\widetilde{\bm{K}}_{D}$ 
defined above, it is non-trivial to find suitable parameter transformations $\theta_{i}\mapsto\hat{\theta}_{i}$, $\tau\mapsto\hat{\tau}$ which makes the transformed Fisher information $|\hat{\bm{F}}(\hat{\bm{\theta}},\hat{\tau})|$ a constant. This can be seen by retracing the steps in the computation of $|\hat{\bm{F}}(\hat{\bm{\theta}},\hat{\tau})|$ shown in the appendix. We present a workaround on this problem below. Define
\begin{align}
&\widetilde{t}_{k}^{2}\mathdef\sum_{i=1}^{N}\frac{\widehat{\widetilde{|\bm{w}}}_{k}[i]|^{2}}{|\widehat{u}[i]|^{2}},\ 1\leq k\leq n\label{def_noise_weights}
\intertext{and the change of variables}
&\widetilde{\bm{x}}\mathdef\widetilde{\bm{W}}^{T}\bm{x},\ \widetilde{\bm{\theta}}\mathdef\widetilde{\bm{W}}^{T}\bm{\theta},\nonumber\\
&\widetilde{\widetilde{\bm{x}}}\mathdef(\tau\widetilde{\bm{K}}_{D})^{-1/2}\widetilde{\bm{x}}=\text{\bf{diag}}(\widetilde{t}_{i}^{-1})_{1\leq i\leq d}\ \widetilde{\bm{x}},\nonumber\\
&\widetilde{\widetilde{\bm{\theta}}}\mathdef(\tau\widetilde{\bm{K}}_{D})^{-1/2}\widetilde{\bm{\theta}}=\text{\bf{diag}}(\widetilde{t}_{i}^{-1})_{1\leq i\leq d}\ \widetilde{\bm{\theta}}\label{def_transformed_data}
\end{align}
We may now use our previous results Theorem \ref{theorem1}, Corollary \ref{corollary_new_marginal} to compute an approximation to 
the marginal distribution $m_{\gamma_{d}}(\bm{x})$ of the deconvolved data $\bm{x}$ in (\ref{discrete_deconvolution}). It is easy to verify by
inspection of the proof of Theorem \ref{theorem1} that the approximation result for the marginal density provided in Theorem \ref{theorem1}
applies to the transformed data $\widetilde{\widetilde{\bm{x}}}$ and parameters $\widetilde{\widetilde{\bm{\theta}}}$ with
minor adjustments. However, there are some important remarks to be made here:

\begin{enumerate}
\item As before, the parameters $\widetilde{\theta}_{i},\ 1\leq i\leq d$ are modelled as identically and independently GGD distributed parameters
with density $\pi_{\lambda}(\widetilde{\theta}_{i})$. We note that the empirical research on the modeling of image wavelet coefficients in the litterature 
\cite{Moulin-Liu:1999} concerns pure wavelet bases, not mirror wavelet bases as in the current context, but we will here use the GGD model also 
for the case of mirror wavelet bases.
\item The transformed parameters $\widetilde{\widetilde{\theta}}_{i},\ 1\leq i\leq d$ are independently distributed, but {\em not} 
identically distributed. By the coordinate transformations defined in (\ref{def_transformed_data}) we see that
\begin{align}
&\widetilde{\theta_{i}}\sim\pi_{\lambda}\left(\widetilde{\theta}_{i}\right)\Longrightarrow\widetilde{\widetilde{\theta}}_{i}\sim\pi_{\widetilde{\lambda}_{i}}\left(\widetilde{\widetilde{\theta}}_{i}\right),\ 1\leq i\leq d,\label{GGD_transformed_theta}
\intertext{where}
&\widetilde{\lambda}_{i}\mathdef\lambda\widetilde{t}_{i}^{2}\label{def_tilde_lambda_i}
\intertext{with $\widetilde{t}_{i}$ as defined in (\ref{def_noise_weights}). Furthermore, we note that}
&\widetilde{\lambda}_{i}^{1/2}\widetilde{\widetilde{\theta}}_{i}=\lambda^{1/2}\widetilde{\theta}_{i}.\label{nice_relation}
\end{align} 
\item The proper definition on the SNR $\widetilde{\Omega}\left(\{\widetilde{\lambda}_{i}\}_{i=1}^{d},\tau\right)$ 
in the current case of transformed data $\widetilde{\widetilde{\bm{x}}}$ and parameters $\widetilde{\widetilde{\bm{\theta}}}$ is
\begin{align}
&\widetilde{\Omega}\left(\{\widetilde{\lambda}_{i}\}_{i=1}^{d},\tau\right)\mathdef\frac{\sum_{i=1}^{d}\frac{1}{\widetilde{\lambda}_{i}}}{n\frac{1}{\tau}}
=\frac{\sum_{i=1}^{d}\widetilde{t}_{i}^{-2}\frac{1}{\lambda}}{n\frac{1}{\tau}}\nonumber\\
&=\frac{\sum_{i=1}^{d}\widetilde{t}_{i}^{-2}}{d}\Omega(\lambda,\tau),\text{ where }\Omega(\lambda,\tau)\mathdef\frac{d\lambda^{-1}}{n\tau^{-1}}.\label{def_tilde_Omega}
\end{align}
\end{enumerate}
By applying the same mappings $\psi(\hat{\tau})$ and $\phi(\hat{\theta}_{i},\hat{\tau})$ defined in (\ref{def_phi_psi}) to the current 
choice of coordinates $\widetilde{\widetilde{\bm{x}}}$, $\widetilde{\widetilde{\bm{\theta}}}$
and going through the proof of Theorem \ref{theorem1} provided in the appendix, replacing $x_{i}$ by $\widetilde{\widetilde{x}}_{i}$ and $\theta_{i}$ 
by $\widetilde{\widetilde{\theta}}_{i}$, we see that our previous results generalize straightforwardly to the current case of nonwhite gaussian noise 
through the whitening transformation defined in (\ref{def_transformed_data}). Because the MAP-estimator $\theta^{*}$ defined in (\ref{MAP_estimator_theta1}) is nonlinear, 
some care has to be taken to estimate the parameter $\lambda$ which is 
needed to estimate the $\widetilde{\lambda}_{i}$ and the $\widetilde{\widetilde{\theta}}_{i},\ 1\leq i\leq d$: 
Given an initial model index vector $\gamma_{d}^{(0)}$ we may define $\widetilde{\bm{x}}_{\parallel}$ and thus initial
estimates $\lambda^{*}_{0}$, $\tau^{*}_{0}$ and $\widetilde{\bm{\theta}}^{*}_{0}$. We define
$\widetilde{\lambda}_{i}^{*}\mathdef\lambda^{*}\widetilde{t}_{i}^{2}$ and this may be used to compute the MAP estimates 
$\widetilde{\widetilde{\theta^{*}}}_{i}$. Applying the model selection principle given in Proposition \ref{INMDL_Selection_Principle} 
to the whitened data $\widetilde{\widetilde{\bm{x}}}$ and their corresponding parameter estimates $\widetilde{\widetilde{\theta_{i}}}^{*}$ 
will provide us with an updated model index vector $\gamma_{d}$. Then the same iterative procedure applied previously in the case of 
white gaussian noise may be used to compute successive estimates $\lambda^{*}$, $\tau^{*}$, $\widetilde{\bm{\theta}}^{*}$
and through these we compute new estimates $\widetilde{\lambda}_{i}^{*}$ and $\widetilde{\widetilde{\bm{\theta}^{*}}}$.

The prior distribution $\pi_{\widetilde{\lambda}_{i}}(\widetilde{\widetilde{\bm{\theta}}})$ in the current case of 
independently distributed $\widetilde{\widetilde{\theta}}_{i},\ 1\leq i\leq d$ where $\pi_{\lambda}(\theta_{i})$ 
is given in (\ref{form_GGD_density}), becomes
\begin{align}
&\pi_{\widetilde{\lambda}_{i}}\left(\widetilde{\widetilde{\bm{\theta}}}\right)\mathdef\prod_{i=1}^{d}\pi_{\widetilde{\lambda}_{i}}\left(\widetilde{\widetilde{\theta}}_{i}\right)\nonumber\\
&=\prod_{i=1}^{d}\left(\frac{\nu\eta(\nu)\widetilde{\lambda}_{i}^{1/2}}{2\Gamma(1/\nu)}\right)\exp\left(-\eta(\nu)^{\nu}\sum_{i=1}^{d}\left|\widetilde{\lambda}_{i}^{1/2}\widetilde{\widetilde{\theta}}_{i}\right|^{\nu}\right)\nonumber\\
&=\prod_{i=1}^{d}\left(\frac{\nu\eta(\nu)(\widetilde{t}_{i}^{2}\lambda)^{1/2}}{2\Gamma(1/\nu)}\right)\exp\left(-\eta(\nu)^{\nu}\sum_{i=1}^{d}\left|(\widetilde{t}_{i}^{2}\lambda)^{1/2}\widetilde{\widetilde{\theta}}_{i}\right|^{\nu}\right)\nonumber\\
&=\left(\prod_{i=1}^{d}\widetilde{t}_{i}\right)\left(\frac{\nu\eta(\nu)\lambda^{1/2}}{2\Gamma(1/\nu)}\right)^{d}\exp\left(-\eta(\nu)^{\nu}\sum_{i=1}^{d}\widetilde{t}_{i}^{\nu}\left|\lambda^{1/2}\widetilde{\widetilde{\theta}}_{i}\right|^{\nu}\right)
\end{align}
We note that the factor $\prod_{i=1}^{d}\widetilde{t}_{i}$ vanishes in the formula (\ref{almost_invariant_marginal}) for the marginal density 
$m_{\gamma_{d}}(\bm{x})$ because this factor is also included in the normalization factor $C_{\gamma_{d}}(\bm{x})$.
To evaluate the model selection criterion $C()$ defined in (\ref{def_model_selection_selection_criterion}) in the current case of 
non-IID parameters $\widetilde{\widetilde{\theta}}_{i}$, we will use the MAP-estimator $\lambda^{*}$ 
in (\ref{GGD_lambda_estimator_form1}):
\begin{align}
&\lambda^{*}=\frac{(d+2)^{2/\nu}}{\nu^{2/\nu}\eta(\nu)^{2}\left(\sum_{i=1}^{d}|\widetilde{t}_{i}\widetilde{\widetilde{\theta}}_{i}|^{\nu}\right)^{2/\nu}}
\intertext{and we then get}
&C\left(\widetilde{\widetilde{\bm{x}}}_{\parallel}(i)|S_{i-1}\right)\mathdef-(n-d+2)\frac{|\widetilde{\widetilde{\bm{x}}}_{\parallel}(i)|}{\|\widetilde{\widetilde{\bm{x}}}\|_{2}^{2}-\|\widetilde{\widetilde{\bm{x}}}_{\parallel}\|_{2}^{2}}\nonumber\\
&-\frac{\partial}{\partial|\widetilde{\widetilde{\bm{x}}}_{\parallel}(i)|}\log\left(\frac{\pi_{\widetilde{\lambda}_{i}}\left(\widetilde{\widetilde{\bm{\theta}^{*}}}|\widetilde{\lambda}_{i}^{*}\right)}{|\Psi_{\lambda\lambda}\left(\widetilde{\widetilde{\bm{\theta}^{*}}},\lambda^{*}\right)|^{\frac{1}{2}}}\right)\nonumber\\
&=-(n-d+2)\frac{|\widetilde{\widetilde{\bm{x}}}_{\parallel}(i)|}{\|\widetilde{\widetilde{\bm{x}}}\|_{2}^{2}-\|\widetilde{\widetilde{\bm{x}}}_{\parallel}\|_{2}^{2}}\nonumber\\
&+(d+2)\frac{\widetilde{t}_{i}|\widetilde{t}_{i}\widetilde{\widetilde{\theta^{*}}}_{i}|^{\nu-1}}{\sum_{j=1}^{d}|\widetilde{t}_{j}\widetilde{\widetilde{\theta^{*}}}_{j}|^{\nu}}\frac{\partial|\widetilde{\widetilde{\theta^{*}}}_{i}(\widetilde{\widetilde{\bm{x}}}_{\parallel}(i))|}{\partial|\widetilde{\widetilde{\bm{x}}}_{\parallel}(i)|},\ 1\leq i\leq d.\label{GGD_evaluated_selection_criterion2}
\end{align} 
The current criterion $C(\cdot)$ in (\ref{GGD_evaluated_selection_criterion2}) is not as easy to minimize over the data
$\widetilde{\widetilde{\bm{x}}}$ as in the 
previous case of IID parameters in (\ref{GGD_evaluated_selection_criterion}). The explanation for this is as follows:
Suppose $0<\nu\leq 1$, then even if $|\widetilde{\widetilde{\bm{x}}}_{\parallel}(i)|$ is large, it may still happen that 
$\widetilde{\widetilde{\theta^{*}}}_{i}(\widetilde{\widetilde{\bm{x}}}(i))=0$ yielding 
$C\left(\widetilde{\widetilde{\bm{x}}}_{\parallel}(i)|S_{i-1}\right)=\infty$, because the MAP estimator 
$\widetilde{\widetilde{\theta^{*}}}_{i}$
is a threshold estimator with a threshold which grows with $\widetilde{\lambda}_{i}\mathdef\lambda\widetilde{t}_{i}^{2}$ as may be seen from (\ref{threshold_size}). We have observed that this effect is a real problem in our numerical experiments. As shown in the proof of Theorem \ref{theorem1}, 
our marginal approximation formula is not valid for small parameter estimates $\theta_{i}^{*}$, and so we cannot allow 
the selection of model indices $i$ with $\widetilde{\widetilde{\theta^{*}}}_{i}=0$. 
To overcome this problem, we will adopt a possibly suboptimal model selection
algorithm which we believe/hope is not far from the model selection procedure which minimizes $C(\cdot)$ in (\ref{GGD_evaluated_selection_criterion2}):
We will simply select the indices $i$ with the largest estimates $|\widetilde{\widetilde{\theta^{*}}}_{i}|$. Now, we will skip the 
details on going through the proof of the Theorem \ref{theorem1} and making the necessary adaptations to the current case of non-identically distributed 
$\widetilde{\widetilde{\theta}}_{i}$, $1\leq i\leq d$. However, the changes are straightforward and we list below the 
ones concerning the sufficient conditions
on the numbers $\zeta$, $\kappa$, $\xi$, $X$ under which the Theorem \ref{theorem1} on the marginal approximation and the Proposition 
\ref{Marginal_Renormalization_Constant} concerning the marginal normalization constant, 
still both apply to the current model for the data $\widetilde{\widetilde{\bm{x}}}$ and parameters $\widetilde{\widetilde{\bm{\theta}}}$ defined above.
We need upper bounds on the numbers $\zeta$, $\kappa$, $\xi$, $X$ in order to check the validity of the marginal approximation and the resulting codelength principle in our numerical work. 
Renaming $\zeta\rightarrow\widetilde{\zeta}$, $\kappa\rightarrow\widetilde{\kappa}$, $\xi\rightarrow\widetilde{\xi}$, $X\rightarrow\widetilde{X}$ 
under the current model we have
\begin{align}
&\widetilde{\zeta}\mathdef \sup_{1\leq i\leq d}\left\{\widetilde{t}_{i}^{\nu/2}C_{\nu}\nu|\nu-1|\left(\frac{n}{d}\Omega(\lambda^{*},\tau^{*})\right)^{-\frac{\nu}{2}}|(\tau^{*})^{\frac{1}{2}}\widetilde{\widetilde{\theta^{*}}}_{i}|^{\nu-2}\right\}<1\label{def_zeta_tilde}
\intertext{where $\Omega(\lambda,\tau)$ is a signal to noise ratio (SNR) defined as}
&\Omega(\lambda,\tau)\mathdef\frac{d\lambda^{-1}}{n\tau^{-1}}\label{def_Omega_IID}\\
&\widetilde{X}\mathdef\frac{\widetilde{S}_{\nu}}{n-d+2}\frac{2C_{\nu}^{2}\nu^{2}\left(\frac{n}{d}\Omega(\lambda^{*},\tau^{*})\right)^{-h(\nu)}}{1-\widetilde{\zeta}}<1,\label{det_H_claim2_tilde}
\intertext{where}
&h(\nu)=\left\{\begin{array}{ll}\nu&\text{ if }0<\nu\leq 1\\
\nu/2 & \text{ if } 1<\nu<2.\end{array}\right.\nonumber
\intertext{and}
&\widetilde{S}_{\nu}=\left\{\begin{array}{ll}\sum_{i=1}^{d}\widetilde{t}_{i}^{\nu/2}&\text{ if }0<\nu\leq 1\\
\sum_{i=1}^{d}\widetilde{t}_{i}^{\nu} & \text{ if } 1<\nu<2.\end{array}\right.\nonumber
\intertext{Then under the claims (\ref{def_zeta_tilde}) and (\ref{det_H_claim2_tilde}) we have the following bounds on the normalization constant $C_{\gamma_{d}}(\bm{x})$ and the upper bounds on the error terms $\widetilde{\kappa}$ and $\widetilde{\xi}$ for the marginal expression $m_{\gamma_{d}}(\widetilde{\widetilde{\bm{z}}})$ in Theorem \ref{theorem1}:}
&|\widetilde{\kappa}|< \frac{4}{3}(1+\widetilde{\zeta})\frac{C_{\nu}\nu|\nu-1|\cdot|\nu-2|}{\left(\frac{n}{d}\Omega(\lambda^{*},\tau^{*})\right)^{\frac{\nu}{2}}}\nonumber\\
&\times\left|\sum_{j=1}^{d}\frac{\widetilde{t}_{j}^{\nu/2}\left|(\tau^{*})^{\frac{1}{2}}\theta_{j}^{*}\right|^{\nu-1}\sgn(\widetilde{\widetilde{\theta^{*}}}_{j})\left(1+\frac{2}{\tau^{*}(\widetilde{\widetilde{\theta^{*}}}_{j})^{2}}\right)}{\exp\left(\frac{1}{2}\tau^{*}(\widetilde{\widetilde{\theta^{*}}}_{j})^{2}\right)}\right|\nonumber\\
&+\frac{1}{N(\lambda^{*},\nu,\gamma_{d})}\sum_{j=1}^{d}\frac{\tau^{*}(\widetilde{\widetilde{\theta^{*}}}_{j})^{2}}{\exp\left(\frac{1}{2}\tau^{*}(\widetilde{\widetilde{\theta^{*}}}_{j})^{2}\right)}\nonumber\\
&+\left|\frac{(2\pi)^{-\frac{1}{2}}}{N(\lambda^{*},\nu,\gamma_{d})}\sum_{i,j=1}^{d}\frac{\tau^{*}(\widetilde{\widetilde{\bm{x}}}_{\parallel}(i)-\frac{1}{2}\theta_{i}^{*})(\widetilde{\widetilde{\bm{x}}}_{\parallel}(j)-\frac{1}{2}\widetilde{\widetilde{\theta^{*}}}_{j})}{\exp\left(\frac{1}{2}\tau^{*}\left[(\widetilde{\widetilde{\theta^{*}}}_{i})^{2}+(\widetilde{\widetilde{\theta^{*}}}_{j})^{2}\right]\right)}\right|.\label{def_kappa_tilde}
\intertext{where} 
&N(\lambda^{*},\nu,\gamma_{d})\sim\left\{\begin{array}{ll}\frac{n-d+2}{2},&\text{ if } 0<\nu\leq 1\\
\frac{n-d+2}{2}-\frac{C_{\nu}\nu|\nu-1|}{4}d,&\text{ if } 1<\nu< 2.\end{array}\right.\nonumber\\
&1\leq\widetilde{\xi}+1<\nonumber\\
&\prod_{i=1}^{d}\left\{1+\left[2P_{G}\left(-\tau_{1}^{\frac{1}{2}}\left|\widetilde{\widetilde{\bm{x}}}_{\parallel}(i)\right|\right)\sup_{t\in\mathbb{R}}{\pi_{\lambda=1}(t)}/\pi_{\lambda=1}\left(u_{0,i}\left(\tau_{1}^{\frac{1}{2}}\widetilde{\widetilde{\bm{x}}}_{\parallel}(i)\right)\right)\right]\times\right.\nonumber\\
&\left.\left[1+\erf\left(\tau_{2}^{\frac{1}{2}}|\widetilde{\widetilde{\bm{x}}}_{\parallel}(i)|\right)\frac{\inf_{t\in\left(0,u_{0,i}\left(\tau_{2}^{\frac{1}{2}}\widetilde{\widetilde{\bm{x}}}_{\parallel}(i)\right)\right)}{\pi_{\lambda=1}(t)}}{\pi_{\lambda=1}\left(u_{0,i}\left(\tau_{2}^{\frac{1}{2}}\widetilde{\widetilde{\bm{x}}}_{\parallel}(i)\right)\right)}\right.\right.\nonumber\\
&\left.\left.-\frac{2C_{\nu}\nu}{(2\pi)^{\frac{1}{2}}}\widetilde{L}_{\nu,i}(\tau_{2}^{\frac{1}{2}}\widetilde{\widetilde{\bm{x}}}_{\parallel}(i))\frac{\sup_{t\in\left(u_{0,i}\left(\tau_{2}^{\frac{1}{2}}\widetilde{\widetilde{\bm{x}}}_{\parallel}(i)\right),\infty\right)}{\pi_{\lambda=1}(t)}}{\pi_{\lambda=1}\left(u_{0,i}\left(\tau_{2}^{\frac{1}{2}}\widetilde{\widetilde{\bm{x}}}_{\parallel}(i)\right)\right)}\right]^{-1}\right\}\label{def_xi_tilde}
\intertext{if}
&\frac{\pi_{\lambda=1}\left(u_{0,i}\left(\tau_{2}^{\frac{1}{2}}\widetilde{\widetilde{\bm{x}}}_{\parallel}(i)\right)\right)}{\sup_{t\in\left(u_{0,i}\left(\tau_{2}^{1/2}\widetilde{\widetilde{\bm{x}}}_{\parallel}(i)\right),\infty\right)}{\pi_{\lambda=1}(t)}}> \frac{2C_{\nu}\nu}{(2\pi)^{\frac{1}{2}}}\widetilde{L}_{\nu,i}\left(\tau_{2}^{\frac{1}{2}}\widetilde{\widetilde{\bm{x}}}_{\parallel}(i)\right),\ \forall\ i\in\gamma_{d}\nonumber
\intertext{where $\tau_{1},\tau_{2}\in I_{\tau}$, $u_{0,i}(s)=\left(\widetilde{t}_{i}^{2}\frac{n}{d}\Omega(\lambda^{*},\tau^{*})\right)^{-\frac{1}{2}}|s|$ and}\nonumber\\
&\widetilde{L}_{\nu,i}\left(\tau^{\frac{1}{2}}x\right)\mathdef\left\{\begin{array}{ll}\left|\tau^{\frac{1}{2}}x\right|^{\nu-1}\left(\widetilde{t}_{i}^{2}\frac{n}{d}\Omega(\lambda,\tau)\right)^{-\frac{\nu}{2}} & \text{ if }0<\nu\leq 1\\
\left(\widetilde{t}_{i}^{2}\frac{n}{d}\Omega(\lambda,\tau)\right)^{-\frac{1}{2}}\left(1+\left|\tau^{\frac{1}{2}}x\right|\left(\widetilde{t}_{i}^{2}\frac{n}{d}\Omega(\lambda,\tau)\right)^{-\frac{1}{2}}\right)& \text{ if }1<\nu<2.\end{array}\right.\label{def_L_nu_tilde}
%\item The number $\zeta$ defined in () becomes
%\item The number $\kappa$ defined in () is bounded by
%\item The number $\xi$
%\item The number $u_{0}$ defined in () becomes
\end{align}
We also note that the INMDL-optimal quantization principle given in Proposition \ref{INMDL_optimal_quantization} only applies to the transformed 
parameters $\widetilde{\widetilde{\bm{\theta}}}$ and not $\widetilde{\bm{\theta}}$ because the Laplace approximation used to estimate the marginal distribution 
$m_{\gamma_{d}}(\bm{x})$ in Theorem \ref{theorem1} was deduced under the assumption of IID gaussian noise.  

%STATE THE BOUNDS ON THE MARGINAL NORMALIZATION CONSTANT $C_{\gamma_{d}}$ IN THIS CASE !

%\section{Numerical experiments on estimating GGD-modeled signals and images in colored gaussian noise}
\section{Numerical methods and experiments}
We applied the INMDL-principle to the deconvolution problem defined above for some test images and compared the 
results to the thresholding algorithm proposed in \cite{MallatKalifa:2003_AnnStat}. 
We define the total variation measure $\|\cdot\|_{tv}$
\begin{align}
&\|\bm{\theta}\|_{tv}\mathdef\sum_{m,n=0}^{N-1}\left[(\theta[m,n+1]-\theta[m,n])^{2}+(\theta[m+1,n]-\theta[m,n])^{2}\right]^{1/2}
\label{def_tv_norm2}
\end{align}
for data $\bm{\theta}\in\mathbb{R}^{N\times N}$. It may be used to compare the smoothness of the original, degraded and estimated
datasets.

\begin{enumerate}
\item The noise variance $\tau^{-1}$ is set to $\tau^{-1}=1$ in all the experiments on graylevel images below, with the graylevel 
  values ranging in the integer range $[0,255]$. 
\item The definition of the SNR is the same is in the previous experimental section on estimating in white gaussian noise.
\item The method of thresholding in a mirror wavelet basis \cite{MallatKalifa:2003_AnnStat}, \cite{MallatKalifa:2003_IEEE} is denoted 
  MWT below. We note that the MWT implemented here does not include a shift-invariant estimation as used in the works cited above. 
  This is due to lack of time to implement the required numerical wavelet-routines.
\item We note that a Fourier cutoff frequency $k_{c}$ with $k_{c}=\frac{N}{2}-8$ was used to cut the the deconvolved data in the Fourier domain 
  because it was needed in the MWT algorithm to stabilize the algorithm. We note that the INMDL algorithm was found to yield the same results with no
  Fourier cutoff.
\item The wavelet used was the Symmlet of filter length 20 in all experiments in this section.
 We note that in all the {\em image} experiments below, the numbers: $\widetilde{\kappa}$, $\widetilde{\xi}$, $\widetilde{\zeta}$, 
$\widetilde{\omega}$, $\widetilde{X}$, $\inf_{1\leq i\leq d}|\sqrt{\tau^{*}}\theta_{i}^{*}|$ defined above on which bounds are needed to ensure 
 the validity of our asymptotic marginal formula in Theorem \ref{theorem1}, Corollary \ref{corollary_new_marginal} and 
 the marginal renormalization constant in Proposition \ref{Marginal_Renormalization_Constant}, were found to range in intervals 
 approximately as stated in the previous experiments section.
\end{enumerate}

\begin{figure}[h]
\begin{center}
\includegraphics[scale=0.80]{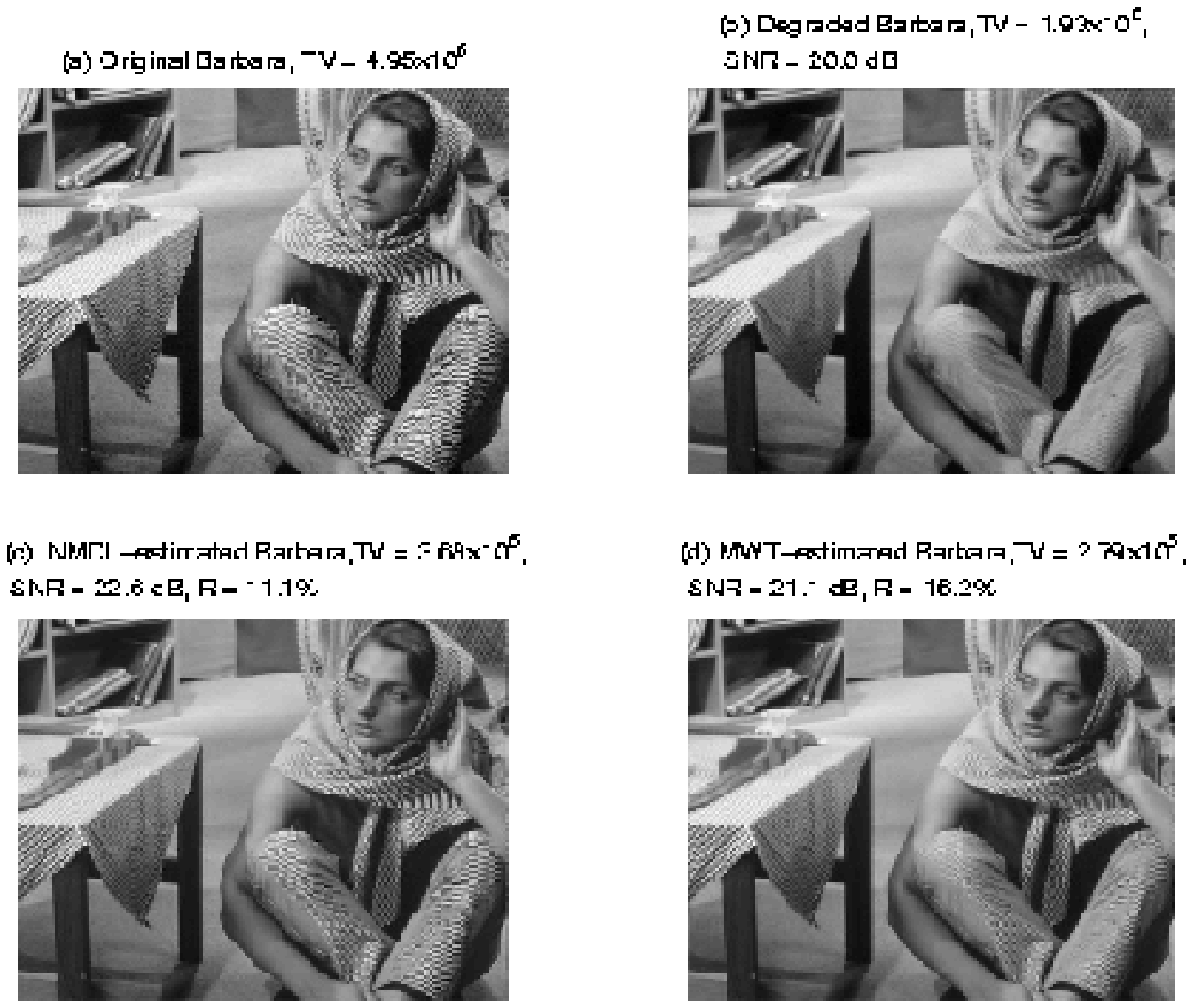}
\end{center}
\caption{Comparison of estimator performance for INMDL and the MWT algorithms on 8-bit grayscale test image barbara with $N= 512\times 512$. The noise variance is $\tau^{-1}=1.0$ and the lowpass filter used 
  to degrade the image was: $\cos^{p_{x}}\left(k_{x}\pi/\sqrt{N}\right)\cos^{p_{y}}\left(k_{y}\pi/\sqrt{N}\right)$ in the Fourier domain 
  with $p_{x}=p_{y}=3.0$. TV denotes the total variation (\ref{def_tv_norm2}) and $R$ is the fraction of nonzero wavelet coefficients in the estimate. The GGD shape parameter was estimated by 
  the INMDL-algorithm to: $\nu=0.606$.}
\label{DeconvolutionPerformance_barbara}
\end{figure}

\begin{figure}[h]
\begin{center}
\includegraphics[scale=0.80]{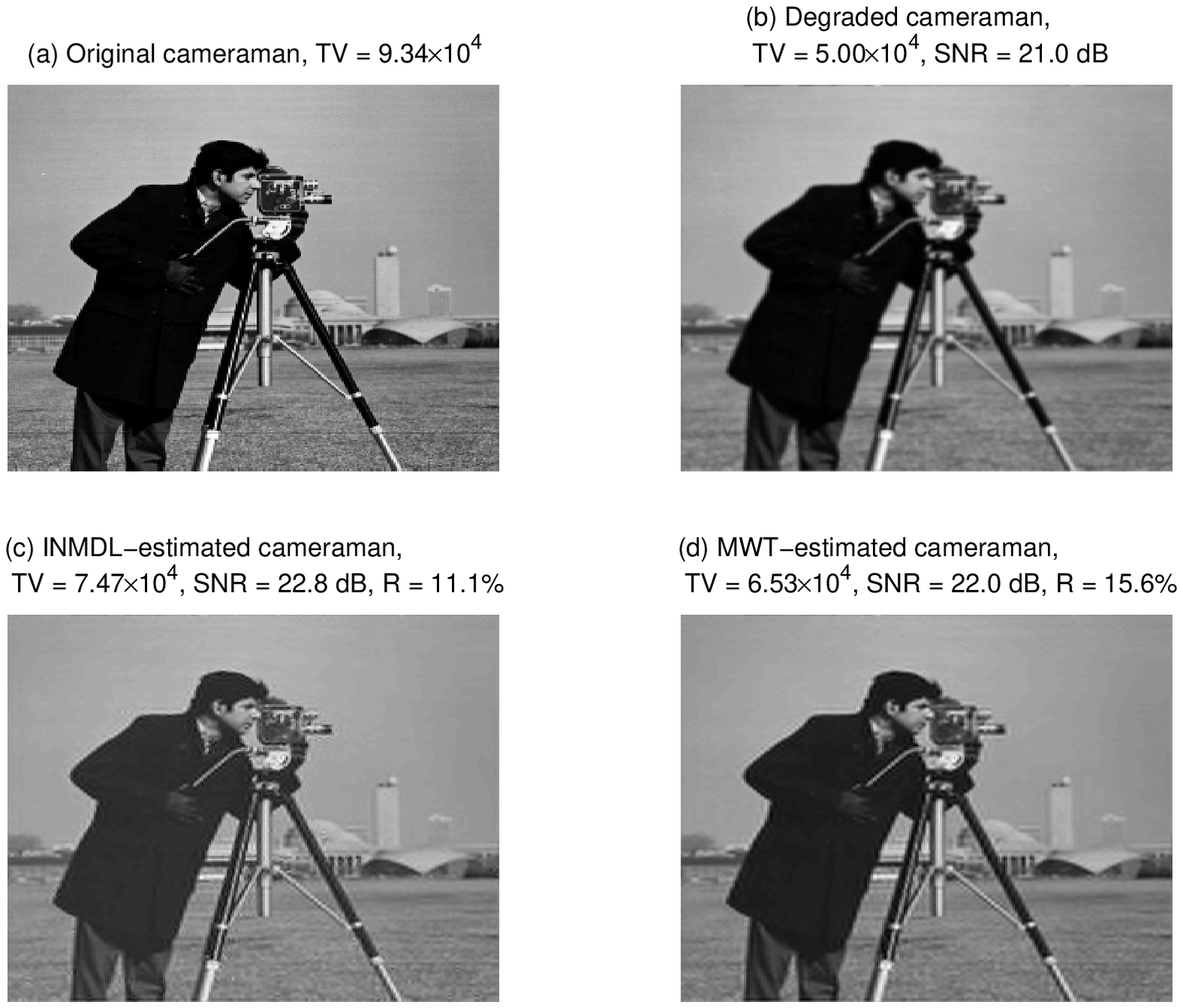}
\end{center}
\caption{Comparison of estimator performance for INMDL and the MWT algorithms on  8-bit grayscale test image cameraman with $N=256\times 256$. The noise variance is $\tau^{-1}=1.0$ and the lowpass filter used 
  to degrade the image was: $\cos^{p_{x}}\left(k_{x}\pi/\sqrt{N}\right)\cos^{p_{y}}\left(k_{y}\pi/\sqrt{N}\right)$ in the Fourier domain 
  with $p_{x}=p_{y}=3.0$. TV denotes the total variation (\ref{def_tv_norm2}) and $R$ is the fraction of nonzero wavelet coefficients in the estimate. The GGD shape parameter was estimated by 
  the INMDL-algorithm to: $\nu=0.556$.}
\label{DeconvolutionPerformance_camera1}
\end{figure}

We wanted to investigate the performance of the INMDL-algorithm using a harder blurring operator, for example operators given by box car convolution filters. Since the frequency response of such a filter is a sinc-function with multiple zeros in the frequency domain, the MWT-algorithm
is not applicable in this case. We have implemented a INMDL-based method which uses an adapted wavelet packet basis $\mathcal{B}$ where the basis is 
adapted to both the degraded input data $y$ in (\ref{deconv_problem}) and the deconvolution filter $u^{-1}$ in (\ref{def_inverse_u}).
The only difference to the INMDL-based deconvolution algorithm defined above for the hyperbolic filters, is that the mirror wavelet basis
is exchanged for a specially chosen wavelet packet basis $\bm{B}^{*}$. We briefly outline below the main ingredients in the process of 
selecting a suitable basis $\bm{B}^{*}$ and refer to \cite{Wickerhauser:1994} and \cite{Mallat:1998} for details on wavelet packet bases.
\begin{enumerate}
\item A wavelet is chosen (Symmlet 20 in our case) and the degraded data $y$ is expanded into some (not full) constrained 
  anisotropic wavelet packet analysis on $\mathbb{R}^{n\times n}$, see \cite{Wickerhauser:1994}. 
  An {\em additive} cost measure \cite{CoifWick:1992} is specified, we used here the entropy-measure $S_{e}$.
  \begin{align}
    S_{e}(\bm{y}) \mathdef -\sum_{i=1}^{n}y_{i}^{2}\log{y_{i}^{2}}.\label{additive_entropy_measure}
  \end{align}
  Then $S_{e}\left(\bm{B}_{i}^{T}\bm{y}\right)$ is evaluated for all the allowed discrete wavelet packet bases $\bm{B}_{i}$ on $\mathbf{R}^{n\times n}$
  for the given wavelet (S20) using the fast ''best basis algorithm'' of \cite{CoifWick:1992}. We note that the total number of different
  wavelet packet bases exceeds $2^{N/2}$, \cite{CoifWick:1992} where $N=n\times n$ for image data. However, the ''best basis algorithm'' 
  ensures that the unique basis minimizing the additive cost measure is found in $\bigo(N\log_{2}{N})$ operations.
\item A constraint is imposed on the search for the optimal wavelet packet basis: Wavelet packet subspaces $\bm{W}$
  spanning a Fourier frequency rectangle $R\mathdef [k_{x}^{(1)},k_{x}^{(2)}]\times[k_{y}^{(1)},k_{y}^{(2)}]$ 
  where the deconvolution filter $\hat{u}^{-1}(k_{x},k_{y})$ ''varies too much'' are marked as not selectable.
  We used here the restriction 
  \begin{align}
  \frac{\sup_{(k_{x},k_{y})\in R}\hat{u}^{-1}(k_{x},k_{y})}{\inf_{(k_{x},k_{y})\in R}\hat{u}^{-1}(k_{x},k_{y})}\leq Q\label{deconv_restrict}
  \end{align}
  with $Q=16$, this value on $Q$ corresponds to the variation factor of the kernel $\cos^{2}(k_{x}\pi/\sqrt{N})\cos^{2}(k_{y}\pi/\sqrt{N})$ 
  (used in \cite{MallatKalifa:2003_IEEE}) inside the different subspaces of the mirror wavelet basis.
  It is easy to realize that there exists wavelet packet bases $\bm{B}$ of which the subspaces $\bm{W}$ satisfies the constraint 
  (\ref{deconv_restrict}) because each wavelet packet basis element has an essential frequency support inside a frequency box 
  $R_{j_{x},m_{x}}\times R_{j_{y},m_{y}}\subset I\mathdef [-\pi,\pi]\times[-\pi,\pi]$ with
  \begin{align}
    &R_{j_{x},m_{x}}\mathdef [-\pi 2^{j_{x}}(m_{x}+1),-\pi 2^{j_{x}}m_{x}]\cup [\pi 2^{j_{x}}m_{x},\pi 2^{j_{x}}(m_{x}+1)],\nonumber\\
     & -log_{2}{n}\leq j_{x},j_{y}<0,\ 0\leq m_{x}<2^{-j_{x}},\ 0\leq m_{y}< 2^{-j_{y}}\label{dyadic_frequancy_tiling}
  \end{align}
  and there exists an injection from the collection of different tilings of the frequency square $I$ by elements 
  $R_{j_{x},m_{x}}\times R_{j_{y},m_{y}}$ into the collection of 
  different discrete wavelet packet bases on $\mathbb{R}^{n\times n}$, \cite{Wickerhauser:1994}, \cite{Mallat:1998}.
\item  A diagonal estimate $\bm{K}^{*}$ of the covariance matrix of the deconvolved noise represented in the selected 
  wavelet packet basis $\bm{B}^{*}$ is computed similarly to the case of the mirror wavelet basis shown above. We note that since 
    the blurring kernels used in the model of the degradation process are separable, the required numbers 
    $\langle\widehat{\bm{K}}\widehat{\bm{w}}_{k},\widehat{\bm{w}}_{k}\rangle$, where $\widehat{\bm{K}}$ is the discrete Fourier representation of
    the covariance $\bm{K}$ of the deconvolved noise and $\widehat{\bm{w}}_{k}$ is the discrete Fourier transform of a wavelet packet basis 
    element $\bm{w}_{k}$, may be computed fast with $\bigo(\sqrt{N})$ operations for each $k$.
\end{enumerate}

\begin{figure}[h]
\begin{center}
\includegraphics[scale=0.80]{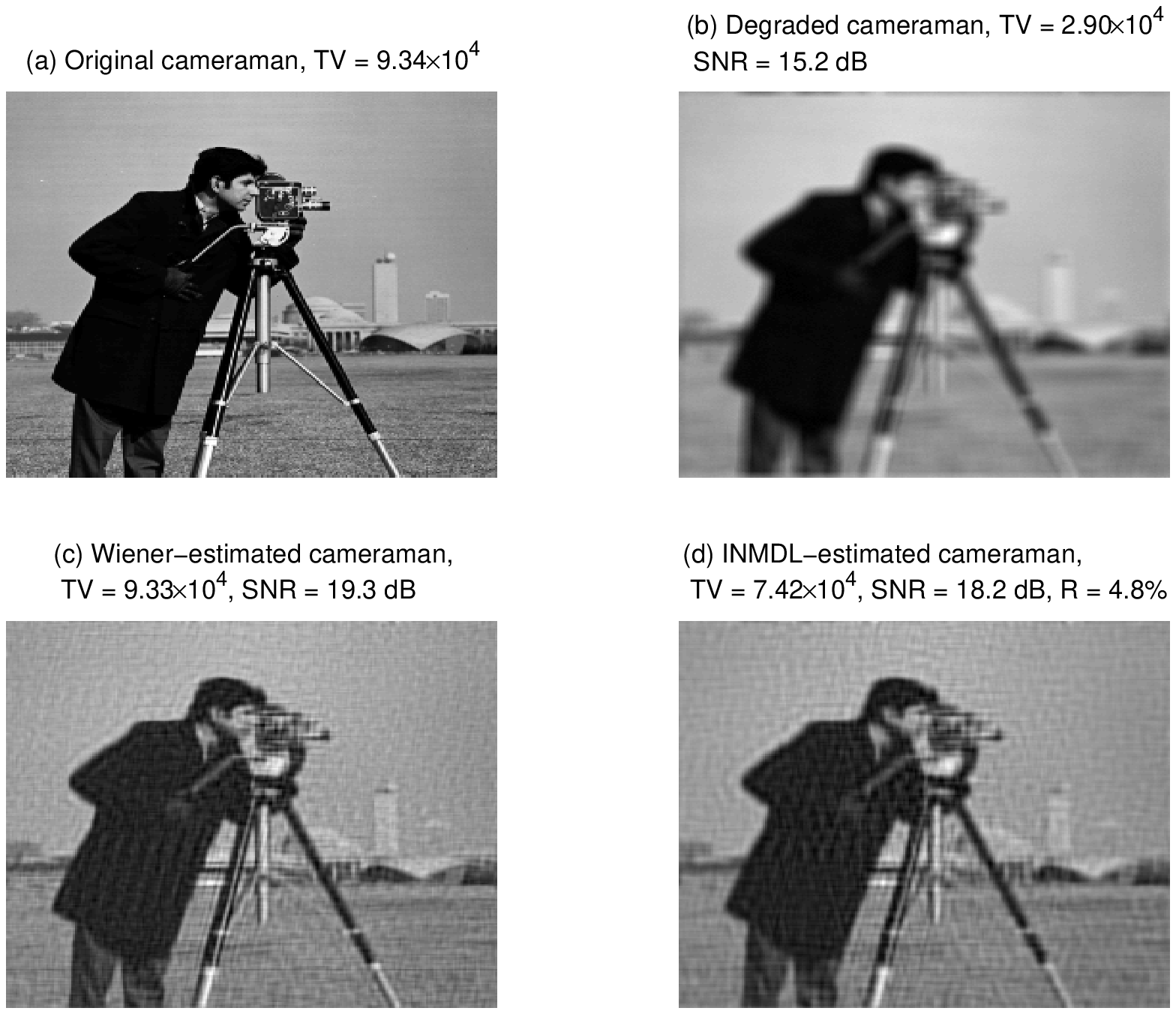}
\end{center}
\caption{Comparison of estimator performance for INMDL and the Wiener filter algorithms on 8-bit grayscale test image cameraman with $N=256\times 256$. The noise variance is $\tau^{-1}=1.0$ and the lowpass filter used 
  to degrade the image was a $9\times9$ box car filter. TV denotes the total variation (\ref{def_tv_norm2}) and $R$ is the fraction of nonzero wavelet coefficients in the estimate. The GGD shape parameter was estimated by 
  the INMDL-algorithm to: $\nu=0.667$. The Wiener filter regularization parameter used was $\alpha=1.0$.}
\label{DeconvolutionPerformance_camera2}
\end{figure}

\begin{figure}[h]
\begin{center}
\includegraphics[scale=0.80]{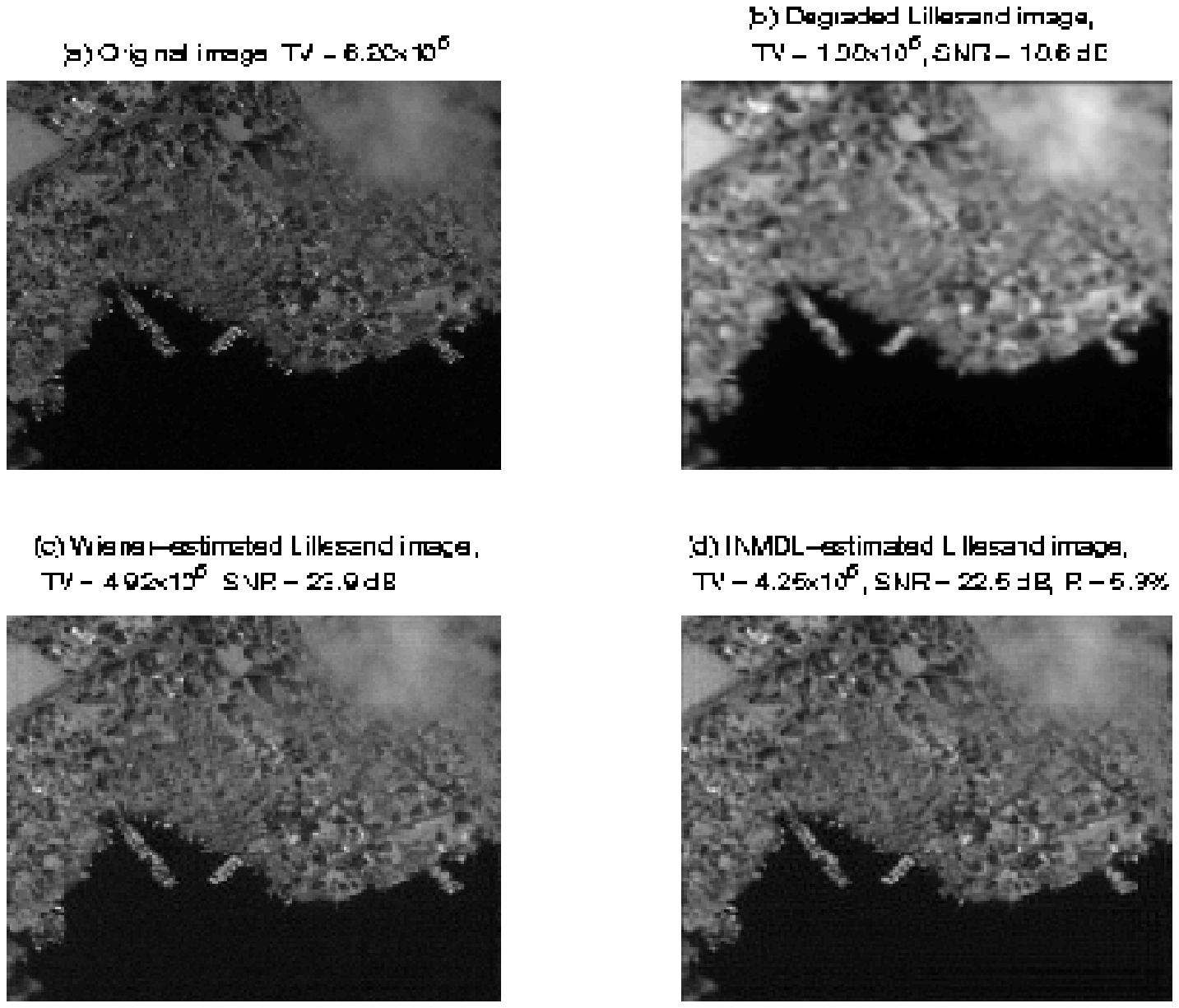}
\end{center}
\caption{Comparison of estimator performance for INMDL and the Wiener filter algorithms on 8-bit grayscale test image Lillesand with $N=512\times 512$. The noise variance is $\tau^{-1}=1.0$ and the lowpass filter used 
  to degrade the image was a $9\times9$ box car filter. TV denotes the total variation (\ref{def_tv_norm2}) and $R$ is the fraction of nonzero wavelet coefficients in the estimate. The GGD shape parameter was estimated by 
  the INMDL-algorithm to: $\nu=0.679$. The Wiener filter regularization parameter used was $\alpha=1.0$. This image dataset 
was provided to us by the Earth Observation Group at NORUT Information Technology Ltd, Tromsø.}
\label{DeconvolutionPerformance_lillesand}
\end{figure}

We compared the INMDL-deconvolution in the adapted basis defined above to the performance of the Wiener filter $R_{\alpha}(k_{x},k_{y})$
\begin{align}
&R_{\alpha}(k_{x},k_{y})\mathdef\frac{1}{1+\alpha\frac{\sigma^{2}|U_{x}(k_{x})|^{-2}|U_{y}(k_{y})|^{-2}}{|P_{\theta}(k_{x},k_{y})|}}\label{def_WienerFilter}
\end{align}  
where $U_{x}$ and $U_{y}$ denote the Fourier transforms of the convolution filters $u_{x}$ and $u_{y}$ 
applied along rows and colums of the image, respectively in the degradation process (\ref{deconv_problem}),
$P_{\theta}$ denotes the power spectrum of the unknown signal $\theta$ in (\ref{deconv_problem}) and
$0<\alpha<\infty$ is a regularization parameter.
We applied the iterative algorithm given in \cite{HilleryChin:1991} to estimate $P_{\theta}$. 
The results on test images are shown in Figure \ref{DeconvolutionPerformance_camera2} and Figure \ref{DeconvolutionPerformance_lillesand}.
We also tried the INMDL-deconvolution algorithm on a high resolution optical gray level image taken by satellite Ikonos, this is
shown in the test image Lillesand in Figure \ref{DeconvolutionPerformance_lillesand}. The bitdepth of the image is 11, and
the pixel resolution is 1 meter. 

\subsection{Discussion of experimental results}

The results shown in Figure \ref{DeconvolutionPerformance_barbara} and Figure \ref{DeconvolutionPerformance_camera1} show that the INMDL-based
restoration algorithm performs slightly better than the MWT-method. However, the blurring of the images imposed by the 
kernels $\cos^{p_{x}}\left(k_{x}\pi/\sqrt{N}\right)\cos^{p_{y}}\left(k_{y}\pi/\sqrt{N}\right)$ is not very hard as may be seen from the
Figure \ref{DeconvolutionPerformance_barbara} and Figure \ref{DeconvolutionPerformance_camera1}. 
Also, much better restoration results using a MWT-method are reported in \cite{MallatKalifa:2003_AnnStat}, \cite{MallatKalifa:2003_IEEE}, but this
difference from our reported results is likely due to a post-processing of the MWT-estimates by the ''spin-cycling''-method \cite{DonCoif:1995} 
yielding a shift-invariant estimate. Unfortunately, we have not had the time to implement this important stage of the 
estimation process, but we have no reason 
to believe that the INMDL-based estimates would not benefit as much from this kind of posterior regularization techniques as is shown to be 
the case for the MWT-method in \cite{MallatKalifa:2003_AnnStat}. 
Therefore, the experimental results obtained here for the INMDL and MWT estimators, although not impressive in performance,
we believe they may be used to compare the (potential) performance of the MWT and the INMDL-based estimators. Our conclusion
is then that the INMDL principle offers an alternative deconvolution technique which compares favourably to the MWT method.

In the case of using the INMDL-principle to restore images degraded by a box car filter, we had to use a basis which approximately diagonalizes the 
covariance matrix $\bm{K}$ of the deconvolved noise. For this purpose we used a certain type of anisotropic discrete wavelet packet analysis 
for $\mathbb{R}^{n\times n}$ \cite{Wickerhauser:1994}, \cite{Mallat:1998} together with the best basis algorithm \cite{CoifWick:1992} 
and some constraints on selectable wavelet packet bases as explained above. 
Figure \ref{DeconvolutionPerformance_camera2} shows that both the Wiener and the INMDL-estimate visually suffers from the same kind of 
global ripple artifacts. The explanation for this in the case of the Wiener filter is of course that the each Fourier basis element
has a support equal the entire spatial (pixel) domain. In the case of the INMDL-estimate the explanation is that the method outlined 
above of not allowing wavelet packet bases with elements possessing a Fourier frequency support over which the deconvolution filters 
$U_{x}^{-1}(k_{x})$ or $U_{y}^{-1}(k_{y})$ are not ''approximately constant'', favours the selection of wavelet packet bases with 
(at least some) basis elements of high frequency resolution (small frequency support) and thus these basis elements must have a large spatial support.

Finally, 
we note that \cite{BarNeelChoi:1999} has reported results obtained with a {\em hybrid} method where one first preprocess the deconvolved data with a collection of adaptive Wiener filters $\{R_{\alpha_{j}}\}_{j\in J},\ 0<\alpha_{j}<1$ in the Fourier domain, and then one estimates the  signal in the wavelet 
domain from the Fourier-regularized data by the universial thresholding scheme \cite{DJ_Ideal:1994}. 
The constructed hybrid estimator is shown in the cited paper to outperform the ordinary Wiener filter $R_{\alpha}$ in experiments. 
Thus, one possible approach to improving the performance of the INMDL-deconvolution principle as defined and tested above, would be to apply some
kind of regularization in the Fourier domain (or possibly in a suitable wavelet packet domain) to the deconvolved data, and then denoise the deconvolved data in a suitable wavelet/wavelet packet basis. To incorporate such a regularization in a codelength principle is a 
topic for future research.

\bibliography{RefDataBase}

\providecommand{\bysame}{\leavevmode\hbox to3em{\hrulefill}\thinspace}
\providecommand{\MR}{\relax\ifhmode\unskip\space\fi MR }
\providecommand{\MRhref}[2]{%
  \href{http://www.ams.org/mathscinet-getitem?mr=#1}{#2}
}
\providecommand{\href}[2]{#2}
\begin{thebibliography}{KMR03}

\bibitem[AS70]{Stegun:1960}
M.~Abramowitz and I.A. Stegun, \emph{Handbook of mathematical functions},
  Dover, 1970.

\bibitem[Bal96]{Bala:1996}
V.~Balasubramanian, \emph{A geometric formulation of occam's razor for
  inference of parametric distributions}, Tech. report, Princeton University,
  Jan. 1996, Princeton preprint PUPT-1588. Available online at: {\tt
  http://schwinger.harvard.edu/${\sim}$vijayb/}.

\bibitem[Bal97]{Bala:1997}
\bysame, \emph{Statistical inference, occam's razor, and statistical mechanics
  on the space of probability distributions}, Neural Computation \textbf{9}
  (1997), no.~2, 349--368, Available online at: {\tt
  http://schwinger.harvard.edu/${\sim}$vijayb/}.

\bibitem[BCN99]{BarNeelChoi:1999}
R.G. Baraniuk, H.~Choi, and R.~Neelmani, \emph{Wavelet-domain regularized
  deconvolution for ill-conditioned systems}, IEEE Image Processing, 1999. ICIP
  99. Proceedings. 1999 International Conference on \textbf{1} (1999),
  204--208, Available online at: {\tt http://citeseer.nj.nec.com}.

\bibitem[BG95a]{BruceGao:1995a}
A.~Bruce and H.~Gao, \emph{Understanding waveshrink: Variance and bias
  estimation}, Tech. report, StatSci Division of MathSoft Inc., 1995, {In
  preparation. Bruce, A. G. and Gao, H.-Y. (1995b). Understanding WaveShrink:
  Variance and Bias Estimation. Technical report, StatSci Division, MathSoft,
  Inc., 1700 Westlake Ave. N, Seattle, WA 98109-9891}.

\bibitem[BG95b]{BruceGao:1995c}
\bysame, \emph{Waveshrink: Shrinkage functions and thresholds}, Tech. report,
  StatSci Division, MathSoft, Inc., 1995, Proc. SPIE, San Diego, CA, 1995.

\bibitem[BG95c]{BruceGao:1995b}
\bysame, \emph{Waveshrink with semisoft shrinkage}, Tech. report, StatSci
  Division of MathSoft Inc., 1995, { Bruce, A. and Gao, H., WaveShrink with
  Semisoft Shrinkage. StaSci Research Report No. 39 (1995) }.

\bibitem[BRY98]{Rissanen:1998a}
A.~Barron, J.~Rissanen, and Bin Yu, \emph{The minimum description length
  principle in coding and modeling}, IEEE Transactions on Information Theory
  \textbf{44} (1998), no.~6, 2743--2760.

\bibitem[CD95]{DonCoif:1995}
R.R. Coifman and D.L. Donoho, \emph{Translation-invariant denoising}, Tech.
  report, Stanford University and Yale University, 1995, Available online at:
  {\tt www-stat.stanford.edu/\verb+~+donoho/reports.html}.

\bibitem[CH91]{HilleryChin:1991}
R.~T. Chin and A.~D. Hillery, \emph{Iterative wiener filters for image
  restoration}, IEEE Transactions on Signal Processing \textbf{39} (1991),
  no.~8, 1892--1899.

\bibitem[CRM98]{CohenMalahRaz:1998}
I.~Cohen, S.~Raz, and D.~Malah, \emph{Mdl-based translation-invariant denoising
  and robust time-frequency representations}, Proc. of the 4th IEEE-SP
  Int.Symposium on Time-Frequency and Time-Scale Analysis, Pittsburgh,
  Pennsylvania, 6--9 Oct. 1998. (1998), Available online at: {\tt
  http://citeseer.nj.nec.com/cohen98mdlbased.html}.

\bibitem[CT91]{CoverThomas:1991}
T.M. Cover and J.A. Thomas, \emph{Elements of information theory}, Wiley, 1991.

\bibitem[CV00]{ChangYuVetterli:2000}
Bin~Yu Chang, S.G. and M.~Vetterli, \emph{Adaptive wavelet thresholding for
  image denoising and compression}, IEEE Transactions on Image Processing
  \textbf{9} (2000), no.~9, 1532--1547.

\bibitem[CW92]{CoifWick:1992}
R.R. Coifman and M.V. Wickerhauser, \emph{Entropy-based algorithms for best
  basis selection}, IEEE Transactions on Information Theory \textbf{38} (1992),
  no.~2, 713--718.

\bibitem[Dau92]{Daubechies:1992}
I.~Daubechies, \emph{Ten lectures on wavelets}, Siam, 1992.

\bibitem[DJ94]{DJ_Ideal:1994}
D.L. Donoho and I.M. Johnstone, \emph{Ideal spatial adaptation by wavelet
  shrinkage}, Biometrika \textbf{81} (1994), no.~3, 425--455.

\bibitem[DJ95]{DJ_Adapt:1995}
\bysame, \emph{Adapting to unknown smoothness via wavelet shrinkage}, Journal
  of the American Statistical Association \textbf{90} (1995), no.~432,
  1200--1224.

\bibitem[DJ98]{DJ_minimax:1998}
\bysame, \emph{Minimax estimation via wavelet shrinkage}, Annals of Statistics
  \textbf{26} (1998), no.~3, 879--921.

\bibitem[DV02]{DoVetterli:2002}
M.N. Do and M.~Vetterli, \emph{Wavelet-based texture retrieval using
  generalized gaussian density and kullback-leibler distance}, IEEE
  Transactions on Image Processing \textbf{11} (2002), 146--158.

\bibitem[GR00]{Gradshteyn:2000}
I.S. Gradshsteyn and I.M. Ryzhic, \emph{Table of integrals, series, and
  products, sixth edition}, Academic Press, 2000.

\bibitem[HY00]{Hansen-Yu:2000}
M.~Hansen and Bin Yu, \emph{Wavelet thresholding via mdl for natural images},
  IEEE Transactions on Information Theory \textbf{46} (2000), no.~5,
  1778--1788.

\bibitem[KM03]{MallatKalifa:2003_AnnStat}
J.~Kalifa and S.~Mallat, \emph{Thresholding estimators for linear inverse
  problems and deconvolutions}, Annals of Statistics \textbf{31} (2003), no.~1,
  58--109, Available online at: {\tt http://projecteuclid.org/}.

\bibitem[KMR03]{MallatKalifa:2003_IEEE}
J.~Kalifa, S.~Mallat, and B.~Rouge, \emph{Deconvolution by thresholding in
  mirror wavelet bases}, IEEE Transactions on Image Processing \textbf{12}
  (2003), no.~4, 446--457, Available online at: {\tt
  www.cs.nyu.edu/cs/faculty/mallat/biblio.html}.

\bibitem[KTK88]{KassKadaneTierney:1988}
R.E. Kass, L.~Tierney, and J.B Kadane, \emph{Asymptotics in bayesian
  computation}, Bayesian Statistics 3, vol.~3, Oxford University Press, 1988,
  pp.~261--278.

\bibitem[Lan01]{Lanterman:2001}
A.D. Lanterman, \emph{Schwarz, wallace, and rissanen: Intertwining themes in
  theories of model order estimation}, International Statistical Review
  \textbf{69} (2001), no.~2, 185--212.

\bibitem[Mal98a]{Mallat:1998b}
S.~Mallat, \emph{Applied mathematics meets signal processing}, 1998, Available
  online at: {\tt www.cs.nyu.edu/cs/faculty/mallat/biblio.html}.

\bibitem[Mal98b]{Mallat:1998}
S.~Mallat, \emph{A wavelet tour of signal processing}, Academic Press, 1998.

\bibitem[ML99]{Moulin-Liu:1999}
P.~Moulin and J.~Liu, \emph{Analysis of multiresolution image denoising schemes
  using generalized gaussian and complexity priors}, IEEE Transactions on
  Information Theory \textbf{45} (1999), no.~3, 909--919.

\bibitem[OB94a]{OliverBaxter:1994b}
J.J. Oliver and R.~Baxter, \emph{Mdl and mml: Similarities and differences},
  Tech. report, Department of Computer Science, Monash University, 1994,
  Available online at: {\tt http://citeseer.nj.nec.com/cs}.

\bibitem[OB94b]{OliverBaxter:1994a}
\bysame, \emph{Mml and bayesianism: Similarities and differences}, Tech.
  report, Department of Computer Science, Monash University, 1994, Available
  online at: {\tt http://citeseer.nj.nec.com/cs}.

\bibitem[OH94]{OliverHand:1994}
J.J. Oliver and D.~Hand, \emph{Introduction to minimum encoding inference},
  Tech. report, Department of Computer Science, Monash University, 1994,
  Available online at: {\tt http://citeseer.nj.nec.com/cs}.

\bibitem[Ris96]{Rissanen:1996}
J.~Rissanen, \emph{Fisher information and stochastic complexity}, IEEE
  Transactions on Information Theory \textbf{42} (1996), no.~1, 40--47.

\bibitem[Ris98]{Rissanen:1998b}
\bysame, \emph{Stochastic complexity in statistical inquiry}, World Scientific,
  1998.

\bibitem[Ris00]{Rissanen:2000}
\bysame, \emph{Mdl denoising}, IEEE Transactions on Information Theory
  \textbf{46} (2000), no.~7, 2537--2543.

\bibitem[Ris01]{Rissanen:2001}
\bysame, \emph{Strong optimality of the normalized ml models as universal codes
  and information in data}, IEEE Transactions on Information Theory \textbf{47}
  (2001), no.~5, 1712--1717.

\bibitem[Sai94]{Saito:1994}
N.~Saito, \emph{Local feature extraction and its applications using a library
  of bases}, Ph.D. thesis, Yale University, Department of Mathematics, 10
  Hillhouse Avenue, P.O. Box 208283 New Haven, CT 06520-8283, december 1994,
  Available online at: {\tt http://www.math.yale.edu/pub/papers/}.

\bibitem[Ste81]{Stein:1981}
C.M. Stein, \emph{Estimation of the mean of a multivariate normal
  distribution}, Annals of Statistics \textbf{9} (1981), no.~6, 1135--1151.

\bibitem[TK86]{TierneyKadane:1986}
L.~Tierney and J.B Kadane, \emph{Accurate approximations for posterior moments
  and marginal densities}, Journal of the American Statistical Association
  \textbf{81} (1986), 82--86.

\bibitem[TKK89]{TierneyKassKadane:1989}
L.~Tierney, R.E. Kass, and J.B. Kadane, \emph{Fully exponential laplace
  approximations for expectations and variances of nonpositive functions},
  Journal of the American Statistical Association \textbf{84} (1989), 710--716.

\bibitem[Vid98]{Vidakovic:1998}
B.~Vidakovic, \emph{Nonlinear wavelet shrinkage with bayes rules and bayes
  factors}, Journal of the American Statistical Association \textbf{93} (1998),
  173--179.

\bibitem[WF87]{WallaceFreeman:1987}
C.S. Wallace and D.M. Freeman, \emph{Estimation and inference by compact
  coding}, Computer Journal \textbf{11} (1987), 185--194.

\bibitem[Wic94]{Wickerhauser:1994}
M.L. Wickerhauser, \emph{Adapted wavelet analysis from theory to software}, A K
  Peters, 1994.

\end{thebibliography}
\bibliographystyle{amsalpha}

%APPENDIX
\appendix
%\input{/home/eirikf/LaTeX/Thesis/appendix1}
%\appendix
\chapter{Notation and definitions}
\begin{enumerate}
\item Let $\mathcal{A}$ be a set, then $\mathcal{A}^{n}$ denotes the collection of all strings with $n$ elements taken from $\mathcal{A}$.
\item $[x]_{u}$ is the physical dimension unit of the real variable $x$, that is: $[2 \text{ meter}]_{u}=\text{ meter }$.
\item $[x]_{v}$ is the number of the real variable $x$, that is: $[2 \text{ meter}]_{v}=2$.
\item $\log{x}$ is the natural logarithm of $x$, that is: $x=\exp\left(\log{x}\right),\ \forall x\in\mathbb{R}_{+}$. 
\item $\log_{a}{x}$ is the logarithm of $x$ in base $a$, that is: $x=a^{\left(\log_{a}{x}\right)},\ \forall x\in\mathbb{R}_{+}$, 
$a>1$. 
\item For $\bm{x}\in\mathbb{C}^{N}$ and $1\leq p<\infty$, define the norm $\|\bm{x}\|_{p}$ by: 
$\|\bm{x}\|_{p}\mathdef\left(\sum_{i=1}^{N}\left|x_{i}\right|^{p}\right)^{1/p}$. 
\item For countable sequences $\{a_{k}\}_{k=-\infty}^{\infty}$ of real or complex numbers, define: 
$\|a_{k}\|_{p}\mathdef\left(\sum_{k=-\infty}^{\infty}\left|a_{k}\right|^{p}\right)^{1/p}$. 
\item For functions $f:\mathbb{R}^{n}\longrightarrow\mathbb{C}$ and $1\leq p<\infty$ define the norm $\|f\|_{p}$ by:
$\|f\|_{p}\mathdef\left(\int_{\bm{x}\in\mathbb{R}^{n}}|f(\bm{x})|^{p}\ d\bm{x}\right)^{1/p}$.
\item Define $\ell^{p}\mathdef\left\{\{a_{k}\}_{k=-\infty}^{\infty}: \|a_{k}\|_{p}<\infty\right\}$.
\item Define $L^{p}(\mathbb{R}^{n})\mathdef\left\{f:\mathbb{R}^{n}\longrightarrow\mathbb{C}:\|f\|_{p}<\infty\right\}$.
\item Let $C^{k}(\mathbb{R}^{n})$ denote the space under addition of functions $f(\bm{x}):\mathbb{R}^{n}\longrightarrow\mathbb{C}$ with
$k$ continuous derivatives. 
\item For functions $f:\mathbb{R}^{n}\longrightarrow\mathbb{C}$ such that $f\in L^{1}(\mathbb{R}^{n})$ define the Fourier transform 
$\mathcal{F}:L^{1}(\mathbb{R}^{n})\longrightarrow C(\mathbb{R}^{n})$ by:
$\hat{f}(\bm{\xi})\mathdef\mathcal{F}(f)(\bm{\xi})\mathdef(2\pi)^{-1/2}\int_{\bm{x}\in\mathbb{R}^{n}}f(\bm{x})\exp(-i\bm{x}\cdot\bm{\xi})\ d\bm{x}$, 
$\bm{\xi}\in\mathbb{R}^{n}$. 
\item For $\bm{x}\in\mathbb{C}^{N}$ define the discrete Fourier transform $\mathcal{F}(\{x_{m}\}_{m=1}^{N})$ 
by: $\hat{x}[k]\mathdef\mathcal{F}(\{x_{m}\}_{m=1}^{N})[k]\mathdef N^{-1/2}\sum_{m=1}^{N}x_{m}\exp(-2\pi imk/N)$, 
$-N/2\leq k<N/2$. This definition is extended to countable sequences $\{a_{m}\}\in\ell^{1}$ by:
 $\hat{a}(\omega)\mathdef\mathcal{F}(\{a_{m}\}_{m=-\infty}^{\infty})(\omega)$\\ $\mathdef$ 
$(2\pi)^{-1/2}\sum_{m=-\infty}^{\infty}a_{m}\exp(-im\omega)$, $\omega\in [-\pi,\pi)$.
\item For functions $f:\mathbb{R}^{n}\longrightarrow\mathbb{C}$ such that $f\in L^{1}(\mathbb{R}^{n})$ define the Fourier transform 
$\mathcal{F}:L^{1}(\mathbb{R}^{n})\longrightarrow C(\mathbb{R}^{n})$ by:
$\hat{f}(\bm{\xi})\mathdef\mathcal{F}(f)(\bm{\xi})\mathdef(2\pi)^{-1/2}\int_{\bm{x}\in\mathbb{R}^{n}}f(\bm{x})\exp(-i\bm{x}\cdot\bm{\xi})\ d\bm{x}$, 
$\bm{\xi}\in\mathbb{R}^{n}$. 
\item For column vectors $\bm{x},\bm{y}\in\mathbb{C}^{N}$ , define the inner product 
$\langle\cdot,\cdot\rangle:\mathbb{C}^{N}\times\mathbb{C}^{N}\longrightarrow\mathbb{C}$ by:
$\langle\bm{x},\bm{y}\rangle\mathdef\overline{\bm{y}}^{T}\bm{x}$ $=\sum_{k=1}^{N}x_{k}\overline{y}_{k}$.
\item For sequences $\{a_{k}\},\{b_{k}\}\in \ell^{2}$ define the inner product $\langle\cdot,\cdot\rangle: \ell^{2}\times \ell^{2}\longrightarrow\mathbb{C}$ by: $\langle a_{k},b_{k}\rangle\mathdef\sum_{n=-\infty}^{\infty}a_{n}\overline{b_{n}}$.
\item For functions $f,g\in L^{2}(\mathbb{R}^{n})$ define the inner product $\langle\cdot,\cdot\rangle: L^{2}(\mathbb{R}^{n})\times L^{2}(\mathbb{R}^{n})\longrightarrow\mathbb{C}$ by: $\langle f,g\rangle\mathdef\int_{\bm{x}\in\mathbb{R}^{n}}f(\bm{x})\overline{g(\bm{x})}\ d\bm{x}$.
\item For sequences $\{a_{k}\},\{b_{k}\}\in\ell^{1}$ define the convolution operator $\ast: \ell^{1}\times\ell^{1}\longrightarrow\ell^{1}$ by:
$(a\ast b)_{k}\mathdef\sum_{n=-\infty}^{\infty}a_{n}b_{k-n}$.
\item For functions $f,g\in L^{1}(\mathbb{R}^{n})$ define the convolution operator 
$\ast: L^{1}(\mathbb{R}^{n})\times L^{1}(\mathbb{R}^{n})\longrightarrow L^{1}(\mathbb{R}^{n})$ by: 
$f\ast g(\bm{x})\mathdef\int_{\bm{y}\in\mathbb{R}^{n}}f(\bm{y})g(\bm{x}-\bm{y})\ d\bm{y}$.
\item $a_{n}=\bigo(b_{n})$ implies the existence of a constant $A>0$ such that $\frac{a_{n}}{b_{n}}\leq A,\forall\ n\geq 1$.
\item $a_{n}=\littleo(b_{n})$ implies that $\lim_{n\rightarrow\infty}\frac{a_{n}}{b_{n}}=0$.
\item $P_{G}$ is the gaussian distribution function:\\ $P_{G}(x)\mathdef\frac{1}{\sqrt{2\pi}}\int_{-\infty}^{x}\exp\left(-\frac{1}{2}x^{2}\right)\ dx$.
\item $\erf$ is the normal error function: $\erf(x)\mathdef\frac{1}{\sqrt{2\pi}}\int_{-x}^{x}\exp\left(-\frac{1}{2}x^{2}\right)\ dx$. 
\item $\Gamma(x)\mathdef\int_{0}^{\infty}t^{x-1}\exp(-t)\ dt,\ x>0$, is the gamma-function.
\item $\mathcal{O}_{n}$ is the set of all orthogonal real $n\times n$ matrices.
\item $\log^{*}{n}\mathdef\log{c}+\log{n}+\log{\log{n}}+\log{\log{\log{n}}}+\cdots$ for $n\in\mathbb{N}$ where the sum includes all positive iterates and $c\approx 2.865$ is a normalization constant. 
\item Let $I\subset\mathbb{R}$ denote an interval, then $I^{d}\mathdef\{\bm{x}=(x_{1},x_{2},...,x_{d})^{T}\in\mathbb{R}^{d}:\ x_{i}\in I\}$.
\end{enumerate}
%\appendix
%\chapter{Wavelet preliminaries}
%HERE SHOULD BE SOME GENERIC STUFF ON CONTINUOUS WAVELETS AND THEIR DISCRETE ANALOGS: \cite{Daubechies:1992}, \cite{Wickerhauser:1994} Should put the presentation of mirror wavelet basis in sections below.
%\subsection{Multiresolution analysis, wavelet and wavelet packet bases}
%Given a conjugate pair of mirror filters $h$, $g$, blah-blah $N$-periodic discrete wavelets are defined.......
%\section{The mirror wavelet basis}
\chapter{The mirror wavelet basis}
The degradation process of data $\bm{\theta}$ is modelled as
\begin{align}
&y=u\ast\theta+\eta\label{deconv_problem2}
\end{align}
where $u$ is a known lowpass filter and $\eta$ is IID gaussian noise and $\ast$ denotes the convolution operator.
Let $\bm{U}\in\mathbb{R}^{N\times N}$ denote the discretized circular convolution operator representing the smoothing
degrading on the data $\bm{\theta}$ by the lowpass filter $u$ 
\begin{equation}
\bm{y}=\bm{U}\bm{\theta}+\bm{\eta}\label{deconv_problem_matrix_form2}
\end{equation}
where $\bm{U}$ is the smoothing matrix representing the smoothing operation performed by lowpass filter $u$, $\bm{\theta}$
is the parameters we want to estimate and $\bm{\eta}$ is white gaussian noise. After deconvolving with inverse operator $\bm{U}^{-1}$
we have
\begin{align}
&\bm{x}\mathdef\bm{U}^{-1}\bm{y}=\bm{\theta}+\bm{U}^{-1}\bm{\eta}\label{deconv_solution_matrix_form}
\intertext{Now, the noise $\bm{Z}\mathdef\bm{U}^{-1}\bm{\eta}$ is non-white gaussian with covariance $\bm{K}$}
&\bm{K}\mathdef\sigma^{2}\bm{U}^{-1}\bm{U}^{-T}.\label{deconv_noise_covariance}
\end{align}
Circular convolution operators $\bm{U}$ are diagonal in the discrete Fourier basis 
$\bm{W}_{F}=[\bm{b}_{k}]_{k=0}^{N-1}$ where $\bm{b}_{k}$ are column vectors with 
\begin{align}
&\bm{b}_{k}[n]\mathdef\frac{1}{\sqrt{N}}\exp\left(i\frac{2\pi kn}{N}\right),\ 0\leq n<N\label{def_fourier_basis}
\intertext{It follows from this fact and (\ref{deconv_noise_covariance}) that the eigenvalues $\sigma_{k}^{2}$ of $\bm{K}$ are given by}
&\sigma^{2}_{k}\mathdef\langle\bm{K}\bm{b}_{k},\bm{b}_{k}\rangle=\left\langle\bm{W}_{F}^{T}\bm{K}\bm{W}_{F}\bm{W}_{F}^{T}\bm{b}_{k},\bm{W}_{F}^{T}\bm{b}_{k}\right\rangle\nonumber\\
&=\left\langle\bm{W}_{F}^{T}\sigma^{2}\bm{U}^{-1}\bm{U}^{-T}\bm{W}_{F}\bm{W}_{F}^{T}\bm{b}_{k},\bm{W}_{F}^{T}\bm{b}_{k}\right\rangle\nonumber\\
&=\sigma^{2}\left\langle\bm{W}_{F}^{T}\bm{U}^{-1}\bm{W}_{F}\bm{W}_{F}^{T}\bm{U}^{-T}\bm{W}_{F}\bm{W}_{F}^{T}\bm{b}_{k},\bm{W}_{F}^{T}\bm{b}_{k}\right\rangle\nonumber\\
&=\sigma^{2}\left\langle\left(\bm{W}_{F}^{T}\bm{U}\bm{W}_{F}\right)^{-1}\left(\bm{W}_{F}^{T}\bm{U}\bm{W}_{F}\right)^{-T}\bm{W}_{F}^{T}\bm{b}_{k},\bm{W}_{F}^{T}\bm{b}_{k}\right\rangle\nonumber\\
&=\frac{\sigma^{2}}{|\widehat{u}[k]|^{2}}.\label{K_eigenvalues}
\end{align}
Now, (\ref{def_fourier_basis}) shows that the Fourier basis elements have full support in the space domain
and therefore this basis, while providing a domain where the noise 
$Z\sim\mathcal{N}\left(0,\text{diag}\left(\frac{|\hat{u}[k]|^{2}}{\sigma^{2}}\bm{Id}\right)\right)$ is IID, 
is not suitable for estimating $\theta$ in (\ref{deconv_problem}). The main idea in  \cite{MallatKalifa:2003_IEEE}, 
\cite{MallatKalifa:2003_AnnStat} is to construct a wavelet packet basis
$\widetilde{\bm{W}}=[\bm{\psi}_{l}]_{l=0}^{N-1}$ of $\mathbb{R}^{N}$, where $\bm{\psi}_{l}[n]$ are 
supported in the space domain $0\leq n\leq N-1$, which {\em approximately} diagonalizes the covariance $\bm{K}$. 
\begin{figure}[h]
\begin{center}
\includegraphics[scale=0.5]{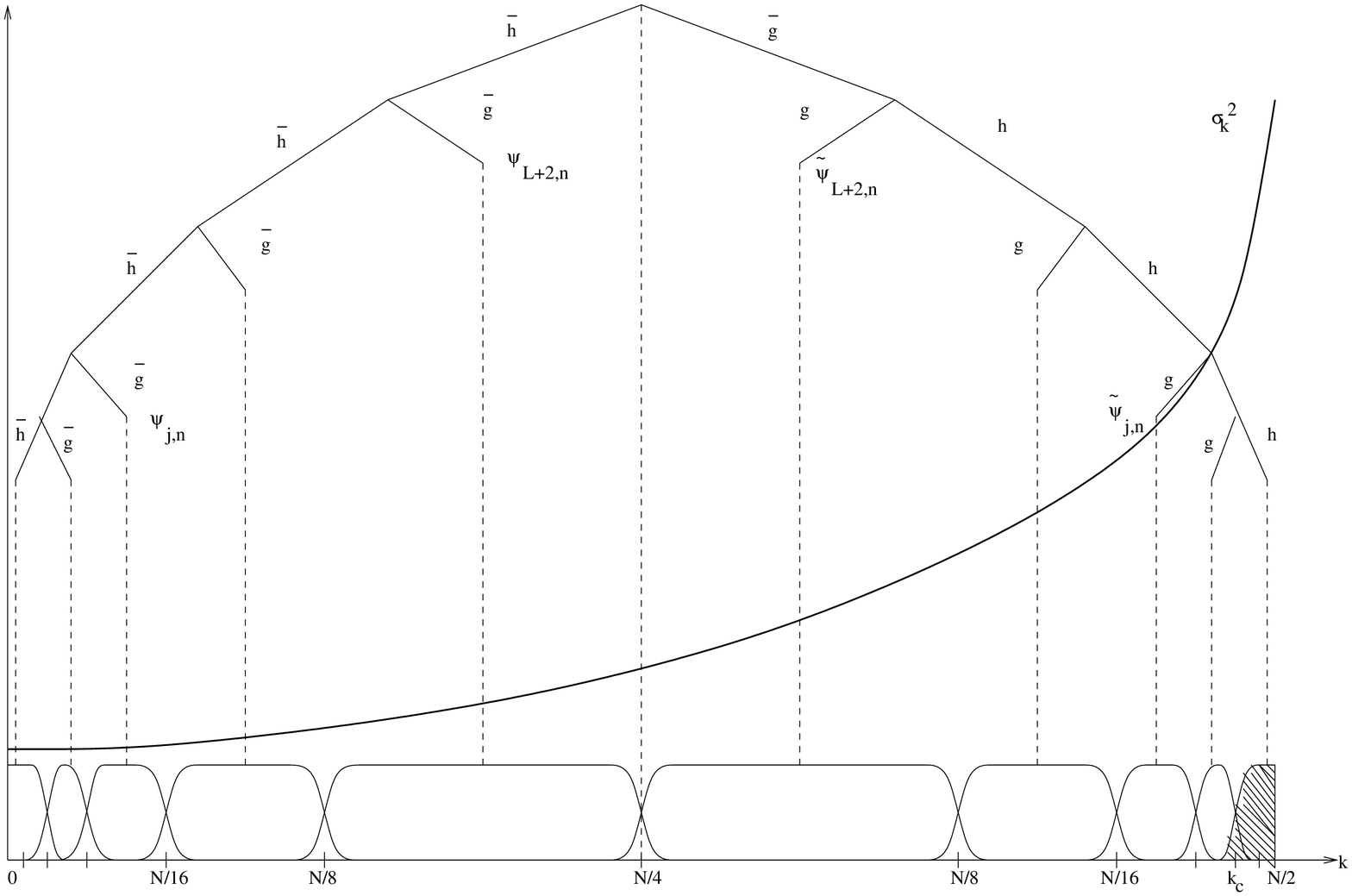}
\end{center}
\caption{Illustration copied from \cite{MallatKalifa:2003_AnnStat} of the mirror wavelet decomposition algorithm and the Fourier support of the mirror wavelet basis.
Each branch in the decomposition tree represents a convolution with a filter $\bar{h}$ or $\bar{g}$ followed by decimation by 2. 
The fourier frequency index $k$ is plotted on the first axis. The curved fat line growing from left to right shows the 
diagonal covariance matrix $\sigma_{k}^{2}$ of the noise in the Fourier domain plotted as a function of the fourier frequency.
The noise variance $\sigma_{k}^{2}$ varies by a bounded factor which do not grow with $N$, on the frequency support of 
each mirror wavelet $\widetilde{\psi}_{j,k}$.
There is a critical frequency $k_{c}$ above which the noise variance $\sigma_{k}^{2}$ is too high for reconstruction 
to be possible.}
\label{MirrorWavelet}
\end{figure}
Assuming the sample space dimension $N$ is a power of 2, we define
\begin{align}
&L\mathdef-\log_{2}{N}.\label{def_L_scale}
\intertext{Given a conjugate pair of mirror filters $h$, $g$, 
$N$-periodic discrete mirror wavelets $\widetilde{\psi}_{j,k}[n]$ are defined from orignal $N$-periodic discrete wavelets $\psi_{j,k}[n]$ by}
&\widetilde{\psi}_{j,k}[n]\mathdef(-1)^{n-1}\psi_{j,k}[1-n]\label{def_mirror_wavelet}
\intertext{where}
&\widehat{\psi}_{j}[n]\mathdef\widehat{g}[2^{j-L-1}n]\hat{\phi}_{j-1}[n]\text{ and }\widehat{\phi}_{j}[n]\mathdef\prod_{l=0}^{j-L-1}\widehat{h}[2^{l}n],\ L<j<1\label{def_discrete_psi_and_phi}
\intertext{and $\widehat{\phi}_{L}[n]\equiv N^{-1/2},\ \forall n\in 0,1,2,...,N-1$, and}
&\psi_{j,k}[n]\mathdef\psi_{j}[n-N2^{j}k],\ 0\leq k< 2^{-j}\label{def_discrete_wavelet_def}
\intertext{and we also define}
&\psi_{1,0}[n]=\widetilde{\psi}_{1,0}[n]\equiv N^{-1/2},\ \forall n\in{0,1,2,...,N-1}.\label{def_coarsest_scale_wavelet}
\intertext{We note that the Fourier support, $\text{supp }\widehat{\psi}_{j}[n]$, satisfies}
&\text{supp }\widehat{\psi}_{j}\approx[2^{-j-1},2^{-j}].\label{Fourier_support_wavelet}
\intertext{The Fourier transform of the mirror wavelets by definition satisfies}
&|\widehat{\widetilde{\psi}}_{j,k}[n]|=|\widehat{\psi}_{j,k}[N/2-n]|\label{fourier_mirror_wavelet}
\intertext{and by (\ref{Fourier_support_wavelet}) we have}
&\text{supp }\widehat{\widetilde{\psi}}_{j}\approx[N/2-2^{-j},N/2-2^{-j-1}].\label{Fourier_support_mirror_wavelet}
\end{align}
The mirror wavelet coefficients $\langle f,\tilde{\psi}_{j,k}\rangle,1>j>L+1$ are calculated from the finest scale 
wavelet coefficients $\langle f,\psi_{L+1,k}\rangle$
by a cascade of convolutions and decimations by 2 with the pair of conjugate filters $h$, $g$ as illustrated in Figure \ref{MirrorWavelet}.
(see appendix). Defining the discrete $N$-periodic mirror wavelet basis $\widetilde{\bm{W}}$ by
\begin{align}
&\widetilde{\bm{W}}\mathdef\left[[\bm{\psi}_{j,k}]_{0\leq k< 2^{-j},L+1<j\leq 1},[\tilde{\bm{\psi}}_{j,k}]_{0\leq k< 2^{-j},L+1<j\leq 1}\right]\label{def_W_tilde}
\intertext{where}
&\bm{\psi}_{j,k}[n]\mathdef\psi_{j,k}[n],\ \widetilde{\bm{\psi}}_{j,k}[n]\mathdef\widetilde{\psi}_{j,k}[n].\label{def_mirror_wavelet_basis}
\end{align}
we have by general properties of wavelet packets \cite{Mallat:1998} that $\widetilde{\bm{W}}$ is an orthonormal basis for $\mathbb{R}^{N}$.
Furthermore, it is proved in \cite{MallatKalifa:2003_AnnStat} that the covariance matrix $\bm{K}$ of the noise defined 
in (\ref{deconv_noise_covariance}) is nearly diagonalized in the mirror wavelet basis $\widetilde{\bm{W}}$ for all $N$ if
the wavelet $\psi$ has $q>p$ vanishing moments where $p$ is the order of the zero of the lowpass smoothing 
filter $\hat{u}[k]$ at the highest Fourier frequency $k=\pm N/2$. In \cite {MallatKalifa:2003_IEEE} one considers smoothing filters $u$ which have
a Fourier transform $\hat{u}$ with a zero of order $p\geq 1$ at highest Fourier frequency $k=\pm N/2$, that is
\begin{align}
&\hat{u}[k]\sim\left|\frac{2|k|}{N}-1\right|^{p},\ p\geq 1\label{generic_smoothing_filter}
\intertext{we will here use the smoothing filter}
&\hat{u}[k]=\cos^{p}\left(\frac{\pi k}{N}\right),\ p\geq 1\label{smoothing_filter}
\end{align}
which have the type of smoothing behaviour that the mirror wavelet basis is designed to work with. We define the pseudo-inverse smoothing 
filter $u^{-1}$ by 
\begin{equation}
\widehat{u^{-1}}[k]\mathdef\left\{\begin{array}{ll}\frac{1}{\hat{u}[k]},& \text{ if } \hat{u}[k]\neq 0\\
0,& \text{ if }\hat{u}[k]=0\end{array}\right.\label{pseudo_inverse_u}
\end{equation}
Define
\begin{align}
&\bm{x}^{(\widetilde{W})}_{j}[k]\mathdef\langle\bm{x},\bm{\psi}_{j,k}\rangle ,\ \
\bm{\theta}^{(\widetilde{W})}_{j}[k]\mathdef\langle\bm{\theta},\bm{\psi}_{j,k}\rangle\label{def_x_W}\\
&\tilde{\bm{x}}^{(\widetilde{W})}_{j}[k]\mathdef\langle\bm{x},\tilde{\bm{\psi}}_{j}[k]\rangle ,\ \
\tilde{\bm{\theta}}^{(\widetilde{W})}_{j}[k]\mathdef\langle\bm{\theta},\tilde{\bm{\psi}}_{j}[k]\rangle\label{def_x_W_tilde}
\end{align}
and observe by (\ref{deconv_solution_matrix_form}) and (\ref{deconv_noise_covariance}) that the data 
$\bm{x}^{\widetilde{\bm{W}}}_{j}[k]$, $\tilde{\bm{x}}^{\widetilde{\bm{W}}}_{j}[k], 0\leq k<2^{-j},L+2\leq j<1$ 
are gaussian random variables with means $\bm{\theta}^{(\widetilde{\bm{W}})}_{j}[k]$, $\tilde{\bm{\theta}}^{(\widetilde{\bm{W}})}_{j}[k]$ 
and variances $\sigma^{2}_{j,k}\mathdef\langle \bm{K}\bm{\psi}_{j,k},\bm{\psi}_{j,k}\rangle$ and $\tilde{\sigma}^{2}_{j,k}\mathdef\langle \bm{K}\tilde{\bm{\psi}}_{j,k},\tilde{\bm{\psi}}_{j,k}\rangle$, respectively. 
We then have by (\ref{K_eigenvalues}), (\ref{Fourier_support_wavelet}) and (\ref{generic_smoothing_filter})
\begin{align}
&\sigma_{j}^{2}\mathdef\langle\bm{K}\bm{\psi}_{j,k},\bm{\psi}_{j,k}\rangle
=\langle\widehat{\bm{K}}\widehat{\bm{\psi}_{j,k}},\widehat{\bm{\psi}_{j,k}}\rangle
=\sigma^{2}\sum_{n=-N/2}^{N/2-1}\frac{|\widehat{\bm{\psi}_{j}}[n]|^{2}}{|\widehat{u}[n]|^{2}}\sim\sigma^{2}.\label{lowpass_variances}
\intertext{Furthermore by (\ref{Fourier_support_mirror_wavelet}), (\ref{generic_smoothing_filter}) we have}
&\tilde{\sigma}_{j}^{2}\mathdef\langle \bm{K}\widetilde{\bm{\psi}}_{j,k},\widetilde{\bm{\psi}}_{j,k}\rangle
=\langle \widehat{\bm{K}}\widehat{\widetilde{\bm{\psi}}_{j,k}},\widehat{\widetilde{\bm{\psi}}_{j,k}}\rangle
=\sigma^{2}\sum_{n=-N/2}^{N/2-1}\frac{|\widehat{\widetilde{\bm{\psi}}}_{j}[n]|^{2}}{|\widehat{u}[n]|^{2}}\sim\sigma^{2}2^{2p(j-L)}.\label{highpass_variances}
\end{align}
We note that the noise variances $\sigma_{j,k}^{2}$, $\tilde{\sigma}_{j,k}^{2}$ do not depend on the translation index $k$ of the 
wavelets $\psi_{j,k}$, $\tilde{\psi}_{j,k}$. Using thresholding estimators as described in 
\cite{DJ_Ideal:1994}, \cite{BruceGao:1995c} the estimators $\hat{\bm{\theta}}^{(\widetilde{\bm{W}})}_{j}[k]$, 
$\hat{\tilde{\bm{\theta}}}^{(\widetilde{\bm{W}})}_{j}[k],\ 0\leq k<2^{-j}$ are given by a thresholding scheme 
on the wavelet expansions $\bm{x}^{(\widetilde{W})}_{j}$, 
$\tilde{\bm{x}}^{(\widetilde{W})}_{j}$ of the deconvolved data $\bm{x}$.
The thresholds $T$, $\widetilde{T}_{j}$ used in \cite{MallatKalifa:2003_AnnStat} on 
$\bm{x}^{(\widetilde{W})}_{j}[k]$, $\tilde{\bm{x}}^{(\widetilde{W})}_{j}[k]$, respectively, are the ideal thresholds described 
in \cite{DJ_Ideal:1994}
\begin{align}
&T\mathdef\sigma\sqrt{2\log_{e}{(N/2)}}\label{def_T_j}\\
&\widetilde{T}_{j}\mathdef\left\{\begin{array}{ll}\tilde{\sigma}_{j}\sqrt{2\log_{e}{(2^{-j})}}, & \text{if }
\tilde{\sigma}_{j}\sqrt{2\log_{e}(2^{-j})}<\tilde{s}_{j}\mathdef\sup_{f\in\Theta}|\langle f,\widetilde{\psi}_{j,k}\rangle|\\
\infty, &\text{otherwise.}\end{array}\right.\label{def_T_j_tilde}
\end{align}
that is the same constant threshold $T$ defined above is used on all of the $N/2$ low frequency wavelet coefficients 
$\bm{x}^{(\widetilde{W})}_{j}[k], 0\leq k<2^{-j},\ L+2\leq j<1$, and the threshold $\widetilde{T}_{j}$ is used on the high frequency mirror 
wavelet coefficients $\tilde{\bm{x}}^{(\widetilde{W})}_{j}[k],0\leq k<2^{-j},\ L+2\leq j<1$, where the noise variance 
$E\left(\tilde{\bm{x}}^{(\widetilde{W})}_{j}[k]-\tilde{\bm{\theta}}^{(\widetilde{W})}_{j}[k]\right)^{2}$ 
may be approximated by $\tilde{\sigma}_{j}^{2}$ defined in (\ref{highpass_variances}) on each subband: 
$\text{span}_{0\leq k<2^{-j}}\widetilde{\bm{\psi}}_{j,k}$.
We note that the hard thresholding function is the MAP-estimator 
for $GGD_{\nu}$-distributed $\bm{\theta}^{(\widetilde{\bm{W}})}$ with $\nu=1$, i.e Laplace-distributed.\\

There are some remarks which should be made on the mirror wavelet deconvolution algorithm as presented above. 
The set $\Theta$ over which the numbers $\tilde{s}_{j}$ defined in (\ref{def_T_j_tilde}) are computed, is in \cite{MallatKalifa:2003_AnnStat} taken to be
the set $\Theta_{\text{tv}}$ of bounded discrete total variation: 
\begin{align}
&\Theta_{\text{tv}}\mathdef\left\{\bm{\theta}:\|\bm{\theta}\|_{\text{tv}}\mathdef\sum_{n=0}^{N-1}\left||\bm{\theta}[n]-\bm{\theta}[n-1]\right|\leq C\right\}\label{Theta_tv}
\intertext{where $C>0$ is some universial constant. Then it is shown in \cite{MallatKalifa:2003_AnnStat} that}
&\tilde{s}_{j}\sim C2^{(L-j)/2}\label{s_j_growth}
\end{align}
The critical scale $2^{c}$ is defined as the smallest scale such that $\widetilde{T}_{j}=\infty$ for all 
scales $2^{j}$ with $2^{j}>2^{c}$. The mirror wavelets $\widetilde{\psi}_{c,k}$ on the critical scale have a Fourier transform
$\widehat{\widetilde{\psi}}_{c,k}$ whose support is essentially at Fourier frequencies $|k|>k_{c}\mathdef N/2-2^{-c}$, this is illustrated 
in Figure \ref{MirrorWavelet}. This implies the existence of a cutoff Fourier frequency $k_{c}$ for thresholding estimators
and so we can replace the pseudo inverse smoothing filter $u^{-1}$ in (\ref{pseudo_inverse_u}) by a truncated pseudo inverse 
$\widetilde{u}^{-1}$ defined by
\begin{align}
&\widehat{\widetilde{u}^{-1}}\mathdef\left\{\begin{array}{ll}\frac{1}{\widehat{u}[k]},& \text{ if } |k|<k_{c}\\
0,& \text{ otherwise }\end{array}\right.\label{truncated_pseudo_inverse_u}
\end{align}
Also, in the numerical experiments in \cite{MallatKalifa:2003_IEEE}, \cite{MallatKalifa:2003_AnnStat} one uses the translation invariant 
thresholding algorithm \cite{DonCoif:1995}, however we have not had the time to implement this algorithm, and so we stick to the 
ordinary thresholding algorithm in our numerical experiments in this thesis.

\begin{figure}[h]
\begin{center}
\includegraphics[scale=0.6]{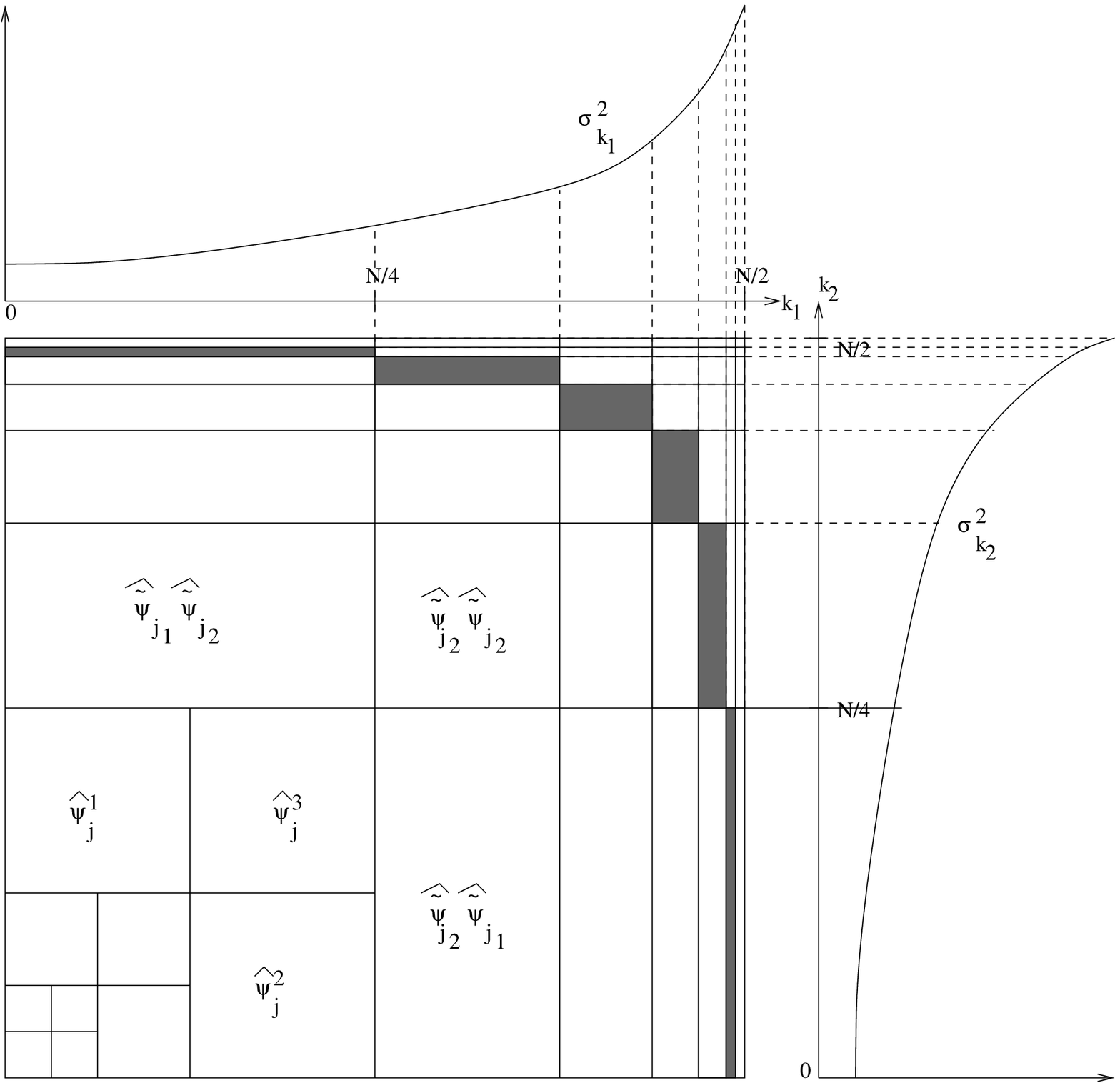}
\end{center}
\caption{Illustration copied from \cite{MallatKalifa:2003_AnnStat} of the separable mirror wavelet basis for functions of two variables and its Fourier support properties.
Note that the mirror wavelet basis segments the frequency plane $(k_{1},k_{2})$ into rectangles over which the noise variance
$\sigma_{k_{1},k_{1}}^{2}=\sigma_{k_{1}}^{2}\sigma_{k_{2}}^{2}$ varies by a bounded factor which do not grow with $N$. The gray rectangles correspond to the critical scales beyond which the thresholding sets all coefficients to zero.}
\label{MirrorWavelet2}
\end{figure}

The restoration algorithm may then be summed up as follows:

\begin{enumerate}
\item Estimate the variance $E(\eta^{2})$ of the white gaussian noise $\eta$ in (\ref{deconv_problem}).
\item Decide on the order $p$ of the smoothing filter $\widehat{u}[k]=\cos^{p}(\pi k/N)$ in (\ref{deconv_problem}) and on 
the numbers $\tilde{s}_{j}\mathdef\sup_{f\in\Theta}|\langle f,\widetilde{\psi}_{j,k}\rangle|$, alternatively decide on a 
critical frequency $k_{c}$.
\item Expand the given data $y$ into the Fourier basis and deconvolve the transformed data $\hat{y}$ in the Fourier domain
by computing $\hat{x}\mathdef\hat{y}[k]\cdot\widehat{\widetilde{u}^{-1}}[k],\ 0\leq k<N$ where 
$\widetilde{u}^{-1}$ is the truncated pseudo inverse smoothing filter defined in (\ref{truncated_pseudo_inverse_u}) 
and apply the inverse Fourier transform on the result to yield $x\mathdef\mathcal{F}^{-1}(\hat{x})$.
\item Calculate the variances $\sigma_{j}^{2}$, $\tilde{\sigma}^{2}_{j},\ L+2\leq j<1$ of the mirror wavelet basis expansion coefficients
$\widetilde{W}^{T}Z$ of the deconvolved noise $Z\mathdef u^{-1}\ast\eta$, or their approximations 
in (\ref{lowpass_variances}), (\ref{highpass_variances}).
\item Expand the deconvolved data $x$ into the mirror wavelet basis $\widetilde{W}$ and apply the thresholding operation $\mathcal{T}$
with thresholds $T$, $\widetilde{T}_{j}$ in (\ref{def_T_j}), (\ref{def_T_j_tilde})  on the transformed data $x^{(\widetilde{W})}\mathdef\widetilde{W}^{T}x$ in the mirror wavelet domain and apply the inverse mirror wavelet transformation 
$\widetilde{W}$ on the thresholded transformed data $\mathcal{T}(\widetilde{W}^{T}x)$ to find the parameter estimate: 
$\hat{\theta}=\widetilde{W}\mathcal{T}(\widetilde{W}^{T}x)$.
\end{enumerate}
The deconvolution estimator described above for signals have a separable extension to image data.
The smoothing filter $u$ in (\ref{deconv_problem}) is here a separable lowpass filter
\begin{align}
&u[n_{1},n_{2}]=u_{1}[n_{1}]u_{2}[n_{2}],\ 0\leq n_{1}<N,\ 0\leq n_{2}<N\label{separable_smoothing_filter}
\intertext{with Fourier transforms $\widehat{u_{1}}[k_{1}]$ and $\widehat{u_{2}}[k_{2}]$ as in (\ref{generic_smoothing_filter}). 
The deconvolved noise has a covariance $\bm{K}$ which is diagonalized in a two-dimesional discrete Fourier basis 
$\bm{W}_{F\otimes F}=[\bm{b}_{k_{1},k_{2}}]_{0\leq k_{1},k_{2}<N}$ and it follows as in (\ref{K_eigenvalues}) that the eigenvalues $\sigma_{k_{1},k_{2}}^{2}$ of $\bm{K}$ are}
&\sigma_{k_{1},k_{2}}^{2}\mathdef\left\langle\bm{K}\bm{b}_{k_{1},k_{2}},\bm{b}_{k_{1},k_{2}}\right\rangle=\frac{\sigma^{2}}{|\widehat{u_{1}}[k_{1}]|^{2}|\widehat{u_{2}}[k_{2}]|^{2}}\nonumber\\
&\sim\sigma^{2}\left|\frac{2|k_{1}|}{N}-1\right|^{-2p_{1}}\left|\frac{2|k_{2}|}{N}-1\right|^{-2p_{2}}.\label{K_eigenvalues2}
\end{align}
A separable discrete mirror wavelet basis of $\mathbb{R}^{N\times N}$ is constructed from the one-dimesional
discrete wavelets $\psi_{j}$ and scaling functions $\phi_{j},\ L<j<1,\ 0\leq k<2^{-j}$ by
\begin{align}
&\psi_{j}^{(1)}[n_{1},n_{2}]\mathdef\phi_{j}[n_{1}]\psi_{j}[n_{2}],\ \psi_{j}^{(2)}[n_{1},n_{2}]\mathdef\psi_{j}[n_{1}]\phi_{j}[n_{2}],\nonumber\\
&\psi_{j}^{(3)}[n_{1},n_{2}]\mathdef\psi_{j}[n_{1}]\psi_{j}[n_{2}],\ \psi_{1}^{(0)}[n_{1},n_{2}]=N^{-1}.\label{def_separable_basis_elements}
\intertext{Defining the translates}
&\psi_{j,m_{1},m_{2}}^{(\alpha)}[n_{1},n_{2}]\mathdef\psi_{j}^{(\alpha)}[n_{1}-2^{j-L}m_{1},n_{2}-2^{j-L}m_{2}],\ \alpha=1,2,3,
\label{def_uniform_translates}
\intertext{then the family}
&\mathcal{B}\mathdef\left\{\psi_{1}^{(0)},\psi^{(1)}_{j,m_{1},m_{2}},\psi^{(3)}_{j,m_{1},m_{2}}\right\}_{L<j<1,0\leq m_{1},m_{2}<2^{-j}}
\label{def_B}
\intertext{is an orthonormal basis of $\mathbb{R}^{N\times N}$. It follows from the definition 
(\ref{def_separable_basis_elements}) that the family $\mathcal{B}_{0}$ of lower frequency wavelets}
&\mathcal{B}_{0}\mathdef\left\{\psi_{1}^{(0)},\psi^{(1)}_{j,m_{1},m_{2}},\psi^{(1)}_{j,m_{1},m_{2}},\psi^{(3)}_{j,m_{1},m_{2}}\right\}_{L+1<j<1,0\leq m_{1},m_{2}<2^{-j}}
\label{def_B0}
\intertext{have Fourier transforms which are essentially supported in the low frequency square $[-N/4,N/4]^{2}$ where the eigenvalues 
$\sigma_{k_{1},k_{2}}^{2}$ of $\bm{K}$ are constant to within a universial constant factor, and therefore the elements of $\mathcal{B}$
are approximate eigenvectors of $\bm{K}$, whereas the family $\mathcal{B}_{1}$ of higher frequency wavelets}
&\mathcal{B}_{1}\mathdef\mathcal{B}\setminus\mathcal{B}_{0}=\left\{\psi^{(1)}_{L+1,m_{1},m_{2}},\psi^{(2)}_{L+1,m_{1},m_{2}},\psi^{(3)}_{m_{1},m_{2}}\right\}_{0\leq m_{1},m_{2}<N/2}\label{def_B1}
\intertext{are not approximate eigenvectors of $\bm{K}$ and these are replaced by the familiy $\widetilde{\mathcal{B}}_{1}$ 
of separable mirror wavelets defined by}
&\widetilde{\mathcal{B}}_{1}\mathdef\left\{\widetilde{\psi}_{j_{1},m_{1}}[n_{1}]\widetilde{\psi}_{j_{2},m_{2}}[n_{2}]\right\}_{\begin{array}{l}L<j_{1},j_{2}<1,(j_{1},j_{2})\neq(L+1,L+1),\\ 0\leq m_{1}<2^{-j_{1}},0\leq m_{2}<2^{-j_{2}}\end{array}}\label{def_tilde_B1}
\intertext{with $\widetilde{\psi}_{j}$ as defined in (\ref{def_mirror_wavelet}). It follows that the family}
&\widetilde{\mathcal{B}}\mathdef\mathcal{B}_{0}\cup\widetilde{\mathcal{B}}_{1}\label{def_separable_mirror_wavelet_basis}
\end{align}
is a discrete separable anisotropic wavelet packet basis for $\mathbb{R}^{N\times N}$ of 
approximate eigenvectors of $\bm{K}$. The tiling of the Fourier frequency plane that results from the separable mirror wavelet basis 
$\widetilde{\mathcal{B}}$ defined in (\ref{def_separable_mirror_wavelet_basis}) is illustrated in Figure \ref{MirrorWavelet2}.
Like in (\ref{lowpass_variances}), (\ref{highpass_variances}) one has
\begin{align}
 &\sigma_{j,\alpha}^{2}\mathdef\left\langle K\psi^{(\alpha)}_{j,m_{1},m_{2}},\psi^{(\alpha)}_{j,m_{1},m_{2}}\right\rangle
=\left\langle\widehat{K}\widehat{\psi^{(\alpha)}_{j,m_{1},m_{2}}},\widehat{\psi^{(\alpha)}_{j,m_{1},m_{2}}}\right\rangle\nonumber\\
&=\sigma^{2}\sum_{n_{1},n_{2}=-N/2}^{N/2-1}\frac{|\widehat{\psi^{(\alpha)}_{j}}[n_{1},n_{2}]|^{2}}{|\widehat{u_{1}}[n_{1}]|^{2}|\widehat{u_{2}}[n_{2}]|^{2}}\sim\sigma^{2},\ \alpha=1,2,3.\label{lowpass_variances2}\\
&\tilde{\sigma}_{j_{1},j_{2}}^{2}\mathdef\langle K\widetilde{\psi}_{j_{1},m_{1}}\widetilde{\psi}_{j_{2},m_{2}},\widetilde{\psi}_{j_{1},m_{1}}\widetilde{\psi}_{j_{2},m_{2}}\rangle=\langle \widehat{K}\widehat{\widetilde{\psi}_{j_{1},m_{1}}}\widehat{\widetilde{\psi}_{j_{2},m_{2}}},\widehat{\widetilde{\psi}_{j_{1},m_{1}}}\widehat{\widetilde{\psi}_{j_{2},m_{2}}}\rangle\nonumber\\
&=\sigma^{2}\sum_{n_{1},n_{2}=-N/2}^{N/2-1}\frac{|\widehat{\widetilde{\psi}}_{j_{1}}[n_{1}]|^{2}|\widehat{\widetilde{\psi}}_{j_{2}}[n_{2}]|^{2}}{|\widehat{u_{1}}[n_{1}]|^{2}|\widehat{u_{2}}[n_{2}]|^{2}}\sim\sigma^{2}2^{2p_{1}(j_{1}-L)}2^{2p_{2}(j_{2}-L)}.\label{highpass_variances2}
\end{align}
Using the same ideal thresholds as in (\ref{def_T_j}), (\ref{def_T_j_tilde}), 
we define the thresholds $T^{(2)}$, $\widetilde{T}^{(2)}_{j_{1},j_{2}}$ by
\begin{align}
&T^{(2)}\mathdef\sigma\sqrt{2\log_{e}{(N^{2}/4)}}\label{def_T2}\\
&\widetilde{T}_{j_{1},j_{2}}\mathdef\left\{\begin{array}{ll}\tilde{\sigma}_{j_{1},j_{2}}\sqrt{2\log_{e}{(2^{-j_{1}-j_{2}})}}, & \text{if }
\tilde{\sigma}_{j_{1},j_{2}}\sqrt{2\log_{e}(2^{-j_{1}-j_{2}})}<\tilde{s}_{j_{1},j_{2}}\\
\infty, &\text{otherwise.}\end{array}\right.\label{def_T2_j1_j2_tilde}
\intertext{where}
&\tilde{s}_{j_{1},j_{2}}\mathdef\sup_{f\in\Theta}\left|\left\langle f,\widetilde{\psi}_{j_{1},m_{1}}\widetilde{\psi}_{j_{2},m_{2}}\right\rangle\right|.
\label{def_s_j1_j2}
\end{align}
Critical scales $2^{c_{1}}$, $2^{c_{2}}$ are defined by: For each scale $j_{1}$, define $2^{c_{2}}$ as the smallest scale such that 
$2^{j_{2}}>2^{c_{2}}$ implies $\widetilde{T}_{j_{1},j_{2}}=\infty$, and for each scale $j_{2}$ define $2^{c_{1}}$ as the smallest scale such that 
$2^{j_{1}}>2^{c_{1}}$ implies $\widetilde{T}_{j_{1},j_{2}}=\infty$. These critical scales are illustrated in Figure \ref{MirrorWavelet2}.
Critical frequencies $k_{c_{1}}$, $k_{c_{2}}$ may then be deduced as in the one-dimensional case by $k_{c_{i}}=N/2-2^{-c_{i}},\ i=1,2$
and so may truncated smoothing filters $\widetilde{u}_{1}$, $\widetilde{u}_{2}$.

%\appendix
%\chapter{Calculation of  Fisher matrix for the reparameterized gaussian likelihood function}
\chapter{Calculation of Fisher matrix for the likelihood function}
We have from (\ref{split_likelihood}) and (\ref{def_phi_psi})  the definitions
\begin{align}
&f(\bm{x}|\bm{\theta},\tau)=\left(\frac{\tau}{2\pi}\right)^{\frac{n}{2}}\exp\left(-\frac{\tau}{2}\|\bm{x}_{\perp}\|^{2}\right)\exp\left(-\frac{\tau}{2}\|\bm{x}_{\parallel}-\bm{\theta}\|^{2}\right).\nonumber\\
&\tau=\psi(\hat{\tau}),\ \psi(0)=\tau_{0},\ \theta_{i}=\phi(\hat{\theta}_{i},\hat{\tau})\mathdef\frac{\bar{\tau}^{1/2}}{\tau^{1/2}}\hat{\theta}_{i}
=\left(\frac{\bar{\tau}}{\psi(\hat{\tau})}\right)^{\frac{1}{2}}\hat{\theta}_{i},\ 1\leq i\leq d.\nonumber
\intertext{Define the reparameterized log-likelihood function $\hat{L}$ as}
&\hat{L}(\bm{x},\hat{\tau},\hat{\bm{\theta}})\mathdef \log{f(\bm{x}|\psi(\hat{\tau}),\bm{\phi}(\hat{\bm{\theta}}))}\nonumber\\
&=\frac{n}{2}\log\psi(\hat{\tau})-\frac{1}{2}\psi(\hat{\tau})\|\bm{x}_{\perp}\|^{2}-\frac{1}{2}\psi(\hat{\tau})
\left\|\bm{x}_{\parallel}-\bar{\tau}^{\frac{1}{2}}\psi(\hat{\tau})^{-1/2}\hat{\bm{\theta}}\right\|^{2}.\nonumber
\intertext{We compute the required partial derivatives of $\hat{L}$ and get}
&\frac{\partial \hat{L}(\hat{\tau},\hat{\bm{\theta}})}{\partial\hat{\theta}_{k}}=\psi(\hat{\tau})\left(\bm{x}_{\parallel}(k)-
\bar{\tau}^{\frac{1}{2}}\psi(\hat{\tau})^{-1/2}\hat{\bm{\theta}}(k)\right)\bar{\tau}^{\frac{1}{2}}\psi(\hat{\tau})^{-1/2}\nonumber\\
&\frac{\partial^{2} \hat{L}(\hat{\tau},\hat{\bm{\theta}})}{\partial\hat{\theta}_{k}^{2}}=-\bar{\tau}\nonumber\\
&\frac{\partial^{2} \hat{L}(\hat{\tau},\hat{\bm{\theta}})}{\partial\hat{\theta}_{k}\partial\hat{\tau}}=
\frac{1}{2}\bar{\tau}^{\frac{1}{2}}\psi(\hat{\tau})^{-1/2}\frac{\partial\psi(\hat{\tau})}{\partial\hat{\tau}}\bm{x}_{\parallel}(k)\nonumber\\
&\frac{\partial \hat{L}(\hat{\tau},\hat{\bm{\theta}})}{\partial\hat{\tau}}=\frac{n/2}{\psi(\hat{\tau})}\frac{\partial\psi(\hat{\tau})}{\partial\hat{\tau}}
-\frac{1}{2}\frac{\partial\psi(\hat{\tau})}{\partial\hat{\tau}}\|\bm{x}_{\perp}\|^{2}
-\frac{1}{2}\frac{\partial\psi(\hat{\tau})}{\partial\hat{\tau}}\left\|\bm{x}_{\parallel}-\bar{\tau}^{\frac{1}{2}}\psi(\hat{\tau})^{-1/2}\hat{\bm{\theta}}\right\|^{2}\nonumber\\
&-\frac{1}{2}\psi(\hat{\tau})\sum_{i=1}^{d}\left(\bm{x}_{\parallel}(i)-\bar{\tau}^{\frac{1}{2}}\psi(\hat{\tau})^{-1/2}\hat{\bm{\theta}}(i)\right)\left(\bar{\tau}^{\frac{1}{2}}\psi(\hat{\tau})^{-3/2}\hat{\bm{\theta}}(i)\frac{\partial\psi(\hat{\tau})}{\partial\hat{\tau}}\right)\nonumber\\
&\frac{\partial^{2} \hat{L}(\hat{\tau},\hat{\bm{\theta}})}{\partial\hat{\tau}^{2}}=-\frac{n/2}{\psi(\hat{\tau})^{2}}\left(\frac{\partial\psi(\hat{\tau})}{\partial\hat{\tau}}\right)^{2}+\frac{n/2}{\psi(\hat{\tau})}\frac{\partial^{2}\psi(\hat{\tau})}{\partial\hat{\tau}^{2}}
-\frac{1}{2}\frac{\partial^{2}\psi(\hat{\tau})}{\partial\hat{\tau}^{2}}\|\bm{x}_{\perp}\|^{2}\nonumber\\
&-\frac{1}{2}\frac{\partial^{2}\psi(\hat{\tau})}{\partial\hat{\tau}^{2}}\left\|\bm{x}_{\parallel}-\bar{\tau}^{\frac{1}{2}}\psi(\hat{\tau})^{-1/2}\hat{\bm{\theta}}\right\|^{2}\nonumber\\
&-\frac{1}{2}\frac{\partial\psi(\hat{\tau})}{\partial\hat{\tau}}\sum_{i=1}^{d}\left(\bm{x}_{\parallel}(i)-\bar{\tau}^{\frac{1}{2}}\psi(\hat{\tau})^{-1/2}\hat{\bm{\theta}}(i)\right)\left(\bar{\tau}^{\frac{1}{2}}\psi(\hat{\tau})^{-3/2}\hat{\bm{\theta}}(i)\frac{\partial\psi(\hat{\tau})}{\partial\hat{\tau}}\right)\nonumber\\
&-\frac{1}{2}\psi(\hat{\tau})\sum_{i=1}^{d}\left(\bar{\tau}^{\frac{1}{2}}\frac{1}{2}\psi(\hat{\tau})^{-\frac{3}{2}}\hat{\bm{\theta}}(i)\frac{\partial\psi(\hat{\tau})}{\partial\hat{\tau}}\right)\left(\bar{\tau}^{\frac{1}{2}}\psi(\hat{\tau})^{-\frac{3}{2}}\hat{\bm{\theta}}(i)\frac{\partial\psi(\hat{\tau})}{\partial\hat{\tau}}\right)\nonumber\\
&-\frac{1}{2}\psi(\hat{\tau})\sum_{i=1}^{d}\left(\bm{x}_{\parallel}(i)-\bar{\tau}^{\frac{1}{2}}\psi(\hat{\tau})^{-\frac{1}{2}}\hat{\bm{\theta}}(i)\right)\left(-\frac{3}{2}\bar{\tau}^{\frac{1}{2}}\psi(\hat{\tau})^{-\frac{5}{2}}\hat{\bm{\theta}}(i)\left(\frac{\partial\psi(\hat{\tau})}{\partial\hat{\tau}}\right)^{2}\right.\nonumber\\
&\left.+\bar{\tau}^{\frac{1}{2}}\psi(\hat{\tau})^{-\frac{3}{2}}\hat{\bm{\theta}}(i)\frac{\partial^{2}\psi(\hat{\tau})}{\partial\hat{\tau}^{2}}\right).\nonumber
\intertext{We take the negative expectation $-E_{\bm{x}}$ of the data $\bm{x}=\bm{x}_{\perp}+\bm{x}_{\parallel}$ with respect to the likelihood $f$. Using
$E_{\bm{x}}[\bm{x}_{\parallel}-\bar{\tau}^{1/2}\psi(\hat{\tau})^{-1/2}\hat{\bm{\theta}}]=\bm{0}$ and $E_{\bm{x}}\|\bm{x}_{\perp}\|^{2}=(n-d)\psi(\hat{\tau})^{-1}$ and 
$E_{\bm{x}}\|\bm{x}_{\parallel}-\bar{\tau}^{1/2}\psi(\hat{\tau})^{-1/2}\hat{\bm{\theta}}\|^{2}=d\psi(\hat{\tau})^{-1}$ we get}
&c\mathdef-E_{\bm{x}}\left[\frac{\partial^{2} \hat{L}(\hat{\tau},\hat{\bm{\theta}})}{\partial\hat{\theta}_{k}^{2}}\right]=\bar{\tau}.\label{def_c}\\
&b_{k}\mathdef-E_{\bm{x}}\left[\frac{\partial^{2} \hat{L}(\hat{\tau},\hat{\bm{\theta}})}{\partial\hat{\theta}_{k}\partial\hat{\tau}}\right]=
-\frac{1}{2}\bar{\tau}\psi(\hat{\tau})^{-1}\frac{\partial\psi(\hat{\tau})}{\partial\hat{\tau}}\hat{\theta}_{k},\ 1\leq k\leq d.\label{def_b}\\
&a\mathdef-E_{\bm{x}}\left[\frac{\partial^{2} \hat{L}(\hat{\tau},\hat{\bm{\theta}})}{\partial\hat{\tau}^{2}}\right]=\frac{n/2}{\psi(\hat{\tau})^{2}}\left(\frac{\partial\psi(\hat{\tau})}{\partial\hat{\tau}}\right)^{2}\nonumber\\
&+\frac{1}{4}\bar{\tau}\psi(\hat{\tau})^{-2}\left(\frac{\partial\psi(\hat{\tau})}{\partial\hat{\tau}}\right)^{2}\sum_{i=1}^{d}\hat{\theta}_{i}^{2}.\label{def_a}
\intertext{We may now write the Fisher matrix $\hat{\bm{F}}(\hat{\bm{\theta}},\hat{\tau})$ of the reparameterized likelihood 
$f(\bm{x}|\psi(\hat{\tau}),\bm{\phi}(\hat{\bm{\theta}},\hat{\tau}))$ as the $(d+1)\times (d+1)$ matrix}
\hat{\bm{F}}=&\left(\begin{array}{ccccc}a & b_{1} & b_{2} &\cdots & b_{d}\\
b_{1} & c_{1} & 0 & \cdots & 0\\
b_{2} & 0 & c_{2} & \cdots & 0\\
\vdots &\vdots &\vdots & \ddots &\vdots\\
b_{d} & 0 & 0 & \cdots & c_{d} \end{array}\right)\label{form_of_G}\\
\intertext{where the only nonzero elements of $\hat{\bm{F}}$ are located on the first row and the first column and the diagonal. The determinant of the matrix in
(\ref{form_of_G}) is easily verified to be}
&\det\hat{\bm{F}}=\left(\prod_{i=1}^{d}c_{i}\right)\left(a-\sum_{j=1}^{d}\frac{b_{j}^{2}}{c_{j}}\right)\label{det_form_of_general_G}
\intertext{and since in this case $c_{i}=c,\ 1\leq i\leq d$, we get}
&\det\hat{\bm{F}}=ac^{d}-c^{d-1}\sum_{i=1}^{d}b_{i}^{2}.\label{det_form_of_G}
\intertext{Plugging in (\ref{def_c}), (\ref{def_b}), (\ref{def_a}) into (\ref{det_form_of_G}) we get}
&|\hat{\bm{F}}(\hat{\bm{\theta}},\hat{\tau})|=\bar{\tau}^{d}\left(\frac{n/2}{\psi(\hat{\tau})^{2}}\left(\frac{\partial\psi(\hat{\tau})}{\partial\hat{\tau}}\right)^{2}+\frac{1}{4}\bar{\tau}\psi(\hat{\tau})^{-2}\left(\frac{\partial\psi(\hat{\tau})}{\partial\hat{\tau}}\right)^{2}\sum_{i=1}^{d}\hat{\theta}_{i}^{2}\right)\nonumber\\ 
&-\bar{\tau}^{d-1}\sum_{i=1}^{d}\left(-\frac{1}{2}\bar{\tau}\frac{1}{\psi(\hat{\tau})}\frac{\partial\psi(\hat{\tau})}{\partial\hat{\tau}}\hat{\theta}_{i}\right)^{2}
=\bar{\tau}^{d}\frac{n/2}{\psi(\hat{\tau})^{2}}\left(\frac{\partial\psi(\hat{\tau})}{\partial\hat{\tau}}\right)^{2}.\label{reparameterized_fisher_determinant}
\intertext{We finally verify that our calculated reparameterized Fisher matrix $\hat{\bm{F}}$ satisfies the relation 
$|\hat{\bm{F}}(\hat{\bm{\theta}},\hat{\tau})|=|\bm{J}^{T}\bm{F}(\bm{\theta},\tau)\bm{J}|$ where $\bm{J}$ is the jacobi matrix induced by the transformations 
$\hat{\theta}_{i}\mapsto\phi(\hat{\theta}_{i},\hat{\tau}),\ 1\leq i\leq d$ and $\hat{\tau}\mapsto\psi(\hat{\tau})$. The jacobian is}
\bm{J}=&\left(\begin{array}{ccccc}\frac{\partial\psi(\hat{\tau})}{\partial\hat{\tau}} &\frac{\partial\psi(\hat{\tau})}{\partial\hat{\theta}_{1}}  
&\frac{\partial\psi(\hat{\tau})}{\partial\hat{\theta}_{2}} &\cdots & \frac{\partial\psi(\hat{\tau})}{\partial\hat{\theta}_{d}}\\
\frac{\partial\phi(\hat{\theta}_{1},\hat{\tau})}{\partial\hat{\tau}} &\frac{\partial\phi(\hat{\theta}_{1},\hat{\tau})}{\partial\hat{\theta}_{1}}  
& \frac{\partial\phi(\hat{\theta}_{1},\hat{\tau})}{\partial\hat{\theta}_{2}} & \cdots & \frac{\partial\phi(\hat{\theta}_{1},\hat{\tau})}{\partial\hat{\theta}_{d}}\\
\frac{\partial\phi(\hat{\theta}_{2},\hat{\tau})}{\partial\hat{\tau}} &\frac{\partial\phi(\hat{\theta}_{2},\hat{\tau})}{\partial\hat{\theta}_{1}}  
& \frac{\partial\phi(\hat{\theta}_{2},\hat{\tau})}{\partial\hat{\theta}_{2}} & \cdots & \frac{\partial\phi(\hat{\theta}_{2},\hat{\tau})}{\partial\hat{\theta}_{d}}\\
\vdots &\vdots &\vdots & \ddots &\vdots\\
\frac{\partial\phi(\hat{\theta}_{d},\hat{\tau})}{\partial\hat{\tau}} &\frac{\partial\phi(\hat{\theta}_{d},\hat{\tau})}{\partial\hat{\theta}_{1}}  
& \frac{\partial\phi(\hat{\theta}_{d},\hat{\tau})}{\partial\hat{\theta}_{2}} & \cdots & \frac{\partial\phi(\hat{\theta}_{d},\hat{\tau})}{\partial\hat{\theta}_{d}}\end{array}\right)\nonumber\\
\nonumber\\
&=\left(\begin{array}{ccccc}\psi^{\prime}(\hat{\tau}) & 0 & 0 &\cdots & 0\\
t_{1} & s & 0 & \cdots & 0\\
t_{2} & 0 & s & \cdots & 0\\
\vdots &\vdots &\vdots & \ddots &\vdots\\
t_{d} & 0 & 0 & \cdots & s \end{array}\right)\label{form_of_J}
\intertext{where $t_{i}=-\frac{1}{2}\bar{\tau}^{\frac{1}{2}}\delta_{n}\epsilon_{d}\psi(\hat{\tau})^{-1/2}\hat{\theta}_{i},\ 1\leq i\leq d$ and $s=\bar{\tau}^{\frac{1}{2}}\psi(\hat{\tau})^{-\frac{1}{2}}$.
Since $\bm{J}$ is triangular matrix we have $|\bm{J}|=\psi^{\prime}(\hat{\tau})s^{d}$. Now (\ref{split_likelihood}) yields after a trivial computation}
&|\bm{F}(\bm{\theta},\tau)|=\frac{n}{2}\tau^{d-2}.\label{Fisher_determinant} 
\intertext{Thus we have}
&|\bm{J}^{T}\bm{F}(\bm{\theta},\tau)\bm{J}|=\left(\frac{\partial\psi(\hat{\tau})}{\partial\hat{\tau}}\right)^{2}s^{2d}\frac{n}{2}\psi(\hat{\tau})^{d-2}\nonumber\\
&=\left(\frac{\partial\psi(\hat{\tau})}{\partial\hat{\tau}}\right)^{2}\bar{\tau}^{d}\psi(\hat{\tau})^{-d}\frac{n}{2}\psi(\hat{\tau})^{d-2}
=\left(\frac{\partial\psi(\hat{\tau})}{\partial\hat{\tau}}\right)^{2}\bar{\tau}^{d}\psi(\hat{\tau})^{-2}\frac{n}{2}.\label{det_JGJ}
\end{align}
Comparing (\ref{det_JGJ}) and (\ref{reparameterized_fisher_determinant}) we see that $|\hat{\bm{F}}(\hat{\bm{\theta}},\hat{\tau})|=|\bm{J}^{T}\bm{F}(\bm{\theta},\tau)\bm{J}|$
is satisfied.
%\appendix
%\chapter{Proof of the validity of the Laplace approximation formula for the marginal density $m_{\gamma_{d}}(\bm{x})$}
\chapter{The Laplace approximation formula for the marginal}
\noindent Let $\hat{T}_{m}^{(\hat{\tau}^{*},\hat{\bm{\theta}}^{*})}(\bm{x},\hat{\tau},\hat{\bm{\theta}})$ denote the $m$'th degree Taylor polynomial expansion of 
$\hat{\Phi}$ as a function of $\hat{\theta}_{i},\ i=1,...,d$ about the points $\hat{\bm{\theta}}^{*}$ and $\hat{\tau}^{*}$. Since the prior 
$\pi_{\lambda}(\bm{\theta})$ may not be smooth at $\bm{\theta}=\bm{0}$ we will have to claim that $\hat{\bm{\theta}}^{*}$ is nonzero. The $\hat{\bm{\theta}}$ 
integration in (\ref{reparameterized_integral}) will have to be split up into the $2^{d}$ integration areas consisting of 
$\mathbb{R}^{d}_{+}$ and the remaining $2^{d}-1$ ''quadrants'' which union is 
$\mathbb{R}^{d}\setminus\mathbb{R}^{d}_{+}$. We assume that $\hat{\bm{\theta}}^{*}\in\mathbb{R}_{+}^{d}$, and as will become clear below, 
this assumption implies no loss of generality.
Let $\hat{\Phi}_{\alpha_{1},...\alpha_{s}}(\bm{x},\hat{\tau},\hat{\bm{\theta}})$ denote the $s$'th order partial derivative of $\hat{\Phi}$ with respect to the 
ordered list of parameters $\bm{\alpha}=\alpha_{1},...,\alpha_{s}$. We write out the terms of 
$\hat{T}_{2}^{(\hat{\tau}^{*},\hat{\bm{\theta}}^{*})}(\bm{x},\hat{\tau},\hat{\bm{\theta}})$ explicitely below.
Define
\begin{align}
&\hat{T}_{K}^{(\hat{\tau}^{*},\hat{\bm{\theta}}^{*})}(\bm{x},\hat{\tau},\hat{\bm{\theta}})\mathdef\sum_{i=1}^{d}\sum_{j,k=1}^{j+k\leq K}\frac{1}{j!k!}\frac{\partial^{j+k}\hat{\Phi}(\bm{x},\hat{\tau}^{*},\hat{\bm{\theta}}^{*})}{\partial\hat{\tau}^{j}\partial\hat{\theta}_{i}^{k}}(\hat{\tau}-\hat{\tau}^{*})^{j}(\hat{\theta}_{i}-\hat{\theta}_{i}^{*})^{k}\nonumber\\
&\hat{R}^{(\hat{\tau}^{*},\hat{\bm{\theta}}^{*})}_{K}(\bm{x},\hat{\tau},\hat{\bm{\theta}})=\sum_{i=1}^{d}\sum_{j+k\geq K}^{\infty}\frac{1}{j!k!}\frac{\partial^{j+k}\hat{\Phi}(\bm{x},\hat{\tau},\hat{\bm{\theta}}^{*})}{\partial\hat{\tau}^{j}\partial\hat{\theta}_{i}^{k}}(\hat{\tau}-\hat{\tau}^{*})^{j}(\hat{\theta}_{i}-\hat{\theta}_{i}^{*})^{k}\nonumber
\intertext{we may then write}
&\hat{\Phi}(\bm{x},\hat{\tau},\hat{\bm{\theta}})=\hat{T}_{2}^{*}(\bm{x},\hat{\tau}^{*},\hat{\bm{\theta}}^{*})+\hat{R}_{3}^{(\hat{\tau}^{*},\hat{\bm{\theta}}^{*})}(\bm{x},\hat{\tau},\hat{\bm{\theta}})\nonumber
\intertext{where}\nonumber
&\hat{T}^{(\hat{\tau}^{*},\hat{\bm{\theta}}^{*})}_{2}(\bm{x},\hat{\tau},\hat{\bm{\theta}})=\hat{\Phi}(\bm{x},\hat{\tau}^{*},\hat{\bm{\theta}}^{*})
+\hat{\Phi}_{\hat{\tau}}(\bm{x},\hat{\tau}^{*},\hat{\bm{\theta}}^{*})(\hat{\tau}-\hat{\tau}^{*})\nonumber\\
&+\sum_{i=1}^{d}\hat{\Phi}_{\hat{\theta}_{i}}(\bm{x},\hat{\tau}^{*},\hat{\bm{\theta}}^{*})(\hat{\theta}_{i}-\hat{\theta}_{i}^{*})
+\frac{1}{2}\hat{\Phi}_{\hat{\tau},\hat{\tau}}(\bm{x},\hat{\tau}^{*},\hat{\bm{\theta}}^{*})(\hat{\tau}-\hat{\tau}^{*})^{2}\nonumber\\
&+\frac{1}{2}\sum_{i=1}^{d}\hat{\Phi}_{\hat{\theta}_{i},\hat{\theta}_{i}}(\bm{x},\hat{\tau}^{*},\hat{\bm{\theta}}^{*})(\hat{\theta}_{i}-\hat{\theta}_{i}^{*})^{2}+\sum_{i=1}^{d}\hat{\Phi}_{\hat{\tau},\hat{\theta}_{i}}(\bm{x},\hat{\tau}^{*},\hat{\bm{\theta}}^{*})(\hat{\tau}-\hat{\tau}^{*})(\hat{\theta}_{i}-\hat{\theta}_{i}^{*})\label{T2_Taylor}
\end{align}
where we can safely omit cross derivative terms of type $\hat{\Phi}_{\hat{\theta}_{i},\hat{\theta}_{j}},\ i\neq j$ 
because of our IID modeling assumptions on the $\theta_{i}$ and the functional relation $\theta_{i}=\phi(\hat{\theta}_{i},\hat{\tau})$. 
We note that the pure first order terms in $\hat{\theta}_{i}$ and $\hat{\tau}$
will vanish because $\hat{\Phi}_{\hat{\theta}_{i}}(\bm{x},\hat{\tau}^{*},\hat{\bm{\theta}}^{*})=0,\ 1\leq i\leq d$, and 
$\hat{\Phi}_{\hat{\tau}}(\bm{x},\hat{\tau}^{*},\hat{\bm{\theta}}^{*})=0$ by definition of $\hat{\tau}^{*}$ and $\hat{\bm{\theta}}^{*}$.
We will approximate the innermost integral in (\ref{reparameterized_integral}) by completing the squares in the parameters $\hat{\theta}_{i}$ in 
$\hat{T}_{2}^{(\hat{\tau}^{*},\hat{\bm{\theta}}^{*})}(\bm{x},\hat{\tau}^{*},\hat{\bm{\theta}}^{*})$ and integrate the resulting shifted quadratic exponential 
against a remainder polynomial $\hat{P}(\hat{\tau},\hat{\bm{\theta}})$ over 
the parameter manifold $\hat{\Theta}_{d}$. To prove that this method is sound, we have to show that the error due to the terms in 
$\hat{R}^{(\hat{\tau}^{*},\hat{\bm{\theta}}^{*})}_{3}(\bm{x},\hat{\tau},\hat{\bm{\theta}})$ not included in the remainder 
polynomial $\hat{P}(\hat{\tau},\hat{\bm{\theta}})$ can be made small enough. To do this we will have to analyse the relative magnitudes of the coefficients of 
$\hat{R}_{3}^{(\hat{\tau}^{*},\hat{\bm{\theta}}^{*})}$ to find the leading order terms. In order to explicitely express the dependency of terms of $\hat{T}_{2}^{(\hat{\tau}^{*},\hat{\bm{\theta}}^{*})}$ on $\bar{\tau}$ and $\bar{\bar{\tau}}$ we use equation (\ref{def_Phi_hat}) and the chain rule to rewrite partial derivatives of $\hat{\Phi}$ with respect to the parameters $\hat{\theta}_{i}$ and $\hat{\tau}$ as combinations of partial derivatives of $\Phi$.
%From equations (\ref{O_tau})-(\ref{O_tau_tau_tau}) we see that the coefficients of $\hat{T}_{2}^{(\hat{\tau}^{*},\hat{\bm{\theta}}^{*})}$ depend strongly on our choice of values on the free 
%parameters $\bar{\tau}$ and $\bar{\bar{\tau}}$. We have to choose these to make the integral (\ref{reparameterized_integral}) converge sufficiently fast. To be able to state bounds on the possible values on $\bar{\tau}$ and $\bar{\bar{\tau}}$ we express the integral (\ref{reparameterized_integral}) by the terms in $\hat{T}_{2}^{(\hat{\tau}^{*},\hat{\bm{\theta}}^{*})}$ and leave the discussion of which terms to include and neglect until we have calculated this expression. 
By completing squares in $\hat{\tau}$ and $\hat{\theta}_{i},\ 1\leq i\leq d$ and omitting the zero first order terms 
we may rewrite $\hat{T}_{2}^{(\hat{\tau}^{*},\hat{\bm{\theta}}^{*})}$ as
\begin{align}
&\hat{T}_{2}^{(\hat{\tau}^{*},\hat{\bm{\theta}}^{*})}(\bm{x},\hat{\tau},\hat{\bm{\theta}})=\hat{\Phi}(\bm{x},\hat{\tau}^{*},\hat{\bm{\theta}}^{*})
+\frac{1}{2}\hat{\Phi}_{\hat{\tau},\hat{\tau}}(\bm{x},\hat{\tau}^{*},\hat{\bm{\theta}}^{*})\left(\hat{\tau}-\hat{\tau}^{*}\right)^{2}\nonumber\\
&+\sum_{i=1}^{d}\left\{\frac{1}{2}\left[\hat{\Phi}_{\hat{\theta}_{i},\hat{\theta}_{i}}(\bm{x},\hat{\tau}^{*},\hat{\bm{\theta}}^{*})\right]\times\right.\nonumber\\
&\left.\left(\hat{\theta}_{i}-\hat{\theta}_{i}^{*}+\frac{\hat{\Phi}_{\hat{\tau},\hat{\theta}_{i}}(\bm{x},\hat{\tau}^{*},\hat{\bm{\theta}}^{*})(\hat{\tau}-\hat{\tau}^{*})}{\hat{\Phi}_{\hat{\theta}_{i},\hat{\theta}_{i}}(\bm{x},\hat{\tau}^{*},\hat{\bm{\theta}}^{*})}\right)^{2}-\frac{1}{2}\frac{\hat{\Phi}^{2}_{\hat{\tau},\hat{\theta}_{i}}(\bm{x},\hat{\tau}^{*},\hat{\bm{\theta}}^{*})(\hat{\tau}-\hat{\tau}^{*})^{2}}{\hat{\Phi}_{\hat{\theta}_{i},\hat{\theta}_{i}}(\bm{x},\hat{\tau}^{*},\hat{\bm{\theta}}^{*})}\right\}\nonumber\\
\intertext{Now we may write (\ref{reparameterized_integral}) on the form}
&m_{\gamma_{d}}(\bm{x})=I_{1}+I_{2}
\intertext{where}
&I_{1}\mathdef\bar{\bar{\tau}}^{\frac{d}{2}}\exp\left(-\hat{\Phi}(\bm{x},\hat{\tau}^{*},\hat{\bm{\theta}}^{*})\right)\times\nonumber\\
&\int_{\hat{\tau}\in\hat{I}_{\hat{\tau}}}d\hat{\tau}\ \exp\left(-\frac{1}{2}\left[\hat{\Phi}_{\hat{\tau},\hat{\tau}}(\bm{x},\hat{\tau}^{*},\hat{\bm{\theta}}^{*})-\sum_{i=1}^{d}\frac{\hat{\Phi}^{2}_{\hat{\tau},\hat{\theta}_{i}}(\bm{x},\hat{\tau}^{*},\hat{\bm{\theta}}^{*})}{\hat{\Phi}_{\hat{\theta}_{i},\hat{\theta}_{i}}(\bm{x},\hat{\tau}^{*},\hat{\bm{\theta}}^{*})}\right](\hat{\tau}-\hat{\tau}^{*})^{2}
\right)\times\nonumber\\
&\int_{\hat{\bm{\theta}}\in\mathbb{R}^{d}_{+}}\exp\left(-\frac{1}{2}\sum_{i=1}^{d}\left[\hat{\Phi}_{\hat{\theta}_{i},\hat{\theta}_{i}}(\bm{x},\hat{\tau}^{*},\hat{\bm{\theta}}^{*})\right]\times\right.\nonumber\\
&\left.\left(\hat{\theta}_{i}-\hat{\theta}_{i}^{*}+\frac{\hat{\Phi}_{\hat{\tau},\hat{\theta}_{i}}(\bm{x},\hat{\tau}^{*},\hat{\bm{\theta}}^{*})(\hat{\tau}-\hat{\tau}^{*})}{\hat{\Phi}_{\hat{\theta}_{i},\hat{\theta}_{i}}(\bm{x},\hat{\tau}^{*},\hat{\bm{\theta}}^{*})}\right)^{2}\right)
\exp\left(-\hat{R}_{3}^{(\hat{\tau}^{*},\hat{\bm{\theta}}^{*})}(\bm{x},\hat{\tau},\hat{\bm{\theta}})\right)\ d\hat{\bm{\theta}}.\label{def_I1}\\
&I_{2}\mathdef\bar{\bar{\tau}}^{\frac{d}{2}}\int_{\hat{\tau}\in\hat{I}_{\hat{\tau}}}d\hat{\tau}\int_{\hat{\bm{\theta}}\in\mathbb{R}^{d}\setminus\mathbb{R}^{d}_{+}}\exp\left(-\hat{\Phi}(\bm{x},\hat{\tau},\hat{\bm{\theta}})\right)\ d\hat{\bm{\theta}}.\label{def_I2}
\end{align}
We begin with the calculation of the $\hat{\bm{\theta}}$-part of the integral $I_{1}$. We will first introduce som notation and some 
claims. Define
\begin{align}
&\Upsilon_{\lambda}(\theta)\mathdef-\log[\pi_{\lambda}(\theta)]_{v}=-\log\left[C\lambda^{\frac{1}{2}}\exp\left(-f(\lambda^{\frac{1}{2}}\theta)\right)\right]_{v}\label{def_Upsilon}
\intertext{where $C>0$ is a normalization constant independent of $\lambda$. We claim that $f(\theta)$ is an integrable, symmetric, function of $\theta$, such that}
&\lim_{|\theta|\rightarrow\infty}f(\lambda^{\frac{1}{2}}\theta)=\infty,\label{integrable_claim_prior}
\intertext{and such that there exist numbers $0<\nu<2$, $B^{\prime}_{\nu}\leq B_{\nu}\in\mathbb{R}$, $C_{\nu}>0$ with the properties}
&B^{\prime}_{\nu}\leq f(\lambda^{\frac{1}{2}}\theta)\leq B_{\nu}+C_{\nu}|\lambda^{\frac{1}{2}}\theta|^{\nu},\forall\ [\theta]_{v}\in\mathbb{R},\label{tail_claim_prior1}
\intertext{and}
&\left|\left[\frac{\partial^{k}}{\partial\theta^{k}}f(\lambda^{\frac{1}{2}}\theta)\right]_{v}\right|\leq C_{\nu}\left|\left[\frac{\partial^{k}}{\partial\theta^{k}}|\lambda^{\frac{1}{2}}\theta|^{\nu}\right]_{v}\right|,\  1\leq k<\infty.\label{tail_claim_prior2}
\intertext{Because of (\ref{tail_claim_prior2}) we have}
&\left|\left[\frac{\partial^{2}}{\partial\theta^{2}}\Upsilon_{\lambda}(\theta)\right]_{v}\right|\leq\left|\left[C_{\nu}\nu(\nu-1)\lambda|\lambda^{\frac{1}{2}}\theta|^{v-2}\right]_{v}\right|,\ \forall\ [\theta]_{v}\in\mathbb{R}.\label{upper_bound_second_derivative_neg_log_prior}
\intertext{Define the signal to noise ratio (SNR) $\Omega$ by the power ratio in the data model (\ref{noise_model})}
&\Omega(\lambda,\tau)\mathdef \frac{d\frac{1}{\lambda}}{n\frac{1}{\tau}}\text{\hspace{1cm} (signal to noise ratio)}\label{def_SNR}
\intertext{we may deduce from (\ref{upper_bound_second_derivative_neg_log_prior}), (\ref{def_SNR}) and the fact that the likelihood is gaussian, the inequalities}
&[\tau^{*}]_{v}\leq \Phi_{\theta_{i},\theta_{i}}(\bm{x},\tau^{*},\bm{\theta}^{*})\nonumber\\
&\leq [\tau^{*}]_{v}\left(1+\frac{d}{n}\Omega^{-1}(\lambda,\tau^{*})|\lambda^{\frac{1}{2}}\theta_{i}^{*}|^{\nu-2}C_{\nu}\nu(\nu-1)\right),\ \text{ if }1\leq\nu<2\label{nu_bounds1}\\
\intertext{and}
&[\tau^{*}]_{v}\geq \Phi_{\theta_{i},\theta_{i}}(\bm{x},\tau^{*},\bm{\theta}^{*})\nonumber\\
&\geq [\tau^{*}]_{v}\left(1+\frac{d}{n}\Omega^{-1}(\lambda,\tau^{*})|\lambda^{\frac{1}{2}}\theta_{i}^{*}|^{\nu-2}C_{\nu}\nu(\nu-1)\right),\ \text{ if } 0<\nu< 1.\label{nu_bounds2}
\intertext{We define}
&\mu_{\lambda,\nu}(\tau,\theta_{i})\mathdef\frac{d}{n}\Omega^{-1}(\lambda,\tau)|\lambda^{\frac{1}{2}}\theta_{i}|^{\nu-2}C_{\nu}\nu(\nu-1)\label{def_mu}\\
&=\left(\frac{n}{d}\Omega(\lambda,\tau)\right)^{-\frac{\nu}{2}}|\tau^{\frac{1}{2}}\theta_{i}|^{\nu-2}C_{\nu}\nu(\nu-1)\label{rewritten_mu}
\intertext{and we claim there exists a number $0<\zeta_{\mu_{\lambda,\nu}}<1$ such that}
&\left|\mu_{\lambda,\nu}(\tau^{*},\theta_{i}^{*})\right|\leq\zeta_{\mu_{\lambda,\nu}}< 1,\ 1\leq i\leq d.\label{mu_claim1}
\intertext{We will investigate this claim further below. By (\ref{nu_bounds1}) and (\ref{nu_bounds2}) we see that if the SNR-value $\Omega(\lambda,\tau)$ is high enough and the relative model size $\frac{d}{n}$ small enough, then the value of $\Phi_{\theta_{i},\theta_{i}}(\bm{x},\tau^{*},\bm{\theta}^{*})$ may be approximated by $[\tau^{*}]_{v}$ for all practical purposes for 
the actual value of $\nu$ under consideration and ''reasonable'' $|\lambda^{\frac{1}{2}}\theta_{i}^{*}|$. We will discuss this question at the end of the proof.
Proceeding analogously to the steps above, one may show}
&\Phi_{\theta_{i},\theta_{i},\theta_{i}}(\bm{x},\tau^{*},\bm{\theta}^{*})=[\tau^{*}]_{v}^{\frac{3}{2}}\sigma_{\lambda,\nu}(\tau^{*},\theta_{i}^{*})\label{order_Phi_theta_theta_theta}
\intertext{where}
&\sigma_{\lambda,\nu}(\tau,\theta_{i})\mathdef\left(\frac{d}{n}\right)^{\frac{3}{2}}\Omega^{-\frac{3}{2}}(\lambda,\tau^{*})|\lambda^{\frac{1}{2}}\theta_{i}|^{\nu-3}C_{\nu}\nu(\nu-1)(\nu-2)\sgn(\theta_{i})\label{def_sigma}\\
&=\left(\frac{n}{d}\Omega(\lambda,\tau)\right)^{-\frac{\nu}{2}}|\tau^{\frac{1}{2}}\theta_{i}|^{\nu-3}C_{\nu}\nu(\nu-1)(\nu-2)\sgn(\theta_{i})\label{rewritten_sigma}
\intertext{and}
&\Phi_{\theta_{i},\theta_{i},\theta_{i},\theta_{i}}(\bm{x},\tau^{*},\bm{\theta}^{*})=[\tau^{*}]_{v}^{2}\kappa_{\lambda,\nu}(\tau^{*},\theta_{i}^{*})\label{order_Phi_theta_theta_theta_theta}
\intertext{where}
&\kappa_{\lambda,\nu}(\tau,\theta_{i})\mathdef\left(\frac{d}{n}\right)^{2}\Omega^{-2}(\lambda,\tau)|\lambda^{\frac{1}{2}}\theta_{i}|^{\nu-4}C_{\nu}\nu(\nu-1)(\nu-2)(\nu-3)\label{def_kappa}\\
&=\left(\frac{n}{d}\Omega(\lambda,\tau)\right)^{-\frac{\nu}{2}}|\tau^{\frac{1}{2}}\theta_{i}|^{\nu-4}C_{\nu}\nu(\nu-1)(\nu-2)(\nu-3).\label{rewritten_kappa}
\end{align}
We define
\begin{align}
&\Delta_{\lambda}(\theta_{i}^{*})\mathdef\frac{\partial^{2}}{\partial\theta_{i}^{2}}\Upsilon_{\lambda}(\bm{\theta})\label{def_Delta}
\intertext{and thus}
&\Phi_{\theta_{i},\theta_{i}}(\bm{x},\tau^{*},\bm{\theta}^{*})=\tau^{*}+\left.\frac{\partial^{2}}{\partial\theta_{i}^{2}}\Upsilon_{\lambda}(\bm{\theta})\right|_{\bm{\theta}=\bm{\theta}^{*}}=\tau^{*}+\Delta_{\lambda}(\theta_{i}^{*}).\label{Phi_by_Delta}
\intertext{By (\ref{tail_claim_prior2}) we deduce}
&\left|\Delta_{\lambda}(\theta_{i}^{*})\right|\leq\tau^{*}\left|\mu_{\lambda,\nu}(\tau^{*},\theta_{i}^{*})\right|.\label{order_of_delta}
\end{align}
We now continue with the calculation of the integral in (\ref{def_I1}).
\begin{align}
&\exp\left(-\hat{R}_{3}^{(\hat{\tau}^{*},\hat{\bm{\theta}}^{*})}(\bm{x},\hat{\tau},\hat{\bm{\theta}})\right)=\hat{P}(\hat{\tau},\hat{\bm{\theta}})+\hat{E}(\hat{\tau},\hat{\bm{\theta}})\nonumber\\
\intertext{where}
&\hat{P}(\hat{\tau},\hat{\bm{\theta}})\mathdef 1.\label{def_P}\\
&\hat{E}(\hat{\tau},\hat{\bm{\theta}})\mathdef\sum_{k=1}^{\infty}\frac{(-1)^{k}}{k!}\left(\hat{R}_{3}^{(\hat{\tau}^{*},\hat{\bm{\theta}}^{*})}(\bm{x},\hat{\tau},\hat{\bm{\theta}})\right)^{k}.\label{def_E}
\intertext{Define}
&P_{G}(x)\mathdef\frac{1}{\sqrt{2\pi}}\int_{-\infty}^{x}\exp\left(-\frac{1}{2}t^{2}\right)\ dt.\label{def_P_G}
\intertext{We may then proceed to write}
&I_{1}=\hat{U}\left(\bm{x},\hat{\tau}^{*},\hat{\bm{\theta}}^{*}\right)+\hat{W}\left(\bm{x},\hat{\tau}^{*},\hat{\bm{\theta}}\right)\nonumber
\intertext{where}
&\hat{U}\left(\bm{x},\hat{\tau}^{*},\hat{\bm{\theta}}^{*}\right)\mathdef
\bar{\bar{\tau}}^{\frac{d}{2}}\exp\left(-\hat{\Phi}(\bm{x},\hat{\tau}^{*},\hat{\bm{\theta}}^{*})\right)\frac{(2\pi)^{\frac{d}{2}}}{\prod_{i=1}^{d}\hat{\Phi}^{\frac{1}{2}}_{\hat{\theta}_{i},\hat{\theta}_{i}}(\bm{x},\hat{\tau}^{*},\hat{\bm{\theta}}^{*})}
\times\nonumber\\
&\int_{\hat{\tau}\in\hat{I}_{\hat{\tau}}}d\hat{\tau}\ \exp\left(-\frac{1}{2}\left[\hat{\Phi}_{\hat{\tau},\hat{\tau}}(\bm{x},\hat{\tau}^{*},\hat{\bm{\theta}}^{*})-\sum_{i=1}^{d}\frac{\hat{\Phi}^{2}_{\hat{\tau},\hat{\theta}_{i}}(\bm{x},\hat{\tau}^{*},\hat{\bm{\theta}}^{*})}{\hat{\Phi}_{\hat{\theta}_{i},\hat{\theta}_{i}}(\bm{x},\hat{\tau}^{*},\hat{\bm{\theta}}^{*})}\right](\hat{\tau}-\hat{\tau}^{*})^{2}
\right)\times\nonumber\\
&\prod_{i=1}^{d}P_{G}\left(\hat{\Phi}^{\frac{1}{2}}_{\hat{\theta}_{i},\hat{\theta}_{i}}(\bm{x},\hat{\tau}^{*},\hat{\bm{\theta}}^{*})\left(\hat{\theta}_{i}^{*}-\frac{\hat{\Phi}_{\hat{\tau},\hat{\theta}_{i}}(\bm{x},\hat{\tau}^{*},\hat{\bm{\theta}}^{*})(\hat{\tau}-\hat{\tau}^{*})}{\hat{\Phi}_{\hat{\theta}_{i},\hat{\theta}_{i}}(\bm{x},\hat{\tau}^{*},\hat{\bm{\theta}}^{*})}\right)\right)\label{def_U}\\
\intertext{and}
&\hat{W}\left(\bm{x},\hat{\tau}^{*},\hat{\bm{\theta}}^{*}\right)\mathdef
\bar{\bar{\tau}}^{\frac{d}{2}}\exp\left(-\hat{\Phi}(\bm{x},\hat{\tau}^{*},\hat{\bm{\theta}}^{*})\right)\times\nonumber\\
&\int_{\hat{\tau}\in\hat{I}_{\hat{\tau}}}d\hat{\tau}\ \exp\left(-\frac{1}{2}\left[\hat{\Phi}_{\hat{\tau},\hat{\tau}}(\bm{x},\hat{\tau}^{*},\hat{\bm{\theta}}^{*})-\sum_{i=1}^{d}\frac{\hat{\Phi}^{2}_{\hat{\tau},\hat{\theta}_{i}}(\bm{x},\hat{\tau}^{*},\hat{\bm{\theta}}^{*})}{\hat{\Phi}_{\hat{\theta}_{i},\hat{\theta}_{i}}(\bm{x},\hat{\tau}^{*},\hat{\bm{\theta}}^{*})}\right](\hat{\tau}-\hat{\tau}^{*})^{2}
\right)\times\nonumber\\
&\int_{\hat{\bm{\theta}}\in\mathbb{R}^{d}_{+}}\exp\left(-\frac{1}{2}\sum_{i=1}^{d}\left[\hat{\Phi}_{\hat{\theta}_{i},\hat{\theta}_{i}}(\bm{x},\hat{\tau}^{*},\hat{\bm{\theta}}^{*})\right]\times\right.\nonumber\\
&\left.\left(\hat{\theta}_{i}-\hat{\theta}_{i}^{*}+\frac{\hat{\Phi}_{\hat{\tau},\hat{\theta}_{i}}(\bm{x},\hat{\tau}^{*},\hat{\bm{\theta}}^{*})(\hat{\tau}-\hat{\tau}^{*})}{\hat{\Phi}_{\hat{\theta}_{i},\hat{\theta}_{i}}(\bm{x},\hat{\tau}^{*},\hat{\bm{\theta}}^{*})}\right)^{2}\right)\hat{E}(\hat{\tau},\hat{\bm{\theta}})\ d\hat{\bm{\theta}}.\label{def_W}
\intertext{We first investigate the term $\hat{W}\left(\bm{x},\hat{\tau}^{*},\hat{\bm{\theta}}\right)$. The lowest order term of 
$\hat{E}(\hat{\tau},\hat{\bm{\theta}})$ in (\ref{def_E}) is $\hat{R}^{(\hat{\tau}^{*},\hat{\bm{\theta}}^{*})}_{3}(\bm{x},\hat{\tau},\hat{\bm{\theta}})$, and is given by}
&\hat{R}^{(\hat{\tau}^{*},\hat{\bm{\theta}}^{*})}_{3}(\bm{x},\hat{\tau},\hat{\bm{\theta}})=
\frac{1}{6}\Phi_{\hat{\tau},\hat{\tau},\hat{\tau}}(\bm{x},\hat{\tau}^{*},\hat{\bm{\theta}}^{*})(\hat{\tau}-\hat{\tau}^{*})^{3}+\nonumber\\
&+\frac{1}{2}\sum_{i=1}^{d}\hat{\Phi}_{\hat{\tau},\hat{\tau},\hat{\theta}_{i}}(\bm{x},\hat{\tau}^{*},\hat{\bm{\theta}}^{*})(\hat{\tau}-\hat{\tau}^{*})^{2}(\hat{\theta}_{i}-\hat{\theta}_{i}^{*})\nonumber\\
&+\frac{1}{2}\sum_{i=1}^{d}\hat{\Phi}_{\hat{\tau},\hat{\theta}_{i},\hat{\theta}_{i}}(\bm{x},\hat{\tau}^{*},\hat{\bm{\theta}}^{*})(\hat{\tau}-\hat{\tau}^{*})(\hat{\theta}_{i}-\hat{\theta}_{i}^{*})^{2}\nonumber\\
&+\frac{1}{6}\sum_{i=1}^{d}\hat{\Phi}_{\hat{\theta}_{i},\hat{\theta}_{i},\hat{\theta}_{i}}(\bm{x},\hat{\tau}^{*},\hat{\bm{\theta}}^{*})(\hat{\theta}_{i}-\hat{\theta}_{i}^{*})^{3}+\text{ higher order }.\label{truncated_R3_Taylor}
\end{align}
We will now make a few observations and claims which together will imply that it suffices to consider the part of $\hat{E}(\hat{\tau},\hat{\bm{\theta}})$ given
by the third order terms listed in (\ref{truncated_R3_Taylor}) to compute $\hat{W}(\bm{x},\hat{\tau}^{*},\hat{\bm{\theta}}^{*})$ to leading order.
As will become clear from considerations below, the interval $\hat{I}_{\hat{\tau}}$ will have to include the point
$\hat{\tau}^{*}$ in order to get convergence of the $\hat{\tau}-$integration step. Furthermore, it will become clear that 
$\hat{\tau}-\hat{\tau}^{*}$ must be bounded below. In fact we will see below that we must have
\begin{align}
&\hat{I}_{\hat{\tau}}=\left[\hat{\tau}^{*}-a_{\hat{\tau}}\delta_{n}^{-1}\epsilon_{d}^{-1},\hat{\tau}^{*}+b_{\hat{\tau}}\delta_{n}^{-1}\epsilon_{d}^{-1}\right],\ 0<a_{\hat{\tau}}\ll 1,\ a_{\hat{\tau}}\leq b_{\hat{\tau}}<\infty.\label{I_tau_claim1}
\end{align}
where $a_{\hat{\tau}}$ and $b_{\hat{\tau}}$ are to be chosen large enough to make the integral of $\exp\left(-\hat{T}_{2}(\bm{x},\hat{\tau},\hat{\bm{\theta}})\right)$ 
converge with respect to the integration in $\hat{\tau}$. Furthermore, we will see below that we may choose $a_{\hat{\tau}}=b_{\hat{\tau}}$, thus making the interval 
$\hat{I}_{\hat{\tau}}$ symmetric about $\hat{\tau}^{*}$. This fact will be used to simplify the computations below. Next, we observe
\begin{align}
&\hat{\Phi}_{\hat{\theta}_{i},\hat{\tau}}(\bm{x},\hat{\tau}^{*},\hat{\bm{\theta}}^{*})=
\bar{\tau}^{\frac{1}{2}}\delta_{n}\epsilon_{d}\left(-(\bm{x}_{\parallel}(i)-\theta_{i}^{*})(\tau^{*})^{\frac{1}{2}}\right.\nonumber\\
&\left.-\frac{1}{2}(\tau^{*}+\Delta_{\lambda}(\theta_{i}^{*}))(\tau^{*})^{-\frac{1}{2}}\theta_{i}^{*}\right) \nonumber\\
&\in\left[\left(-\frac{\bm{x}_{\parallel}(i)}{\theta_{i}^{*}}+1-\frac{1}{2}(1+\mu_{\lambda,\nu}(\tau^{*},\theta_{i}^{*}))\right)\bar{\tau}^{\frac{1}{2}}\delta_{n}\epsilon_{d}(\tau^{*})^{\frac{1}{2}}\theta^{*}_{i},\right.\nonumber\\
&\left.\left(-\frac{1}{2}(1-\mu_{\lambda,\nu}(\tau^{*},\theta_{i}^{*}))\right)\bar{\tau}^{\frac{1}{2}}\delta_{n}\epsilon_{d}(\tau^{*})^{\frac{1}{2}}\theta^{*}_{i}\right]\subset\mathbb{R}_{-}\nonumber\\
&\text{ by (\ref{O_tau_theta}), (\ref{order_of_delta}), (\ref{mu_claim1}). }\label{hat_Phi_theta_tau2}\\
&\hat{\Phi}_{\hat{\tau},\hat{\tau}}(\bm{x},\hat{\tau}^{*},\hat{\bm{\theta}}^{*})=\delta_{n}^{2}\epsilon_{d}^{2}\left(\frac{n-d+2}{2}+\frac{1}{4}\|(\tau^{*})^{\frac{1}{2}}\bm{\theta}^{*}\|_{2}^{2}+\right.\nonumber\\
&\left.+\sum_{i=1}^{d}\tau^{*}(\theta_{i}^{*})^{2}\mu_{\lambda,\nu}(\tau^{*},\theta_{i}^{*})
+\sum_{i=1}^{d}\tau^{*}(\bm{x}_{\parallel}(i)-\theta^{*}_{i})\theta_{i}^{*}\right)\text { by (\ref{O_tau_tau})}.\label{hat_Phi_tau_tau2}\\
&\hat{\Phi}_{\hat{\theta}_{j},\hat{\theta}_{j}}(\bm{x},\hat{\tau}^{*},\hat{\bm{\theta}}^{*})=\Phi_{\theta_{i},\theta_{i}}(\bm{x},\tau^{*},\bm{\theta}^{*})\bar{\tau}(\tau^{*})^{-1}\nonumber\\
&=\left(1+\littleo\left(\mu_{\lambda,\nu}(\tau^{*},\theta_{i}^{*})\right)\right)\bar{\tau}\text{ by (\ref{O_theta_theta}), (\ref{Phi_by_Delta}), (\ref{order_of_delta}). }
\label{hat_Phi_theta_theta2}\\
&\hat{\Phi}_{\hat{\theta}_{i},\hat{\theta}_{i},\hat{\theta}_{i}}(\bm{x},\hat{\tau}^{*},\hat{\bm{\theta}}^{*})=
\Phi_{\theta_{i},\theta_{i},\theta_{i}}(\bm{x},\tau^{*},\bm{\theta}^{*})\bar{\tau}^{\frac{3}{2}}(\tau^{*})^{-\frac{3}{2}}\nonumber\\
&=\bar{\tau}^{\frac{3}{2}}\sigma_{\lambda,\nu}(\tau^{*},\theta_{i}^{*}) \text{ by (\ref{O_theta_theta_theta}) and (\ref{order_Phi_theta_theta_theta}). } \label{hat_Phi_theta_theta_theta2}\\
&\hat{\Phi}_{\hat{\tau},\hat{\tau},\hat{\tau}}(\bm{x},\hat{\tau}^{*},\hat{\bm{\theta}}^{*})=\delta_{n}^{3}\epsilon_{d}^{3}\left(\frac{n-d+2}{2}+
\frac{3}{8}\|(\tau^{*})^{\frac{1}{2}}\bm{\theta}^{*}\|_{2}^{2}+\right.\nonumber\\ 
&\left.+\frac{1}{8}C_{\nu}\nu(\nu-1)(\nu-2)\sum_{i=1}^{d}|\lambda^{\frac{1}{2}}\theta^{*}_{i}|^{\nu}\right)\nonumber\\
&+\delta_{n}^{3}\epsilon_{d}^{3}\littleo\left(\|(\tau^{*})^{\frac{1}{2}}\bm{\theta}^{*}\|_{2}^{2}\right)
\text{ by (\ref{O_tau_tau_tau}), (\ref{order_Phi_theta_theta_theta})-(\ref{def_sigma}). } \label{hat_Phi_tau_tau_tau2}\\
&\hat{\Phi}_{\hat{\tau},\hat{\tau},\hat{\theta}_{i}}(\bm{x},\hat{\tau}^{*},\hat{\bm{\theta}}^{*})=
\delta_{n}^{2}\epsilon_{d}^{2}\bar{\tau}^{\frac{1}{2}}(\tau^{*})^{\frac{1}{2}}\theta_{i}^{*}\left\{-\frac{1}{4}+
\frac{3}{4}\littleo\left(\mu_{\lambda,\nu}(\tau^{*},\theta^{*}_{i})\right)\right.\nonumber\\
&\left.+\frac{1}{4}\sigma_{\lambda,\nu}(\tau^{*},\theta_{i}^{*})(\tau^{*})^{\frac{1}{2}}\theta_{i}^{*}\right\}
\text{ by (\ref{O_tau_tau_theta}), (\ref{order_Phi_theta_theta_theta}). } \label{hat_Phi_tau_tau_theta2}\\
&\hat{\Phi}_{\hat{\tau},\hat{\theta}_{i},\hat{\theta}_{i}}(\bm{x},\hat{\tau}^{*},\hat{\bm{\theta}}^{*})=
-\frac{1}{2}\delta_{n}\epsilon_{d}\bar{\tau}(\tau^{*})^{-1}\Phi_{\theta_{i},\theta_{i},\theta_{i}}(\bm{x},\tau^{*},\bm{\theta}^{*})\theta_{i}^{*}\nonumber\\
&=-\frac{1}{2}\delta_{n}\epsilon_{d}\bar{\tau}\sigma_{\lambda,\nu}(\tau^{*},\theta_{i}^{*})(\tau^{*})^{\frac{1}{2}}\theta_{i}^{*}
\text{ by (\ref{O_theta_theta_tau}), (\ref{order_Phi_theta_theta_theta}). } \label{hat_Phi_theta_theta_tau2}\\
&\hat{\Phi}_{\hat{\tau},\hat{\tau},\hat{\theta}_{i},\hat{\theta}_{i}}(\bm{x},\hat{\tau}^{*},\hat{\bm{\theta}}^{*})
=\delta_{n}^{2}\epsilon_{d}^{2}\bar{\tau}\mu_{\lambda,\nu}(\tau^{*},\theta_{i}^{*})\nonumber\\
&+\delta_{n}^{2}\epsilon_{d}^{2}\bar{\tau}\sigma_{\lambda,\nu}(\tau^{*},\theta_{i}^{*})(\tau^{*})^{\frac{1}{2}}\theta^{*}_{i}
+\delta_{n}^{2}\epsilon_{d}^{2}\bar{\tau}\frac{1}{4}\kappa_{\lambda,\nu}(\tau^{*},\theta_{i}^{*})\tau^{*}(\theta_{i}^{*})^{2}\nonumber\\
&\text{ by (\ref{O_theta_theta_tau_tau}), (\ref{order_Phi_theta_theta_theta})-(\ref{Phi_by_Delta}). } \label{hat_Phi_theta_theta_tau_tau2}\\
&\hat{\Phi}_{\hat{\theta}_{i},\hat{\theta}_{i},\hat{\theta}_{i},\hat{\theta}_{i}}(\bm{x},\hat{\tau}^{*},\hat{\bm{\theta}}^{*})=
\Phi_{\theta_{i},\theta_{i},\theta_{i},\theta_{i}}(\bm{x},\tau^{*},\bm{\theta}^{*})\bar{\tau}^{2}(\tau^{*})^{-2}\nonumber\\
&=\bar{\tau}^{2}\kappa_{\lambda,\nu}(\tau^{*},\theta_{i}^{*})\text{ by (\ref{hat_Phi_hat_theta_hat_theta_hat_theta_hat_theta}) and (\ref{def_kappa}).}\nonumber
\end{align}                                                      
We will begin with considering the terms in $\hat{E}(\hat{\tau},\hat{\bm{\theta}})$ that are \\ $\hat{\Phi}_{\hat{\theta}_{i},\hat{\theta}_{i},\hat{\theta}_{i}}(\bm{x},\hat{\tau}^{*},\hat{\bm{\theta}}^{*})\left(\hat{\theta}_{i}-\hat{\theta}_{i}^{*}\right)^{3}$. We define

%\begin{figure}[h]
%\begin{center}
%\includegraphics[scale=0.6]{./Cum_Gauss}
%\end{center}
%\caption{Plot of the gaussian distribution function $P_{G}(\xi)$.}
%\label{plot_cum_gauss}
%\end{figure}
We may now express (\ref{def_I1}) on the form
\begin{align}
&I_{1}=\hat{W}(\hat{\tau}^{*},\hat{\bm{\theta}}^{*})+\hat{U}(\hat{\tau}^{*},\hat{\bm{\theta}}^{*})\nonumber\\
&=\frac{\bar{\bar{\tau}}^{\frac{d}{2}}(2\pi)^{\frac{d}{2}}\exp\left(-\hat{\Phi}(\bm{x},\hat{\tau}^{*},\hat{\bm{\theta}}^{*})\right)}{\prod_{i=1}^{d}\hat{\Phi}^{\frac{1}{2}}_{\hat{\theta}_{i},\hat{\theta}_{i}}(\bm{x},\hat{\tau}^{*},\hat{\bm{\theta}}^{*})}
\left[\hat{\Phi}_{\hat{\tau},\hat{\tau}}(\bm{x},\hat{\tau}^{*},\hat{\bm{\theta}}^{*})-\sum_{i=1}^{d}\frac{\hat{\Phi}^{2}_{\hat{\tau},\hat{\theta}_{i}}(\bm{x},\hat{\tau}^{*},\hat{\bm{\theta}}^{*})}{\hat{\Phi}_{\hat{\theta}_{i},\hat{\theta}_{i}}(\bm{x},\hat{\tau}^{*},\hat{\bm{\theta}}^{*})}\right]^{-\frac{1}{2}}\nonumber\\
&\times\prod_{i=1}^{d}P_{G}\left((\tau^{*})^{\frac{1}{2}}\theta_{i}^{*}\left(1+\zeta\right)\right)\times\nonumber\\
&\left\{1+\sum_{j=1}^{d}\frac{\tau^{*}(\theta_{j}^{*})^{2}}{N(\lambda,\nu,\gamma_{d})}\exp\left(-\frac{1}{2}\tau^{*}(\theta_{j}^{*})^{2}\right)+\frac{4}{3}(1+\zeta)\frac{C_{\nu}\nu|\nu-1|\cdot|\nu-2|}{\left(\frac{n}{d}\Omega(\tau^{*},\lambda^{*})\right)^{\frac{\nu}{2}}}\times\right.\nonumber\\
&\left.\sum_{j=1}^{d}\left[\frac{\left|(\tau^{*})^{\frac{1}{2}}\theta_{j}^{*}\right|^{\nu-1}\sgn(\theta_{j}^{*})\left(1+\frac{2}{\tau^{*}(\theta_{j}^{*})^{2}}\right)+\bigo\left(\left[\frac{k_{\hat{\tau}}\log{N(\lambda,\nu,\gamma_{d})}}{N^{k_{\hat{\tau}}}(\lambda,\nu,\gamma_{d})}\right]^{\frac{1}{2}}\right)}{\exp\left(\frac{1}{2}\tau^{*}(\theta^{*}_{j})^{2}\right)}\right]\right.\nonumber\\
&\left.-\frac{(2\pi)^{-\frac{1}{2}}}{N(\lambda,\nu,\gamma_{d})}\sum_{i,j=1}^{d}\frac{\tau^{*}(\bm{x}_{\parallel}(i)-\frac{1}{2}\theta_{i}^{*})(\bm{x}_{\parallel}(j)-\frac{1}{2}\theta_{j}^{*})}{\exp\left(\frac{1}{2}\tau^{*}\left[(\theta_{i}^{*})^{2}+(\theta_{j}^{*})^{2}\right]\right)}\right.\nonumber\\
&\left.+\bigo\left(\frac{1}{N(\lambda,\nu,\gamma_{d})}\sum_{j=1}^{d}\frac{\tau^{*}(\theta_{j}^{*})^{2}}{\exp\left(\frac{1}{2}\tau^{*}(\theta_{j}^{*})^{2}\right)}\right)\right\}.\label{I1_form3}
\end{align}
Now, utilizing (\ref{form_of_G}), (\ref{det_form_of_G}) we recognize the determinant of the Hessian $\hat{\bm{H}}(\bm{x},\hat{\tau}^{*},\hat{\bm{\theta}}^{*})$ of 
$\hat{\Phi}(\bm{x},\hat{\tau}^{*},\hat{\bm{\theta}}^{*})$ with respect to parameters $\hat{\tau}$,$\hat{\bm{\theta}}$ inside expression (\ref{I1_form3}). Assuming $N(\lambda,\nu,\gamma_{d})$ is large we may write
\begin{align}
&I_{1}=\bar{\bar{\tau}}^{\frac{d}{2}}\exp\left(-\hat{\Phi}(\bm{x},\hat{\tau}^{*},\hat{\bm{\theta}}^{*})\right)\frac{(2\pi)^{\frac{d+1}{2}}}{|\hat{\bm{H}}(\bm{x},
\hat{\tau}^{*},\hat{\bm{\theta}}^{*})|^{\frac{1}{2}}}\prod_{i=1}^{d}P_{G}\left((\tau^{*})^{\frac{1}{2}}\theta_{i}^{*}\left(1+\zeta\right)\right)\nonumber\\
&\times\left\{1+\sum_{j=1}^{d}\frac{\tau^{*}(\theta_{j}^{*})^{2}}{N(\lambda,\nu,\gamma_{d})}\exp\left(-\frac{1}{2}\tau^{*}(\theta_{j}^{*})^{2}\right)\right.\nonumber\\
&\left.+\frac{4}{3}(1+\zeta)\frac{C_{\nu}\nu|\nu-1|\cdot|\nu-2|}{\left(\frac{n}{d}\Omega(\tau^{*},\lambda^{*})\right)^{\frac{\nu}{2}}}
\sum_{j=1}^{d}\frac{\left|(\tau^{*})^{\frac{1}{2}}\theta_{j}^{*}\right|^{\nu-1}\sgn(\theta_{j}^{*})\left(1+\frac{2}{\tau^{*}(\theta_{j}^{*})^{2}}\right)}{\exp\left(\frac{1}{2}\tau^{*}(\theta^{*}_{j})^{2}\right)}\right.\nonumber\\
&\left.-\frac{(2\pi)^{-\frac{1}{2}}}{N(\lambda,\nu,\gamma_{d})}\sum_{i,j=1}^{d}\frac{\tau^{*}(\bm{x}_{\parallel}(i)-\frac{1}{2}\theta_{i}^{*})(\bm{x}_{\parallel}(j)-\frac{1}{2}\theta_{j}^{*})}{\exp\left(\frac{1}{2}\tau^{*}\left[(\theta_{i}^{*})^{2}+(\theta_{j}^{*})^{2}\right]\right)}\right\}.\label{I1_form4}
\intertext{Using (\ref{def_Phi_hat}), and the relations $|\bm{J}_{\psi,\bm{\phi}}^{T}\bm{H}\bm{J}_{\psi,\bm{\phi}}|=|\hat{\bm{H}}|$ and $|\bm{J}_{\psi,\bm{\phi}}^{T}\bm{F}\bm{J}_{\psi,\bm{\phi}}|=|\hat{\bm{F}}|=\bar{\bar{\tau}}^{d}$, where $\bm{J}_{\psi,\bm{\phi}}$ is the jacobi matrix of the transformations $\psi(\hat{\tau}),\phi(\theta_{i}),\ 1\leq i\leq d$, and $\bm{H}(\bm{x},\tau,\bm{\theta})$ is the Hessian of $\bm{\Phi}(\bm{x},\tau,\bm{\theta})$,  we finally get}
&I_{1}=\frac{(2\pi)^{\frac{d+1}{2}}}{|\bm{H}(\bm{x},\tau^{*},\bm{\theta}^{*})|^{\frac{1}{2}}}f(\bm{x}|\tau^{*},\bm{\theta}^{*})\pi_{\lambda}(\bm{\theta}^{*})
\prod_{i=1}^{d}P_{G}\left((\tau^{*})^{\frac{1}{2}}\theta_{i}^{*}\left(1+\zeta\right)\right)\times\nonumber\\
&\left\{1+\sum_{j=1}^{d}\frac{\tau^{*}(\theta_{j}^{*})^{2}}{N(\lambda,\nu,\gamma_{d})}\exp\left(-\frac{1}{2}\tau^{*}(\theta_{j}^{*})^{2}\right)\right.\nonumber\\
&\left.+\frac{4}{3}(1+\zeta)\frac{C_{\nu}\nu|\nu-1|\cdot|\nu-2|}{\left(\frac{n}{d}\Omega(\tau^{*},\lambda^{*})\right)^{\frac{\nu}{2}}}
\sum_{j=1}^{d}\frac{\left|(\tau^{*})^{\frac{1}{2}}\theta_{j}^{*}\right|^{\nu-1}\sgn(\theta_{j}^{*})\left(1+\frac{2}{\tau^{*}(\theta_{j}^{*})^{2}}\right)}{\exp\left(\frac{1}{2}\tau^{*}(\theta^{*}_{j})^{2}\right)}\right.\nonumber\\
&\left.-\frac{(2\pi)^{-\frac{1}{2}}}{N(\lambda,\nu,\gamma_{d})}\sum_{i,j=1}^{d}\frac{\tau^{*}(\bm{x}_{\parallel}(i)-\frac{1}{2}\theta_{i}^{*})(\bm{x}_{\parallel}(j)-\frac{1}{2}\theta_{j}^{*})}{\exp\left(\frac{1}{2}\tau^{*}\left[(\theta_{i}^{*})^{2}+(\theta_{j}^{*})^{2}\right]\right)}\right\}.\label{I1_final}
\end{align}
There are some observations to be remarked upon in connection with the result (\ref{I1_final}).\\
\begin{enumerate}
\item $a_{\hat{\tau}}$ and $b_{\hat{\tau}}$ has to be chosen large enough to make the $\hat{\tau}$-integrals $\int_{\hat{I}_{\hat{\tau}}}(\cdot)\ d\hat{\tau}$ in (\ref{def_U}) and (\ref{def_W}) 
converge, that is $\int_{\hat{I}_{\hat{\tau}}}(\cdot)\ d\hat{\tau}$ $\approx\int_{-\infty}^{\infty}(\cdot)\ d\hat{\tau}$.
\item We note that if $\theta^{*}$ is the hard threshold estimator used by Donoho and Johnstone in \cite{DJ_Ideal:1994}, we have for all nonzero $\hat{\theta}^{*}$ that $(\tau^{*})^{\frac{1}{2}}\theta^{*}\geq\sqrt{2\log{n}}$ and so: 
$1\geq\prod_{i=1}^{d}P_{G}\left((\tau^{*})^{\frac{1}{2}}\theta_{i}^{*}\right)$ $\geq P^{d}_{G}\left(\sqrt{2\log{n}}\right)$.
\item We note that the result in \cite{Rissanen:2000} concerning IID signal in additive white gaussian noise, which in our setting 
coincides with a prior density equal the Fisher information, (which for IID gaussian likelihood is the uniform density in $\theta_{i},\ 1\leq i\leq d$), 
yields asymptotically for large $n$ that $\inf_{1\leq i\leq d}(\tau^{*})^{\frac{1}{2}}|\bm{x}_{\parallel}(i)|=\sqrt{\log{n}+\littleo(\log{n})}$.\\
\end{enumerate}
We proceed to estimate the size of $N(\lambda,\nu,\gamma_{d})$.
By equation (\ref{O_tau_tau}), (\ref{def_Phi}), (\ref{def_delta}) and the fact that the likelihood $f$ is gaussian we have
\begin{align}
&\delta_{n}^{-2}\epsilon_{d}^{-2}\hat{\Phi}_{\hat{\tau},\hat{\tau}}(\bm{x},\hat{\tau}^{*},\hat{\bm{\theta}}^{*})=\frac{1}{2}(n-d+2)
+\sum_{i=1}^{d}\left(\bm{x}_{\parallel}(i)-\theta_{i}^{*}\right)\theta_{i}^{*}\tau^{*}\nonumber\\
&+\frac{1}{4}\sum_{i=1}^{d}\left(\tau^{*}+\Delta_{\lambda}(\theta_{i}^{*})\right)(\theta_{i}^{*})^{2}.\label{term1}
\intertext{By (\ref{hat_Phi_theta_tau2}), (\ref{hat_Phi_theta_theta2}) we have}
&\delta_{n}^{-2}\epsilon_{d}^{-2}\sum_{i=1}^{d}\frac{\hat{\Phi}^{2}_{\hat{\tau},\hat{\theta}_{i}}(\bm{x},\hat{\tau}^{*},\hat{\bm{\theta}}^{*})}{\hat{\Phi}_{\hat{\theta}_{i},\hat{\theta}_{i}}(\bm{x},\hat{\tau}^{*},\hat{\bm{\theta}}^{*})}=\sum_{i=1}^{d}\left(1+\littleo\left(\mu_{\lambda,\nu}(\tau^{*},\theta_{i}^{*})\right)\right)\times\nonumber\\
&\left((\bm{x}_{\parallel}(i)-\theta_{i}^{*})(\tau^{*})^{\frac{1}{2}}-\frac{1}{2}(\tau^{*}+\Delta_{\lambda}(\theta_{i}^{*})(\tau^{*})^{-\frac{1}{2}}\theta_{i}^{*}\right)^{2}.\label{term2}
\intertext{Combining (\ref{term1}) and (\ref{term2}) we get}
&N(\lambda,\nu,\gamma_{d})\mathdef\delta_{n}^{-2}\epsilon_{d}^{-2}\left\{\hat{\Phi}_{\hat{\tau},\hat{\tau}}(\bm{x},\hat{\tau}^{*},\hat{\bm{\theta}}^{*})-\sum_{i=1}^{d}\frac{\hat{\Phi}^{2}_{\hat{\tau},\hat{\theta}_{i}}(\bm{x},\hat{\tau}^{*},\hat{\bm{\theta}}^{*})}{\hat{\Phi}_{\hat{\theta}_{i},\hat{\theta}_{i}}(\bm{x},\hat{\tau}^{*},\hat{\bm{\theta}}^{*})}\right\}\label{def_N}\\
&=\frac{1}{2}(n-d+2)+\sum_{i=1}^{d}\left(\bm{x}_{\parallel}(i)-\theta_{i}^{*}\right)\theta_{i}^{*}\tau^{*}-\sum_{i=1}^{d}\left(\bm{x}_{\parallel}(i)-\theta_{i}^{*}\right)^{2}\tau^{*}+\nonumber\\
&+\sum_{i=1}^{d}\left\{\Delta_{\lambda}(\theta_{i}^{*})(\tau^{*})^{-\frac{1}{2}}\theta_{i}^{*}-\frac{1}{4}(\theta_{i}^{*})^{2}\Delta_{\lambda}(\theta_{i}^{*})-\frac{1}{4}\Delta^{2}_{\lambda}(\theta_{i}^{*})(\tau^{*})^{-1}(\theta_{i}^{*})^{2}\right\}\nonumber\\
&-\sum_{i=1}^{d}\littleo\left(\mu_{\lambda,\nu}(\tau^{*},\theta_{i}^{*})\right)\left((\bm{x}_{\parallel}(i)-\theta_{i}^{*})(\tau^{*})^{\frac{1}{2}}-\frac{1}{2}(\tau^{*}+\Delta_{\lambda}(\theta_{i}^{*}))(\tau^{*})^{-\frac{1}{2}}\theta_{i}^{*}\right)^{2}\nonumber
\intertext{ using (\ref{order_of_delta}) and rewriting a bit we get}
&N(\lambda,\nu,\gamma_{d})=\frac{1}{2}(n-d+2)+\sum_{i=1}^{d}\tau^{*}\left(\bm{x}_{\parallel}(i)-\theta_{i}^{*}\right)\left(2\theta_{i}^{*}-\bm{x}_{\parallel}(i)\right)+\nonumber\\
&+\sum_{i=1}^{d}\left\{\mu_{\lambda,\nu}(\tau^{*},\theta_{i}^{*})(\tau^{*})^{\frac{1}{2}}\theta_{i}^{*}-\frac{1}{4}\mu_{\lambda,\nu}(\tau^{*},\theta_{i}^{*})\tau^{*}(\theta_{i}^{*})^{2}+\right.\nonumber\\
&\left.-\frac{1}{4}\mu^{2}_{\lambda,\nu}(\tau^{*},\theta_{i}^{*})\tau^{*}(\theta_{i}^{*})^{2}\right\}\nonumber\\
&-\sum_{i=1}^{d}\littleo\left(\mu_{\lambda,\nu}(\tau^{*},\theta_{i}^{*})\right)\left((\bm{x}_{\parallel}(i)-\theta_{i}^{*})(\tau^{*})^{\frac{1}{2}}-\frac{1}{2}(\tau^{*}+\Delta_{\lambda}(\theta_{i}^{*}))(\tau^{*})^{-\frac{1}{2}}\theta_{i}^{*}\right)^{2}\label{tau_det1}\\
\intertext{ rewriting $\mu_{\lambda,\nu}(\tau^{*},\theta_{i}^{*})$ we get}
&N(\lambda,\nu,\gamma_{d})=\frac{1}{2}(n-d+2)+\sum_{i=1}^{d}\tau^{*}\left(\bm{x}_{\parallel}(i)-\theta_{i}^{*}\right)\left(2\theta_{i}^{*}-\bm{x}_{\parallel}(i)\right)+\nonumber\\
&+\sum_{i=1}^{d}C_{\nu}\nu(\nu-1)\left[\left(\frac{n}{d}\Omega(\tau^{*},\lambda^{*})\right)^{-\frac{\nu}{2}}\left|(\tau^{*})^{\frac{1}{2}}\theta_{i}^{*}\right|^{\nu-1}\sgn(\theta_{i}^{*})-\frac{1}{4}\left|\lambda^{\frac{1}{2}}\theta_{i}^{*}\right|^{\nu}\right]\nonumber\\
&-\frac{1}{4}\sum_{i=1}^{d}\mu^{2}_{\lambda,\nu}(\tau^{*},\theta_{i}^{*})\tau^{*}(\theta_{i}^{*})^{2}\nonumber\\
&-\sum_{i=1}^{d}\littleo\left(\mu_{\lambda,\nu}(\tau^{*},\theta_{i}^{*})\right)\left((\bm{x}_{\parallel}(i)-\theta_{i}^{*})(\tau^{*})^{\frac{1}{2}}-\frac{1}{2}(\tau^{*}+\Delta_{\lambda}(\theta_{i}^{*}))(\tau^{*})^{-\frac{1}{2}}\theta_{i}^{*}\right)^{2}\nonumber\\
\label{tau_det2}
\end{align}
Now, it is reasonable to claim that the sum $\sum_{i=1}^{d}\tau^{*}\left(\bm{x}_{\parallel}(i)-\theta_{i}^{*}\right)\times$ $\left(2\theta_{i}^{*}-\bm{x}_{\parallel}(i)\right)$ is either positive, or failing that, very small in absolute value compared to $\frac{1}{2}(n-d+2)$. 
By (\ref{def_mu}) we see that $\mu_{\lambda,1}(\tau^{*},\theta_{i}^{*})\equiv 0$ and $\mu_{\lambda,\nu}(\tau^{*},\theta_{i}^{*})<0$, when $\ 0<\nu<1$. 
By (\ref{mu_relation_SNR1}) we have $|\mu_{\lambda,\nu}(\tau^{*},\theta_{i})|<\zeta_{\mu_{\lambda,\nu}}<1,$ when  $0<\nu<2$, therefore to leading order it suffices to consider the terms linear in $\mu_{\lambda,\nu}(\tau^{*},\theta_{i})$ in the expression (\ref{tau_det1}). In the case  $0<\nu\leq 1$ we note that the 
$|\lambda^{\frac{1}{2}}\theta_{i}^{*}|$-terms contribute
positively to the right hand side of expression (\ref{tau_det1}) and because of the claim (\ref{mu_relation_SNR1}) the $\left(\frac{n}{d}\Omega(\tau^{*},\lambda^{*})\right)^{-\frac{\nu}{2}}\left|(\tau^{*})^{\frac{1}{2}}\theta_{i}^{*}\right|^{\nu-1}\sgn(\theta_{i}^{*})$-terms are bounded in absolute value by 
$\zeta|(\tau^{*})^{\frac{1}{2}}\theta_{i}^{*}|$ for some positive number $\zeta<1$, and since the $\theta_{i}$ are modelled as zero mean parameters, we may expect a cancellation effect to make the number value of the sum of $d$ such terms small compared to $d$. Alternatively: $\tau^{*}(\theta_{i}^{*})^{2}\geq 1$ and $\left(\frac{n}{d}\Omega(\lambda^{*},\tau^{*})\right)^{-\frac{\nu}{2}}\ll 1$. In the case $1<\nu<2$ we observe that
\begin{align}
&\|\bm{z}\|_{\nu}\leq K(d,\nu)\|\bm{z}\|_{2},\ \forall\bm{z}\in\mathbb{R}^{d},\ \nu\geq 1\label{lp_norm_ineq}\\
\intertext{where}
&K(d,\nu)\mathdef\sup_{\|\bm{z}\|_{2}=1}\frac{\|\bm{z}\|_{\nu}}{\|\bm{z}\|_{2}}=\max\left(1,d^{\frac{1}{\nu}-\frac{1}{2}}\right),\ \nu\geq 1.
\intertext{Then, using the estimator $\lambda^{*}$ for $\lambda$ given in (\ref{lambda_estimator1}) we may write}
&\sum_{i=1}^{d}\left|(\lambda^{*})^{\frac{1}{2}}\theta_{i}^{*}\right|^{\nu}=d^{\frac{\nu}{2}}\|\bm{\theta}^{*}\|_{2}^{-\nu}\|\bm{\theta}\|_{\nu}^{\nu}\leq d^{\frac{\nu}{2}}\|\bm{\theta}^{*}\|_{2}^{-\nu}\left(d^{\frac{1}{\nu}-\frac{1}{2}}\|\bm{\theta}\|_{2}\right)^{\nu}\nonumber\\
&\leq d,\ \forall\ \nu\in\left(1,2\right).\label{ineq_lambda1}
\intertext{Inserting our results from the discussion above in (\ref{tau_det2}) we may write}
&N(\lambda^{*},\nu,\gamma_{d})=\delta_{n}^{-2}\epsilon_{d}^{-2}\left\{\hat{\Phi}_{\hat{\tau},\hat{\tau}}(\bm{x},\hat{\tau}^{*},\hat{\bm{\theta}}^{*})-\sum_{i=1}^{d}\frac{\hat{\Phi}^{2}_{\hat{\tau},\hat{\theta}_{i}}(\bm{x},\hat{\tau}^{*},\hat{\bm{\theta}}^{*})}{\hat{\Phi}_{\hat{\theta}_{i},\hat{\theta}_{i}}(\bm{x},\hat{\tau}^{*},\hat{\bm{\theta}}^{*})}\right\}\nonumber\\
&\geq\frac{1}{2}(n-d+2)-\frac{1}{4}C_{\nu}\nu|\nu-1|d
+\sum_{i=1}^{d}\tau^{*}\left(\bm{x}_{\parallel}(i)-\theta_{i}^{*}\right)\left(2\theta_{i}^{*}-\bm{x}_{\parallel}(i)\right)\nonumber\\
&-\littleo\left(\sum_{i=1}^{d}\mu^{2}_{\lambda,\nu}(\tau^{*},\theta_{i}^{*})\tau^{*}(\theta_{i}^{*})^{2}\right),\text{ when }1<\nu< 2,\label{tau_det3}
\intertext{and}
&N(\lambda^{*},\nu,\gamma_{d})=\frac{1}{2}(n-d+2)+\sum_{i=1}^{d}\tau^{*}\left(\bm{x}_{\parallel}(i)-\theta_{i}^{*}\right)\left(2\theta_{i}^{*}-\bm{x}_{\parallel}(i)\right)\nonumber\\
&-\littleo\left(\sum_{i=1}^{d}\mu^{2}_{\lambda,\nu}(\tau^{*},\theta_{i}^{*})\tau^{*}(\theta_{i}^{*})^{2}\right),\text{ when }0<\nu\leq 1.\label{tau_det4}
\end{align}
We note that the result in (\ref{tau_det3}) also holds when using the estimator $\lambda^{*}$ given in (\ref{lambda_estimator2}).
Alternatively, we may just evaluate the expression
(\ref{tau_det2}) for a given dataset $\bm{x}$, estimator $\lambda^{*}$ and corresponding model as indexed by $\gamma_{d}$ to get the exact value of $N(\lambda^{*},\nu,\gamma_{d})$.
Now we consider the integral $I_{2}$ in (\ref{def_I2}). It is difficult to evaluate as we have no natural center about which to do a Taylor expansion.
Instead we will show that $I_{2}\ll I_{1}$, by an indirect approach. 
Since we will simply bring $I_{1}$ and $I_{2}$ onto forms that are easily compared, we will keep to the coordinates $\bm{\theta}$, $\tau$ for simplicity. 
We need to compare
\begin{align}
&I_{1}=\int_{\tau\in I_{\tau}}d\tau\ \int_{\bm{\theta}\in\mathbb{R}_{+}^{d}}d\bm{\theta}\ \exp\left(-\Phi(\bm{x},\tau,\bm{\theta})\right)
|\bm{F}(\bm{\theta},\tau)|^{\frac{1}{2}}\nonumber\\
&=\int_{\tau\in I_{\tau}}d\tau\ \int_{\bm{\theta}\in\mathbb{R}_{+}^{d}}d\bm{\theta}\ \prod_{i=1}^{d}g_{\tau}(x_{i}-\theta_{i})\pi_{\lambda}(\theta_{i})\label{I1_compare1}
\intertext{and}
&I_{2}=\int_{\tau\in I_{\tau}}d\tau\ \int_{\bm{\theta}\in\mathbb{R}^{d}\setminus\mathbb{R}_{+}^{d}}d\bm{\theta}\ \exp\left(-\Phi(\bm{x},\tau,\bm{\theta})\right)
|\bm{F}(\bm{\theta},\tau)|^{\frac{1}{2}}\nonumber\\
&=\int_{\tau\in I_{\tau}}d\tau\ \int_{\bm{\theta}\in\mathbb{R}^{d}\setminus\mathbb{R}_{+}^{d}}d\bm{\theta}\ \prod_{i=1}^{d}g_{\tau}(x_{i}-\theta_{i})\pi_{\lambda}(\theta_{i})\label{I2_compare1}
\intertext{where we have defined}
&g_{\tau}(x)\mathdef\frac{\tau^{1/2}}{\sqrt{2\pi}}\exp\left(-\frac{1}{2}\tau x^{2}\right).\nonumber
\intertext{Now consider the integral}
&Q_{2}(x,\tau)\mathdef\int_{-\infty}^{0}\pi_{\lambda}(\theta)g_{\tau}\left(x-\theta\right)\ d\theta\label{def_Q2}
\intertext{changing variables $u\mathdef\lambda^{\frac{1}{2}}\theta$, recalling the definition of SNR $\Omega(\lambda,\tau)$ in (\ref{def_SNR}), we get}
&Q_{2}(x,\tau)=\tau^{1/2}\int_{-\infty}^{0}\pi_{1}(u)g_{1}\left(\left(\frac{n}{d}\Omega(\lambda,\tau)\right)^{\frac{1}{2}}u-\tau^{\frac{1}{2}}x\right)\ du\nonumber\\
&\leq\tau^{1/2}\left(\sup_{u\in\mathbb{R}_{-}}\pi_{1}(u)\right)\left(\frac{n}{d}\Omega(\lambda,\tau)\right)^{-\frac{1}{2}}P_{G}\left(-\tau^{\frac{1}{2}}x\right).\label{Q1_bound}
\intertext{We continue with}
&Q_{1}(x,\tau)\mathdef\int_{0}^{\infty}\pi_{\lambda}(\theta)g_{\tau}\left(x-\theta\right)\ d\theta\label{i2_form1}\\
&=\tau^{1/2}\int_{0}^{\infty}\pi_{1}(u)g_{1}\left(\left(\frac{n}{d}\Omega(\lambda,\tau)\right)^{\frac{1}{2}}u-\tau^{\frac{1}{2}}x\right)\ du.
\intertext{We define}
&u_{0}(\tau^{\frac{1}{2}}x)\mathdef\tau^{\frac{1}{2}}x\left(\frac{n}{d}\Omega(\lambda,\tau)\right)^{-\frac{1}{2}}\label{def_u0}
\intertext{and we then write}
&Q_{1}(x,\tau)=\tau^{1/2}\int_{0}^{u_{0}}\pi_{1}(u)g_{1}\left(\left(\frac{n}{d}\Omega(\lambda,\tau)\right)^{\frac{1}{2}}\left(u-u_{0}\right)\right)\ du\nonumber\\
&+\tau^{1/2}\int_{u_{0}}^{\infty}\pi_{1}(u)g_{1}\left(\left(\frac{n}{d}\Omega(\lambda,\tau)\right)^{\frac{1}{2}}\left(u-u_{0}\right)\right)\ du.\label{i2_form2}
\intertext{Now we have}
&\int_{0}^{u_{0}}\pi_{1}(u)g_{1}\left(\left(\frac{n}{d}\Omega(\lambda,\tau)\right)^{\frac{1}{2}}\left(u-u_{0}\right)\right)\ du\nonumber\\
&\geq\inf_{t\in(0,u_{0})}\pi_{1}(t)\int_{0}^{u_{0}}g_{1}\left(\left(\frac{n}{d}\Omega(\lambda,\tau)\right)^{\frac{1}{2}}\left(u-u_{0}\right)\right)\ du\nonumber\\
&=\inf_{t\in(0,u_{0})}\pi_{1}(t)\frac{1}{2}\left(\frac{n}{d}\Omega\left(\lambda,\tau\right)\right)^{-\frac{1}{2}}\erf(\tau^{\frac{1}{2}}x).\label{i2_form3_part1}
\intertext{Taylor expanding $\pi_{1}(u)$ to first order about $u_{0}$ yields}
&\int_{u_{0}}^{\infty}\pi_{1}(u)g_{1}\left(\left(\frac{n}{d}\Omega(\lambda,\tau)\right)^{\frac{1}{2}}\left(u-u_{0}\right)\right)\ du\nonumber\\
&=\int_{u_{0}}^{\infty}du\ \left(\pi_{1}(u_{0})+\pi_{1}^{\prime}(\xi_{u_{0}})(u-u_{0})\right)g_{1}\left(\left(\frac{n}{d}\Omega(\lambda,\tau\right)^{\frac{1}{2}}\left(u-u_{0}\right)\right)\nonumber\\
&=\frac{1}{2}\pi_{1}(u_{0})\left(\frac{n}{d}\Omega(\lambda,\tau\right)^{-\frac{1}{2}}\nonumber\\
&+\int_{u_{0}}^{\infty}\pi_{1}^{\prime}(\xi_{u_{0}})(u-u_{0})g_{1}\left(\left(\frac{n}{d}\Omega(\lambda,\tau\right)^{\frac{1}{2}}\left(u-u_{0}\right)\right)\ du\label{i2_form3_part2a}
\intertext{where $\xi_{u_{0}}$ is some number such that $\xi_{u_{0}}\in(u_{0},u)$. Using the bound (\ref{tail_claim_prior2}) we may write}
&\sup_{0<u_{0}<\xi_{u_{0}}<u}\left|\pi_{1}^{\prime}(\xi_{u_{0}})\right|\leq\sup_{t\in(u_{0},\infty)}{\pi_{1}(t)}\cdot\left\{\begin{array}{ll}C_{\nu}\cdot\nu\cdot u_{0}^{\nu-1} & \text{ if }0<\nu\leq 1 \\
C_{\nu}\cdot\nu\cdot u^{\nu-1}& \text{ if }1<\nu<2.\end{array}\right.\label{prior_bound_on_some_interval}
\intertext{We may then in the case $0<\nu\leq 1$ write}
&\left|\int_{u_{0}}^{\infty}\pi_{1}^{\prime}(\xi_{u_{0}})(u-u_{0})g_{1}\left(\left(\frac{n}{d}\Omega(\lambda,\tau)\right)^{\frac{1}{2}}(u-u_{0})\right)\ du\right|\nonumber\\
&\leq C_{\nu}\cdot\nu\cdot\sup_{t\in(u_{0},\infty)}{\pi_{1}(t)}\cdot u_{0}^{\nu-1}\int_{u_{0}}^{\infty}(u-u_{0})g_{1}\left(\left(\frac{n}{d}\Omega(\lambda,\tau)\right)^{\frac{1}{2}}(u-u_{0})\right)\ du\nonumber\\
&= C_{\nu}\cdot\nu\cdot u_{0}^{\nu-1}\sup_{t\in(u_{0},\infty)}{\pi_{1}(t)}\cdot\left(\frac{n}{d}\Omega(\lambda,\tau)\right)^{-1}(2\pi)^{-\frac{1}{2}}\nonumber\\
&= \frac{C_{\nu}\cdot\nu}{(2\pi)^{\frac{1}{2}}}\cdot\sup_{t\in(u_{0},\infty)}{\pi_{1}(t)}\cdot\left(\tau^{\frac{1}{2}}x\right)^{\nu-1}\left(\frac{n}{d}\Omega(\lambda,\tau)\right)^{-\frac{\nu+1}{2}}.\label{i2_form3_part2b}
\intertext{In the case $1<\nu<2$ we have}
&\left|\int_{u_{0}}^{\infty}\pi_{1}^{\prime}(\xi_{u_{0}})(u-u_{0})g_{1}\left(\left(\frac{n}{d}\Omega(\lambda,\tau)\right)^{\frac{1}{2}}(u-u_{0})\right)\ du\right|\nonumber\\
&\leq C_{\nu}\cdot\nu\cdot\sup_{t\in(u_{0},\infty)}{\pi_{1}(t)}\int_{u_{0}}^{\infty}(u-u_{0})u^{\nu-1}g_{1}\left(\left(\frac{n}{d}\Omega(\lambda,\tau)\right)^{\frac{1}{2}}(u-u_{0})\right)\ du.\label{i2_form3_part2c}
%\intertext{Now, by differentiating the expression (\ref{i2_form3_part2c}) with respect to $\nu$ we see that it has a maximum for $\nu=1$, and so by the previous calculations we may write}
\intertext{Now, by maximizing the integrand in (\ref{i2_form3_part2c}) with respect to $\nu\in(1,2)$ for each of the cases $u_{0}\leq 1$ and $u_{0}>1$, 
we find that the expression (\ref{i2_form3_part2c}) may be bounded from above for all $u_{0}>0$ by}
&1<\nu<2\Rightarrow\left|\int_{u_{0}}^{\infty}du\ \pi_{1}^{\prime}(\xi_{u_{0}})(u-u_{0})g_{1}\left(\left(\frac{n}{d}\Omega(\lambda,\tau)\right)^{\frac{1}{2}}(u-u_{0})\right)\right|\nonumber\\
%&<\frac{C_{\nu}\cdot\nu}{(2\pi)^{\frac{1}{2}}}\cdot\sup_{t\in(u_{0},\infty)}\pi_{1}(t)\cdot\left(\frac{n}{d}\Omega(\lambda,\tau)\right)^{-1}.\label{i2_form3_part2d}
&<\frac{C_{\nu}\cdot\nu}{(2\pi)^{\frac{1}{2}}}\cdot\sup_{t\in(u_{0},\infty)}\pi_{1}(t)\cdot\left(\frac{n}{d}\Omega(\lambda,\tau)\right)^{-1}
\left(1+u_{0}+\frac{\sqrt{2\pi}}{2}\left(\frac{n}{d}\Omega(\lambda,\tau)\right)^{-1/2}\right).\label{i2_form3_part2d}
\intertext{Then by the estimates (\ref{i2_form3_part1}), (\ref{i2_form3_part2a}), (\ref{i2_form3_part2b}), (\ref{i2_form3_part2d}) we may bound $Q_{1}(x,\tau)$ from below as follows}
&Q_{1}(x,\tau)>\frac{1}{2}\tau^{1/2}\left(\frac{n}{d}\Omega(\lambda,\tau)\right)^{-\frac{1}{2}}\sup_{t\in(u_{0},\infty)}\pi_{1}(t)\left\{\frac{\pi_{1}(u_{0})}{\sup_{t\in(u_{0},\infty)}{\pi_{1}(t)}}\right.\nonumber\\
&\left.+\erf(\tau^{\frac{1}{2}}x)\frac{\inf_{t\in(0,u_{0})}{\pi_{1}(t)}}{\sup_{t\in(u_{0},\infty)}{\pi_{1}(t)}}-\frac{2C_{\nu}\nu}{(2\pi)^{\frac{1}{2}}}L_{\nu}(\tau^{\frac{1}{2}}x)\right\}\label{Q2_bound}
\intertext{where}
&L_{\nu}(\tau^{\frac{1}{2}}x)\mathdef\left\{\begin{array}{ll}\left(\tau^{\frac{1}{2}}x\right)^{\nu-1}\left(\frac{n}{d}\Omega(\lambda,\tau)\right)^{-\frac{\nu}{2}} & \text{ if }0<\nu\leq 1\\
\left(\frac{n}{d}\Omega(\lambda,\tau)\right)^{-\frac{1}{2}}\left(1+\tau^{\frac{1}{2}}x\left(\frac{n}{d}\Omega(\lambda,\tau)\right)^{-\frac{1}{2}}\right)& \text{ if }1<\nu<2.\end{array}\right.\label{def_L_nu}
\intertext{and $u_{0}(\tau^{\frac{1}{2}}x)=\left(\frac{n}{d}\Omega(\lambda,\tau)\right)^{-\frac{1}{2}}\tau^{\frac{1}{2}}x$ and we have implicitely made the assumption: $\pi_{1}(u_{0})/(\sup_{t\in\mathbb{R}}{\pi_{1}(t)})> 2C_{\nu}\nu(2\pi)^{-\frac{1}{2}}L_{\nu}(\tau^{\frac{1}{2}}x)$. We may then by (\ref{Q1_bound}), (\ref{def_u0}), (\ref{Q2_bound}), (\ref{def_L_nu}) write}
&\frac{Q_{2}(x,\tau)}{Q_{1}(x,\tau)}\leq\frac{2P_{G}\left(-\tau^{\frac{1}{2}}x\right)\sup_{t\in\mathbb{R}_{-}}{\pi_{1}(t)}}{\pi_{1}(u_{0})\left[1+\erf(\tau^{\frac{1}{2}}x)\frac{\inf_{t\in(0,u_{0})}{\pi_{1}(t)}}{\pi_{1}(u_{0})}-\frac{2C_{\nu}\nu}{(2\pi)^{\frac{1}{2}}}\frac{\sup_{t\in(u_{0},\infty)}{\pi_{1}(t)}}{\pi_{1}(u_{0})}L_{\nu}(\tau^{\frac{1}{2}}x)\right]}.\label{compare_Q1_Q2_form1}
\intertext{We note that if $\pi_{\lambda}(\theta)$ is a monotone decreasing function of $|\theta|$ we may write}
&\frac{Q_{2}(x,\tau)}{Q_{1}(x,\tau)}\leq\frac{2P_{G}\left(-\tau^{\frac{1}{2}}x\right)\pi_{1}(0)/\pi_{1}\left(\left[\frac{\tau x^{2}}{\frac{n}{d}\Omega(\lambda,\tau)}\right]^{\frac{1}{2}}\right)}{1+\erf(\tau^{\frac{1}{2}}x)-\frac{2C_{\nu}\nu}{(2\pi)^{\frac{1}{2}}}L_{\nu}(\tau^{\frac{1}{2}}x)}.\label{compare_Q1_Q2_form2}
\intertext{We may write}
&\frac{I_{2}}{I_{1}}=\frac{I_{1}+I_{2}-I_{1}}{I_{1}}=-1+\frac{I_{1}+I_{2}}{I_{1}}\nonumber\\
&=-1+\frac{\int_{\tau\in I_{\tau}}\prod_{i=1}^{d}\left(Q_{1}(\bm{x}_{\parallel}(i),\tau)+Q_{2}(\bm{x}_{\parallel}(i),\tau)\right)\ d\tau}{\int_{\tau\in I_{\tau}}\prod_{j=1}^{d}Q_{1}(\bm{x}_{\parallel}(j),\tau)\ d\tau}\nonumber
\intertext{using the integral mean value theorem we get}
&=-1+\frac{|I_{\tau}|\prod_{i=1}^{d}\left(Q_{1}(\bm{x}_{\parallel}(i),\tau_{1})+Q_{2}(\bm{x}_{\parallel}(i),\tau_{1})\right)}{|I_{\tau}|\prod_{j=1}^{d}Q_{1}(\bm{x}_{\parallel}(j),\tau_{2})},\text{ for some }\tau_{1},\tau_{2}\in I_{\tau}\nonumber\\
&=-1+\frac{\prod_{i=1}^{d}Q_{1}(\bm{x}_{\parallel}(i),\tau_{1})\left(1+\frac{Q_{2}(\bm{x}_{\parallel}(i),\tau_{1})}{Q_{1}(\bm{x}_{\parallel}(i),\tau_{1})}\right)}{\prod_{j=1}^{d}Q_{1}(\bm{x}_{\parallel}(j),\tau_{2})}.\label{approx_frac1}\\
\intertext{Now, if $|I_{\tau}|$ is chosen small enough, then $\tau_{1}\approx\tau_{2}$, and using the bounds on $Q_{1}(x,\tau)$ and $Q_{2}(x,\tau)$ calculated above, we may write (\ref{approx_frac1}) as}
%&\prod_{i=1}^{d}\left\{\frac{|I_{\tau}|Q_{2}\left(\bm{x}_{\parallel}(i),\tau_{2}\right)}{|I_{\tau}|Q_{1}\left(\bm{x}_{\parallel}(i),\tau_{1}\right)}\right\}\leq\nonumber\\
&\frac{I_{2}}{I_{1}}\approx-1+\prod_{i=1}^{d}\left(1+\frac{Q_{2}(\bm{x}_{\parallel}(i),\tau_{1})}{Q_{1}(\bm{x}_{\parallel}(i),\tau_{1})}\right)\label{approx_frac2}\\
&\leq-1+\prod_{i=1}^{d}\left\{1+\right.\nonumber\\
&\left.\frac{2P_{G}\left(-\tau_{1}^{\frac{1}{2}}\bm{x}_{\parallel}(i)\right)\sup_{t\in\mathbb{R}_{-}}{\pi_{1}(t)}/\pi_{1}\left(u_{0}(\tau_{1}^{1/2}\bm{x}_{\parallel}(i))\right)}{1+\erf(\tau_{1}^{1/2}\bm{x}_{\parallel}(i))\frac{\inf_{t\in\left(0,u_{0}\left(\tau_{1}^{1/2}\bm{x}_{\parallel}(i)\right)\right)}{\pi_{1}(t)}}{\pi_{1}\left(u_{0}(\tau_{1}^{1/2}\bm{x}_{\parallel}(i))\right)}
-\frac{2C_{\nu}\nu}{(2\pi)^{\frac{1}{2}}}L_{\nu}(\tau_{1}^{\frac{1}{2}}\bm{x}_{\parallel}(i))\frac{\sup_{t\in(u_{0},\infty)}{\pi_{1}(t)}}{\pi_{1}\left(u_{0}(\tau_{1}^{1/2}\bm{x}_{\parallel}(i))\right)}}
\right\}\label{I2_compare_I1}
\intertext{where we have by (\ref{I_tau_claim3}) that}
&\tau_{1},\tau_{2}\in I_{\tau}\subset\left(\tau^{*}\exp\left[-\left(\frac{\log^{2}{N(\lambda,\nu,\gamma_{d})}}{N(\lambda,\nu,\gamma_{d})}\right)^{\frac{1}{2}}\right],\right.\nonumber\\
&\left.\tau^{*}\exp\left[\left(\frac{\log^{2}{N(\lambda,\nu,\gamma_{d})}}{N(\lambda,\nu,\gamma_{d})}\right)^{\frac{1}{2}}\right]\right).\nonumber
%&\tau^{*}\exp\left(-\sqrt{\frac{\log^{2}{N(\lambda,\nu,\gamma_{d})}}{N(\lambda,\nu,\gamma_{d})}}\right)<\tau_{1},\tau_{2}<\tau^{*}\exp\left(\sqrt{\frac{\log^{2}{N(\lambda,\nu,\gamma_{d})}}{N(\lambda,\nu,\gamma_{d})}}\right)\nonumber
\intertext{with $N(\lambda,\nu,\gamma_{d})$ as given in (\ref{tau_det3}) and (\ref{tau_det4}).}\nonumber
\end{align}
%We therefore conclude that under the claims made in this proof, the contribution from $I_{2}$ to $I_{1}+I_{2}$ may be neglected 
%when $\bm{x}_{\parallel}\in\mathbb{R}^{d}_{+}$.

We did begin the proof under the assumption that $\hat{\bm{\theta}}^{*}\in\mathbb{R}^{d}_{+}$. But when considering the calculations leading to 
the expression in (\ref{def_U}), we see that if $\hat{\theta}_{j}^{*}<0$
for some $1\leq i\leq d$, then by changing the domain of integration from $\mathbb{R}_{+}$ to $\mathbb{R}_{-}$ on the $\hat{\theta}_{j}$-axis in 
(\ref{def_I1}) and correspondingly changing the domain of integration for $I_{2}$ in (\ref{def_I2}) so that the union of integration domains in $I_{1}$ and $I_{2}$ is
$\mathbb{R}^{d}$, what we get in formula (\ref{I1_final}) is simply that $\hat{\theta}_{j}$ changes to $-\hat{\theta}_{j}^{*}$ inside the 
$P_{G}(\cdot)$-expression. 
Thus, if we replace $\theta_{i}^{*}$ by $|\theta_{i}^{*}|$ inside the $P_{G}(\cdot)$-expression  in (\ref{I1_final}), 
and likewise replace  $\bm{x}_{\parallel}(i)$ by $|\bm{x}_{\parallel}(i)|$ inside the $P_{G}(\cdot)$-expressions  in (\ref{I2_compare_I1}),
we see that our proof of the formulas (\ref{I1_final}) and (\ref{I2_compare_I1}) is invariant of sign changes on $\bm{x}_{\parallel}(i)$ and $\theta_{i}^{*}$.

There remains one question that need to be answered before the proof can be said to be complete, that is the problem of estimating the parameter $\lambda$. Since $\lambda^{-1/2}$ is the second order moment of the prior density $\pi_{\lambda}(\bm{\theta})$,
we could simply define
\begin{align}
&\frac{1}{\lambda^{*}}\mathdef\frac{1}{d}\sum_{i=1}^{d}(\theta_{i}^{*})^{2}\label{lambda_estimator1}
\intertext{This would lead to}
&\Omega(\lambda^{*},\tau^{*})=\frac{d\frac{1}{\lambda^{*}}}{n\frac{1}{\tau^{*}}}=\frac{\frac{d}{d}\sum_{i=1}^{d}(\theta_{i}^{*})^{2}}{n\frac{1}{\tau^{*}}}
=\frac{1}{n}\sum_{i=1}^{d}\tau^{*}(\theta_{i}^{*})^{2}.\label{Omega_estimator1}
\intertext{Alternatively}
&\frac{1}{\lambda^{*}}\mathdef\left\{\frac{1}{d}\sum_{i=1}^{d}\bm{x}^{2}_{\parallel}(i)-\frac{1}{\tau^{*}}\right\}\label{lambda_estimator2}
\intertext{leading to}
&\Omega(\lambda^{*},\tau^{*})=\frac{d\frac{1}{\lambda^{*}}}{n\frac{1}{\tau^{*}}}=\frac{1}{n}\left\{\sum_{i=1}^{d}\tau^{*}\bm{x}^{2}_{\parallel}(i)-d\right\}.\label{Omega_estimator2}
\intertext{Another way to proceed which is more in line with the philosophy of the MDL-principle would be to define}
&\lambda^{*}\mathdef\text{arg min}_{\lambda>0}\left\{-\log{m_{\gamma_{d}}(\bm{x})}\right\}\label{lambda_estimator4}
\intertext{That is we select the value of $\lambda$ minimizing the codelength of our dataset $\bm{x}$ given the model $\gamma_{d}$. 
We will generally prefer this maximum likelihood form of the moment estimator $\lambda^{*}$ because of its codelength 
optimality and because it also simplifies computations. 
A special case of interest to us is $\pi_{\lambda}(\theta)$ belongs to the class of priors known as 
''Generalized Gaussian Distributions'' (GGD) which may be expressed on the form, \cite{Moulin-Liu:1999}}
&\pi_{\lambda}(\theta)\mathdef\frac{\nu\eta(\nu)}{2\Gamma(1/\nu)}\lambda^{\frac{1}{2}}
\exp\left(-\eta(\nu)^{\nu}\left|\lambda^{\frac{1}{2}}\theta\right|^{\nu}\right),\ \eta(\nu)\mathdef\left(\frac{\Gamma(3/\nu)}{\Gamma(1/\nu)}\right)^{\frac{1}{2}}\label{def_GGD}
\intertext{By (\ref{integrable_claim_prior})-(\ref{tail_claim_prior2}) and because $-\log\pi_{\lambda}(\theta)$ is taken to be a symmetric, nonnegative function of $\theta$ with a decay limit as stated in (\ref{tail_claim_prior1}), we see that the family of priors under consideration in this proof includes the family of GGD-distributions. 
In the special case of a GGD prior the Maximum Likelihood (ML) estimator $\lambda^{*}$ for $\lambda$ is}
&\frac{1}{[\lambda^{*}]_{v}}\mathdef\left(\frac{\nu\eta(\nu)^{\nu}}{d}\sum_{i=1}^{d}|[\theta^{*}_{i}]_{v}|^{\nu}\right)^{\frac{2}{\nu}}\label{lambda_estimator3}
\intertext{This leads to}
&\Omega(\lambda^{*},\tau^{*})=\frac{d\frac{1}{\lambda^{*}}}{n\frac{1}{\tau^{*}}}=\left(\frac{\nu\eta(\nu)^{\nu}}{n^{\frac{\nu}{2}}d^{1-\frac{\nu}{2}}}\sum_{i=1}^{d}|(\tau^{*})^{\frac{1}{2}}\theta^{*}_{i}|^{\nu}\right)^{\frac{2}{\nu}}.\label{Omega_estimator3}
\end{align}
The second question is the claim (\ref{mu_claim1}) which is
\begin{align}
&\left|\mu_{\lambda,\nu}(\tau^{*},\theta_{i}^{*})\right|\leq \zeta_{\mu_{\lambda,\nu}}<1,\ 1\leq i\leq d.\label{t1}
\intertext{Rewriting (\ref{def_mu}) yields}
&\left|\mu_{\lambda,\nu}(\tau^{*},\theta_{i}^{*})\right|=
\left|\frac{d}{n}\Omega^{-1}(\lambda,\tau^{*})|(\lambda)^{\frac{1}{2}}\theta_{i}^{*}|^{\nu-2}C_{\nu}\nu(\nu-1)\right|\nonumber\\
&= C_{\nu}\nu|\nu-1|\frac{\lambda}{\tau^{*}}\left|\frac{(\lambda)^{\frac{1}{2}}}{(\tau^{*})^{\frac{1}{2}}}(\tau^{*})^{\frac{1}{2}}\theta_{i}^{*}\right|^{\nu-2}\nonumber\\
&=C_{\nu}\nu|\nu-1|\left(\frac{n}{d}\Omega(\lambda,\tau^{*})\right)^{-\frac{\nu}{2}}\left|(\tau^{*})^{\frac{1}{2}}\theta_{i}^{*}\right|^{\nu-2},\ 0<\nu< 2\label{mu_relation_SNR1}
\intertext{The claim (\ref{t1}) may then be expressed as}
&\sup_{1\leq i\leq d}\left\{C_{\nu}\nu|\nu-1|\left(\frac{n}{d}\Omega(\lambda,\tau^{*})\right)^{-\frac{\nu}{2}}\left|(\tau^{*})^{\frac{1}{2}}\theta_{i}^{*}\right|^{\nu-2}\right\}\mathdef \zeta_{\mu_{\lambda,\nu}}<1,\label{mu_relation_SNR2}
\intertext{where $0<\nu<2$.}\nonumber
\end{align}
It is now clear by considering (\ref{mu_relation_SNR2})
that (\ref{t1}) will be satisfied for ''reasonable'' values on the SNR $\Omega(\lambda^{*},\tau^{*})$, the relative model size $\frac{d}{n}$, 
the tail parameter $0<\nu<2$ and the tail constant $C_{\nu}$ on the prior distribution $\pi_{\lambda}$.
The proof is now complete.

%\section{Unadressed questions}
%\begin{itemize}
%\item Elaborate on the size of $N(\lambda,\nu,\gamma_{d})$.
%\item Elaborate on the choice of $k_{\hat{\tau}}$ and its implications on the error terms.....
%\end{itemize}

%\section{Unresolved questions}
%\begin{itemize}
%\item Is it possible to formulate ''general'' conditions in terms of properties of the prior distribution and the data, under which the 
%claim $\prod_{i=1}^{d}P_{G}\left((\tau^{*})^{\frac{1}{2}}\theta_{i}^{*}\right)\approx 1$ may be said to be true?
%\end{itemize}
%\appendix
\chapter{The marginal normalization $C_{\gamma_{d}}$}
We must address the problem of calculating the normalizing constant $C_{\gamma_{d}}$ defined in (\ref{def_C_gamma_d}). Clearly, $C_{\gamma_{d}}$ will depend on our choice of the domain $Y\ni\bm{z}$ on which $m_{\gamma_{d}}(\bm{z})$ is normalized to be a density. We will take care in choosing this region $Y$ as it will possibly have significant influence on the model selection principle we will end up with. The given data set $\bm{x}$ must be contained in the region $Y$. The geometry of the region $Y$ is determined canonically by the model index vector $\gamma_{d}$ and the form of the estimators $\tau^{*}$ and $\lambda^{*}$, as will be demonstrated below. We will concentrate on the generic case of priors $\pi_{\lambda}$ defined in Theorem \ref{theorem1}. 

Given a data set $\bm{x}$, a model as indexed by $\gamma_{d}$ and assuming the conditions in Theorem \ref{theorem1}. We then need to calculate
\begin{align}
&C_{\gamma_{d}}=\frac{(2\pi)^{\frac{d+2}{2}}}{|I_{\tau}|\cdot|I_{\lambda}|}\int_{\bm{z}\in Y}d\bm{z}\ \frac{f(\bm{z}|\tau^{*},\bm{\theta}^{*})\pi(\bm{\theta}^{*}|\lambda^{*})}{|\bm{H}(\bm{z},\tau^{*},\bm{\theta}^{*})|^{\frac{1}{2}}|\Psi_{\lambda\lambda}(\bm{\theta}^{*},\lambda^{*})|^{\frac{1}{2}}}\times\nonumber\\
&\left\{\prod_{i=1}^{d}P_{G}\left((\tau^{*})^{\frac{1}{2}}|\theta_{i}^{*}|\left\{1+\littleo(\zeta)\right\}^{\frac{1}{2}}\right)\right\}\label{C_NMDL1}
\intertext{where $\bm{\theta}^{*}=\bm{\theta}(\bm{z}_{\parallel})$, $\tau^{*}=\tau^{*}(\bm{z})$ are the MAP-estimators defined in (\ref{def_invariant_estimators}) and 
$\lambda^{*}$ is the estimator for the parameter $\lambda$ defined in (\ref{def_invariant_lambda_estimator}). 
Using (\ref{def_Phi}) and exploiting the orthogonal 
decomposition $Y=Y_{\perp}\oplus Y_{\parallel}$ induced by the model $\gamma_{d}$, we may express the invariant MAP-estimator $\tau^{*}$ for the noise as}
&\frac{1}{\tau^{*}(\bm{z})}=\frac{1}{n-d+2}\left(\|\bm{z}_{\perp}\|_{2}^{2}+\|\bm{z}_{\parallel}-\bm{\theta}^{*}(\bm{z}_{\parallel})\|_{2}^{2}\right),\ \bm{z}_{\perp}\in Y_{\perp},\ \bm{z}_{\parallel}\in Y_{\parallel}.\label{invariant_noise_estimator}
\intertext{We evaluate the determinant of Hessian $\bm{H}(\bm{z},\tau^{*},\bm{\theta}^{*})$ of $\Phi(\bm{z},\tau,\bm{\theta})$ by means of (\ref{def_Phi}), (\ref{form_of_G}), (\ref{det_form_of_general_G}), (\ref{def_Delta}), (\ref{order_of_delta}) and we get}
&|\bm{H}(\bm{z},\tau^{*},\bm{\theta}^{*})|=\frac{n-d+2}{2}(\tau^{*})^{d-2}
\left\{\prod_{i=1}^{d}\left[1+\littleo\left(\mu_{\lambda,\nu}(\tau^{*},\theta_{i}^{*})\right)\right]\right\}\times\nonumber\\
&\left[1-\frac{2\tau^{*}}{n-d+2}\sum_{j=1}^{d}\frac{\left(\bm{z}_{\parallel}(j)-\theta_{j}^{*}\right)^{2}}{1+\littleo\left(\mu_{\lambda,\nu}(\tau^{*},\theta_{j}^{*})\right)
}\right]\label{explicit_det_Hessian}
\intertext{where}
&\mu_{\lambda,\nu}(\tau^{*},\theta_{j}^{*})\mathdef C_{\nu}\nu(\nu-1)\left(\frac{n}{d}\Omega(\lambda^{*},\tau^{*})\right)^{-\frac{\nu}{2}}\left|(\tau^{*})^{\frac{1}{2}}\theta_{j}^{*}\right|^{\nu-2}.\label{def_mu2}
\intertext{We note that the matrix $\bm{H}(\bm{z},\tau^{*},\bm{\theta}^{*})$ is singular for $\tau^{*}=0$. 
By (\ref{explicit_det_Hessian}), (\ref{invariant_noise_estimator}) and under the conditions in Theorem \ref{theorem1} we may write}
&|\bm{H}(\bm{z},\tau^{*},\bm{\theta}^{*})|=\frac{1}{2}(n-d+2)(\tau^{*})^{d-2}\exp\left(d\cdot\littleo(\zeta)\right)\times\nonumber\\
&\left[1-\frac{2\tau^{*}}{n-d+2}\frac{\|\bm{z}_{\parallel}-\bm{\theta}^{*}\|_{2}^{2}}{1+\littleo\left(\zeta\right)
}\right]\label{explicit_det_Hessian2}.
\intertext{We note that by expression (\ref{explicit_det_Hessian2}) the matrix $\bm{H}(\bm{z},\tau^{*},\bm{\theta}^{*})$ also becomes singular when $\tau^{*}\|\bm{z}_{\parallel}-\bm{\theta}^{*}\|_{2}^{2}$ increases from zero and becomes large enough. 
%Now, let $Y=Y_{\perp}\cup Y_{\parallel}$ be the splitting of $X$ into mutually disjoint sets induced by the decomposition 
%$\mathbb{R}^{n}=V_{\perp}\oplus V_{\parallel}$. 
Inserting (\ref{invariant_noise_estimator}) into (\ref{split_likelihood}) and exploiting the orthogonal decomposition $\bm{z}=\bm{z}_{\perp}+\bm{z}_{\parallel}$, we get}
&C_{\gamma_{d}}=\frac{(2\pi)^{\frac{d+2}{2}}}{|I_{\tau}|\cdot|I_{\lambda}|}\exp\left(-\frac{n-d+2}{2}\right)\int_{\bm{z}_{\parallel}\in Y_{\parallel},\ \bm{z}_{\perp}\in Y_{\perp}}d\bm{z}_{\parallel}\ d\bm{z}_{\perp}\ \left(\frac{\tau^{*}}{2\pi}\right)^{\frac{n}{2}}\times\nonumber\\
&\frac{\pi(\bm{\theta}^{*}|\lambda^{*})}{\left|\bm{H}(\bm{z},\tau^{*},\bm{\theta}^{*})\right|^{\frac{1}{2}}|\Psi_{\lambda\lambda}(\bm{\theta}^{*},\lambda^{*})|^{\frac{1}{2}}}
\left\{\prod_{i=1}^{d}P_{G}\left((\tau^{*})^{\frac{1}{2}}\left|\theta_{i}^{*}\right|\left\{1+\littleo(\zeta)\right\}^{\frac{1}{2}}\right)\right\}\nonumber\\
\intertext{Exploiting the spherical symmetry of $Y_{\perp}$ as induced by the form of the estimator $\tau^{*}$ in (\ref{invariant_noise_estimator}), we change to polar coordinates in $\bm{z}_{\perp}$, that is we set}
&R_{\perp}^{2}\mathdef\|\bm{z}_{\perp}\|_{2}^{2},\label{def_R_perp}\\
&S_{k}(r)\mathdef\frac{\pi^{\frac{k}{2}}k}{\Gamma(\frac{k}{2}+1)}r^{k-1},\label{def_S_k}
\intertext{where $S_{k}(r)$ is the surface area of a $k$-dimensional hyper sphere of radius $r$. We find it convenient to
use $r^{2}$ instead of $r$ as a integration variable, thus we make a change of variables $r\rightarrow r^{2}$ which gives}
&S_{k}(1)r^{k-1}\ dr\mapsto \frac{1}{2}S_{k}(1)(r^{2})^{\frac{k-2}{2}}\ dr^{2}.\label{r_mapsto_R_square}
\intertext{We may then write}
&C_{\gamma_{d}}=\frac{\exp\left(-\frac{n-d+2}{2}\right)}{2(2\pi)^{\frac{n-d-2}{2}}}\frac{S_{n-d}(1)}{|I_{\tau}|\cdot|I_{\lambda}|}\int_{\bm{z}_{\parallel}\in Y_{\parallel},\ R^{2}_{\perp}\in J}d\bm{z}_{\parallel}\ dR^{2}_{\perp}\ (\tau^{*})^{\frac{n}{2}}\times\nonumber\\
&\frac{\pi(\bm{\theta}^{*}|\lambda^{*})\left(R_{\perp}^{2}\right)^{\frac{n-d-2}{2}}}{\left|\bm{H}(\bm{z},\tau^{*},\bm{\theta}^{*})\right|^{\frac{1}{2}}|\Psi_{\lambda\lambda}(\bm{\theta}^{*},\lambda^{*})|^{\frac{1}{2}}}
\left\{\prod_{i=1}^{d}P_{G}\left((\tau^{*})^{\frac{1}{2}}\left|\theta_{i}^{*}\right|\left\{1+\littleo(\zeta)\right\}^{\frac{1}{2}}\right)\right\}.\label{C_form2}
\intertext{where $J\subset \mathbb{R}_{+}$ is an interval. We will find it convenient to change integration variables in the integral (\ref{C_form2}) from $(R_{\perp}^{2},\bm{z}_{\parallel}^{T})$
to $(\tau^{*},(\bm{\theta}^{*})^{T})$. We define}
&\bm{\beta}\mathdef(\tau,\bm{\theta})^{T},\ \ \bm{\beta}^{*}\mathdef(\tau^{*},\bm{\theta}^{*})^{T},\ \ \bm{y}\mathdef (R_{\perp}^{2},\bm{z}_{\parallel}^{T})^{T}.\label{def_beta_R_perp_y}
\intertext{Using (\ref{def_Phi}) we define the gradient vector $\bm{Q}$ as}
&\bm{Q}(\bm{y},\bm{\beta})\mathdef\left.\frac{\partial\Phi(\bm{z},\bm{\alpha})}{\partial\bm{\alpha}}\right|_{\bm{\alpha}=\bm{\beta}}=\bm{0}.\label{def_Q}
\intertext{By (\ref{def_Q}) the total differential of $\bm{Q}(\bm{y},\bm{\beta})$ along $\bm{\beta}=\bm{\beta}^{*}$ may be expressed formally as}
&\bm{0}=d\bm{Q}=\left.\frac{\partial\bm{Q}(\bm{y},\bm{\beta})}{\partial\bm{y}}\right|_{\bm{\beta}=\bm{\beta}^{*}}d\bm{y}
+\left.\frac{\partial\bm{Q}(\bm{y},\bm{\beta})}{\partial\bm{\beta}}\right|_{\bm{\beta}=\bm{\beta}^{*}}d\bm{\beta}^{*}\label{total_diff_Q}
\intertext{This yields the formal expression for Jacobian $\frac{\partial\bm{y}}{\partial\bm{\beta}^{*}}$ as}
&\frac{\partial\bm{y}}{\partial\bm{\beta}^{*}}=-\left(\left.\frac{\partial\bm{Q}(\bm{y},\bm{\beta})}{\partial\bm{y}}\right|_{\bm{\beta}=\bm{\beta}^{*}}\right)^{-1}\left.\frac{\partial\bm{Q}(\bm{y},\bm{\beta})}{\partial\bm{\beta}}\right|_{\bm{\beta}=\bm{\beta}^{*}}.\label{def_jacobian_z_to_theta}
\intertext{Now, we observe that}
&\left.\frac{\partial\bm{Q}(\bm{y},\bm{\beta})}{\partial\bm{\beta}}\right|_{\bm{\beta}=\bm{\beta}^{*}}=\bm{H}(\bm{z},\tau^{*},\bm{\theta}^{*})\label{dQ_dbeta}
\intertext{and}
&\left.\frac{\partial\bm{Q}(\bm{y},\bm{\beta})}{\partial\bm{y}}\right|_{\bm{\beta}=\bm{\beta}^{*}}=\left(\begin{array}{ccccc}
\frac{1}{2} & a_{1} & a_{2} &\cdots & a_{d}\\
0 & -\tau^{*} & 0 & \cdots & 0\\
0 & 0 & -\tau^{*}  & \cdots & 0\\
\vdots & \vdots & \vdots & \ddots & \vdots\\
0 & 0 & \cdots & 0 & -\tau^{*}\end{array}\right)\label{dQ_dy}
\intertext{where $a_{j}\mathdef-(\bm{x}_{\parallel}(j)-\theta_{j}^{*}), \ 1\leq j\leq d$. The Jacobi-determinant for the change of variables 
$\bm{y}\rightarrow\bm{\beta}^{*}$ then becomes}
&\left|\frac{\partial\bm{y}}{\partial\bm{\beta}^{*}}\right|=\left|\left.\left(\frac{\partial\bm{Q}(\bm{y},\bm{\beta})}{\partial\bm{y}}\right|_{\bm{\beta}=\bm{\beta}^{*}}\right)^{-1}
\left.\frac{\partial\bm{Q}(\bm{y},\bm{\beta})}{\partial\bm{\beta}}\right|_{\bm{\beta}=\bm{\beta}^{*}}\right|\nonumber\\
&=\left(\frac{1}{2}(\tau^{*})^{d}\right)^{-1}|\bm{H}(\bm{z},\tau^{*},\bm{\theta}^{*})|.\label{jacobian_z_to_beta}
\intertext{Define the inverse function $\bm{\theta}^{*}_{-1}(\bm{\alpha})$ of $\bm{\alpha}=\bm{\theta}^{*}(\bm{z}_{\parallel})$ by}
&\bm{\theta}^{*}_{-1}\left(\bm{\theta}^{*}(\bm{z}_{\parallel})\right)\mathdef\bm{z}_{\parallel}\label{def_estimator_inverse}
\intertext{assuming such an inverse exists ($\theta^{*}_{i}\neq 0,\ 1\leq i\leq d$). Using (\ref{invariant_noise_estimator}), (\ref{jacobian_z_to_beta}) and the relation $\bm{z}_{\parallel}(\bm{\theta}^{*})=\bm{\theta}^{*}_{-1}(\bm{\theta}^{*})$ , then (\ref{C_form2}) becomes}
&C_{\gamma_{d}}=\frac{\exp\left(-\frac{n-d+2}{2}\right)}{2(2\pi)^{\frac{n-d-2}{2}}}S_{n-d}(1)(n-d+2)^{\frac{n-d-2}{2}}2|I_{\tau}|^{-1}|I_{\lambda}|^{-1}\times\nonumber\\
&\int_{\bm{\theta}^{*}\in\Theta^{*},\ \tau^{*}\in J_{\tau}^{*}}d\bm{\theta}^{*}\ d\tau^{*}\ (\tau^{*})^{\frac{-d+2}{2}}
\pi(\bm{\theta}^{*}|\lambda^{*})\left|\bm{H}(\bm{z},\tau^{*},\bm{\theta}^{*})\right|^{\frac{1}{2}}|\Psi_{\lambda\lambda}(\bm{\theta}^{*},\lambda^{*})|^{-\frac{1}{2}}\times\nonumber\\
&\left(1-\frac{\tau^{*}\|\bm{z}_{\parallel}(\bm{\theta}^{*})-\bm{\theta}^{*}\|_{2}^{2}}{n-d+2}\right)^{\frac{n-d-2}{2}}\times\nonumber\\
&\left\{\prod_{i=1}^{d}P_{G}\left((\tau^{*})^{\frac{1}{2}}\left|\theta_{i}^{*}\right|\left\{1+\littleo(\zeta)\right\}^{\frac{1}{2}}\right)\right\}\label{C_form3}
\intertext{where $\Theta^{*}\subset\mathbb{R}^{d}\setminus\left\{\bm{0}\right\}$ is some set still to be chosen subject to the constraints of containing the MAP estimate $\bm{\theta}^{*}$ and minimizing the total codelength expression (\ref{def_nice_codelength_part2}) while $\lambda^{*}$ is constant on the boundary $\partial\Theta^{*}$ of $\Theta^{*}$.  Also, $J_{\tau}^{*}\subset\mathbb{R}_{+}$ is a bounded interval. Inserting (\ref{explicit_det_Hessian2}) into (\ref{C_form3}) yields}
&C_{\gamma_{d}}=\frac{\exp\left(-\frac{n-d+2}{2}\right)}{\sqrt{2}(2\pi)^{\frac{n-d-2}{2}}}\frac{S_{n-d}(1)}{|I_{\tau}|\cdot|I_{\lambda}|}
(n-d+2)^{\frac{n-d-1}{2}}\exp\left(\frac{d}{2}\littleo(\zeta)\right)\times\nonumber\\
&\int_{\bm{\theta}^{*}\in\Theta^{*},\ \tau^{*}\in J_{\tau}^{*}}d\bm{\theta}^{*}\ d\tau^{*}\
\frac{\pi(\bm{\theta}^{*}|\lambda^{*})}{|\Psi_{\lambda\lambda}(\bm{\theta}^{*},\lambda^{*})|^{\frac{1}{2}}}
\left(1-\frac{\tau^{*}\|\bm{z}_{\parallel}(\bm{\theta}^{*})-\bm{\theta}^{*}\|_{2}^{2}}{n-d+2}\right)^{\frac{n-d-2}{2}}\times\nonumber\\
&\left(1-\frac{2}{1+\littleo(\zeta)}\frac{\tau^{*}\|\bm{z}_{\parallel}(\bm{\theta}^{*})-\bm{\theta}^{*}\|_{2}^{2}}{n-d+2}\right)\times\nonumber\\
&\left\{\prod_{i=1}^{d}P_{G}\left((\tau^{*})^{\frac{1}{2}}\left|\theta_{i}^{*}\right|\left\{1+\littleo(\zeta)\right\}^{\frac{1}{2}}\right)\right\}.\label{C_form4}
\intertext{We will need bounds on $\|\bm{z}_{\parallel}-\bm{\theta}^{*}\|_{2}^{2}$. Using the notation
from Theorem \ref{theorem1} we have}
&\pi(\theta|\lambda)=C\cdot\lambda^{\frac{1}{2}}\exp\left(-f(\lambda^{\frac{1}{2}}\theta)\right)\nonumber
\intertext{for some constant $C>0$. The MAP-estimator $\theta^{*}$ is given by} 
&\theta^{*}=\text{arg min}_{\ \theta\in\mathbb{R}}\left[\frac{1}{2}\tau(x-\theta)^{2}+f(\lambda^{\frac{1}{2}}\theta)\right]\label{def_MAP_estimator}
\intertext{which yields the solution $\theta^{*}$ expressed by}
&x-\theta^{*}=\frac{1}{\tau}\left.
\frac{d}{d\theta}f(\lambda^{\frac{1}{2}}\theta)\right|_{\theta=\theta^{*}}.\label{MAP_estimator}
\intertext{Using the bound on $f^{\prime}$ stated in (\ref{C3}) in Theorem \ref{theorem1} together with the expression (\ref{MAP_estimator}), we get}
&0\leq\|\bm{z}_{\parallel}-\bm{\theta}^{*}\|_{2}^{2}\leq \frac{\lambda}{\tau^{2}}C_{\nu}^{2}\nu^{2}\sum_{i=1}^{d}\left|\lambda^{\frac{1}{2}}\theta_{i}^{*}\right|^{2\nu-2}.\label{MAP_estimator_bounds}
\intertext{To bound the righthand side of (\ref{MAP_estimator_bounds}) from above we will make use of the claim $\tau^{*}(\theta_{i}^{*})^{2}\geq 1,\ \forall\ i\in\gamma_{d},\ \forall\ \tau\in I_{\tau}$ in Theorem \ref{theorem1} together with the norm inequality relation for $\ell^{p}$ norms on $\mathbb{R}^{d}$}
&\|\bm{x}\|_{p}\leq K(d,p)\|\bm{x}\|_{2},\ \bm{x}\in\mathbb{R}^{d},\ 1\leq p<\infty\label{lp_norm_ineq2}
\intertext{where}
& K(d,p)\mathdef\sup_{\|\bm{x}\|_{2}=1}\frac{\|\bm{x}\|_{p}}{\|\bm{x}\|_{2}}=\max\left(1,d^{\frac{1}{p}-\frac{1}{2}}\right).\label{def_lp_norm_constant}
\intertext{First we consider the case $0<\nu\leq 1$. Recalling the definition on the SNR (\ref{def_SNR}) we write}
&\|\bm{z}_{\parallel}-\bm{\theta}^{*}\|_{2}^{2}\leq\frac{\lambda}{\tau^{2}}C_{\nu}^{2}\nu^{2}\sum_{i=1}^{d}\left|\frac{\lambda^{\frac{1}{2}}}{\tau^{\frac{1}{2}}}\tau^{\frac{1}{2}}\theta_{i}^{*}\right|^{2\nu-2}=\frac{1}{\tau}C_{\nu}^{2}\nu^{2}\left(\frac{\lambda}{\tau}\right)^{\nu}\sum_{i=1}^{d}\left|\tau^{\frac{1}{2}}\theta_{i}^{*}\right|^{2\nu-2}\nonumber\\
&=\frac{1}{\tau}C_{\nu}^{2}\nu^{2}\left(\frac{n}{d}\Omega(\lambda,\tau)\right)^{-\nu}\sum_{i=1}^{d}\left|\tau^{\frac{1}{2}}\theta_{i}^{*}\right|^{2\nu-2}\nonumber\\
&\leq\frac{1}{\tau}C_{\nu}^{2}\nu^{2}\left(\frac{n}{d}\Omega(\lambda,\tau)\right)^{-\nu}d,\ \forall\ \nu\in\left(0,1\right]\label{norm_ineq1}
\intertext{where in the last inequality in (\ref{norm_ineq1}) we used that $\tau(\theta_{i}^{*})^{2}\geq 1$. In the case $1<\nu<2$ we may write}
&\|\bm{z}_{\parallel}-\bm{\theta}^{*}\|_{2}^{2}\leq\frac{\lambda}{\tau^{2}}C_{\nu}^{2}\nu^{2}\sum_{i=1}^{d}\left|\lambda^{\frac{1}{2}}\theta_{i}^{*}\right|^{\nu}\left|\frac{\lambda^{\frac{1}{2}}}{\tau^{\frac{1}{2}}}\tau^{\frac{1}{2}}\theta_{i}^{*}\right|^{\nu-2}\nonumber\\
&=\frac{1}{\tau}C_{\nu}^{2}\nu^{2}\left(\frac{n}{d}\Omega(\lambda,\tau)\right)^{-\frac{\nu}{2}}\sum_{i=1}^{d}\left|\lambda^{\frac{1}{2}}\theta_{i}^{*}\right|^{\nu}\left|\tau^{\frac{1}{2}}\theta_{i}^{*}\right|^{\nu-2}\nonumber\\
&\leq\frac{1}{\tau}C_{\nu}^{2}\nu^{2}\left(\frac{n}{d}\Omega(\lambda,\tau)\right)^{-\frac{\nu}{2}}\sum_{i=1}^{d}\left|\lambda^{\frac{1}{2}}\theta_{i}^{*}\right|^{\nu}\nonumber\\
&\leq\frac{1}{\tau}C_{\nu}^{2}\nu^{2}\left(\frac{n}{d}\Omega(\lambda,\tau)\right)^{-\frac{\nu}{2}}\left(K(d,\nu)\|\lambda^{\frac{1}{2}}\bm{\theta}^{*}\|_{2}\right)^{\nu}\nonumber\\
&=\frac{1}{\tau}C_{\nu}^{2}\nu^{2}\left(\frac{n}{d}\Omega(\lambda,\tau)\right)^{-\frac{\nu}{2}}d^{1-\nu/2}\left(\frac{d^{\frac{1}{2}}}{\|\bm{\theta}^{*}\|_{2}}\|\bm{\theta}^{*}\|_{2}\right)^{\nu}\nonumber\\
&=\frac{1}{\tau}C_{\nu}^{2}\nu^{2}\left(\frac{n}{d}\Omega(\lambda,\tau)\right)^{-\frac{\nu}{2}}d,\ \forall\ \nu\in\left(1,2\right)\label{norm_ineq2}
\intertext{where we have made use of the expression (\ref{lambda_estimator1}) for the size of $\lambda$. We will have to choose the regions $\Theta^{*}$ and $J_{\tau}^{*}$ of integration subject to the claim that the Theorem \ref{theorem1} is valid. Thus, we will have to ensure the Hessian $\bm{H}(\bm{z},\tau^{*},\bm{\theta}^{*})$ is non-singular over the region of integration. Using (\ref{norm_ineq1}), (\ref{norm_ineq2}) we may now write}
&\tau^{*}\|\bm{z}_{\parallel}(\bm{\theta}^{*})-\bm{\theta}^{*}\|_{2}^{2}\leq C_{\nu}^{2}\nu^{2}d\left(\frac{n}{d}\Omega(\lambda^{*},\tau^{*})\right)^{-h(\nu)},\nonumber\\
&\forall\ \bm{\theta^{*}}\in \Theta^{*},\ \forall\ \tau^{*}\in I_{\tau}\subset\mathbb{R}_{+}\label{tau_theta_claim1}
\intertext{where}
&h(\nu)\mathdef\left\{\begin{array}{ll}\nu &\text{ if } 0<\nu\leq 1\\
\nu/2 &\text{ if } 1<\nu<2.\end{array}\right.\label{def_h}
\intertext{By combining (\ref{explicit_det_Hessian2}) and (\ref{tau_theta_claim1}) we get} 
&\det\bm{H}(\bm{z},\tau^{*},\bm{\theta}^{*})>\frac{1}{2}(n-d+2)(\tau^{*})^{d-2}\exp\left(-d\zeta\right)\times\nonumber\\
&\left(1-\frac{d}{n-d+2}\frac{2C_{\nu}^{2}\nu^{2}\left(\frac{n}{d}\Omega(\lambda^{*},\tau^{*})\right)^{-h(\nu)}}{1-\zeta}\right) \label{lower_bound_det_Hessian}
\intertext{which will always be a positive number if}
&\frac{d}{n-d+2}\cdot\frac{2C_{\nu}^{2}\nu^{2}\left(\frac{n}{d}\Omega(\lambda^{*},\tau^{*})\right)^{-h(\nu)}}{1-\zeta}<1,\nonumber\\
&\text{and } 0<\zeta<1, \text{ and }\tau^{*}>0.\label{det_H_claim}
\intertext{We observe that the integral (\ref{C_form4}) diverges in $\tau^{*}$ at infinity. The description length as given by $-\log\left(m_{\gamma_{d}}(\bm{x})/C_{\gamma_{d}}\right)$ 
decreases with decreasing $C_{\gamma_{d}}$. The expression (\ref{def_mu2}) tells us that $\mu_{\lambda,\nu}(\tau^{*},\theta_{j}^{*})\rightarrow\infty$ as $\tau^{*}\rightarrow 0$.  Because of the claim (\ref{C4}) in Theorem \ref{theorem1} the left end of $J_{\tau}^{*}$ must not be ''too near'' zero, unless $\mu_{\lambda,\nu}(\tau^{*},\theta_{i}^{*}) = 0$ which is the case for priors flat in $\bm{\theta}$. However, by equation (\ref{invariant_noise_estimator}) 
we see that $\tau^{*}(\bm{z})\rightarrow\infty$ as $\|\bm{z}_{\perp}\|_{2}\rightarrow 0$ 
and $\|\bm{z}_{\parallel}-\bm{\theta}^{*}\|_{2}\rightarrow 0$ and $\tau^{*}(\bm{z})$ is bounded from below by a positive number when 
$\bm{z}_{\perp}\rightarrow\bm{x}_{\perp}$ and $\bm{z}_{\parallel}\rightarrow\bm{x}_{\parallel}$. 
%Therefore we will define}
%&J_{\tau}^{*}\mathdef \left(\tau^{*}(\bm{x}),T\right)\label{def_I_star}
%\intertext{where $T\gg\tau^{*}(\bm{x})$ is some large number. 
First we discuss the case $\|\bm{z}_{\parallel}-\bm{\theta}^{*}(\bm{z}_{\parallel})\|_{2}^{2}\equiv 0$. This means that the estimator $\bm{\theta}^{*}$ is the ML-estimator $\bm{\theta}^{*}(\bm{z}_{\parallel})=\bm{z}_{\parallel}$ corresponding to the choice of a prior distribution $\pi_{\lambda}(\bm{\theta})$ which is uniform (flat) in $\bm{\theta}$ and is centered in the origin. This is the case discussed in \cite{Rissanen:1998a}, \cite{Rissanen:2001}. In this case we have $\zeta\equiv 0$ and because this distribution is infinitely differentiable at the origin, the term $\prod_{i=1}^{d}P_{G}\left((\tau^{*})^{\frac{1}{2}}\theta_{i}^{*}\right)$ in (\ref{C_form4}) may be replaced by $1$. We then have}
&C_{\gamma_{d}}=\frac{\exp\left(-\frac{n-d+2}{2}\right)}{\sqrt{2}(2\pi)^{\frac{n-d-2}{2}}}S_{n-d}(1)(n-d+2)^{\frac{n-d-1}{2}}|I_{\tau}|^{-1}|I_{\lambda}|^{-1}\times\nonumber\\
&\int_{\bm{\theta}^{*}\in\Theta^{*},\ \tau^{*}\in J_{\tau}^{*}}d\bm{\theta}^{*}\ d\tau^{*}\
\frac{\pi(\bm{\theta}^{*}|\lambda^{*})}{|\Psi_{\lambda\lambda}(\bm{\theta}^{*},\lambda^{*})|^{\frac{1}{2}}}.\label{C_form5}
\intertext{We continue with the case of priors non-flat in $\bm{\theta}$ and flat (constant) in $\tau$. 
By (\ref{tau_theta_claim1}) we have the following bounds} 
&1\geq\left(1-\frac{\tau^{*}\|\bm{z}_{\parallel}-\bm{\theta}^{*}\|_{2}^{2}}{n-d+2}\right)^{\frac{n-d-2}{2}}\left(1-\frac{2}{1+\littleo(\zeta)}\frac{\tau^{*}\|\bm{z}_{\parallel}-\bm{\theta}^{*}\|_{2}^{2}}{n-d+2}\right)\nonumber\\
&\geq\left(1-\frac{d}{n-d+2}C_{\nu}^{2}\nu^{2}\left(\frac{n}{d}\Omega(\lambda^{*},\tau^{*})\right)^{-h(\nu)}\right)^{\frac{n-d-2}{2}}\times\nonumber\\
&\left(1-\frac{2d}{n-d+2}\frac{C_{\nu}^{2}\nu^{2}\left(\frac{n}{d}\Omega(\lambda^{*},\tau^{*})\right)^{-h(\nu)}}{1-\zeta}\right).\label{bound_problem_terms}
\intertext{Using the integral version of the mean value theorem on the $P_{G}(\cdot)$-part of the integrand we may state the following bounds}
&\frac{\exp\left(-\frac{n-d+2}{2}\right)}{\sqrt{2}(2\pi)^{\frac{n-d-2}{2}}}\frac{S_{n-d}(1)}{|I_{\lambda}||I_{\tau}|}(n-d+2)^{\frac{n-d-1}{2}}\exp\left(\frac{d}{2}\littleo(\zeta)\right)\times\nonumber\\
&\left(1-\frac{d}{n-d+2}C_{\nu}^{2}\nu^{2}\left(\frac{n}{d}\Omega(\lambda^{*},\tau^{*})\right)^{-h(\nu)}\right)^{\frac{n-d-2}{2}}\times\nonumber\\
&\left(1-\frac{2d}{n-d+2}\frac{C_{\nu}^{2}\nu^{2}\left(\frac{n}{d}\Omega(\lambda^{*},\tau^{*})\right)^{-h(\nu)}}{1-\zeta}\right)\times\nonumber\\
&\left\{\prod_{i=1}^{d}P_{G}\left((\xi)^{\frac{1}{2}}\left|\alpha_{i}\right|\left\{1+\littleo(\zeta)\right\}^{\frac{1}{2}}\right)\right\}
\int_{\bm{\theta}^{*}\in\Theta^{*},\tau^{*}\in J_{\tau}^{*}}d\bm{\theta}^{*}\ d\tau^{*}\ 
\frac{\pi(\bm{\theta}^{*}|\lambda^{*})}{|\Psi_{\lambda\lambda}(\bm{\theta}^{*},\lambda^{*})|^{\frac{1}{2}}}\nonumber\\
&\leq C_{\gamma_{d}}\nonumber\\
&\leq\frac{\exp\left(-\frac{n-d+2}{2}\right)}{\sqrt{2}(2\pi)^{\frac{n-d-2}{2}}}\frac{S_{n-d}(1)}{|I_{\lambda}||I_{\tau}|}(n-d+2)^{\frac{n-d-1}{2}}\exp\left(\frac{d}{2}\littleo(\zeta)\right)
\times\nonumber\\
&\left\{\prod_{i=1}^{d}P_{G}\left((\xi)^{\frac{1}{2}}\left|\alpha_{i}\right|\left\{1+\littleo(\zeta)\right\}^{\frac{1}{2}}\right)\right\}
\int_{\bm{\theta}^{*}\in\Theta^{*},\tau^{*}\in J_{\tau}^{*}}d\bm{\theta}^{*}\ d\tau^{*}\
\frac{\pi(\bm{\theta}^{*}|\lambda^{*})}{|\Psi_{\lambda\lambda}(\bm{\theta}^{*},\lambda^{*})|^{\frac{1}{2}}}\label{C_form7}
\intertext{for some $\xi\in J_{\tau}^{*}$ and some $\bm{\alpha}\in\Theta^{*}$. Now, applying the Stirling approximation \cite{Stegun:1960} to the $\Gamma$-function in (\ref{def_S_k}), we may write}
&\Gamma\left(\frac{n-d}{2}+1\right)=\Gamma\left(\frac{n-d+2}{2}\right)\nonumber\\
&=(2\pi)^{\frac{1}{2}}\left(\frac{n-d+2}{2}\right)^{\frac{n-d+1}{2}}\exp\left(-\frac{n-d+2}{2}\right)\left(1+\littleo\left(\frac{1}{n-d+2}\right)\right)\label{stirling_approximation}
\intertext{and using the bounds: $\frac{1}{2}=P_{G}(0)\leq P_{G}(x)\leq 1,\ \forall\ x\geq 0$ ,
%(I'm still uncertain about the sharpness of the claim $\tau^{*}(\theta_{i}^{*})^{2}\geq 1$), 
we may now bound $C_{\gamma_{d}}$ as follows:}
&\frac{n-d}{n-d+2}\exp\left(\frac{d}{2}\littleo(\zeta)\right)P_{G}^{d}\left((\tau^{*})^{1/2}\inf_{1\leq i\leq d}|\theta_{i}^{*}|\right)\frac{\sqrt{2\pi}}{|I_{\lambda}||I_{\tau}|}\times\nonumber\\
&\left(1-\frac{d}{n-d+2}C_{\nu}^{2}\nu^{2}\left(\frac{n}{d}\Omega(\lambda^{*},\tau^{*})\right)^{-h(\nu)}\right)^{\frac{n-d-2}{2}}\times\nonumber\\
&\left(1-\frac{2d}{n-d+2}\frac{C_{\nu}^{2}\nu^{2}\left(\frac{n}{d}\Omega(\lambda^{*},\tau^{*})\right)^{-h(\nu)}}{1-\zeta}\right)\times\nonumber\\
&\int_{\bm{\theta}^{*}\in\Theta^{*},\tau^{*}\in J_{\tau}^{*}}d\bm{\theta}^{*}\ d\tau^{*}\ 
\frac{\pi(\bm{\theta}^{*}|\lambda^{*})}{|\Psi_{\lambda\lambda}(\bm{\theta}^{*},\lambda^{*})|^{\frac{1}{2}}}\nonumber\\
&\leq C_{\gamma_{d}}\nonumber\\
&\leq\frac{n-d}{n-d+2}\exp\left(\frac{d}{2}\littleo(\zeta)\right)\frac{\sqrt{2\pi}}{|I_{\lambda}||I_{\tau}|}\int_{\bm{\theta}^{*}\in\Theta^{*},\tau^{*}\in J_{\tau}^{*}}d\bm{\theta}^{*}\ d\tau^{*}
\frac{\pi(\bm{\theta}^{*}|\lambda^{*})}{|\Psi_{\lambda\lambda}(\bm{\theta}^{*},\lambda^{*})|^{\frac{1}{2}}}.\label{C_bounds}
\end{align}\\
The result in Propostion \ref{Marginal_Renormalization_Constant} follows.
\chapter{The partial derivatives of $\hat{\Phi}(\bm{x},\hat{\tau},\hat{\bm{\theta}})$ up to order 3}
\noindent Define 
\begin{align}
&\tau^{*}\mathdef\psi(\hat{\tau}^{*}),\ \text{ and } \bm{\theta}^{*}\mathdef\bm{\phi}(\hat{\bm{\theta}}^{*},\hat{\tau}).\nonumber\\
\intertext{Using the independency of the parameters $\theta_{i},\ 1\leq i\leq d$ and the functional relations $\tau=\psi(\hat{\tau})$ and 
$\theta_{i}=\phi(\hat{\theta}_{i},\hat{\tau})$ as given in (\ref{def_phi_psi}) we may write}
&\hat{\Phi}_{\hat{\tau}}(\bm{x},\hat{\tau}^{*},\hat{\bm{\theta}}^{*})=\Phi_{\tau}(\bm{x},\tau^{*},\bm{\theta}^{*})
\left.\frac{\partial\psi(\hat{\tau})}{\partial\hat{\tau}}\right|_{\hat{\tau}=\hat{\tau}^{*},\hat{\bm{\theta}}=\hat{\bm{\theta}}^{*}}\nonumber\\
&+\sum_{i=1}^{d}\Phi_{\theta_{i}}(\bm{x},\tau^{*},\bm{\theta}^{*})\left.\frac{\partial\phi(\hat{\theta}_{i},\hat{\tau})}{\partial\hat{\tau}}\right|_{\hat{\tau}=\hat{\tau}^{*},\hat{\bm{\theta}}=\hat{\bm{\theta}}^{*}}.\label{hat_Phi_hat_tau}\\ 
\nonumber\\
&\hat{\Phi}_{\hat{\theta}_{k}}(\bm{x},\hat{\tau}^{*},\hat{\bm{\theta}}^{*})=\Phi_{\theta_{k}}(\bm{x},\tau^{*},\bm{\theta}^{*})\left.\frac{\partial\phi(\hat{\theta}_{k},\hat{\tau})}{\partial\hat{\theta}_{k}}\right|_{\hat{\tau}=\hat{\tau}^{*},\hat{\bm{\theta}}=\hat{\bm{\theta}}^{*}}.\label{hat_Phi_hat_theta}\\
\nonumber\\
&\hat{\Phi}_{\hat{\tau},\hat{\tau}}(\bm{x},\hat{\tau}^{*},\hat{\bm{\theta}}^{*})=\Phi_{\tau,\tau}(\bm{x},\tau^{*},\bm{\theta}^{*})
\left(\left.\frac{\partial\psi(\hat{\tau})}{\partial\hat{\tau}}\right|_{\hat{\tau}=\hat{\tau}^{*},\hat{\bm{\theta}}=\hat{\bm{\theta}}^{*}}\right)^{2}\nonumber\\
&+\Phi_{\tau}(\bm{x},\tau^{*},\bm{\theta}^{*})\left.\frac{\partial^{2}\psi(\hat{\tau})}{\partial\hat{\tau}^{2}}\right|_{\hat{\tau}=\hat{\tau}^{*},\hat{\bm{\theta}}=\hat{\bm{\theta}}^{*}}\nonumber\\
&+2\sum_{i=1}^{d}\Phi_{\theta_{i},\tau}(\bm{x},\tau^{*},\bm{\theta}^{*})\left.\frac{\partial\phi(\hat{\theta}_{i},\hat{\tau})}{\partial\hat{\tau}}\right|_{\hat{\tau}=\hat{\tau}^{*},\hat{\theta}_{i}=\hat{\theta}_{i}^{*}}\left.\frac{\partial\psi(\hat{\tau})}{\partial\hat{\tau}}\right|_{\hat{\tau}=\hat{\tau}^{*},\hat{\bm{\theta}}=\hat{\bm{\theta}}^{*}}\nonumber\\
&+\sum_{i=1}^{d}\Phi_{\theta_{i}}(\bm{x},\tau^{*},\bm{\theta}^{*})\left.\frac{\partial^{2}\phi(\hat{\theta}_{i},\hat{\tau})}{\partial\hat{\tau}^{2}}\right|_{\hat{\tau}=\hat{\tau}^{*},\hat{\bm{\theta}}=\hat{\bm{\theta}}^{*}}\nonumber\\
&+\sum_{i=1}^{d}\Phi_{\theta_{i},\theta_{i}}(\bm{x},\tau^{*},\bm{\theta}^{*})\left(\left.\frac{\partial\phi(\hat{\theta}_{i},\hat{\tau})}{\partial\hat{\tau}}\right|_{\hat{\tau}=\hat{\tau}^{*},\hat{\bm{\theta}}=\hat{\bm{\theta}}^{*}}\right)^{2}.\label{hat_Phi_hat_tau_hat_tau}\\
\nonumber\\
&\hat{\Phi}_{\hat{\theta}_{k},\hat{\theta}_{k}}(\bm{x},\hat{\tau}^{*},\hat{\bm{\theta}}^{*})=\Phi_{\theta_{k},\theta_{k}}(\bm{x},\tau^{*},\bm{\theta}^{*})\left(\left.\frac{\partial\phi(\hat{\theta}_{k},\hat{\tau})}{\partial\hat{\theta}_{k}}\right|_{\hat{\tau}=\hat{\tau}^{*},\hat{\bm{\theta}}=\hat{\bm{\theta}}^{*}}\right)^{2}.\label{hat_Phi_hat_theta_hat_theta}\\
\nonumber\\
&\hat{\Phi}_{\hat{\theta}_{k},\hat{\tau}}(\bm{x},\hat{\tau}^{*},\hat{\bm{\theta}}^{*})=\Phi_{\theta_{k}}(\bm{x},\tau^{*},\bm{\theta}^{*})\left.\frac{\partial^{2}\phi(\hat{\theta}_{k},\hat{\tau})}{\partial\hat{\tau}\partial\hat{\theta_{k}}}\right|_{\hat{\tau}=\hat{\tau}^{*},\hat{\bm{\theta}}=\hat{\bm{\theta}}^{*}}\nonumber\\
&+\Phi_{\theta_{k},\tau}(\bm{x},\tau^{*},\bm{\theta}^{*})\left.\frac{\partial\phi(\hat{\theta}_{k},\hat{\tau})}{\partial\hat{\theta}_{k}}\right|_{\hat{\tau}=\hat{\tau}^{*},\hat{\bm{\theta}}=\hat{\bm{\theta}}^{*}}\left.\frac{\partial\psi(\hat{\tau})}{\partial\hat{\tau}}\right|_{\hat{\tau}=\hat{\tau}^{*},\hat{\bm{\theta}}=\hat{\bm{\theta}}^{*}}\nonumber\\
&+\Phi_{\theta_{k},\theta_{k}}(\bm{x},\tau^{*},\bm{\theta}^{*})\left.\frac{\partial\phi(\hat{\theta}_{k},\hat{\tau})}{\partial\hat{\theta_{k}}}\right|_{\hat{\tau}=\hat{\tau}^{*},\hat{\bm{\theta}}=\hat{\bm{\theta}}^{*}}\left.\frac{\partial\phi(\hat{\theta}_{k},\hat{\tau})}{\partial\hat{\tau}}\right|_{\hat{\tau}=\hat{\tau}^{*},\hat{\bm{\theta}}=\hat{\bm{\theta}}^{*}}.\label{hat_Phi_hat_theta_hat_tau}\\
\nonumber\\
&\hat{\Phi}_{\hat{\theta}_{k},\hat{\theta}_{k},\hat{\tau}}(\bm{x},\hat{\tau}^{*},\hat{\bm{\theta}}^{*})=2\Phi_{\theta_{k},\theta_{k}}(\bm{x},\tau^{*},\bm{\theta}^{*})\left.\frac{\partial\phi(\hat{\theta}_{k},\hat{\tau})}{\partial\hat{\theta}_{k}}\right|_{\hat{\tau}=\hat{\tau}^{*},\hat{\bm{\theta}}=\hat{\bm{\theta}}^{*}}\left.\frac{\partial^{2}\phi(\hat{\theta}_{k},\hat{\tau})}{\partial\hat{\tau}\partial\hat{\theta}_{k}}\right|_{\hat{\tau}=\hat{\tau}^{*},\hat{\bm{\theta}}=\hat{\bm{\theta}}^{*}}\nonumber\\
&+\Phi_{\theta_{k},\theta_{k},\tau}(\bm{x},\tau^{*},\bm{\theta}^{*})\left(\left.\frac{\partial\phi(\hat{\theta}_{k},\hat{\tau})}{\partial\hat{\theta}_{k}}\right|_{\hat{\tau}=\hat{\tau}^{*},\hat{\bm{\theta}}=\hat{\bm{\theta}}^{*}}\right)^{2}\left.\frac{\partial\psi(\hat{\tau})}{\partial\hat{\tau}}\right|_{\hat{\tau}=\hat{\tau}^{*},\hat{\bm{\theta}}=\hat{\bm{\theta}}^{*}}\nonumber\\
&+\Phi_{\theta_{k},\theta_{k},\theta_{k}}(\bm{x},\tau^{*},\bm{\theta}^{*})\left(\left.\frac{\partial\phi(\hat{\theta}_{k},\hat{\tau})}{\partial\hat{\theta}_{k}}\right|_{\hat{\tau}=\hat{\tau}^{*},\hat{\bm{\theta}}=\hat{\bm{\theta}}^{*}}\right)^{2}\left.\frac{\partial\phi(\hat{\theta}_{k},\hat{\tau})}{\partial\hat{\tau}}\right|_{\hat{\tau}=\hat{\tau}^{*},\hat{\bm{\theta}}=\hat{\bm{\theta}}^{*}}.\label{hat_Phi_hat_theta_hat_theta_hat_tau}\\
\nonumber\\
&\hat{\Phi}_{\hat{\tau},\hat{\tau},\hat{\theta}_{k}}(\bm{x},\hat{\tau}^{*},\hat{\bm{\theta}}^{*})=\Phi_{\tau,\tau,\theta_{k}}(\bm{x},\tau^{*},\bm{\theta}^{*})
\left(\left.\frac{\partial\psi(\hat{\tau})}{\partial\hat{\tau}}\right|_{\hat{\tau}=\hat{\tau}^{*},\hat{\bm{\theta}}=\hat{\bm{\theta}}^{*}}\right)^{2}
\left.\frac{\partial\phi(\hat{\theta}_{k},\hat{\tau})}{\partial\hat{\theta}_{k}}\right|_{\hat{\tau}=\hat{\tau}^{*},\hat{\bm{\theta}}=\hat{\bm{\theta}}^{*}}\nonumber\\
&+\Phi_{\tau,\theta_{k}}(\bm{x},\tau^{*},\bm{\theta}^{*})\left.\frac{\partial^{2}\psi(\hat{\tau})}{\partial\hat{\tau}^{2}}\right|_{\hat{\tau}=\hat{\tau}^{*},\hat{\bm{\theta}}=\hat{\bm{\theta}}^{*}}\left.\frac{\partial\phi(\hat{\theta}_{k},\hat{\tau})}{\partial\hat{\theta}_{k}}\right|_{\hat{\tau}=\hat{\tau}^{*},\hat{\bm{\theta}}=\hat{\bm{\theta}}^{*}}\nonumber\\
&+2\Phi_{\theta_{k},\tau}(\bm{x},\tau^{*},\bm{\theta}^{*})\left.\frac{\partial^{2}\phi(\hat{\theta}_{k},\hat{\tau})}{\partial\hat{\theta}_{k}\partial\hat{\tau}}\right|_{\hat{\tau}=\hat{\tau}^{*},\hat{\theta}_{i}=\hat{\theta}_{i}^{*}}\left.\frac{\partial\psi(\hat{\tau})}{\partial\hat{\tau}}\right|_{\hat{\tau}=\hat{\tau}^{*},\hat{\bm{\theta}}=\hat{\bm{\theta}}^{*}}\nonumber\\
&+2\Phi_{\theta_{k},\theta_{k},\tau}(\bm{x},\tau^{*},\bm{\theta}^{*})\left.\frac{\partial\phi(\hat{\theta}_{k},\hat{\tau})}{\partial\hat{\tau}}\right|_{\hat{\tau}=\hat{\tau}^{*},\hat{\bm{\theta}}=\hat{\bm{\theta}}^{*}}\left.\frac{\partial\phi(\hat{\theta}_{k},\hat{\tau})}{\partial\hat{\theta}_{k}}\right|_{\hat{\tau}=\hat{\tau}^{*},\hat{\bm{\theta}}=\hat{\bm{\theta}}^{*}}\left.\frac{\partial\psi(\hat{\tau})}{\partial\hat{\tau}}\right|_{\hat{\tau}=\hat{\tau}^{*},\hat{\bm{\theta}}=\hat{\bm{\theta}}^{*}}\nonumber\\
&+\Phi_{\theta_{k}}(\bm{x},\tau^{*},\bm{\theta}^{*})\left.\frac{\partial^{3}\phi(\hat{\theta}_{k},\hat{\tau})}{\partial\hat{\theta}_{k}\partial\hat{\tau}^{2}}\right|_{\hat{\tau}=\hat{\tau}^{*},\hat{\bm{\theta}}=\hat{\bm{\theta}}^{*}}\nonumber\\
&+\Phi_{\theta_{k},\theta_{k}}(\bm{x},\tau^{*},\bm{\theta}^{*})\left.\frac{\partial^{2}\phi(\hat{\theta}_{k},\hat{\tau})}{\partial\hat{\tau}^{2}}\right|_{\hat{\tau}=\hat{\tau}^{*},\hat{\bm{\theta}}=\hat{\bm{\theta}}^{*}}\left.\frac{\partial\phi(\hat{\theta}_{k},\hat{\tau})}{\partial\hat{\theta}_{k}}\right|_{\hat{\tau}=\hat{\tau}^{*},\hat{\bm{\theta}}=\hat{\bm{\theta}}^{*}}\nonumber\\
&+2\Phi_{\theta_{k},\theta_{k}}(\bm{x},\tau^{*},\bm{\theta}^{*})\left.\frac{\partial\phi(\hat{\theta}_{k},\hat{\tau})}{\partial\hat{\tau}}\right|_{\hat{\tau}=\hat{\tau}^{*},\hat{\bm{\theta}}=\hat{\bm{\theta}}^{*}}\left.\frac{\partial^{2}\phi(\hat{\theta}_{k},\hat{\tau})}{\partial\hat{\theta}_{k}\partial\hat{\tau}}\right|_{\hat{\tau}=\hat{\tau}^{*},\hat{\bm{\theta}}=\hat{\bm{\theta}}^{*}}\nonumber\\
&+\Phi_{\theta_{k},\theta_{k},\theta_{k}}(\bm{x},\tau^{*},\bm{\theta}^{*})\left(\left.\frac{\partial\phi(\hat{\theta}_{k},\hat{\tau})}{\partial\hat{\tau}}\right|_{\hat{\tau}=\hat{\tau}^{*},\hat{\bm{\theta}}=\hat{\bm{\theta}}^{*}}\right)^{2}\left.\frac{\partial\phi(\hat{\theta}_{k},\hat{\tau})}{\partial\hat{\theta}_{k}}\right|_{\hat{\tau}=\hat{\tau}^{*},\hat{\bm{\theta}}=\hat{\bm{\theta}}^{*}}\label{hat_Phi_hat_tau_hat_tau_hat_theta}\\
\nonumber\\
&\hat{\Phi}_{\hat{\theta}_{k},\hat{\theta}_{k},\hat{\theta}_{k}}(\bm{x},\hat{\tau}^{*},\hat{\bm{\theta}}^{*})=\Phi_{\theta_{k},\theta_{k},\theta_{k}}(\bm{x},\tau^{*},\bm{\theta}^{*})\left(\left.\frac{\partial\phi(\hat{\theta}_{k},\hat{\tau})}{\partial\hat{\theta}_{k}}\right|_{\hat{\tau}=\hat{\tau}^{*},\hat{\bm{\theta}}=\hat{\bm{\theta}}^{*}}\right)^{3}\label{hat_Phi_hat_theta_hat_theta_hat_theta}\\
\nonumber\\
&\hat{\Phi}_{\hat{\tau},\hat{\tau},\hat{\tau}}(\bm{x},\hat{\tau}^{*},\hat{\bm{\theta}}^{*})=2\Phi_{\tau,\tau}(\bm{x},\tau^{*},\bm{\theta}^{*})\left.\frac{\partial\psi(\hat{\tau})}{\partial\hat{\tau}}\right|_{\hat{\tau}=\hat{\tau}^{*},\hat{\bm{\theta}}=\hat{\bm{\theta}}^{*}}\left.\frac{\partial^{2}\psi(\hat{\tau})}{\partial\hat{\tau}^{2}}\right|_{\hat{\tau}=\hat{\tau}^{*},\hat{\bm{\theta}}=\hat{\bm{\theta}}^{*}}\nonumber\\
&+\Phi_{\tau,\tau,\tau}(\bm{x},\tau^{*},\bm{\theta}^{*})
\left(\left.\frac{\partial\psi(\hat{\tau})}{\partial\hat{\tau}}\right|_{\hat{\tau}=\hat{\tau}^{*},\hat{\bm{\theta}}=\hat{\bm{\theta}}^{*}}\right)^{3}\nonumber\\
&+\sum_{i=1}^{d}\Phi_{\tau,\tau,\theta_{i}}(\bm{x},\tau^{*},\bm{\theta}^{*})
\left(\left.\frac{\partial\psi(\hat{\tau})}{\partial\hat{\tau}}\right|_{\hat{\tau}=\hat{\tau}^{*},\hat{\bm{\theta}}=\hat{\bm{\theta}}^{*}}\right)^{2}
\left.\frac{\partial\phi(\hat{\theta}_{i},\hat{\tau})}{\partial\hat{\tau}}\right|_{\hat{\tau}=\hat{\tau}^{*},\hat{\bm{\theta}}=\hat{\bm{\theta}}^{*}}\nonumber\\
&+\Phi_{\tau}(\bm{x},\tau^{*},\bm{\theta}^{*})\left.\frac{\partial^{3}\psi(\hat{\tau})}{\partial\hat{\tau}^{3}}\right|_{\hat{\tau}=\hat{\tau}^{*},\hat{\bm{\theta}}=\hat{\bm{\theta}}^{*}}\nonumber\\
&+\Phi_{\tau,\tau}(\bm{x},\tau^{*},\bm{\theta}^{*})\left.\frac{\partial^{2}\psi(\hat{\tau})}{\partial\hat{\tau}^{2}}\right|_{\hat{\tau}=\hat{\tau}^{*},\hat{\bm{\theta}}=\hat{\bm{\theta}}^{*}}\left.\frac{\partial\psi(\hat{\tau})}{\partial\hat{\tau}}\right|_{\hat{\tau}=\hat{\tau}^{*},\hat{\bm{\theta}}=\hat{\bm{\theta}}^{*}}\nonumber\\
&+\sum_{i=1}^{d}\Phi_{\tau,\theta_{i}}(\bm{x},\tau^{*},\bm{\theta}^{*})\left.\frac{\partial^{2}\psi(\hat{\tau})}{\partial\hat{\tau}^{2}}\right|_{\hat{\tau}=\hat{\tau}^{*},\hat{\bm{\theta}}=\hat{\bm{\theta}}^{*}}\left.\frac{\partial\phi(\hat{\theta}_{i},\hat{\tau})}{\partial\hat{\tau}}\right|_{\hat{\tau}=\hat{\tau}^{*},\hat{\bm{\theta}}=\hat{\bm{\theta}}^{*}}\nonumber\\
&+2\sum_{i=1}^{d}\Phi_{\theta_{i},\tau}(\bm{x},\tau^{*},\bm{\theta}^{*})\left.\frac{\partial^{2}\phi(\hat{\theta}_{i},\hat{\tau})}{\partial\hat{\tau}^{2}}\right|_{\hat{\tau}=\hat{\tau}^{*},\hat{\theta}_{i}=\hat{\theta}_{i}^{*}}\left.\frac{\partial\psi(\hat{\tau})}{\partial\hat{\tau}}\right|_{\hat{\tau}=\hat{\tau}^{*},\hat{\bm{\theta}}=\hat{\bm{\theta}}^{*}}\nonumber\\
&+2\sum_{i=1}^{d}\Phi_{\theta_{i},\tau}(\bm{x},\tau^{*},\bm{\theta}^{*})\left.\frac{\partial\phi(\hat{\theta}_{i},\hat{\tau})}{\partial\hat{\tau}}\right|_{\hat{\tau}=\hat{\tau}^{*},\hat{\theta}_{i}=\hat{\theta}_{i}^{*}}\left.\frac{\partial^{2}\psi(\hat{\tau})}{\partial\hat{\tau}^{2}}\right|_{\hat{\tau}=\hat{\tau}^{*},\hat{\bm{\theta}}=\hat{\bm{\theta}}^{*}}\nonumber\\
&+2\sum_{i=1}^{d}\Phi_{\theta_{i},\tau,\tau}(\bm{x},\tau^{*},\bm{\theta}^{*})\left.\frac{\partial\phi(\hat{\theta}_{i},\hat{\tau})}{\partial\hat{\tau}}\right|_{\hat{\tau}=\hat{\tau}^{*},\hat{\theta}_{i}=\hat{\theta}_{i}^{*}}\left(\left.\frac{\partial\psi(\hat{\tau})}{\partial\hat{\tau}}\right|_{\hat{\tau}=\hat{\tau}^{*},\hat{\bm{\theta}}=\hat{\bm{\theta}}^{*}}\right)^{2}\nonumber\\
&+2\sum_{i=1}^{d}\Phi_{\theta_{i},\theta_{i},\tau}(\bm{x},\tau^{*},\bm{\theta}^{*})\left(\left.\frac{\partial\phi(\hat{\theta}_{i},\hat{\tau})}{\partial\hat{\tau}}\right|_{\hat{\tau}=\hat{\tau}^{*},\hat{\theta}_{i}=\hat{\theta}_{i}^{*}}\right)^{2}\left.\frac{\partial\psi(\hat{\tau})}{\partial\hat{\tau}}\right|_{\hat{\tau}=\hat{\tau}^{*},\hat{\bm{\theta}}=\hat{\bm{\theta}}^{*}}\nonumber\\
&+\sum_{i=1}^{d}\Phi_{\theta_{i}}(\bm{x},\tau^{*},\bm{\theta}^{*})\left.\frac{\partial^{3}\phi(\hat{\theta}_{i},\hat{\tau})}{\partial\hat{\tau}^{3}}\right|_{\hat{\tau}=\hat{\tau}^{*},\hat{\bm{\theta}}=\hat{\bm{\theta}}^{*}}\nonumber\\
&+\sum_{i=1}^{d}\Phi_{\theta_{i},\tau}(\bm{x},\tau^{*},\bm{\theta}^{*})\left.\frac{\partial^{2}\phi(\hat{\theta}_{i},\hat{\tau})}{\partial\hat{\tau}^{2}}\right|_{\hat{\tau}=\hat{\tau}^{*},\hat{\bm{\theta}}=\hat{\bm{\theta}}^{*}}\left.\frac{\partial\psi(\hat{\tau})}{\partial\hat{\tau}}\right|_{\hat{\tau}=\hat{\tau}^{*},\hat{\bm{\theta}}=\hat{\bm{\theta}}^{*}}\nonumber\\
&+\sum_{i=1}^{d}\Phi_{\theta_{i},\theta_{i}}(\bm{x},\tau^{*},\bm{\theta}^{*})\left.\frac{\partial^{2}\phi(\hat{\theta}_{i},\hat{\tau})}{\partial\hat{\tau}^{2}}\right|_{\hat{\tau}=\hat{\tau}^{*},\hat{\bm{\theta}}=\hat{\bm{\theta}}^{*}}\left.\frac{\partial\phi(\hat{\theta}_{i},\hat{\tau})}{\partial\hat{\tau}}\right|_{\hat{\tau}=\hat{\tau}^{*},\hat{\bm{\theta}}=\hat{\bm{\theta}}^{*}}\nonumber\\
&+2\sum_{i=1}^{d}\Phi_{\theta_{i},\theta_{i}}(\bm{x},\tau^{*},\bm{\theta}^{*})\left.\frac{\partial\phi(\hat{\theta}_{i},\hat{\tau})}{\partial\hat{\tau}}\right|_{\hat{\tau}=\hat{\tau}^{*},\hat{\bm{\theta}}=\hat{\bm{\theta}}^{*}}\left.\frac{\partial^{2}\phi(\hat{\theta}_{i},\hat{\tau})}{\partial\hat{\tau}^{2}}\right|_{\hat{\tau}=\hat{\tau}^{*},\hat{\bm{\theta}}=\hat{\bm{\theta}}^{*}}\nonumber\\
&+\sum_{i=1}^{d}\Phi_{\theta_{i},\theta_{i},\tau}(\bm{x},\tau^{*},\bm{\theta}^{*})\left(\left.\frac{\partial\phi(\hat{\theta}_{i},\hat{\tau})}{\partial\hat{\tau}}\right|_{\hat{\tau}=\hat{\tau}^{*},\hat{\bm{\theta}}=\hat{\bm{\theta}}^{*}}\right)^{2}\left.\frac{\partial\psi(\hat{\tau})}{\partial\hat{\tau}}\right|_{\hat{\tau}=\hat{\tau}^{*},\hat{\bm{\theta}}=\hat{\bm{\theta}}^{*}}\nonumber\\
&+\sum_{i=1}^{d}\Phi_{\theta_{i},\theta_{i},\theta_{i}}(\bm{x},\tau^{*},\bm{\theta}^{*})\left(\left.\frac{\partial\phi(\hat{\theta}_{i},\hat{\tau})}{\partial\hat{\tau}}\right|_{\hat{\tau}=\hat{\tau}^{*},\hat{\bm{\theta}}=\hat{\bm{\theta}}^{*}}\right)^{3}.\label{hat_Phi_hat_tau_hat_tau_hat_tau}\\
\intertext{Now we need to compute all nonzero partial derivatives up to third order of the parameter mappings $\psi(\hat{\tau})$ and $\phi(\theta_{i},\hat{\tau})$
to get the desired order estimates of the coefficients of $\hat{T}^{*}_{3}$. Recall the definitions of $\psi(\hat{\tau})$ and $\phi(\hat{\theta_{i}},\hat{\tau})$
in (\ref{def_phi_psi}) and let the dimensionless numbers $\alpha$, $\delta_{n}$ and $\epsilon_{d}$ be defined as}
&\alpha^{-1/2}\mathdef\left(\frac{2}{n}\right)^{\frac{1}{2}}\left(\frac{\bar{\bar{\tau}}}{\bar{\tau}}\right)^{\frac{d}{2}}
\text{ and } \epsilon_{d}\mathdef\left(\frac{\bar{\bar{\tau}}}{\bar{\tau}}\right)^{\frac{d}{2}}\text{ and }\delta_{n}\mathdef\left(\frac{2}{n}\right)^{\frac{1}{2}}\label{def_epsilon_and_alpha_and_s}
\intertext{where $\bar{\tau}$ and $\bar{\bar{\tau}}$ are dimensionless positive real numbers. We claim $1\leq d<n$ and $0<\bar{\bar{\tau}}$ and $0<\bar{\tau}$ and 
$0<\delta_{n}\leq 1$. We may then write}
%COMMENTS: r was switched to \epsilon_{d}
&\frac{\partial\psi(\hat{\tau})}{\partial\hat{\tau}}=\alpha^{-\frac{1}{2}}\psi(\hat{\tau})=\delta_{n}\epsilon_{d}\psi(\hat{\tau})=\delta_{n}\epsilon_{d}\tau.\label{a1}\\
&\frac{\partial^{2}\psi(\hat{\tau})}{\partial\hat{\tau}^{2}}=\alpha^{-1}\psi(\hat{\tau})=\delta_{n}^{2}\epsilon_{d}^{2}\psi(\hat{\tau})
=\delta_{n}^{2}\epsilon_{d}^{2}\tau.\label{a2}\\
&\frac{\partial^{3}\psi(\hat{\tau})}{\partial\hat{\tau}^{3}}=\alpha^{-\frac{3}{2}}\psi(\hat{\tau})=\delta_{n}^{3}\epsilon_{d}^{3}\psi(\hat{\tau})
=\delta_{n}^{3}\epsilon_{d}^{3}\tau.\label{a3}\\
&\frac{\partial\phi(\hat{\theta}_{k},\hat{\tau})}{\partial\hat{\theta}_{k}}=\bar{\tau}^{\frac{1}{2}}\psi^{-\frac{1}{2}}(\hat{\tau})=\bar{\tau}^{\frac{1}{2}}\tau^{-\frac{1}{2}}.\label{c1}\\
&\frac{\partial\phi(\hat{\theta}_{k},\hat{\tau})}{\partial\hat{\tau}}=-\frac{1}{2}\bar{\tau}^{\frac{1}{2}}\alpha^{-\frac{1}{2}}\hat{\theta}_{k}\psi^{-\frac{1}{2}}(\hat{\tau})
=-\frac{1}{2}\bar{\tau}^{\frac{1}{2}}\delta_{n}\epsilon_{d}\hat{\theta}_{k}\psi^{-\frac{1}{2}}(\hat{\tau})\nonumber\\
&=-\frac{1}{2}\delta_{n}\epsilon_{d}\theta_{k}.\label{c2}\\
&\frac{\partial^{2}\phi(\hat{\theta}_{k},\hat{\tau})}{\partial\hat{\theta}_{k}\partial\hat{\tau}}=-\frac{1}{2}\bar{\tau}^{\frac{1}{2}}\alpha^{-\frac{1}{2}}\psi^{-\frac{1}{2}}(\hat{\tau})=-\frac{1}{2}\bar{\tau}^{\frac{1}{2}}\delta_{n}\epsilon_{d}\psi^{-\frac{1}{2}}(\hat{\tau})\nonumber\\
&=-\frac{1}{2}\bar{\tau}^{\frac{1}{2}}\delta_{n}\epsilon_{d}\tau^{-\frac{1}{2}}.\label{c3}\\
&\frac{\partial^{2}\phi(\hat{\theta}_{k},\hat{\tau})}{\partial\hat{\tau}^{2}}=\frac{1}{4}\bar{\tau}^{\frac{1}{2}}\alpha^{-1}\hat{\theta}_{k}\psi^{-\frac{1}{2}}(\hat{\tau})=\frac{1}{4}\bar{\tau}^{\frac{1}{2}}\delta_{n}^{2}\epsilon_{d}^{2}\hat{\theta}_{k}\psi^{-\frac{1}{2}}(\hat{\tau})= \frac{1}{4}\delta_{n}^{2}\epsilon_{d}^{2}\theta_{k}.\label{c4}\\
&\frac{\partial^{3}\phi(\hat{\theta}_{k},\hat{\tau})}{\partial\hat{\theta}_{k}\partial\hat{\tau}^{2}}=\frac{1}{4}\bar{\tau}^{\frac{1}{2}}\alpha^{-1}\psi^{-\frac{1}{2}}(\hat{\tau})=\frac{1}{4}\bar{\tau}^{\frac{1}{2}}\delta_{n}^{2}\epsilon_{d}^{2}\psi^{-\frac{1}{2}}(\hat{\tau})=\frac{1}{4}\bar{\tau}^{\frac{1}{2}}\delta_{n}^{2}\epsilon_{d}^{2}\tau^{-\frac{1}{2}}.\label{c5}\\
&\frac{\partial^{3}\phi(\hat{\theta}_{k},\hat{\tau})}{\partial\hat{\tau}^{3}}=-\frac{1}{8}\bar{\tau}^{\frac{1}{2}}\alpha^{-\frac{3}{2}}\hat{\theta}_{k}\psi^{-\frac{1}{2}}(\hat{\tau})=-\frac{1}{8}\bar{\tau}^{\frac{1}{2}}\delta_{n}^{3}\epsilon_{d}^{3}\hat{\theta}_{k}\psi^{-\frac{1}{2}}(\hat{\tau})\nonumber\\
&=-\frac{1}{8}\delta_{n}^{3}\epsilon_{d}^{3}\theta_{k}.\label{c6}
\end{align}
We may now combine the results in (\ref{a1})-(\ref{c6}) above with the calculated partial derivatives of $\hat{\Phi}(\bm{x},\hat{\tau},\hat{\bm{\theta}})$ with
respect to the parameters $\hat{\tau}$ and $\hat{\theta}_{i},\ 1\leq i\leq d$ in (\ref{hat_Phi_hat_tau})-(\ref{hat_Phi_hat_tau_hat_tau_hat_tau}). We then get
\begin{align}
&\hat{\Phi}_{\hat{\tau}}(\bm{x},\hat{\tau}^{*},\hat{\bm{\theta}}^{*})=\Phi_{\tau}(\bm{x},\tau^{*},\bm{\theta}^{*})\delta_{n}\epsilon_{d}\tau^{*}-
\frac{1}{2}\delta_{n}\epsilon_{d}\sum_{i=1}^{d}\Phi_{\theta_{i}}(\bm{x},\tau^{*},\bm{\theta}^{*})\theta_{i}^{*}.\label{O_tau}\\
\nonumber\\
&\hat{\Phi}_{\hat{\theta}_{k}}(\bm{x},\hat{\tau}^{*},\hat{\bm{\theta}}^{*})=\Phi_{\theta_{k}}(\bm{x},\tau^{*},\bm{\theta}^{*})\bar{\tau}^{\frac{1}{2}}(\tau^{*})^{-\frac{1}{2}}.\label{O_theta}\\
\nonumber\\
&\hat{\Phi}_{\hat{\tau},\hat{\tau}}(\bm{x},\hat{\tau}^{*},\hat{\bm{\theta}}^{*})=\Phi_{\tau,\tau}(\bm{x},\tau^{*},\bm{\theta}^{*})\delta_{n}^{2}\epsilon_{d}^{2}(\tau^{*})^{2}
+\Phi_{\tau}(\bm{x},\tau^{*},\bm{\theta}^{*})\delta_{n}^{2}\epsilon_{d}^{2}\tau^{*}\nonumber\\
&-\delta_{n}^{2}\epsilon_{d}^{2}\tau^{*}\sum_{i=1}^{d}\Phi_{\theta_{i},\tau}(\bm{x},\tau^{*},\bm{\theta}^{*})\theta_{i}^{*}+\frac{1}{4}\delta_{n}^{2}\epsilon_{d}^{2}\sum_{i=1}^{d}\Phi_{\theta_{i}}(\bm{x},\tau^{*},\bm{\theta}^{*})\theta_{i}^{*}\nonumber\\
&+\frac{1}{4}\delta_{n}^{2}\epsilon_{d}^{2}\sum_{i=1}^{d}\Phi_{\theta_{i},\theta_{i}}(\bm{x},\tau^{*},\bm{\theta}^{*})(\theta_{i}^{*})^{2}.\label{O_tau_tau}\\
\nonumber\\
&\hat{\Phi}_{\hat{\theta}_{k},\hat{\theta}_{k}}(\bm{x},\hat{\tau}^{*},\hat{\bm{\theta}}^{*})=\Phi_{\theta_{k},\theta_{k}}(\bm{x},\tau^{*},\bm{\theta}^{*})\bar{\tau}(\tau^{*})^{-1}.\label{O_theta_theta}\\
\nonumber\\
&\hat{\Phi}_{\hat{\theta}_{k},\hat{\tau}}(\bm{x},\hat{\tau}^{*},\hat{\bm{\theta}}^{*})=-\frac{1}{2}\Phi_{\theta_{k}}(\bm{x},\tau^{*},\bm{\theta}^{*})\bar{\tau}^{\frac{1}{2}}\delta_{n}\epsilon_{d}(\tau^{*})^{-\frac{1}{2}}\nonumber\\
&+\Phi_{\theta_{k},\tau}(\bm{x},\tau^{*},\bm{\theta}^{*})\bar{\tau}^{\frac{1}{2}}\delta_{n}\epsilon_{d}(\tau^{*})^{\frac{1}{2}}-\frac{1}{2}\Phi_{\theta_{k},\theta_{k}}(\bm{x},\tau^{*},\bm{\theta}^{*})\bar{\tau}^{\frac{1}{2}}\delta_{n}\epsilon_{d}(\tau^{*})^{-\frac{1}{2}}\theta_{k}^{*}.\label{O_tau_theta}\\
\nonumber\\
&\hat{\Phi}_{\hat{\theta}_{k},\hat{\theta}_{k},\hat{\tau}}(\bm{x},\hat{\tau}^{*},\hat{\bm{\theta}}^{*})=
-\Phi_{\theta_{k},\theta_{k}}(\bm{x},\tau^{*},\bm{\theta}^{*})\bar{\tau}\delta_{n}\epsilon_{d}(\tau^{*})^{-1}\nonumber\\
&+\Phi_{\theta_{k},\theta_{k},\tau}(\bm{x},\tau^{*},\bm{\theta}^{*})\delta_{n}\epsilon_{d}\bar{\tau}
-\frac{1}{2}\delta_{n}\epsilon_{d}\bar{\tau}(\tau^{*})^{-1}\Phi_{\theta_{k},\theta_{k},\theta_{k}}(\bm{x},\tau^{*},\bm{\theta}^{*})\theta_{k}^{*}.\label{O_theta_theta_tau}\\
\nonumber\\
&\hat{\Phi}_{\hat{\tau},\hat{\tau},\hat{\theta}_{k}}(\bm{x},\hat{\tau}^{*},\hat{\bm{\theta}}^{*})=
\Phi_{\tau,\tau,\theta_{k}}(\bm{x},\tau^{*},\bm{\theta}^{*})\bar{\tau}^{\frac{1}{2}}(\tau^{*})^{\frac{3}{2}}\delta_{n}^{2}\epsilon_{d}^{2}\nonumber\\
&-\Phi_{\theta_{k},\theta_{k},\tau}(\bm{x},\tau^{*},\bm{\theta}^{*})\delta_{n}^{2}\epsilon_{d}^{2}\bar{\tau}^{\frac{1}{2}}(\tau^{*})^{\frac{1}{2}}\theta_{k}^{*}\nonumber\\
&+\frac{1}{4}\Phi_{\theta_{k}}(\bm{x},\tau^{*},\bm{\theta}^{*})\delta_{n}^{2}\epsilon_{d}^{2}\bar{\tau}^{\frac{1}{2}}(\tau^{*})^{-\frac{1}{2}}\nonumber\\
&+\frac{3}{4}\Phi_{\theta_{k},\theta_{k}}(\bm{x},\tau^{*},\bm{\theta}^{*})\delta_{n}^{2}\epsilon_{d}^{2}\bar{\tau}^{\frac{1}{2}}(\tau^{*})^{-\frac{1}{2}}\theta_{k}^{*}\nonumber\\
&+\frac{1}{4}\Phi_{\theta_{k},\theta_{k},\theta_{k}}(\bm{x},\tau^{*},\bm{\theta}^{*})\delta_{n}^{2}\epsilon_{d}^{2}\bar{\tau}^{\frac{1}{2}}(\tau^{*})^{-\frac{1}{2}}(\theta_{k}^{*})^{2}.\label{O_tau_tau_theta}\\
\nonumber\\
&\hat{\Phi}_{\hat{\theta}_{k},\hat{\theta}_{k},\hat{\theta}_{k}}(\bm{x},\hat{\tau}^{*},\hat{\bm{\theta}}^{*})=\Phi_{\theta_{k},\theta_{k},\theta_{k}}(\bm{x},\tau^{*},\bm{\theta}^{*})\bar{\tau}^{\frac{3}{2}}(\tau^{*})^{-\frac{3}{2}}.\label{O_theta_theta_theta}\\
\nonumber\\
&\hat{\Phi}_{\hat{\tau},\hat{\tau},\hat{\tau}}(\bm{x},\hat{\tau}^{*},\hat{\bm{\theta}}^{*})=
3\Phi_{\tau,\tau}(\bm{x},\tau^{*},\bm{\theta}^{*})\delta_{n}^{3}\epsilon_{d}^{3}(\tau^{*})^{2}\nonumber\\
&+\Phi_{\tau,\tau,\tau}(\bm{x},\tau^{*},\bm{\theta}^{*})\delta_{n}^{3}\epsilon_{d}^{3}(\tau^{*})^{3}
-\sum_{i=1}^{d}\Phi_{\tau,\tau,\theta_{i}}(\bm{x},\tau^{*},\bm{\theta}^{*})\delta_{n}^{3}\epsilon_{d}^{3}\theta_{i}^{*}(\tau^{*})^{2}\nonumber\\
&+\Phi_{\tau}(\bm{x},\tau^{*},\bm{\theta}^{*})\delta_{n}^{3}\epsilon_{d}^{3}\tau^{*}
-\frac{3}{4}\sum_{i=1}^{d}\Phi_{\tau,\theta_{i}}(\bm{x},\tau^{*},\bm{\theta}^{*})\delta_{n}^{3}\epsilon_{d}^{3}\tau^{*}\theta_{i}^{*}\nonumber\\
&+\frac{3}{4}\sum_{i=1}^{d}\Phi_{\tau,\theta_{i},\theta_{i}}(\bm{x},\tau^{*},\bm{\theta}^{*})\delta_{n}^{3}\epsilon_{d}^{3}\tau^{*}(\theta_{i}^{*})^{2}
-\frac{1}{8}\sum_{i=1}^{d}\Phi_{\theta_{i}}(\bm{x},\tau^{*},\bm{\theta}^{*})\delta_{n}^{3}\epsilon_{d}^{3}\theta_{i}^{*}\nonumber\\
&-\frac{3}{8}\sum_{i=1}^{d}\Phi_{\theta_{i},\theta_{i}}(\bm{x},\tau^{*},\bm{\theta}^{*})\delta_{n}^{3}\epsilon_{d}^{3}(\theta_{i}^{*})^{2}
+\frac{1}{8}\sum_{i=1}^{d}\Phi_{\theta_{i},\theta_{i},\theta_{i}}(\bm{x},\tau^{*},\bm{\theta}^{*})\delta_{n}^{3}\epsilon_{d}^{3}(\theta_{i}^{*})^{3}.\label{O_tau_tau_tau}
\end{align}

%%% Local Variables: 
%%% mode: latex
%%% TeX-master: "master"
%%% End: 
%\appendix
\chapter{Some fourth order partial derivatives of $\hat{\Phi}(\bm{x},\tau^{*},\bm{\theta}^{*})$}
\noindent{Differentiating the expression (\ref{hat_Phi_hat_theta_hat_theta_hat_tau}) with respect to $\hat{\tau}$ we get
\begin{align}
&\hat{\Phi}_{\hat{\theta}_{k},\hat{\theta}_{k},\hat{\tau},\hat{\tau}}(\bm{x},\hat{\tau}^{*},\hat{\bm{\theta}}^{*})=
2\Phi_{\theta_{k},\theta_{k}}(\bm{x},\tau^{*},\bm{\theta}^{*})\left(\left.\frac{\partial^{2}\phi(\hat{\theta}_{k},\hat{\tau})}{\partial\hat{\theta}_{k}\partial\hat{\tau}}\right]\right)
%_{\hat{\tau}=\hat{\tau}^{*},\hat{\bm{\theta}}=\hat{\bm{\theta}}^{*}}
\left(\left.\frac{\partial^{2}\phi(\hat{\theta}_{k},\hat{\tau})}{\partial\hat{\tau}\partial\hat{\theta}_{k}}\right]\right)
%_{\hat{\tau}=\hat{\tau}^{*},\hat{\bm{\theta}}=\hat{\bm{\theta}}^{*}}
\nonumber\\
&+2\Phi_{\theta_{k},\theta_{k}}(\bm{x},\tau^{*},\bm{\theta}^{*})\left(\left.\frac{\partial\phi(\hat{\theta}_{k},\hat{\tau})}{\partial\hat{\theta}_{k}}\right]\right)
%_{\hat{\tau}=\hat{\tau}^{*},\hat{\bm{\theta}}=\hat{\bm{\theta}}^{*}}
\left(\left.\frac{\partial^{3}\phi(\hat{\theta}_{k},\hat{\tau})}{\partial\hat{\tau}^{2}\partial\hat{\theta}_{k}}\right]\right)
%_{\hat{\tau}=\hat{\tau}^{*},\hat{\bm{\theta}}=\hat{\bm{\theta}}^{*}}
\nonumber\\
&+2\Phi_{\theta_{k},\theta_{k},\hat{\tau}}(\bm{x},\tau^{*},\bm{\theta}^{*})\left(\left.\frac{\partial\phi(\hat{\theta}_{k},\hat{\tau})}{\partial\hat{\theta}_{k}}\right]\right)
%_{\hat{\tau}=\hat{\tau}^{*},\hat{\bm{\theta}}=\hat{\bm{\theta}}^{*}}
\left(\left.\frac{\partial^{2}\phi(\hat{\theta}_{k},\hat{\tau})}{\partial\hat{\tau}\partial\hat{\theta}_{k}}\right]\right)
%_{\hat{\tau}=\hat{\tau}^{*},\hat{\bm{\theta}}=\hat{\bm{\theta}}^{*}}
\left(\left.\frac{\partial\psi(\hat{\tau})}{\partial\hat{\tau}}\right]\right)
%_{\hat{\tau}=\hat{\tau}^{*},\hat{\bm{\theta}}=\hat{\bm{\theta}}^{*}}
\nonumber\\
&+2\Phi_{\theta_{k},\theta_{k},\hat{\theta}_{k}}(\bm{x},\tau^{*},\bm{\theta}^{*})\left(\left.\frac{\partial\phi(\hat{\theta}_{k},\hat{\tau})}{\partial\hat{\theta}_{k}}\right]\right)
%_{\hat{\tau}=\hat{\tau}^{*},\hat{\bm{\theta}}=\hat{\bm{\theta}}^{*}}
\left(\left.\frac{\partial^{2}\phi(\hat{\theta}_{k},\hat{\tau})}{\partial\hat{\tau}\partial\hat{\theta}_{k}}\right]\right)
%_{\hat{\tau}=\hat{\tau}^{*},\hat{\bm{\theta}}=\hat{\bm{\theta}}^{*}}
\left(\left.\frac{\partial\phi(\hat{\theta}_{k},\hat{\tau})}{\partial\hat{\tau}}\right]\right)
%_{\hat{\tau}=\hat{\tau}^{*},\hat{\bm{\theta}}=\hat{\bm{\theta}}^{*}}
\nonumber\\
&+2\Phi_{\theta_{k},\theta_{k},\tau}(\bm{x},\tau^{*},\bm{\theta}^{*})\left(\left.\frac{\partial\phi(\hat{\theta}_{k},\hat{\tau})}{\partial\hat{\theta}_{k}}\right]\right)
%_{\hat{\tau}=\hat{\tau}^{*},\hat{\bm{\theta}}=\hat{\bm{\theta}}^{*}}
\left(\left.\frac{\partial^{2}\phi(\hat{\theta}_{k},\hat{\tau})}{\partial\hat{\theta}_{k}\partial\hat{\tau}}\right]\right)
%_{\hat{\tau}=\hat{\tau}^{*},\hat{\bm{\theta}}=\hat{\bm{\theta}}^{*}}
\left(\left.\frac{\partial\psi(\hat{\tau})}{\partial\hat{\tau}}\right]\right)
%_{\hat{\tau}=\hat{\tau}^{*},\hat{\bm{\theta}}=\hat{\bm{\theta}}^{*}}
\nonumber\\
&+\Phi_{\theta_{k},\theta_{k},\tau}(\bm{x},\tau^{*},\bm{\theta}^{*})\left(\left.\frac{\partial\phi(\hat{\theta}_{k},\hat{\tau})}{\partial\hat{\theta}_{k}}\right]
%_{\hat{\tau}=\hat{\tau}^{*},\hat{\bm{\theta}}=\hat{\bm{\theta}}^{*}}
\right)^{2}\left(\left.\frac{\partial^{2}\psi(\hat{\tau})}{\partial\hat{\tau}^{2}}\right]\right)
%_{\hat{\tau}=\hat{\tau}^{*},\hat{\bm{\theta}}=\hat{\bm{\theta}}^{*}}
\nonumber\\
&+\Phi_{\theta_{k},\theta_{k},\tau,\tau}(\bm{x},\tau^{*},\bm{\theta}^{*})\left(\left.\frac{\partial\phi(\hat{\theta}_{k},\hat{\tau})}{\partial\hat{\theta}_{k}}\right]
%_{\hat{\tau}=\hat{\tau}^{*},\hat{\bm{\theta}}=\hat{\bm{\theta}}^{*}}
\right)^{2}\left(\left.\frac{\partial\psi(\hat{\tau})}{\partial\hat{\tau}}\right]
%_{\hat{\tau}=\hat{\tau}^{*},\hat{\bm{\theta}}=\hat{\bm{\theta}}^{*}}
\right)^{2}\nonumber\\
&+\Phi_{\theta_{k},\theta_{k},\theta_{k},\tau}(\bm{x},\tau^{*},\bm{\theta}^{*})\left(\left.\frac{\partial\phi(\hat{\theta}_{k},\hat{\tau})}{\partial\hat{\theta}_{k}}\right]
%_{\hat{\tau}=\hat{\tau}^{*},\hat{\bm{\theta}}=\hat{\bm{\theta}}^{*}}
\right)^{2}\left(\left.\frac{\partial\psi(\hat{\tau})}{\partial\hat{\tau}}\right]\right)
%_{\hat{\tau}=\hat{\tau}^{*},\hat{\bm{\theta}}=\hat{\bm{\theta}}^{*}}
\left(\left.\frac{\partial\phi(\hat{\theta}_{k},\hat{\tau})}{\partial\hat{\tau}}\right]\right)
%_{\hat{\tau}=\hat{\tau}^{*},\hat{\bm{\theta}}=\hat{\bm{\theta}}^{*}}
\nonumber\\
&+2\Phi_{\theta_{k},\theta_{k},\theta_{k}}(\bm{x},\tau^{*},\bm{\theta}^{*})\left(\left.\frac{\partial\phi(\hat{\theta}_{k},\hat{\tau})}{\partial\hat{\theta}_{k}}\right]\right)
%_{\hat{\tau}=\hat{\tau}^{*},\hat{\bm{\theta}}=\hat{\bm{\theta}}^{*}}
\left(\left.\frac{\partial^{2}\phi(\hat{\theta}_{k},\hat{\tau})}{\partial\hat{\theta}_{k}\partial\hat{\tau}}\right]\right)
%_{\hat{\tau}=\hat{\tau}^{*},\hat{\bm{\theta}}=\hat{\bm{\theta}}^{*}}
\left(\left.\frac{\partial\phi(\hat{\theta}_{k},\hat{\tau})}{\partial\hat{\tau}}\right]\right)
%_{\hat{\tau}=\hat{\tau}^{*},\hat{\bm{\theta}}=\hat{\bm{\theta}}^{*}}
\nonumber\\
&+\Phi_{\theta_{k},\theta_{k},\theta_{k}}(\bm{x},\tau^{*},\bm{\theta}^{*})\left(\left.\frac{\partial\phi(\hat{\theta}_{k},\hat{\tau})}{\partial\hat{\theta}_{k}}\right]
%_{\hat{\tau}=\hat{\tau}^{*},\hat{\bm{\theta}}=\hat{\bm{\theta}}^{*}}
\right)^{2}\left(\left.\frac{\partial^{2}\phi(\hat{\theta}_{k},\hat{\tau})}{\partial\hat{\tau}^{2}}\right]\right)
%_{\hat{\tau}=\hat{\tau}^{*},\hat{\bm{\theta}}=\hat{\bm{\theta}}^{*}}
\nonumber\\
&+\Phi_{\theta_{k},\theta_{k},\theta_{k},\tau}(\bm{x},\tau^{*},\bm{\theta}^{*})\left(\left.\frac{\partial\phi(\hat{\theta}_{k},\hat{\tau})}{\partial\hat{\theta}_{k}}\right]
%_{\hat{\tau}=\hat{\tau}^{*},\hat{\bm{\theta}}=\hat{\bm{\theta}}^{*}}
\right)^{2}\left(\left.\frac{\partial\phi(\hat{\theta}_{k},\hat{\tau})}{\partial\hat{\tau}}\right]\right)
%_{\hat{\tau}=\hat{\tau}^{*},\hat{\bm{\theta}}=\hat{\bm{\theta}}^{*}}
\left(\left.\frac{\partial\psi(\hat{\tau})}{\partial\hat{\tau}}\right]\right)
%_{\hat{\tau}=\hat{\tau}^{*},\hat{\bm{\theta}}=\hat{\bm{\theta}}^{*}}
\nonumber\\
&+\Phi_{\theta_{k},\theta_{k},\theta_{k},\theta_{k}}(\bm{x},\tau^{*},\bm{\theta}^{*})\left(\left.\frac{\partial\phi(\hat{\theta}_{k},\hat{\tau})}{\partial\hat{\theta}_{k}}\right]
%_{\hat{\tau}=\hat{\tau}^{*},\hat{\bm{\theta}}=\hat{\bm{\theta}}^{*}}
\right)^{2}\left(\left.\frac{\partial\phi(\hat{\theta}_{k},\hat{\tau})}{\partial\hat{\tau}}\right]\right)
%_{\hat{\tau}=\hat{\tau}^{*},\hat{\bm{\theta}}=\hat{\bm{\theta}}^{*}}
\left(\left.\frac{\partial\phi(\hat{\theta}_{k},\hat{\tau})}{\partial\hat{\tau}}\right]\right)
%_{\hat{\tau}=\hat{\tau}^{*},\hat{\bm{\theta}}=\hat{\bm{\theta}}^{*}}
\label{hat_Phi_hat_theta_hat_theta_hat_tau_hat_tau}
\intertext{where $\left.\right]$ means evaluating the derivatives in $\hat{\tau}=\hat{\tau}^{*},\hat{\bm{\theta}}=\hat{\bm{\theta}}^{*}$. By differentiating (\ref{hat_Phi_hat_theta_hat_theta_hat_theta}) with respect to $\hat{\theta}_{k}$ we get}
&\hat{\Phi}_{\hat{\theta}_{k},\hat{\theta}_{k},\hat{\theta}_{k},\hat{\theta}_{k}}(\bm{x},\hat{\tau}^{*},\hat{\bm{\theta}}^{*})=
\Phi_{\theta_{k},\theta_{k},\theta_{k},\theta_{k}}(\bm{x},\tau^{*},\bm{\theta}^{*})\bar{\tau}^{2}(\tau^{*})^{-2}.\label{hat_Phi_hat_theta_hat_theta_hat_theta_hat_theta}
\intertext{By differentiating (\ref{hat_Phi_hat_theta_hat_theta_hat_theta}) with respect to $\hat{\tau}$ we get}
&\hat{\Phi}_{\hat{\theta}_{k},\hat{\theta}_{k},\hat{\theta}_{k},\hat{\tau}}(\bm{x},\hat{\tau}^{*},\hat{\bm{\theta}}^{*})=
\Phi_{\theta_{k},\theta_{k},\theta_{k},\theta_{k}}(\bm{x},\tau^{*},\bm{\theta}^{*})\left(\left.\frac{\partial\phi(\hat{\theta}_{k},\hat{\tau})}{\partial\hat{\tau}}\right]\right)
%_{\hat{\tau}=\hat{\tau}^{*},\hat{\bm{\theta}}=\hat{\bm{\theta}}^{*}}
\left(\left.\frac{\partial\phi(\hat{\theta}_{k},\hat{\tau})}{\partial\hat{\theta}_{k}}\right]
%_{\hat{\tau}=\hat{\tau}^{*},\hat{\bm{\theta}}=\hat{\bm{\theta}}^{*}}
\right)^{3}\nonumber\\
&+\Phi_{\theta_{k},\theta_{k},\theta_{k},\tau}(\bm{x},\tau^{*},\bm{\theta}^{*})\left(\left.\frac{\partial\psi(\hat{\tau})}{\partial\hat{\tau}}\right]\right)
%_{\hat{\tau}=\hat{\tau}^{*},\hat{\bm{\theta}}=\hat{\bm{\theta}}^{*}}
\left(\left.\frac{\partial\phi(\hat{\theta}_{k},\hat{\tau})}{\partial\hat{\theta}_{k}}\right]
%_{\hat{\tau}=\hat{\tau}^{*},\hat{\bm{\theta}}=\hat{\bm{\theta}}^{*}}
\right)^{3}\nonumber\\
&+3\Phi_{\theta_{k},\theta_{k},\theta_{k}}(\bm{x},\tau^{*},\bm{\theta}^{*})\left(\left.\frac{\partial^{2}\phi(\hat{\theta}_{k},\hat{\tau})}{\partial\hat{\theta}_{k}\partial\hat{\tau}}\right]\right)
%_{\hat{\tau}=\hat{\tau}^{*},\hat{\bm{\theta}}=\hat{\bm{\theta}}^{*}}
\left(\left.\frac{\partial\phi(\hat{\theta}_{k},\hat{\tau})}{\partial\hat{\theta}_{k}}\right]
%_{\hat{\tau}=\hat{\tau}^{*},\hat{\bm{\theta}}=\hat{\bm{\theta}}^{*}}
\right)^{2}\label{hat_Phi_hat_theta_hat_theta_hat_theta_hat_tau}
\intertext{In the case of a gaussian likelihood function, the expression (\ref{hat_Phi_hat_theta_hat_theta_hat_tau_hat_tau}) reduces to}
&\hat{\Phi}_{\hat{\theta}_{k},\hat{\theta}_{k},\hat{\tau},\hat{\tau}}(\bm{x},\hat{\tau}^{*},\hat{\bm{\theta}}^{*})=
\Phi_{\theta_{k},\theta_{k}}(\bm{x},\tau^{*},\bm{\theta}^{*})\delta_{n}^{2}\epsilon_{d}^{2}\bar{\tau}(\tau^{*})^{-1}\nonumber\\
&-\Phi_{\theta_{k},\theta_{k},\tau}(\bm{x},\tau^{*},\bm{\theta}^{*})\delta_{n}^{2}\epsilon_{d}^{2}\bar{\tau}
+\Phi_{\theta_{k},\theta_{k},\theta_{k}}(\bm{x},\tau^{*},\bm{\theta}^{*})\delta_{n}^{2}\epsilon_{d}^{2}\bar{\tau}\theta_{k}^{*}(\tau^{*})^{-1}\nonumber\\
&+\frac{1}{4}\Phi_{\theta_{k},\theta_{k},\theta_{k},\theta_{k}}(\bm{x},\tau^{*},\bm{\theta}^{*})\delta_{n}^{2}\epsilon_{d}^{2}\bar{\tau}(\theta_{k}^{*})^{2}(\tau^{*})^{-1}
\label{O_theta_theta_tau_tau}
\intertext{and the expression (\ref{hat_Phi_hat_theta_hat_theta_hat_theta_hat_tau}) reduces to}
&\hat{\Phi}_{\hat{\theta}_{k},\hat{\theta}_{k},\hat{\theta}_{k},\hat{\tau}}(\bm{x},\hat{\tau}^{*},\hat{\bm{\theta}}^{*})=
-\frac{1}{2}\Phi_{\theta_{k},\theta_{k},\theta_{k},\theta_{k}}(\bm{x},\tau^{*},\bm{\theta}^{*})\delta_{n}\epsilon_{d}\bar{\tau}^{\frac{3}{2}}(\tau^{*})^{-\frac{3}{2}}\theta_{k}\nonumber\\
&-\frac{3}{2}\Phi_{\theta_{k},\theta_{k},\theta_{k}}(\bm{x},\tau^{*},\bm{\theta}^{*})\delta_{n}\epsilon_{d}\bar{\tau}^{\frac{3}{2}}(\tau^{*})^{-\frac{3}{2}}.
\end{align}
\chapter{Numerical results}

\begin{table}[h]
\caption{The results from the test image Barbara, $N=512\times 512$, wavelet used: Symmlet 16. 
The first row in each entry in the table is the scaled RMSE measure in (\ref{def_RMSE}), the second row is 
the SNR of the reconstruction, see (\ref{SNR_experiments2}), the third row shows the proportion of nonzero 
wavelet coefficient estimates $\theta_{i}^{*}$ as a fraction of the sample size $N$
and the fourth row shows the estimated value of the GGD shape parameter $\nu$ provided by the INMDL algorithm.}
\begin{tabular}{|l|c|c|c|c|c|c|}\hline\hline 
  SNR & RiskShrink & SureShrink & $T_{MAP}^{(0.70)}$ & $T_{MAP}^{(1.0)}$ & NML & INMDL \\ \hline
  1.0 dB & 24.3\% & 20.0\% & 23.4\% & 24.2\% & 93.4\% & 33.5\%\\
         & 13.1 dB &  14.8 dB & 13.4 dB & 12.9 dB & 3.88 dB &  10.6 dB \\
         & 0.0961\% & 0.790\% & 0.155\% & 0.810\% & 39.7\% & 1.34\%\\ 
         & & & & & & 1.02 \\\hline
  5.0 dB & 34.0\%& 28.5\% & 32.1\% & 35.6\% &81.1\% & 37.8\%\\
      & 14.2 dB & 15.8 dB & 14.8 dB & 13.9 dB & 7.60 dB & 13.5 dB \\
      & 0.188\% & 1.64\% & 1.52\% & 18.8\% & 19.9\% &  1.35\%\\ 
      & & & & & &  0.913\\\hline
  10.0 dB & 51.2\% & 40.3\% & 52.4\% & 62.5\% & 59.2\% & 47.9\%\\
           & 15.7 dB &  17.8 dB & 15.6 dB &  14.2 dB & 14.6 dB & 16.4  dB \\
           & 0.434\% & 8.83\% & 11.1\% & 55.6\% & 5.47\% & 1.92\%\\ 
           & & & & & &  0.842\\\hline
  15.0 dB & 72.6\% & 49.4\% & 67.7\% & 79.4\% & 59.8\% & 58.6\%\\ 
          & 17.7 dB & 21.1 dB & 18.4 dB & 17.1 dB & 19.5 dB & 19.6 dB\\
          & 1.19\% &  15.9\% & 27.7\% & 77.1\% & 4.94\% & 3.20\%\\
          & & & & & & 0.824 \\ \hline
  20.0 dB & 102\% & 66.3\% & 79.3\% & 88.8\% & 75.0\% & 78.7\%\\
         &  19.8 dB &  23.5 dB & 22.0 dB & 21.1 dB & 22.5 dB & 22.1 dB \\
         & 2.81\% & 24.1\% & 44.9\% & 87.9\% & 7.22\% & 5.67\%\\ 
         & & & & & & 0.815 \\\hline
  25.0 dB & 122\% & 75.3\% & 85.6\% & 93.4\% & 89.6\% & 95.1\%\\ 
          & 23.3 dB & 27.4 dB & 26.4 dB & 25.6 dB & 25.9 dB & 25.4 dB \\
          & 5.14\% & 30.0\% & 57.1\% & 92.9\% & 9.4\% & 8.10\%\\
          & & & & & & 0.791 \\ \hline
  50.0 dB & 971\% & 441\% & 129\% & 100\% & 797\% & 873\%\\
          & 30.2 dB & 37.1 dB & 47.8 dB & 50.0 dB & 32.0 dB & 31.2 dB\\
          & 14.0\% & 51.0\% & 74.2\% & 97.4\% & 17.4\% & 15.8\%\\
          & & & & & & 0.703 \\ \hline\hline
\end{tabular}
\label{table_barbara}
\end{table}

\begin{table}[h]
\caption{The results from the test image Lena, $N=512\times 512$, wavelet used: Symmlet 16. 
The first row in each entry in the table is the scaled RMSE measure in (\ref{def_RMSE}), the second row is 
the SNR of the reconstruction, see (\ref{SNR_experiments2}), the third row shows the proportion of nonzero
wavelet coefficient estimates $\theta_{i}^{*}$ as a fraction of the sample size $N$
and the fourth row shows the estimated value of the GGD shape parameter $\nu$ provided by the INMDL algorithm.}
\begin{tabular}{|l|c|c|c|c|c|c|}\hline\hline 
  SNR & RiskShrink & SureShrink & $T_{MAP}^{(0.70)}$ & $T_{MAP}^{(1.0)}$ & NML & INMDL \\ \hline
  1.0 dB & 17.9\% & 14.1\% & 17.1\% & 17.9\% & 93.0\% & 31.8\%\\
         & 15.9 dB &  18.0 dB &  16.2 dB & 15.7 dB & 3.91 dB & 11.2 dB \\
         & 0.0755\% & 0.544 \% & 0.122\% & 0.764\% & 38.9\% & 1.17\%\\ 
         & & & & & & 1.02 \\\hline
  5.0 dB & 24.0\%& 19.1\% & 24.8\% & 31.0\% & 78.5\% & 32.6\%\\
      & 17.3 dB & 19.3 dB & 17.1 dB & 15.2 dB & 7.87 dB & 14.8 dB \\
      & 0.138\% & 0.935\% & 1.24\% & 18.4\% & 17.8\% &  1.21\%\\ 
      & & & & & &  0.901\\\hline
  10.0 dB & 34.2\% & 26.1\% & 46.1\% & 60.5\% & 48.1\% & 35.7\%\\
         &  19.3 dB &  21.6 dB & 16.8 dB & 14.5 dB & 16.4 dB & 18.9 dB \\
         & 0.300\% & 2.39\% & 10.0\% & 55.3\% & 3.52\% & 1.50\%\\ 
         & & & & & &  0.807\\\hline
  15.0 dB & 47.0\% & 34.3\% & 65.0\% & 79.0\% & 44.5\% & 41.7\%\\ 
          & 21.5 dB & 24.3 dB & 18.8 dB & 17.1 dB & 22.0 dB & 22.6 dB\\
          & 0.658\% &  5.65\% & 26.6\% & 77.3\% & 2.42\% & 1.94\%\\
          & & & & & & 0.742 \\ \hline
  20.0 dB & 62.8\% & 45.9\% & 77.6\% & 88.8\% & 53.1\% & 52.4\%\\
         &  24.0 dB &  26.7 dB &  22.2 dB & 21.1 dB & 25.5 dB & 25.6 dB \\
         & 1.34\% & 10.6\% & 44.2\% & 88.3\% & 3.05\% & 2.89\%\\ 
         & & & & & & 0.700 \\\hline
  25.0 dB & 84.4\% & 60.2\% & 93.8\% & 85.6\% & 70.0\% & 69.7\%\\ 
          & 26.5 dB & 29.4 dB & 25.6 dB & 26.4 dB & 28.1 dB & 28.1 dB \\
          & 2.42\% & 17.3\% & 93.5\% & 58.5\% & 4.31\% & 4.26\%\\
          & & & & & &  0.657\\ \hline
  50.0 dB & 738\% & 381\% & 110\% & 99.3\% & 705\% & 689\%\\
          & 32.6 dB & 38.4 dB & 49.2 dB & 50.1 dB & 33.0 dB & 33.2 dB\\
          & 8.9\% & 48.5\% & 82.3\% & 98.6\% & 9.73\% & 10.2\%\\
          & & & & & & 0.588 \\ \hline\hline
\end{tabular}
\label{table_lena}
\end{table}

\begin{table}[h]
\caption{The results from the test image Baboon, $N=512\times 512$, wavelet used: Symmlet 16. 
The first row in each entry in the table is the scaled RMSE measure in (\ref{def_RMSE}), the second row is 
the SNR of the reconstruction, see (\ref{SNR_experiments2}), the third row shows the 
 proportion of nonzero 
wavelet coefficient estimates $\theta_{i}^{*}$ as a fraction of the sample size $N$
and the fourth row shows the estimated value of the GGD shape parameter $\nu$ provided by the INMDL algorithm.}
\begin{tabular}{|l|c|c|c|c|c|c|}\hline\hline 
  SNR & RiskShrink & SureShrink & $T_{MAP}^{(0.70)}$ & $T_{MAP}^{(1.0)}$ & NML & INMDL \\ \hline
  1.0 dB & 25.1\% & 22.9\% & 24.7\% & 24.9\% & 93.7\% & 35.3\%\\
         &  12.8 dB &  13.6 dB & 12.9 dB & 12.8 dB & 3.86 dB & 10.1 dB \\
         & 0.0420\% & 0.580\% & 0.0801\% & 0.731\% & 40.3\% & 1.31\%\\ 
         & & & & & & 1.04 \\\hline
  5.0 dB & 37.5\% & 34.3\% & 37.2\% & 38.9\% & 83.2\% & 42.9\%\\
      & 13.3 dB & 14.1 dB & 13.4 dB & 13.1 dB & 7.40 dB & 12.3 dB \\
      & 0.0946\% & 1.39\% & 1.45\% & 19.1\% & 21.2\% &  1.29\%\\ 
      & & & & & &  0.95\\\hline
  10.0 dB & 62.0\% & 48.7\% & 58.4\% & 64.6\% & 66.9\% & 58.4\%\\
         &  14.0 dB &  16.2 dB & 14.7 dB & 13.9 dB & 13.6 dB & 14.6 dB \\
         & 0.263\% & 12.1\% & 11.9\% & 55.9\% & 6.21\% & 1.90\%\\ 
         & & & & & & 0.904 \\\hline
  15.0 dB & 97.9\% & 66.1\% & 74.7\% & 80.9\% & 81.1\% & 83.8\%\\ 
          & 15.1 dB & 18.5 dB & 17.5 dB & 16.9 dB & 16.8 dB & 16.5 dB\\
          & 0.929\% & 23.6\% & 29.7\% & 77.2\% & 5.93\% & 3.31\%\\
          & & & & & &  0.912\\ \hline
  20.0 dB & 146\% & 84.8\% & 85.0\% & 89.1\% & 110\% & 123\%\\
         &  16.7 dB &  21.4 dB & 21.4 dB & 21.0 dB & 19.1 dB & 18.1 dB \\
         & 2.65\% & 36.5\% & 45.8\% & 87.2\% & 8.70\% & 5.48\%\\ 
         & & & & & &  0.931\\\hline
  25.0 dB & 214\% & 105\% & 92.3\% & 93.3\% & 159\% & 188\%\\ 
          & 18.3 dB & 24.5 dB & 25.7 dB & 25.6 & 20.9 dB & 19.5dB \\
          & 5.27\% & 48.0\% & 56.7\% & 91.7\% & 11.7\% & 5.27\%\\
          & & & & & &  0.954\\ \hline
  50.0 dB & 2788\% & 1030\% & 425\% & 165\% & 2271\% & 2892\%\\
          & 21.1 dB & 29.7 dB & 37.4 dB & 45.6 dB & 22.9 dB & 20.7 dB\\
          & 10.7\% & 59.7\% & 67.4\% & 94.9\% & 15.4\% & 9.96\%\\
          & & & & & &  0.941\\ \hline\hline
\end{tabular}
\label{table_baboon}
\end{table}

\begin{table}[h]
\caption{The results from the test image Goldhill,  $N=512\times 512$, wavelet used: Symmlet 16. 
The first row in each entry in the table is the scaled RMSE measure in (\ref{def_RMSE}), the second row is 
the SNR of the reconstruction, see (\ref{SNR_experiments2}), the third row shows the 
 proportion of nonzero 
wavelet coefficient estimates $\theta_{i}^{*}$ as a fraction of the sample size $N$
 and the fourth row shows the estimated value of the GGD shape parameter $\nu$ provided by the INMDL algorithm.}
\begin{tabular}{|l|c|c|c|c|c|c|}\hline\hline 
  SNR & RiskShrink & SureShrink & $T_{MAP}^{(0.70)}$ & $T_{MAP}^{(1.0)}$ & NML & INMDL \\ \hline
  1.0 dB & 21.2\% & 16.6\% & 19.8\% & 20.6\% & 93.2\% & 32.7\%\\
         &  14.3 dB &  16.5 dB & 14.9 dB & 14.4 dB & 3.90 dB & 10.9 dB \\
         & 0.0763\% & 0.795\% & 0.140\% & 0.795\% & 39.4\% & 1.15\%\\ 
         & & & & & &  0.999\\\hline
  5.0 dB & 28.5\% & 23.0\% & 28.3\% & 33.1\% & 79.9\% & 35.2\%\\
      & 15.8 dB & 17.7 dB & 15.9 dB & 14.6 dB & 7.73 dB & 14.1 dB \\
      & 0.161\% & 1.45\% & 1.34\% & 18.6\% & 18.9\% &  1.26\%\\ 
      & & & & & &  0.891 \\\hline
  10.0 dB & 41.9\% & 33.6\% & 49.5\% & 61.6\% & 53.4\% & 41.6\%\\
         &  17.5 dB &  19.4 dB &  16.1 dB & 14.3 dB & 15.5 dB & 17.6 dB \\
         & 0.354\% & 3.16\% & 10.6\% & 55.7\% & 4.21\% & 1.61\%\\ 
         & & & & & & 0.804 \\\hline
  15.0 dB & 61.6\% & 45.8\% & 68.1\% & 79.9\% & 56.5\% & 54.2\%\\ 
          & 19.2 dB & 21.7 dB & 18.4 dB & 17.0 dB & 20.0 dB & 20.3 dB\\
          & 0.80\% &  9.63\% & 28.3\% & 77.7\% & 3.29\% & 2.34\%\\
          & & & & & &  0.758\\ \hline
  20.0 dB & 88.1\% & 59.5\% & 80.5\% & 89.4\% & 73.6\% & 74.5\%\\
          &  21.1 dB &  24.5 dB & 21.9 dB & 21.0 dB & 22.6 dB &22.5  dB \\
          & 1.78\% & 23.2\% & 46.6\% & 88.5\% & 4.48\% & 3.78\%\\ 
          & & & & & &  0.741\\\hline
  25.0 dB & 124\% & 76.1\% & 88.1\% & 94.2\% & 104\% & 107\%\\ 
          & 23.1 dB & 27.4 dB & 26.10 dB & 25.5 dB & 24.7dB & 24.4 dB \\
          & 3.56\% & 35.4\% & 60.9\% & 93.6\% & 6.50\% & 5.73\%\\
          & & & & & &  0.732\\ \hline
  50.0 dB & 1309\% & 521\% & 152\% & 103\% & 1278\% & 1352\%\\
          & 27.7 dB & 35.6 dB & 46.3dB & 49.8 dB & 27.9 dB & 27.4 dB\\
          & 10.9\% & 58.8\% & 78.6\% & 97.9\% & 11.4\% & 10.3\%\\
          & & & & & &  0.710\\ \hline\hline
\end{tabular}
\label{table_goldhill}
\end{table}

\begin{table}[h]
\caption{The results from the test image Bridge, $N=512\times 512$, wavelet used: Symmlet 16. 
The first row in each entry in the table is the scaled RMSE measure in (\ref{def_RMSE}), the second row is 
the SNR of the reconstruction, see (\ref{SNR_experiments2}), the third row shows the 
 proportion of nonzero 
wavelet coefficient estimates $\theta_{i}^{*}$ as a fraction of the sample size $N$
and the fourth row shows the estimated value of the GGD shape parameter $\nu$ provided by the INMDL algorithm.}
\begin{tabular}{|l|c|c|c|c|c|c|}\hline\hline 
  SNR & RiskShrink & SureShrink & $T_{MAP}^{(0.70)}$ & $T_{MAP}^{(1.0)}$ & NML & INMDL \\ \hline
  1.0 dB & 27.2\% & 23.2\% & 26.2\% & 26.7\% & 93.5\% & 35.0\%\\
         &  12.1 dB &  13.5 dB & 12.3 dB &12.0  dB & 3.87 dB & 10.2 dB \\
         & 0.0931\% & 0.813\% & 0.160\% & 0.849\% & 39.8\% & 1.37\%\\ 
         & & & & & &  1.02\\\hline
  5.0 dB & 38.5\%& 31.0\% & 36.2\% & 38.3\% & 82.3\% & 41.5\%\\
      & 13.1 dB & 15.0 dB & 13.7 dB & 13.2 dB & 7.48 dB & 12.6 dB \\
      & 0.200\% & 2.90\% & 1.62\% & 18.9\% & 24.5\% &  1.42\%\\ 
      & & & & & &  0.928\\\hline
  10.0 dB & 58.7\% & 44.6\% & 56.4\% & 63.9\% & 63.9\% & 54.0\%\\
         &  14.5 dB & 16.9 dB &  15.0 dB &  13.9 dB &  13.9 dB &15.3  dB \\
         & 0.503\% & 8.38\% & 11.7\% & 55.9\% & 5.86\% & 1.97\%\\ 
         & & & & & & 0.870 \\\hline
  15.0 dB & 88.3\% & 62.0\% & 73.4\% & 80.9\% & 75.6\% & 77.1\%\\ 
          & 16.0 dB & 19.1 dB & 17.7 dB & 16.9 dB & 17.4 dB & 17.2 dB\\
          & 1.23\% &  19.8\% & 30.1\% & 77.9\% & 5.46\% & 3.13\%\\
          & & & & & &  0.848\\ \hline
  20.0 dB & 131\% & 79.3\% & 84.2\% & 89.6\% & 104\% & 113\%\\
          &  17.6 dB & 22.0 dB &  21.5 dB &  21.0 dB & 19.7 dB & 18.9 dB \\
          & 2.74\% & 34.7\% & 47.5\% & 88.0\% & 7.64 \% & 5.06\%\\ 
          & & & & & & 0.841 \\\hline
  25.0 dB & 189\% & 95.9 \% & 91.0\% & 93.9\% & 152\% & 173\%\\ 
          & 19.4 dB & 25.3 dB & 25.8 dB & 25.5 dB & 21.4 dB & 20.2 dB \\
          & 5.39\% & 49.8\% & 60.2\% & 92.9\% & 10.3\% & 7.06\%\\
          & & & & & &  0.844\\ \hline
  50.0 dB & 2126\% & 644\% & 244\% & 119\% & 2165\% & 2660\%\\
          & 23.4 dB & 33.8 dB & 42.3 dB & 48.5 dB & 23.3 dB & 21.5 dB\\
          & 14.3\% & 68.2\% & 74.9\% & 96.7\% & 13.9\% & 9.30\%\\
          & & & & &  &  0.829\\ \hline\hline
\end{tabular}
\label{table_bridge}
\end{table}

\begin{table}[h]
\caption{The results from the test image Boat, $N=512\times 512$, wavelet used: Symmlet 16. 
The first row in each entry in the table is the scaled RMSE measure in (\ref{def_RMSE}), the second row is 
the SNR of the reconstruction, see (\ref{SNR_experiments2}), the third row shows the 
 proportion of nonzero 
wavelet coefficient estimates $\theta_{i}^{*}$ as a fraction of the sample size $N$ 
and the fourth row shows the estimated value of the GGD shape parameter $\nu$ provided by the INMDL algorithm.}
\begin{tabular}{|l|c|c|c|c|c|c|}\hline\hline 
  SNR & RiskShrink & SureShrink & $T_{MAP}^{(0.70)}$ & $T_{MAP}^{(1.0)}$ & NML & INMDL \\ \hline
  1.0 dB & 20.7\% & 17.4\% & 19.8\% & 20.4\% & 93.2\% & 33.3\%\\
         &  14.6 dB & 16.1 dB &  14.9 dB & 14.5 dB & 3.90 dB & 10.8 dB \\
         & 0.0687\% & 0.713\% & 0.124\% & 0.792\% & 39.3\% & 1.18\%\\ 
         & & & & & & 1.02 \\\hline
  5.0 dB & 28.9\%& 24.0\% & 28.8\% & 33.4\% & 79.9\% & 35.7\%\\
      & 15.7 dB & 17.3 dB & 15.8 dB & 14.5 dB & 7.72 dB & 14.0 dB \\
      & 0.139\% & 1.34\% & 1.37\% & 18.8\% & 18.9\% &  1.29\%\\ 
      & & & & & &  0.920\\\hline
  10.0 dB & 43.4\% & 32.7\% & 49.3\% & 61.8\% & 53.6\% & 41.5\%\\
         &  17.2 dB & 19.7 dB & 16.2 dB & 14.3 dB & 15.5 dB & 17.6  dB \\
         & 0.333\% & 4.74\% & 10.8\% & 55.9\% & 4.43\% & 1.71\%\\ 
         & & & & & & 0.842 \\\hline
  15.0 dB & 61.1\% & 44.9\% & 67.4\% & 79.7\% & 54.6\% & 51.9\%\\ 
          & 19.2 dB & 21.9 dB & 18.5 dB & 17.0dB & 20.3 dB & 20.7 dB\\
          & 0.872\% & 9.2\% & 28.0\% & 77.7\% & 3.56\% & 2.55\%\\
          & & & & & &  0.812\\ \hline
  20.0 dB & 83.4\% & 55.8\% & 79.3\% & 89.0\% & 68.3\% & 69.0\%\\
          &  21.6 dB &25.1 dB & 22.0 dB &  21.0 dB & 23.3 dB & 23.2 dB \\
          & 1.96\% & 20.5\% & 45.3\% & 88.2\% & 4.68\% & 3.99\%\\ 
          & & & & & &  0.789\\\hline
  25.0 dB & 115\% & 73.5\% & 86.5\% & 93.6\% & 92.0\% & 94.5\%\\ 
          & 23.7 dB & 27.7 dB & 26.3 dB & 25.6dB & 25.7 dB & 25.5 dB \\
          & 3.59\% & 26.8\% & 58.1\% & 93.2\% & 6.68\% & 6.05\%\\
          & & & & & &  0.771\\ \hline
  50.0 dB & 1289\% & 649\% & 183\% & 105\% & 1070\% & 1110\%\\
          & 27.8 dB & 33.7 dB & 44.8 dB & 49.6 dB & 29.4 dB & 29.1 dB\\
          & 8.41\% & 43.3\% & 70.6\% & 96.7\% & 11.7\% & 11.0\%\\
          & & & & & &  0.718\\ \hline\hline
\end{tabular}
\label{table_boat}
\end{table}

\begin{table}[h]
\caption{The results from the test image Tank, $N=512\times 512$, wavelet used: Symmlet 16. 
The first row in each entry in the table is the scaled RMSE measure in (\ref{def_RMSE}), the second row is the SNR of the rconstruction, see (\ref{SNR_experiments2}), the third row shows the 
 proportion of nonzero 
wavelet coefficient estimates $\theta_{i}^{*}$ as a fraction of the sample size $N$ 
and the fourth row shows the estimated value of the GGD shape parameter $\nu$ provided by the INMDL algorithm.}
\begin{tabular}{|l|c|c|c|c|c|c|}\hline\hline 
  SNR & RiskShrink & SureShrink & $T_{MAP}^{(0.70)}$ & $T_{MAP}^{(1.0)}$ & NML & INMDL \\ \hline
   1.0 dB & 19.4\% & 16.4\% & 18.9\% & 19.0\% & 93.2\% & 33.6\%\\
         &  15.1 dB &  16.6 dB &  15.4 dB & 15.3 dB &  3.89 dB & 10.7 dB \\
         & 0.0278\% & 0.593\% & 0.0618\% & 0.667\% & 39.2\% & 1.15\%\\ 
         & & & & & &  1.04 \\\hline
  5.0 dB & 28.5\% & 23.4\% & 29.1\% & 33.4\% & 79.8\% & 36.7\%\\
      & 15.8 dB & 17.5 dB & 15.7 dB & 14.5 dB & 7.73 dB & 13.8 dB \\
      & 0.0698\% & 1.14\% & 1.23\% & 18.3\% & 18.5\% &  1.27\%\\ 
      & & & & & &  0.951\\\hline
  10.0 dB & 44.5\% & 34.8\% & 50.9\% & 61.9\% & 55.1\% & 44.8\%\\
         &  17.0 dB &  19.1 dB &  15.9 dB &  14.3 dB & 15.2 dB & 17.0 dB \\
         & 0.220\% & 3.26\% & 10.3\% & 55.3\% & 3.93\% & 1.60\%\\ 
         & & & & & &  0.885\\\hline
  15.0 dB & 67.6\% & 49.8 \% & 69.9\% & 80.0\% & 62.3\% & 60.5\%\\ 
          & 18.3 dB & 21.0 dB & 18.1 dB & 17.0 dB & 19.1 dB & 19.3 dB\\
          & 0.633\% & 9.68\% & 27.9\% & 77.5\% & 3.04\% & 2.24\%\\
          & & & & & &  0.856\\ \hline
  20.0 dB & 101\% & 68.7\% & 82.4\% & 89.4\% & 87.9\% & 88.9\%\\
          &  19.8 dB &  23.2 dB & 21.7 dB & 21.0  dB & 21.1 dB &  21.0 dB \\
          & 1.52\% & 24.0\% & 46.4\% & 88.4\% & 4.00\% & 3.41\%\\ 
          & & & & & & 0.850 \\\hline
  25.0 dB & 154\% & 93.0\% & 90.3\% & 94.0\% & 134\% & 138\%\\ 
          & 21.2 dB & 25.6 dB & 25.9 dB & 25.5 dB & 22.4 dB & 22.2 dB \\
          & 2.94\% & 36.4\% & 60.0\% & 93.3\% & 5.56\% & 4.85\%\\
          & & & & & &  0.848 \\ \hline
  50.0 dB & 2155\% & 858\% & 288\% & 126\% & 2039\% & 2139\%\\
          & 23.3 dB & 31.3 dB & 40.8 dB & 48.0 dB & 23.8 dB & 23.4 dB\\
          & 6.64\% & 58.3\% & 73.6\% & 96.6\% & 7.86\% & 6.81\%\\
          & & & & & &  0.842\\ \hline\hline
\end{tabular}
\label{table_tank}
\end{table}

\begin{table}[h]
\caption{The results from the test signal Blocks, $N=1024$, wavelet used: Haar wavelet. 
The first row in each entry in the table is the scaled RMSE measure in (\ref{def_RMSE}), the second row is 
the SNR of the reconstruction, see (\ref{SNR_experiments2}), the third row shows the 
 proportion of nonzero 
wavelet coefficient estimates $\theta_{i}^{*}$ as a fraction of the sample size $N$ 
and the fourth row shows the fixed  value of the GGD shape parameter $\nu$ used by the INMDL algorithm.}
\begin{tabular}{|l|c|c|c|c|c|c|}\hline\hline 
  SNR & RiskShrink & SureShrink & $T_{MAP}^{(0.5)}$ & $T_{MAP}^{(1.0)}$ & NML & INMDL \\ \hline
  1.0 dB & 43.4\% & 35.2\% & 42.1\% & 41.1\% & 93.0\% & 54.0\%\\
         & 8.11 dB & 9.45 dB & 7.82 dB & 6.99 dB & 3.89 dB & 6.91 dB \\
         & 1.66\% & 3.22\% & 1.66\% & 7.32\% & 44.3\% & 7.23\%\\ 
         & & & & & &  1.00\\\hline
  5.0 dB & 42.0\% & 39.4\% & 42.4\% & 48.6\% & 78.9\% & 52.9\%\\
         & 12.5 dB & 12.7 dB & 12.2 dB & 10.8 dB & 7.81 dB & 10.7 dB \\
         & 2.83\% & 5.47\% & 3.61\% & 34.3\% & 21.1\% & 7.23\%\\ 
         & & & & & &  1.00\\\hline
  10.0 dB & 48.5\% & 44.9\% & 45.4\% & 69.0\% & 57.5\% & 46.9\%\\
          & 16.3 dB & 16.7 dB & 16.8 dB & 13.2 dB & 14.9 dB & 16.6 dB \\
          & 3.71\% & 11.8\% & 7.81\% & 66.4\% & 9.18\% &  6.83\%\\ 
          & & & & & &  1.00 \\\hline
  15.0 dB & 39.8\% & 44.9\% & 53.4\% & 82.0\% & 46.1\% & 41.8\%\\ 
          &  23.0dB &  21.8 dB &  20.5 dB &  16.8 dB & 21.8 dB & 22.6 dB\\
          & 5.08\% & 14.9\% & 12.7\% & 81.6\% & 7.23\% & 6.64\%\\
          & & & & & &  1.00\\ \hline
  20.0 dB & 30.5\% & 46.3\% & 62.9\% & 89.4\% & 38.9\% & 38.5\%\\
          &  30.3 dB & 26.6 dB & 24.0 dB & 21.0  dB & 28.2 dB & 28.3 dB \\
          & 5.96\% & 15.0\% & 20.6\% & 90.7\% & 6.64\% & 6.64\%\\ 
          & & & & & &  1.00 \\\hline
  25.0 dB & 28.1\% & 44.4\% & 71.2\% & 93.2\% & 35.0\% & 33.8\%\\ 
          & 36.0 dB & 32.0 dB & 28.0 dB & 25.6 dB & 34.1 dB & 34.4 dB \\
          & 6.15\% & 19.2\% & 29.1\% & 94.4\% & 6.64\% & 6.54\%\\
          & & & & & &   1.00\\ \hline
  50.0 dB & 21.9\% & 50.2\% & 80.4\% & 98.3\% & 25.9\% & 25.9\%\\
          & 63.2 dB & 56.0 dB & 51.9 dB & 50.2 dB &  61.7 dB &  61.7 dB\\
          & 5.96\% & 8.39\% & 42.7\% & 99.4\% & 6.05\% & 6.05\%\\
          & & & & & &  1.00\\ \hline\hline
\end{tabular}
\label{table_blocks}
\end{table}

\begin{table}[h]
\caption{The results from the test signal Bumps, $N=1024$, wavelet used: Symmlet 12. 
The first row in each entry in the table is the scaled RMSE measure in (\ref{def_RMSE}), the second row is 
the SNR of the reconstruction, see (\ref{SNR_experiments2}), the third row shows the
 proportion of nonzero 
wavelet coefficient estimates $\theta_{i}^{*}$ as a fraction of the sample size $N$ 
and the fourth row shows the fixed value of the GGD shape parameter $\nu$ used by the INMDL algorithm.}
\begin{tabular}{|l|c|c|c|c|c|c|}\hline\hline 
  SNR & RiskShrink & SureShrink & $T_{MAP}^{(0.5)}$ & $T_{MAP}^{(1.0)}$ & NML & INMDL \\ \hline
  1.0 dB & 44.7\% & 37.2\% & 47.5\% & 56.6\% & 91.0\% & 55.2\%\\
         & 7.59 dB & 8.58 dB & 6.01 dB & 2.15 dB & 3.92 dB & 6.57 dB \\
         & 1.76\% & 5.57\% & 1.56\% & 2.15\% & 38.9\% & 7.81 \%\\ 
         & & & & & &  1.00 \\\hline
  5.0 dB & 55.3\% & 40.1\% & 52.3\% & 49.2\% & 76.9\% & 55.0\%\\
         & 9.83 dB & 12.2 dB & 9.97 dB & 9.94 dB & 7.94 dB & 10.2 dB \\
         & 2.34\% & 7.62\% & 2.73\% & 15.0\% & 20.1\% &  7.42\%\\ 
         & & & & & &  1.00 \\\hline
  10.0 dB & 57.9\% & 47.2\% & 46.7\% & 61.7\% & 63.1\% & 52.8\%\\
          & 14.6 dB & 16.1dB & 16.4 dB & 14.0 dB & 14.1 dB & 15.5 dB \\
          & 4.39\% & 8.79\% & 6.35\% & 53.8\% & 11.1\% & 7.62\%\\ 
          & & & & & &  1.00 \\\hline
  15.0 dB & 55.6\% & 56.5\% & 52.7\% & 78.5\% & 52.8\% & 51.2\%\\ 
          &  20.1 dB &  19.7 dB &  20.5 dB &  17.1 dB & 20.5 dB & 20.8 dB\\
          & 6.15\% & 10.8\% & 11.5\% & 78.1\% & 9.38\% & 8.79\%\\
          & & & & & &  1.00\\ \hline
  20.0 dB & 53.8\% & 56.0\% & 63.0\% & 87.9\% & 56.2\% & 54.8\%\\
          &  25.4 dB &  24.9 dB & 24.0 dB & 21.1 dB & 25.0 dB &25.2  dB \\
          & 8.20\% & 15.7\% & 18.8\% & 89.6\% & 9.38\% & 9.18\%\\ 
          & & & & & & 1.00 \\\hline
  25.0 dB & 62.7\% & 63.1\% & 71.7\% & 92.7\% & 57.1\% & 57.4\%\\ 
          & 29.0dB & 28.9 dB & 27.9 dB & 25.7 dB & 29.9 dB & 29.8 dB \\
          & 9.87\% & 18.6\% & 28.0\% & 93.8\% & 10.9\% & 10.5\%\\
          & & & & & & 1.00  \\ \hline
  50.0 dB & 79.7\% & 83.6\% & 91.5\% & 98.0\% & 80.5\% & 80.5\%\\
          & 52.0 dB & 51.6 dB &  50.8 dB &  50.2 dB & 51.9 dB & 51.9 dB\\
          & 22.0\% & 45.7\% & 68.8\% & 99.6\% & 21.9\% & 21.9\%\\
          & & & & & &  1.00 \\ \hline\hline
\end{tabular}
\label{table_bumps}
\end{table}

\begin{table}[h]
\caption{The results from the test signal Heavisine, $N=1024$, wavelet used: Symmlet 12. 
The first row in each entry in the table is the scaled RMSE measure in (\ref{def_RMSE}), the second row is 
the SNR of the reconstruction, see (\ref{SNR_experiments2}), the third row shows the 
 proportion of nonzero 
wavelet coefficient estimates $\theta_{i}^{*}$ as a fraction of the sample size $N$ 
and the fourth row shows the fixed value of the GGD shape parameter $\nu$ used by the INMDL algorithm.}
\begin{tabular}{|l|c|c|c|c|c|c|}\hline\hline 
  SNR & RiskShrink & SureShrink & $T_{MAP}^{(0.5)}$ & $T_{MAP}^{(1.0)}$ & NML & INMDL \\ \hline
  1.0 dB & 14.7\% & 15.6\% & 19.2\% & 24.3\% & 89.9\% & 50.6\%\\
         & 18.0 dB & 16.7 dB & 15.2 dB & 11.8 dB & 4.28 dB & 7.81 dB \\
         & 0.684\% & 0.977\% & 0.586\% & 1.27\% & 36.9\% &  5.86\%\\ 
         & & & & & &  1.00\\\hline
  5.0 dB & 21.9\% & 17.1\% & 23.2\% & 29.4\% & 72.2\% & 48.2\%\\
         & 18.3 dB & 20.0 dB & 17.7 dB & 15.4 dB & 8.66 dB & 11.8 dB \\
         & 0.684\% & 1.37\% & 0.781\% & 16.4\% & 14.6\% & 5.27\%\\ 
         & & & & & &  1.00\\\hline
  10.0 dB & 22.5\% & 23.5\% & 30.3\% & 57.4\% & 49.4\% & 46.7\%\\
          & 23.1 dB & 22.4 dB & 20.5 dB & 14.9 dB & 16.3 dB &16.8 dB \\
          & 1.07\% & 1.56\% & 2.44\% & 55.0\% & 4.69\% &  4.49\%\\ 
          & & & & & &  1.00\\\hline
  15.0 dB & 31.7\% & 30.3\% & 42.8\% & 76.2\% & 39.6\% & 39.0\%\\ 
          &  25.0 dB &25.3  dB &  22.4 dB & 17.5 dB & 23.1 dB & 23.2 dB\\
          & 1.47\% & 2.25\% & 5.66\% & 76.2\% & 3.13\% & 3.13\%\\
          & & & & & &  1.00\\ \hline
  20.0 dB & 36.1\% & 35.0\% & 54.2\% & 86.3\% & 35.4\% & 36.4\%\\
          &  28.9 dB &  29.0 dB & 25.4 dB & 21.3  dB & 29.1 dB & 28.8 dB \\
          & 1.95\% & 4.30\% & 12.5\% & 88.0\% & 2.93\% & 3.42\%\\ 
          & & & & & &  1.00\\\hline
  25.0 dB & 32.6\% & 40.2\% & 64.4\% & 91.7\% & 34.4\% & 34.3\%\\ 
          & 34.8 dB & 32.9 dB & 28.9 dB & 25.8 dB & 34.3 dB & 34.3 dB \\
          & 2.93\% & 7.62\% & 20.2\% & 92.7\% & 3.71\% & 3.71\%\\
          & & & & & &   1.00\\ \hline
  50.0 dB & 38.4\% & 51.5\% & 90.5\% & 97.9\% & 38.4\% & 38.4\%\\
          & 58.3 dB & 55.8 dB & 50.9 dB &  50.2 dB &  58.3 dB &  58.3 dB\\
          & 7.03 \% & 19.9\% & 65.5\% & 99.6\% & 7.03\% & 7.03\%\\
          & & & & & &  1.00\\ \hline\hline
\end{tabular}
\label{table_heavisine}
\end{table}

\begin{table}[h]
\caption{The results from the test signal Doppler, $N=1024$, wavelet used: Symmlet 12. 
The first row in each entry in the table is the scaled RMSE measure in (\ref{def_RMSE}), the second row is 
the SNR of the reconstruction, see (\ref{SNR_experiments2}), the third row shows the 
 proportion of nonzero 
wavelet coefficient estimates $\theta_{i}^{*}$ as a fraction of the sample size $N$ 
and the fourth row shows the fixed value of the GGD shape parameter $\nu$ used by the INMDL algorithm.}
\begin{tabular}{|l|c|c|c|c|c|c|}\hline\hline 
  SNR & RiskShrink & SureShrink & $T_{MAP}^{(0.5)}$ & $T_{MAP}^{(1.0)}$ & NML & INMDL \\ \hline
  1.0 dB & 33.8\% & 35.1\% & 41.6\% & 51.8\% & 92.3\% & 52.6\%\\
         & 9.98 dB & 8.98 dB & 7.30 dB & 3.06 dB & 3.82 dB & 7.01 dB \\
         & 1.37\% & 2.83\% & 1.17\% & 1.56\% & 42.8\% &  7.42\%\\ 
         & & & & & &  1.00\\\hline
  5.0 dB & 38.4\% & 39.5\% & 37.5\% & 41.1\% & 76.0\% & 50.6\%\\
         & 13.1 dB & 12.4 dB & 13.0 dB & 11.7 dB & 7.98 dB & 11.0 dB \\
         & 1.95\% & 3.81\% & 2.34\% & 15.0\% & 18.2\% &  6.64\%\\ 
         & & & & & &  1.00\\\hline
  10.0 dB & 42.7\% & 45.5\% & 39.4\% & 57.3\% & 58.7\% & 52.0\%\\
g          & 17.3dB & 16.4 dB & 17.9 dB & 14.6 dB & 14.7 dB & 15.7 dB \\
          & 2.64\% & 5.18\% & 4.30\% & 50.1\% & 8.98\% & 7.03\%\\ 
          & & & & & &  1.00\\\hline
  15.0 dB & 42.3\% & 53.2\% & 48.8\% & 75.7\% & 52.9\% & 49.4\%\\ 
          &  22.4 dB &  20.3 dB &  21.2 dB & 17.4  dB & 20.5 dB & 21.1 dB\\
          & 4.00\% & 9.57\% & 8.11\% & 74.3\% & 7.03\% & 6.25\%\\
          & & & & & &  1.00\\ \hline
  20.0 dB & 47.7\% & 57.2\% & 56.5\% & 86.0\% & 49.8\% & 47.1\%\\
          &  26.4 dB & 24.7 dB &  24.9 dB & 21.3 dB & 26.0 dB & 26.5 dB \\
          & 5.27\% & 12.7\% & 12.7\% & 87.7\% & 7.23\% & 6.83\%\\ 
          & & & & & & 1.00 \\\hline
  25.0 dB & 57.8\% & 64.0\% & 64.2\% & 91.6\% & 50.1\% & 49.0\%\\ 
          & 29.8 dB & 28.8 dB & 28.8 dB & 25.8 dB & 31.0 dB & 31.2 dB \\
          & 6.25\% & 14.9\% & 21.0\% & 92.1\% & 8.00\% & 7.81\%\\
          & & & & & &  1.00 \\ \hline
  50.0 dB & 60.0\% & 67.6\% & 90.6\% & 97.9\% & 60.0\% & 60.0\%\\
          & 54.4 dB & 53.4 dB &  50.9 dB & 50.2 dB &  54.4 dB &  54.4 dB\\
          & 12.3\% & 30.6\% & 65.5\% & 99.8\% & 12.3\% & 12.3\%\\
          & & & & & &  1.00\\ \hline\hline
\end{tabular}
\label{table_doppler}
\end{table}

%\bibliography{RefDataBase}
%\bibliographystyle{alpha}

\end{document}